\RequirePackage{fix-cm}
\documentclass[natbib,smallextended]{svjour3}       
\smartqed  
\usepackage{graphicx}
\usepackage{amsmath,amssymb}
\usepackage{xspace}
\usepackage{xcolor}
\usepackage[colorlinks=true,citecolor=blue,urlcolor=blue,breaklinks]{hyperref}
\usepackage{sidecap}
\usepackage{booktabs}
\usepackage[normalem]{ulem}

\newcommand{\rv}{\textcolor{black}}
\newcommand{\TRISTAN}{\texttt{TRISTAN}\xspace}
\newcommand{\TRISTANMPI}{\texttt{TRISTAN-MPI}\xspace}
\newcommand{\Vlasiator}{\texttt{Vlasiator}\xspace}
\newcommand{\DEPOSIT}{\texttt{DEPOSIT}\xspace}
\newcommand{\SPLIT}{\texttt{SPLIT}\xspace}
\newcommand{\DENSITY}{\texttt{DENSITY}\xspace}

\newcommand{\id}[1]{\textcolor{purple}{\texttt{ID: #1}}}

\definecolor{purple}{rgb}{0.65, 0.12, 0.82}

\journalname{Living Reviews in Computational Astrophysics}
\begin{document}

\title{PIC methods in astrophysics: Simulations of relativistic jets {and kinetic physics 
in astrophysical systems}}



\author{Kenichi Nishikawa,  Ioana Du\c{t}an, Christoph K\"ohn, Yosuke Mizuno 
}


\institute{K. Nishikawa \at
Department of Physics,
V. Murry Chambers Building,
Normal, AL  35762, USA \\
\email{kenichi.nishikawa@aamu.edu}           
\and
I. Du\c{t}an \at
Institute of Space Science, Atomistilor 409, RO-077125 Bucharest-Magurele,
Romania \\
\email{idutan@spacescience.ro}               
\and
C. K\"ohn \at
Technical University of Denmark, National Space Institute (DTU Space), 
Elektrovej 328, 2800 Kgs Lyngby, Denmark\\
\email{koehn@space.dtu.dk}
\and
Y. Mizuno \at
Tsung-Dao Lee Institute and School of Physics and Astronomy, Shanghai 
Jiao Tong University, 800 Dongchuan Road, Shanghai, 200240, People's Republic of China \\
\email{mizuno@sjtu.edu.cn}
\and
\at 
Institut f\"ur Theoretische Physik, Goethe Universit\"at, Max-von-Laue Str. 1, D-60438 Frankfurt am Main, Germany
}

\date{Received: date / Accepted: date}

\maketitle

\begin{abstract}
The Particle-In-Cell (PIC) method has been developed by Oscar Buneman, Charles Birdsall, Roger W. Hockney, and John Dawson in the 1950s and, with the advances of computing power, has been further developed for several fields such as astrophysical, magnetospheric as well as solar plasmas and recently also for atmospheric and laser-plasma physics. 
Currently more than 15 semi-public PIC codes are available \rv{which we discuss in this review}. 
Its applications have grown extensively with increasing computing power available on high performance computing facilities around the world.   
These systems allow the study of various topics of astrophysical plasmas, such as magnetic reconnection, pulsars and black hole magnetosphere, non-relativistic and relativistic shocks, relativistic jets, and laser-plasma physics. 
\rv{We review a plethora of
astrophysical phenomena such as relativistic jets, instabilities, magnetic reconnection, pulsars, 
as well as PIC simulations of laser-plasma physics (until 2021) emphasizing the physics involved 
in the simulations. Finally, we give an outlook of the future simulations 
of jets associated to neutron stars, black holes and their merging and discuss the future of PIC
simulations in the light of petascale and exascale computing.}

\end{abstract}

\setcounter{tocdepth}{3}
\tableofcontents

\section{Introduction}
\label{sec:1}

Plasma is one of the four fundamental states of matter, and ubiquitous in the universe. 
Astrophysical plasmas are observed in compact objects like black holes and neutron stars. 
Plasma is also associated with ejection of material in astrophysical jets, which have been observed in systems like accreting black holes and merging neutron stars. 
It consists of a totally or partially ionized gas (of electrons, ions, or positrons) the particles of which can be, in principle, described by Lorentz' equation and Maxwell's equations Eqs. (\ref{maxwB},\ref{maxwE},\ref{cur},\ref{lor}). 
In order to investigate kinetic (i.e., microscopic) processes, these equations need to be solved. For a multitude of particles one way to solve them is to employ Particle-In-Cell (PIC) simulations. 


Extensively, astrophysical plasma phenomena have been investigated by numerical magnetohydrodynamics (MHD) in both non-relativistic and relativistic (RMHD) regimes, 
extensively \citep[e.g.,][]{marti15,Baumgarte10NR,Rezzolla13}. 
In MHD, plasma is taken as a fluid and handles macroscopic processes. The differences between macroscopic and microscopic processes are described in Sect.~\ref{sec:4}.

Another numerical approach for the investigation of astrophysical plasma phenomena is (collisionless) PIC simulations. In these simulations, the plasma charged particles interact only with the electromagnetic fields that are produced by the particles themselves (in their motion), and the PIC method is employed to solve plasma kinetic (microscopic) processes.
Recently, with the advance of supercomputer facilities and PIC algorithms, PIC simulations have become a compliment to the plasma fluid method with the addition of kinetic processes. 
In particular, particle acceleration has been investigated in shocks, magnetic reconnections, and other systems. 
Whilst fluid models calculate densities, concentrations, or generally averaged quantities, the advantage of PIC codes---sometimes combined with Monte Carlo methods---is that they trace individual particles and are therefore able to capture rare events which would not be seen in fluid simulations \citep{rubino_2009}.


The usage of PIC simulations started in the 1950s by a few scientists such as Buneman, Hockney, Birdsall, and Dawson\citep{buneman93}.
Nowadays, the power of supercomputers increases exponentially, and although in general PIC simulations require a very large memory, many extensive 3D PIC simulations have been performed \citep[e.g.,][]{Daughton11,liu18}.



There are several complimentary reviews related to astrophysical plasma and related physics are available. \cite{Sironi15r} reviewed the physics of relativistic shocks in pulsar wind nebulae (PWNe), gamma-ray bursts (GRBs), and active galactic nuclei (AGN) jets where non-thermal particles (i.e., they have a non-thermal  or power-law distribution) are found to be the sources of observed radiation and, possibly, of ultra-high energy cosmic-rays.
In weakly magnetized or quasi-parallel shocks (i.e., where the magnetic field is nearly aligned with the flow), particle acceleration is efficient. 
The accelerated particles stream ahead of the shock, where they generate strong magnetic waves which in turn scatter the particles back and forth across the shock, mediating their acceleration. 
In contrast, in strongly magnetized quasi-perpendicular shocks, the efficiencies of both particle acceleration and magnetic field generation are suppressed. Particle acceleration, when efficient, modifies the turbulence around the shock on a long time scale, and the accelerated particles have a characteristic energy spectral index of $s_{\gamma} \simeq 2.2$ in the ultra-relativistic limit.
They have unveiled the most relevant plasma instabilities that mediate injection and acceleration in relativistic shocks; and they have summarized recent results of large-scale PIC simulations concerning the efficiency and rate of particle acceleration in relativistic shocks, and the long-term evolution of the self-generated magnetic turbulence.

\rv{Several review articles about the applications of plasma physics to astrophysical phenomena have recently been published. For example,} \cite{Pohl2020r} have reviewed particle acceleration in collisionless plasma systems which is commonly found in astrophysical plasma and astroparticle physics. 
%
%
%
The PIC method has been developed to study kinetic phenomena in non-linear plasma interactions in computer experiments. 
By tracing individual quasi-particles that stand for many electrons or ions, one can investigate the evolution of collective electromagnetic fields and the distribution function of the particles. 
\rv{The spectrum of accelerated particles achieved with PIC simulations is very informative 
to understand of shocks and particle acceleration, but further investigations will follow.} 


\citet{Marcowith2020} have reviewed a general introduction to the subject of the numerical study of energetic particle acceleration and transport in turbulent astrophysical flows providing an up-to-date status. The subject is also complemented by a short overview of recent progresses obtained in the domain of laser-plasma experiments. 
They summarize the main physical processes at the heart of the production of a non-thermal distribution in both Newtonian and relativistic astrophysical flows, namely the first and second order Fermi acceleration processes. Shock drift and surfing acceleration, these two processes are important in the context of particle injection in shock acceleration.
The details of the PIC approach used to describe particle kinetic processes were analyzed. The main results by PIC simulations of particle acceleration at shocks and in magnetic reconnection events are presented. 
In this article, the solution of Fokker--Planck problems with application to the study of particle acceleration at shocks but also in hot coronal plasmas surrounding compact objects are discussed. 
They also describe recent developments in MHD simulations giving a special emphasis on the way energetic particle dynamics, which can be coupled to MHD solutions either using a multi-fluid calculation or directly coupling kinetic and fluid calculations. 

\rv{Additionally, the books ``Plasma Astrophysics and Space Physics"
edited by J\"org B\"uchner et al. \citep{Buechner1999} and ``Space Plasma Simulation" edited 
by J\"org B\"uchner et al. \citep{Buechner2003} provide useful examples of simulations of astrophysics and space physics by various authors.}

In this review, as an example for astrophysical plasmas, we especially focus on relativistic jets, mainly investigating the effect of microscopic plasma instabilities such as the Weibel instability 
\citep{weibel59}. Jets have been investigated extensively since 2000 and there has been substantial progress within the last couple of years according to the increase of computing power. A recent article written by \citet{nishikawa19gal} has given an overview of PIC simulations, starting with the Weibel instability in slab models of jets, and then focuses 
on global jet evolution in helical magnetic field geometry. In particular, kinetic Kelvin--Helmholtz and mushroom instabilities were addressed.

We provides an introductory discussion of the PIC method mainly based on Buneman's code \citep{buneman93} such that readers are able to understand how PIC method works and what kind of physical problems can be investigated with PIC simulations. Although a few textbooks and reviews on PIC methods are already available, it may be overwhelming to read such them; therefore this review may serve as an introduction into the topic and guide to those textbooks\citep{dawson83,birdsall91,hockney89}. 
Initially, in order to illustrate how PIC simulations work in the view of astrophysicalplasmas, we concentrate on PIC simulations of relativistic jets in a slab model and recent progresses of relativistic jets including velocity-shear instabilities. Since magnetic reconnection plays an important role in particle acceleration, its development in Harris' model is discussed.
In particular, we bring the importance of PIC simulations of relativistic jets containing helical magnetic fields. In these simulations relativistic jets need to be injected into a simulation system and, therefore, a large simulation system should be used. 
It requires a long simulation time to investigate the jet evolution up to the full nonlinear stage of \rv{growing} kinetic instabilities, including particle acceleration in turbulent magnetic fields generated kinetic instabilities and magnetic reconnection.

We outline this review as follows: Brief history {and new developments} of PIC simulations is described in Sect.~\ref{sec:2}. Section \ref{sec:3} addresses Basic methods of PIC simulations. The differences between kinetic (microscopic) and macroscopic processes in plasma are described in Sect.~\ref{sec:4}. The various applications of PIC simulations are presented in Sect.~\ref{sec:5}. We summarize in Sect.~\ref{sec:6}. 

\section{Brief history {and new developments} of PIC simulations}
\label{sec:2}

In principle, classical mechanics or electrodynamics allow for solving one- or two-particle problems exactly. 
However, depending on the applied potentials, analytic solutions might not be possible. Similarly, problems with more than two particles cannot generally be solved analytically. Therefore there is a requirement for computational methods which becomes even more urgent when dealing with relativistic problems because of their increased level of complexity compared to classical physics. When tracing a multitude of particles, making advanced codes such as fluid codes or PIC codes is necessary for their study. 
PIC simulations have been developed in the 1950s by a few scientists, among whom was Oscar Buneman, the prominent scientist in the field of plasma electrodynamics.



\begin{table}[htbp]
\caption{Electromagnetic PIC computational applications.}
\label{tab:wikipedia}
\footnotesize
\begin{center}
\begin{tabular}{ p{2.1cm} p{1.4cm} p{3.2cm} p{3.0cm} }
  \toprule
  \textbf{Computational application} & \textbf{License} & \textbf{Availability} & \textbf{Canonical reference (DOI)} \\
  \midrule
  {SHARP} & Proprietary & & \href{https://doi.org/10.3847/1538-4357/aa6d13}{10.3847/1538-4357/aa6d13} \\
  \midrule
  \href{https://aladyn.github.io/ALaDyn/}{ALaDyn} & GPLv3+& Open Repo  & \href{https://doi.org/10.5281/zenodo.49553}{10.5281/zenodo.49553} \\
  \midrule
  \href{https://warwick.ac.uk/fac/sci/ccpp/research/}{EPOCH} & GPL& Open to academic users but signup required & \href{https://doi.org/10.1088/0741-3335/57/11/113001}{10.1088/0741-3335/57/11/113001} \\
  \midrule
  \href{https://fbpic.github.io/}{FBPIC} & 3-Clause-BSD-LBNL &	Open Repo  &  \href{https://doi.org/10.1016/j.cpc.2016.02.00}{10.1016/j.cpc.2016.02. 007} \\
  \midrule
  \href{https://www.northropgrumman.com/space/pic-code-software/}{LSP}	& Proprietary  & Available from ATK & \href{https://doi.org/10.1016/S0168-9002(01)00024-9}{10.1016/S0168-9002(01)00024-9} \\
  \midrule
  \href{https://www.northropgrumman.com/space/pic-code-software/}{MAGIC}  & Proprietary	&  Available from ATK &	\href{https://doi.org/10.1016/0010-4655(95)00010-D}{10.1016/0010-4655(95)00010-D} \\
  \midrule
  \href{http://picksc.idre.ucla.edu/software/software-production-codes/osiris/}{OSIRIS} & Proprietary	& Closed (Collaborators with MoU)	   & \href{https://doi.org/10.1007/3-540-47789-6_36}{10.1007/3-540-47789-6\_36} \\
  \midrule
  \href{https://aladyn.github.io/piccante/}{PICCANTE} &	   GPLv3+	  &  Open  Repo		&\href{https://doi.org/10.5281/zenodo.48703}{10.5281/zenodo.48703} \\
  \midrule
  {PICLas} & Proprietary	& Available from \href{https://www.iag.uni-stuttgart.de/en/working-groups/numerical-methods/}{IAG} Stuttgart &	\href{https://doi.org/10.1016/j.crme.2014.07.005}{10.1016/j.crme.2014.07. 005} \\
  \midrule
  \href{https://www.hzdr.de/db/Cms?pOid=31887&pNid=3227}{PIConGPU} &	 GPLv3+ &	Open Repo
  & \href{https://doi.org/10.1145/2503210.2504564}{10.1145/2503210.2504564} \\
  \midrule
  \href{http://www.maisondelasimulation.fr/smilei/}{SMILEI} & 		CeCILL-B  &	Open Repo
  &  \href{https://doi.org/10.1016/j.cpc.2017.09.024}{10.1016/j.cpc.2017.09.024} \\
  \midrule
  \href{https://github.com/CmPA/iPic3D}{iPIC3D} &    Apache License 2.0	& Open Repo
  & \href{https://doi.org/10.1016/j.matcom.2009.08.038}{10.1016/j.matcom.2009. 08.038} \\
  \midrule
  \href{http://www2.mpq.mpg.de/lpg/research/RelLasPlas/Rel-Las-Plas.html}{The Virtual Laser Plasma Library} & Proprietary & Unknown  &	\href{https://doi.org/10.1017/S0022377899007515}{10.1017/S00223778-99007515}  \\
  \midrule
  \href{https://esgeetech.com/products/vizgrain-particle-modeling/}{VizGrain} &   Proprietary &	Commercially available from Esgee Technologies Inc. &  \\
  \midrule
  \href{https://github.com/lanl/vpic}{VPIC} & 3-Clause-BSD & Open Repo &	\href{https://doi.org/10.1063/1.2840133}{10.1063/1.2840133} \\
  \midrule
  \href{https://txcorp.com/vsim/}{VSim (Vorpal)} & Proprietary & Available from Tech-X Corporation & \href{https://doi.org/10.1016/j.jcp.2003.11.004}{10.1016/j.jcp.2003.11.004} \\
  \midrule
  \href{http://warp.lbl.gov/}{Warp} & 3-Clause-BSD-LBNL & Open Repo & \href{https://doi.org/10.1063/1.860024}{10.1063/1.860024} \\
  \midrule
  \href{https://ecp-warpx.github.io/}{WarpX} & 3-Clause-BSD-LBNL & Open Repo
  & \href{https://doi.org/10.1016/j.nima.2018.01.035}{10.1016/j.nima.2018.01. 035} \\
  \midrule
  \href{https://picksc.idre.ucla.edu/software/educational/zpic/}{ZPIC} & AGPLv3+ &	Open Repo & 
  \href{https://www.overleaf.com/project/5d4d1fd3ff2d1910daf6b2a7#cite.Bowers2008PhPl}{10.1063/1.2840133}\\
   \midrule
  \href{https://ipag.osug.fr/~ceruttbe/Zeltron/index.html}{Zelton} & Cerutti \& Werner &	Open Repo & \href{https://www.overleaf.com/project/5d4d1fd3ff2d1910daf6b2a7#cite.Cerutti_2014}{10.1088/0004-637x/782/2/104} \\
   \midrule
  \href{https://www.swri.org/industry/ballistics-explosives/epic-dynamic-finite-element-code}{PolyPIC} & CST &	Open Repo & \href{https://www.frontiersin.org/articles/10.3389/fphy.2018.00100/full}{10.3389/fphy.2018.00100} \\
  \bottomrule	
\end{tabular}
\end{center}
\end{table}


In 1993, Buneman introduced \TRISTAN (TRIdimensional STANford code), a relativistic code simulating the interaction between the solar-wind and the Earth magnetosphere \citep{buneman93}. 

\rv{\TRISTAN is available on the website \url{www.terrapub.co.jp/e-library/cspp/text/10.txt}. 
We note that Oscar Buneman originally ran this code on his PC \citep{buneman92}. Later on, a variant code of the \TRISTAN has been developed by using High Performance Fortran (HPF)\citep{cai02,cai03}. In the following years, several PIC codes based on \TRISTAN have been parallelized using the Message Passage Interface (MPI) 
\citep{Messmer2001T,Messmer2002,Spitkovsky2005,spit08a,niemiec08} and a class of these codes - generally known as \texttt{TRISTAN-MPI} - has been used 
extensively and further developed to investigate relativistic global jets containing helical magnetic
fields \citep[e.g.,][]{niemiec08,Melzani2013,Sironi2014,Melzani2014a,Melzani2014b,nishikawa16a,
nishikawa17,nishikawa19gal,Nishikawa2020}.}
Other PIC codes 
are available for \texttt{KEMPO1}.\footnote{\url{http://www.terrapub.co.jp/e-library/cspp/text/09.txt},
and \url{http://www.terrapub.co.jp/e-library/amss/pdf/209.pdf} described by \citet{omura07} \url{http://www.terrapub.co.jp/e-library/amss/pdf/001.pdf}}

Until now, many PIC codes have been developed by researchers independently. 
At the time of writing this review, a number of 19 PIC codes are listed in
\href{https://en.wikipedia.org/wiki/Particle-in-cell}{Wikipedia}, 
where many of them are publicly available. Table~\ref{tab:wikipedia} presents an enlarged list of PIC codes that includes also the data from Wikipedia. The \texttt{TRISTAN-MPI} is not available publicly at the present time; nevertheless, the code is made available upon request.\footnote{Email address for requesting the \texttt{TRISTAN-MPI}: knishika27@gmail.com}




Each PIC code has its strength, being designed for 
specific uses. The main differences come from methods for solving Maxwell's equations which are partial differential equations (PDEs) by their essence.
There are three classes of methods to solve these: \rv{finite difference methods (FDMs), finite element methods (FEMs), and spectral methods}.

{FDMs convert partial differential equations, which may be nonlinear, into a system of linear equations that can be solved by approximating derivatives with finite differences. Both the spatial domain and time interval are discretized, or broken into a finite number of steps, 
and the value of the solution at these discrete points is 
approximated by solving algebraic equations containing finite differences and values from nearby points. For solving PDEs with FDMs, there are three methods: explicit, implicit and hybrid (implicit-explicit) methods, where the implicit one is usually the most accurate scheme for small time steps.}

{The FEM subdivides a large system into smaller, simpler parts that are called finite elements. The PDEs are treated as an eigenvalue problem and initially 
a trial solution is calculated using basis functions which are localized in each element. The final solution is then obtained by optimization until the required accuracy is reached. Various types of finite element methods are 
developed such as the Applied Element Method, extended finite element methods (XFEM), etc.}

{Also spectral methods, such as the fast Fourier transform (FFT), transform the PDEs into an eigenvalue problem, but this time the basis functions are of high order and defined globally over the whole domain. The domain itself is not discretized in this case, but rather remains continuous. Again, a trial solution is found by inserting the basis functions into the eigenvalue equation and subsequently optimized to determine the best values of the initial trial parameters. Recently, new numerical methods and additional physical processes such as radiation, ionization, gravitation, pair creation and others have been implemented in these PIC codes \citep[e.g.,][]{Cerutti_2013,Cerutti2017}}. 


\rv{Two examples of variations of PIC codes are the hybrid-Vlasov model, which is briefly discussed 
in Sect.~\ref{sec:4}, and General Relativistic
PIC codes whose basics are introduced in Sect.~\ref{sec:5.7}.}


\href{https://www.archer.ac.uk/community/eCSE/eCSE03-01/eCSE03-01.php} {\texttt{EPOCH}}
(Extendable PIC Open Collaboration) is a mature laser-plasma MPI PIC simulation code.
PIC moves Monte-Carlo sampled particles through a fixed grid on which field variables are updated. Thus the core scheme is dominated by particle pushes and interpolations of particle-to-field and field-to-particle. 
In EPOCH new refinements have been developed \citep{Arber_2015}. Modern PIC codes tend to add to high-order shape functions for particles Poisson preserving field updates, collisions, ionization, a hybrid scheme for solid density, and high-field quantum electrodynamics (QED) effects. In addition to these physics packages, the increase in computing power now allows simulations with real mass ratios, full 3D dynamics, and multi-speckle interaction. They present a review of the core algorithms used in current laser-plasma specific PIC codes. 
They also report estimates of self-heating rates, convergence of collisional routines, and the test of ionization models which are not readily available elsewhere. They review the status of PIC algorithms and present a summary of recent applications of such codes in laser-plasma physics, concentrating on stimulated Raman scattering (SRS), short-pulse laser-solid interactions, fast-electron transport, and QED effects. 

{\href{https://github.com/lanl/vpic}{\texttt{VPIC} (Vector Particle-In-Cell)} is a general purpose PIC simulation code for modeling kinetic plasmas \citep{Bowers2008PhPl}. It employs a second-order, explicit, leapfrog algorithm to update charged particle positions and velocities in order to solve the relativistic kinetic equation for each species in the plasma, along with a full Maxwell description for the electric and magnetic fields evolved via a second-order finite-difference-time-domain (FDTD) method. The VPIC code is optimized for modern computing architectures and uses MPI calls for multi-node application as well as data parallelism using threads \citep[e.g.,][]{Bird2021x}. 
VPIC employs a variety of short-vector, single-instruction-multiple-data (SIMD) intrinsics for high performance computing and has been designed such that the data structures align with cache boundaries. 
The current feature set for VPIC includes a flexible input deck format capable of treating a wide variety of problems. These include: the ability to treat electromagnetic materials (scalar and tensor dielectric, conductivity, and diamagnetic material properties); multiple emission models, including user-configurable models; arbitrary, user-configurable boundary conditions for particles and fields; user-definable simulation units; a suite of ``standard'' diagnostics, as well as user-configurable diagnostics; a Monte-Carlo treatment of collisional processes capable of treating binary and unary collisions and secondary particle generation; and, flexible checkpoint-restart semantics enabling 
\texttt{VPIC} checkpoint files to be read as input for subsequent simulations. 
\texttt{VPIC} has a native I/O format that interfaces with the high-performance visualization software 
\href{ https://hpc.llnl.gov/software/visualization-software/ensight }{\texttt{Ensight}} and
\href{https://www.paraview.org/}{Paraview}. 
While the common use cases for \texttt{VPIC} employ low-order particles on rectilinear meshes, a framework exists to treat higher-order particles and curvilinear meshes, as well as more advanced field solvers. \texttt{VPIC} has been used for simulations of reconnections \citep{Guo2020PhPl,Zhang_2020}.}

\texttt{Zeltron} is a code that has been developed by
\href{https://ipag.osug.fr/~ceruttbe/Zeltron/index.html}
{Benno\^{i}t Cerutti and Greg Werner} and has extensively been used for various investigations \citep[e.g.,][]{Cerutti_2014,Cerutti17}. 
\texttt{Zeltron} is an explicit 3D relativistic
electromagnetic PIC code, ideally suited for modeling the particle acceleration in astrophysical plasmas \citep{Cerutti_2013}. The code is efficiently parallelized with the MPI, and can be run on a laptop computer or on multiple cores on current supercomputers. 
The radiation reaction force includes synchrotron and inverse Compton back-reaction force (in the Thomson regime only). The Abraham--Lorentz--Dirac equation is solved using a modified Boris push \citep{Tamburini2010}. Since the numerical scheme does not strictly conserve electric charge, the  electric field 
is corrected by solving the Poisson's equation 
using an iterative Gauss--Seidel method.

\href{https://aladyn.github.io/piccante/}{\texttt{PICCANTE}} is another open source, massively parallel, fully relativistic PIC code
which is primarily developed to study laser-plasma interaction \citep{Sgattoni2015x,Sgattoni2015PhRvE}. The diagnostic tool \texttt{Scalasca} is used 
to identify sub-optimal routines. Different output strategies are discussed. The detailed strong and weak scaling behaviour of the improved code are presented in comparison with the original version of the code. While the final goal of the \texttt{PICCANTE} project is 
to provide a general tool for kinetic simulations of plasma physics, in the near term the code is mainly used to simulate the interaction of plasmas with super-intense 
laser pulses in various regimes and for different applications.

\href{https://www.hzdr.de/db/Cms?pOid=31887&pNid=3227}{\texttt{PIConGPU}}: 
The innovative methods on the implementation of the PIC algorithm on graphic processing units (GPUs) can directly be adapted to any many-core parallelization of the particle-mesh method. With these methods the developers see a peak performance of 7.176 PFLOP/s (double-precision) plus 1.449 PFLOP/s (single-precision), an efficiency of 96\% when weakly scaling from 1 to 18432 nodes, an efficiency of 68.92\%, and a speed up of 794 (ideal: 1152) when strongly scaling from 16 to 18432 nodes (factor 1152). 
\cite{Schuchart2020} present a PIC simulation of the relativistic Kelvin--Helmholtz Instability (KHI) that for the first time delivers angularly resolved radiation spectra of the particle dynamics during the formation of 
the KHI. This enables studying the formation of the KHI with unprecedented spatial, angular, and spectral resolution. Their results are of great importance for understanding astrophysical jet formation and comparable 
plasma phenomena by relating the particle motion observed in the KHI to its radiation signature. 

\href{https://fbpic.github.io/}{\texttt{FBPIC}} is a spectral PIC algorithm that is based on the combination of a Hankel transform and a Fourier transform. For physical problems that have close-to-cylindrical symmetry, 
this algorithm can be much faster than full 3D PIC algorithms. In addition, unlike standard finite-difference PIC codes, in vacuum the proposed algorithm is free of spurious numerical dispersion. This algorithm is benchmarked in several situations that are of interest for laser-plasma interactions. 
These benchmarks show that \texttt{FBPIC} avoids a number of numerical artifacts, that would otherwise affect the physics in a standard PIC algorithm -- including the zero-order numerical Cherenkov effect \citep{LEHE2016}.

\href{https://github.com/CmPA/iPic3D}{\texttt{iPIC3D}} is the implicit PIC method for the computer simulation of plasmas, and its implementation in a three-dimensional parallel code. The implicit integration in time of the Vlasov--Maxwell system, removes the numerical stability constraints and enables kinetic plasma simulations at magnetohydrodynamic time scales. Simulations of magnetic reconnection in plasma are presented to show the
effectiveness of the algorithm \citep{MARKIDIS2010}.



As described previously, some PIC codes such as \texttt{VPIC}, \texttt{EPOCH}, and others are developed for exa-scale computing using CPU/GPU. Hence, because of the large amount of output data, the visualization of such data is very difficult and requires a special way to handle. One of the more recent developments for exa-scale computation is \href{https://github.com/kokkos/kokkos}{\texttt{Kokkos Core}} which implements a programming model in C++ for writing performance portable applications targeting all major HPC platforms. For that purpose it provides abstractions for both the parallel execution of code and data management. \texttt{Kokkos} is designed to target complex node architectures with N-level memory hierarchies and multiple types of execution resources \citep[e.g.,][]{Bird2021x}. It currently can use CUDA, HPX, OpenMP and Pthreads as backend programming models with several other backends under development.

\rv{Today, PIC simulations are used in a variety of fields, such as, but not limited to astrophysical phenomena, atmospheric electricity, cloud physics or space weather \citep[e.g.,][]{lapenta_2012,koehn_2017,koehn_2018,koehn_2020}.}
\section{Basic methods of PIC simulations}
\label{sec:3}


\rv{In this section, we present the methodology of PIC simulations based on the original code developed by Buneman, which is described in the text of the International School of Space Sciencev\citep{buneman93}. However, the description is condensed so much that understanding the content is quite difficult. Henceforth, a plethora of literature has become available, providing detailed descriptions of PIC implementation using High Performance Fortran (HPF) \citep{cai01,cai02,cai03}}. Here, we describe the original \TRISTAN code based on Buneman's philosophy with additional explanations by
\citet{dawson83,hockney88,birdsall91} making the PIC approach to plasma microphysics, as well as its recent developments, more accessible to readers. 
\rv{We also include \id{\sout{representations} configurations} from \TRISTANMPI \citep{niemiec08}, e.g., for the shape functions  representing a (macro) particle, boundary conditions, etc.}



\begin{figure}[htb]
\includegraphics[scale=0.77]{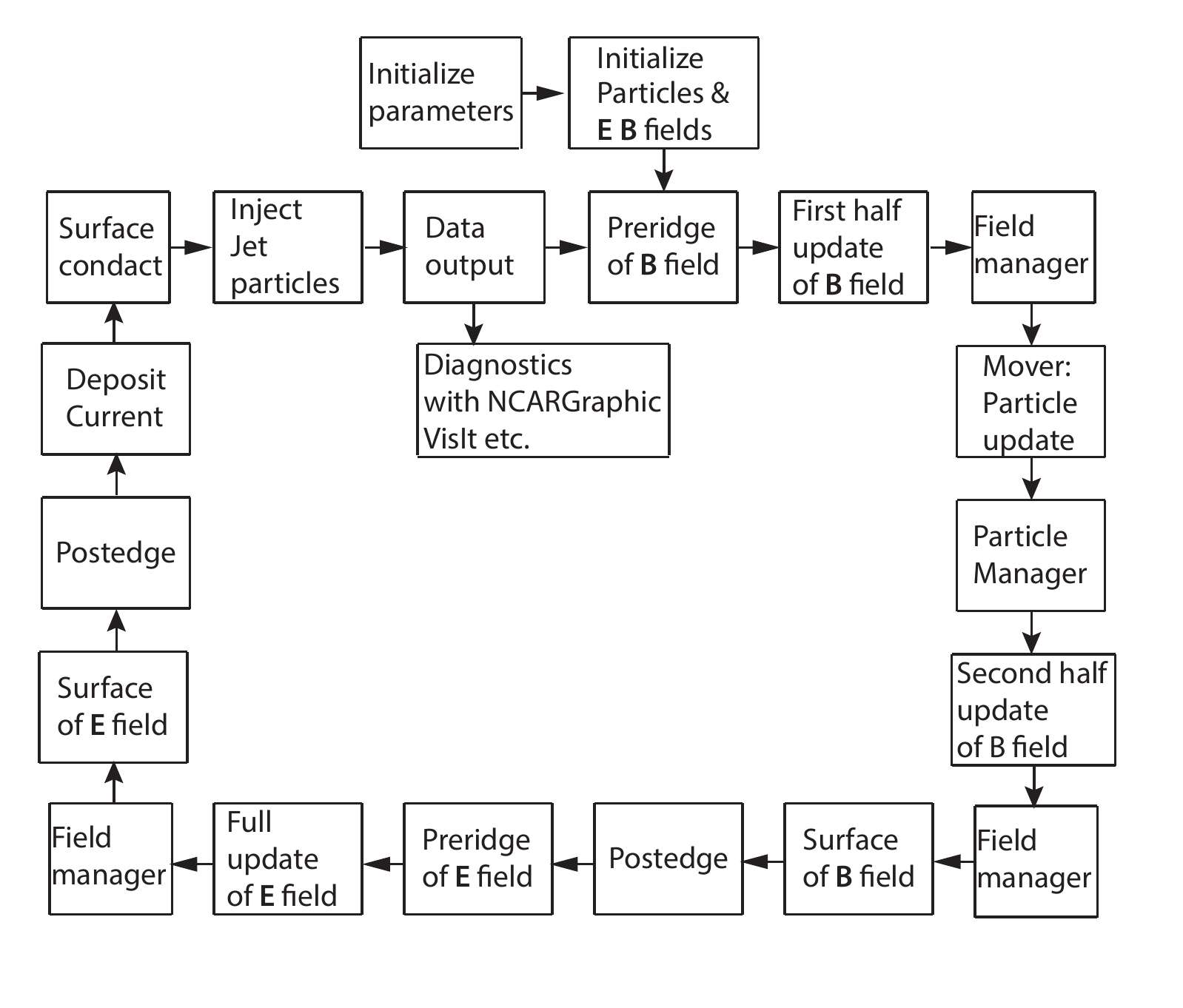}
\caption{Computational cycle of \TRISTANMPI  for jet simulations. This chart shows 
the leaf-frog scheme.}
\label{mpi-TRISTAN}
\end{figure}



\rv{The basic computational cycle of a \TRISTAN-based code is presented in Fig.~\ref{mpi-TRISTAN}.
Details on the sequences of the cycle are provided in the following sections.
In short, the electric and magnetic fields, which are continuous quantities, are discretized on a simulation (computational) grid, which is divided into a finite number of cubic cells (in 3D), such that the fields are stored in the grid point arrays (using a Yee lattice). Charged particles like electrons and ions are modeled by macroparticles. A macroparticle that represents an ensemble of many identical particles is represented in the code through a distribution (or shape) function. (Throughout this review, when referring to PIC calculations the term {\it{particle}} denotes a macroparticle, unless noticed.) During each time step, four main operations are repeated in the same order until the end of the simulation: scatter, solve, gather, and push. The algorithm consist of the following steps:}

\rv{
\begin{itemize}
 \item Initialization: particle data and field (or grid, as the fields are discretized on the grid) data are initialized
 \item Start the main loop: 
 \begin{itemize}
   \item Scatter: calculate particle contributions (using charges or currents as source terms) to the grid via interpolation  
 \item Solve: calculate the fields
 \item Gather: update forces on particle positions via interpolations
 \item Push: move particles to new positions and collect currents
 \end{itemize}
 \item End the main loop
 \item Results: diagnostics
\end{itemize}
}

\rv{Note that the interpolations in scatter (particle-to-field) and gather (field-to-particle) operations should be symmetric; i.e., the same numerical schemes should be used, otherwise the momentum conservation is lost. \TRISTAN uses time-centered and space-centered finite difference schemes to advance the equations in time via the Lorentz force equation and to calculate spatial derivatives. Therefore, the algorithm is second order accurate in space and time.}

\subsection{Basic equations}
\label{sec:3.1}

The basic equations to describe plasmas are Maxwell equations:   

\begin{eqnarray}
\partial \vec{B}/\partial t = - \nabla \times \vec{E}, 
\label{maxwB}
\end{eqnarray}

\begin{eqnarray}
\partial \vec{D}/\partial t = \nabla \times \vec{H} - \vec{J}, 
\label{maxwE}
\end{eqnarray}
where $\vec{H} = \vec{B}/{\mu_0}$, $\vec{D} = \epsilon_{0}\vec{E}$ and 

\begin{eqnarray}
\vec{J}(\vec{r}) = \sum (q_{\rm i}\vec{v}_{\rm i}\delta(\vec{r}-\vec{r}_i) + q_{\rm
e}\vec{v}_{\rm e}\delta(\vec{r}-\vec{r}_e)), 
\label{cur}
\end{eqnarray}
as well as the Newton--Lorentz force 
\begin{eqnarray}
\frac{dm_{\rm i,e}\vec{v}_{\rm i,e}}{dt} = q_{\rm i,e}(\vec{E} + \vec{v}_{\rm
i,e}\times \vec{B}), 
\label{lor}
\end{eqnarray}
where i and e corresponds to ion ($+$ charge) and electron ($-$ charge), respectively.

These equations are the only ones that are implemented into \TRISTAN. 
\rv{Generally, the plasma should also obey the Poisson's equation.} 
However, instead of solving Poisson's equation which is solved numerically in almost all particle simulation codes, \TRISTAN solves only two curls, i.e., Ampere's and Faraday's equations (\ref{maxwB}) and (\ref{maxwE}). 
A rigorous charge conservation method for the current $\vec{J}$ is described in more detail in \citet{villasenor92,umeda03}. 
Particles that are initialized as unmagnetized Maxwell distribution are updated by the leap-frog method using the magnetic and electric fields as an input. In order to get the correct field values, linear interpolation is employed throughout the code. In the  \TRISTANMPI \citep{niemiec08,toggweiler14,Winkel15} higher-order interpolation schemes are employed for solving the Maxwell and Newton-Lorentz equations.

\subsection{Charge distributions of macro-particle}
\label{sec:3.2}

\begin{figure}[htb]
\includegraphics[scale=0.33]{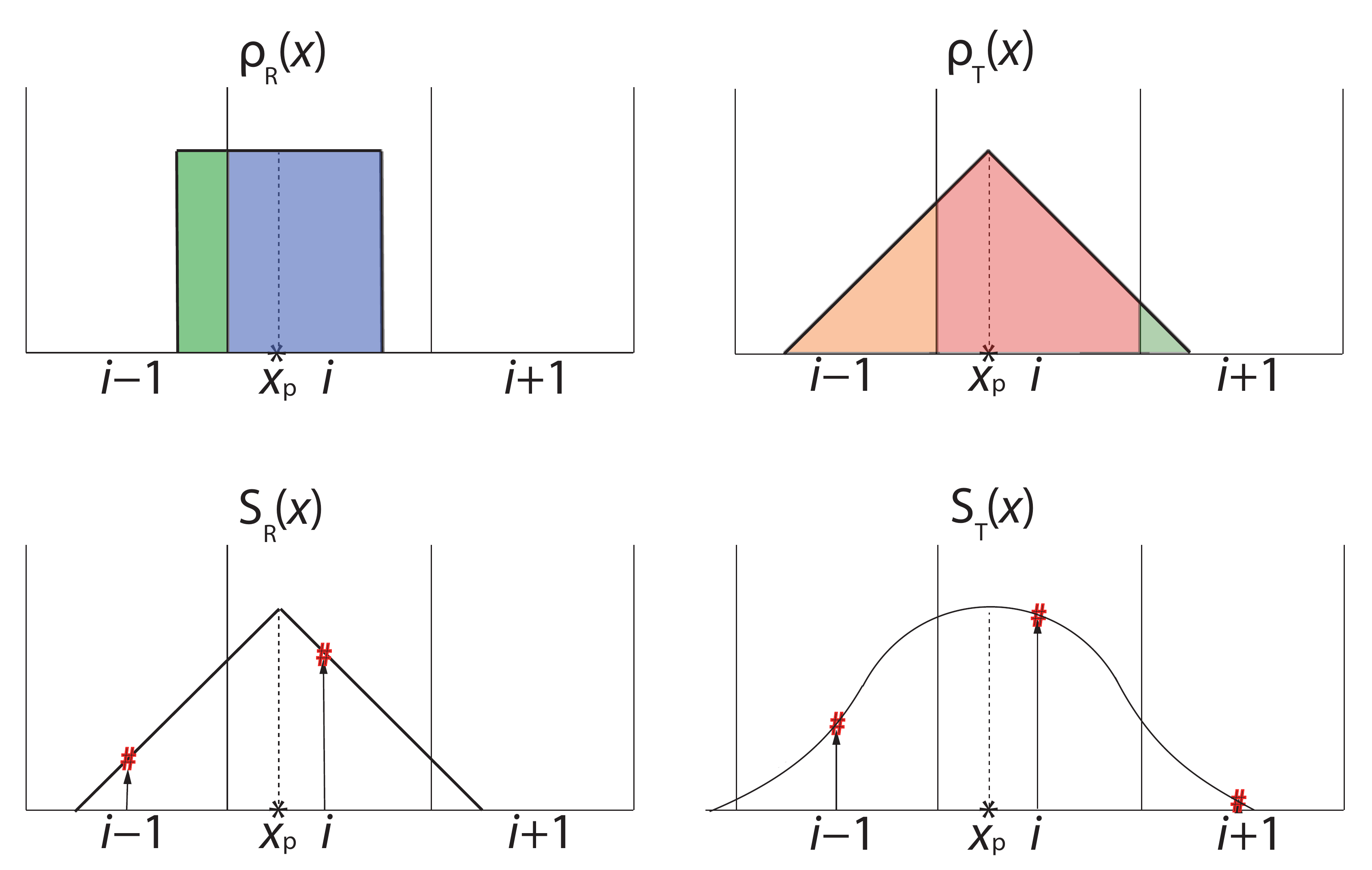}
\caption{The distributions of charge density with two shape factors left: rectangular and right: triangle in a 1D grid. $x_{\rm p}$ denotes the particle location, the rectangular shape with blue and green are allocated at the grid $i$ and $i-1$, respectively. For the triangle shape case the orange, red, and green area are linearly allocated at grid $i-1$, $i$, and $i+1$ respectively.}
\label{charge_density}
\end{figure}

In PIC simulations, to represent a myriad of individual particles, a macro particle is used with some shapes of charge particle such as rectangle, triangle, and Gaussian. Shape functions of cloud-in-cell determine aliasing noise in the corresponding functions in Fourier space \citep[e.g.,][]{dawson83,birdsall91}.
Figure \ref{charge_density} shows how finite-size macro-particle is used to distribute charge on a 1-dimensional (1D) grid. The shape function is schematically presented.

In \TRISTANMPI, the triangle shape is used which is shown in Fig \ref{charge_density}.

On 1D grid with uniform spacing, $\Delta x$, the electric charge $q$ of a particle with triangle shape at location $x_{\rm p}$ makes contributions to three nearest grid points
\begin{eqnarray}
q_{\rm i-1}=0.25\times q \frac{x_{\rm i}-x_{\rm p}+1/2}{\Delta x}, \,\, \, q_{\rm i}= 
q\frac{0.75}{\Delta x}, \,\,\,
q_{\rm i+1}=0.25 \times q\frac{x_{\rm p}-x_{\rm i}+1/2}{\Delta x}.
\end{eqnarray}
Here, the weight function is applied 0.25, 0.75, and 0.25 on grids $i-1$, $i$ and $i+1$, respectively. 

For 3D simulations, the same charge deposition scheme is used for two other dimensions. 
Therefore, 27 weighting factors are also used for calculating current density.

An important requirement for PIC simulations is that the same interpolation scheme is used to calculate force acting on particle as is applied for charge deposition to the grid.  

\subsection{Field update and particle update}
\label{sec:3.3}

\TRISTAN uses $\epsilon_{0} = 1$ such that $\mu_{0}= 1/c^{2}$ and $\vec{E} = \vec{D}=(e_x,e_y,e_z)$. Instead of recording the components of $\vec{B}$ or $\vec{H}$, \TRISTAN records three components $(b_x, b_y, b_z)$ of $c\vec{B}=\vec{H}/c$. This ensures the symmetry for electric field and magnetic field ($\vec{E} \longleftrightarrow \vec{B}$) in Maxwell equations except for the current $\vec{J}$, see below. 
Throughout the code, \TRISTAN uses a cubic grid  with $\delta x = \delta y = \delta z = 1$ and time discretization with $\delta t = 1$ (in \TRISTANMPI we use $\delta t = 0.1$ or less  \rv{to ensure the} numerical stability). Before and after moving (or pushing) the particles, $\vec{B}$ is updated in two half steps, such that it is available at the same time as $\vec{E}$ for the particle update. \rv{As already mentioned, in} \TRISTAN, only two curls of Maxwell equations are solved. 
Subsequently, the current density or charge flux $\vec{J}$ is calculated and subtracted after the particles are moved later in the program employing a rigorous charge conservation method for the current density \citep{villasenor92,Esirkepov2001,umeda03}.

\subsubsection{Magnetic field update}
\label{sec:3.3.1}

\begin{figure}[htb]
\centering
\includegraphics[scale=0.76]{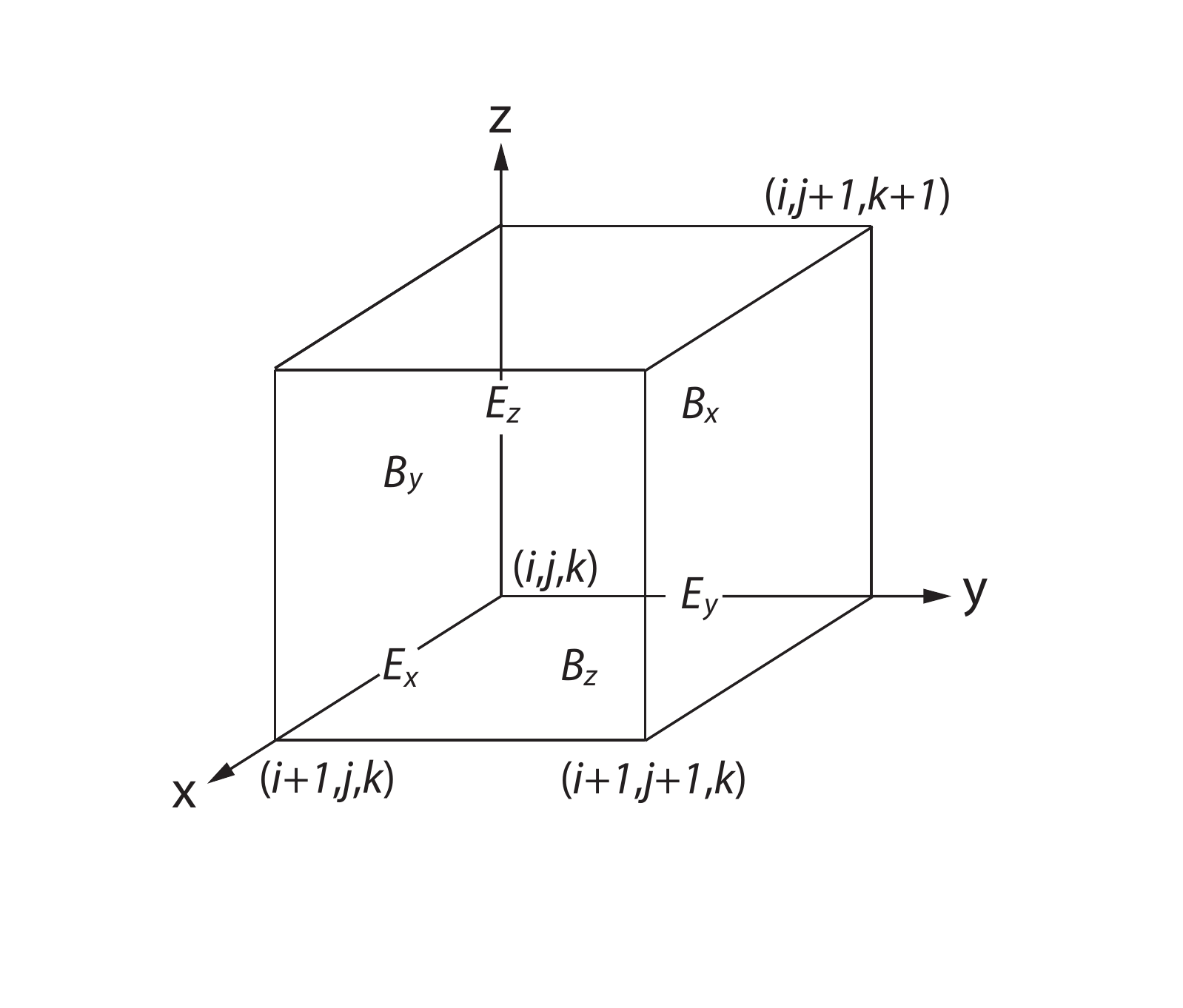}
\caption{Positions of field components. The $E$-components are in the middle of the edges and the $B$-components are in the center of the surfaces.} 
\label{yeelattice}
\end{figure}

In \TRISTAN, the electric and magnetic fields are stored at different locations in the grid cell by using a staggered grid. The staggered mesh system, known in the computational electromagnetic community as Yee-lattice \citep{yee66}, is shown in Fig.~\ref{yeelattice}. 
This \rv{configuration} ensures that the temporal change of $\vec{B}$ through a cell surface equals the negative circulation of $\vec{E}$ around that surface and likewise that the change of $\vec{E}$ equals the circulation of $\vec{B}$ around the surface of the cell containing $\vec{E}$ minus the current through it. Here, $\vec{B}$ and $\vec{E}$ are symmetric except for subtracting the current density $\vec{J}$ in the Ampere equation which is updated and subtracted after all particles have been moved at later positions in the program. 

On the Yee-lattice, $e_{\rm x}, e_{\rm y}, e_{\rm z}, b_{\rm x}, b_{\rm y}$, and $b_{\rm z}$ are staggered and shifted as shown in Fig. \ref{yeelattice} on 0.5 from $(i, j, k)$ and located at the positions as follows
\begin{eqnarray}
e_{\rm x}(i, j, k) &\rightarrow & e_{\rm x}(i + 0.5, j, k), \nonumber \\
e_{\rm y}(i, j, k) &\rightarrow & e_{\rm y}(i, j + 0.5, k), \nonumber \\
e_{\rm z}(i, j, k) &\rightarrow & e_{\rm z}(i, j, k + 0.5),  
\label{efield1}
\end{eqnarray}
and
\begin{eqnarray}
b_{\rm x}(i, j, k) &\rightarrow & b_{\rm x}(i , j+0.5, k+0.5), \nonumber \\
b_{\rm y}(i, j, k) &\rightarrow & b_{\rm y}(i+0.5, j , k+0.5), \nonumber \\
b_{\rm z}(i, j, k) &\rightarrow & b_{\rm z}(i+0.5, j+0.5, k).
\label{bfield1}
\end{eqnarray}

In \rv{\TRISTANMPI,} integer grids \rv{are used}. In both Eqs.~(\ref{efield1}) and (\ref{bfield1}), $i, j, k$ in the right-hand sides correspond to Fortran array indices notations and $i, j, k$ in the left hand sides correspond 
to the real positions in the simulation domains as shown in Fig.~\ref{yeelattice}.  If the value ``0.5'' is added to either $i, j, k$ in the array indices, then the array indices correspond to the real positions in the simulation domains.

Thus, magnetic fields are updated as follows.
The time change of the magnetic flux $\vec{B}$ is expressed through
\begin{eqnarray}
\frac{\partial \vec{B}}{\partial t} &= &-c
\left| \begin{array}{ccc}
\vec{i} & \vec{j} & \vec{k} \\
\frac{\partial}{\partial x} & \frac{\partial}{\partial y} & \frac{\partial}{\partial z} \\
e_{\rm x} & e_{\rm y} & e_{\rm z} \end{array} \right| \nonumber \\
&=& c\left[(\frac{\partial e_{\rm y}}{\partial z}-\frac{\partial e_{\rm z}}{\partial y})
\vec{i} +(\frac{\partial e_{\rm z}}{\partial x}-\frac{\partial e_{\rm x}}{\partial
z})\vec{j}+(\frac{\partial e_{\rm x}}{\partial y}-\frac{\partial e_{\rm y}}{\partial
x})\vec{k}\right] 
\label{ampere}
\end{eqnarray}
allowing to update the magnetic field components $b_{\rm x},b_{\rm y},b_{\rm z}$. The $x$-component 
of Eq.~(\ref{ampere}), $\partial b_x / \partial t = c(\partial e_y / \partial z - \partial e_z / \partial y)$, is translated into 
%
\begin{eqnarray}
(b_{\rm x}^{\rm new}(i , j&+&0.5,
k+0.5)- b_{\rm x}^{\rm old}(i , j+0.5, k+0.5))/\delta t \nonumber \\
& = &c[\{e_{\rm y}(i, j + 0.5, k + 1) - e_{\rm y}(i, j + 0.5, k)\}/\delta z 
\nonumber \\
&   &  -\{e_{\rm z}(i, j + 1, k + 0.5)
- e_{\rm z}(i, j, k + 0.5)\}/\delta y] , 
\label{bupdate}\\
\Rightarrow b_{\rm x}^{\rm new}(i , j, k) &= &b_{\rm x}^{\rm old}(i , j, k)) \nonumber \\
& &+c[\{e_{\rm y}(i, j, k + 1) - e_{\rm y}(i, j, k)\}  \nonumber \\
& &\, \, \, \, \, - \{e_{\rm z}(i, j + 1, k) - e_{\rm z}(i, j, k)\}],
\label{bupdate1}
\end{eqnarray}
with $\delta t = \delta z = \delta y = \delta x = 1$, similarly for $b_{\rm y}$ and $b_{\rm z}$. It should be noted that in \rv{many \TRISTAN-based codes, including \TRISTANMPI,} two-half step calculations are performed as described in \citep{cai03}. 

\subsubsection{Electric field update}
\label{sec:3.3.2}

Similar to the temporal evolution of $\vec{B}$, the temporal change of $\vec{E}$ through a cell surface equals the circulation of $\vec{B}$ around that surface, initially without the current $\vec{J}$. First, the electric field is updated by
\begin{eqnarray}
\frac{\partial \vec{E}}{\partial t} & = & c
\left| \begin{array}{ccc}
\vec{i} & \vec{j} & \vec{k} \\
\frac{\partial}{\partial x} & \frac{\partial}{\partial y} &
\frac{\partial}{\partial z} \\
b_{\rm x} & b_{\rm y} & b_{\rm z} \end{array} \right| \nonumber \\
&=& c\left[(\frac{\partial b_{\rm z}}{\partial y}-\frac{\partial b_{\rm
y}}{\partial z})\vec{i} +(\frac{\partial b_{\rm x}}{\partial z}-\frac{
\partial b_{\rm z}}{\partial x})\vec{j}+(\frac{\partial b_{\rm y}}{
\partial x}-\frac{\partial b_{\rm x}}{\partial y})\vec{k}\right]. 
\label{faraday}
\end{eqnarray}
Thus the electric field components $e_{\rm x}, e_{\rm y}, e_{\rm z}$ are updated by the magnetic field components on the Yee-lattice surface using Eq. (\ref{faraday}).
For example, the $x$ component of the electric fields becomes
\begin{eqnarray}
 (e_{\rm x}^{\rm new}(i&+&0.5 , j, k)- e_{\rm x}^{\rm old}(i+0.5, j, k))/\delta t \nonumber \\
& = &c[\{b_{\rm z}(i + 0.5, j + 0.5, k) - b_{\rm z}(i + 0.5, j - 0.5,
k)\}/\delta y  \nonumber \\
& &-\{b_{\rm y}(i+0.5, j, k + 0.5)
- b_{\rm y}(i+0.5, j, k - 0.5)\}/\delta z], 
\label{eupdate}\\
\Rightarrow e_{\rm x}^{\rm new}(i , j, k)&=& e_{\rm x}^{\rm old}(i, j, k) \nonumber \\
& &+c[\{b_{\rm z}(i, j, k) 
- b_{\rm z}(i, j-1, k)\} \nonumber \\
& &\, \, \, \, \, -\{b_{\rm y}(i, j, k) 
- (b_{\rm y}(i, j, k-1)\}], 
\label{eupdate1}
\end{eqnarray}
with $\delta t = \delta z = \delta y = \delta x = 1$, similarly for $e_{\rm y}$ and $e_{\rm z}$.

After updating the electric field through (\ref{eupdate1}), followed by the update of the position and velocity of particles, the current density $\vec{J}$ is calculated and subtracted from $\partial\vec{E}/\partial t$. See \ref{sec:3.6} Current deposit for details. 

\subsection{Particle update}
\label{sec:3.4}

The time-centered finite difference version of the Newton-Lorentz particle update is given as
\begin{eqnarray}
\vec{v}^{\rm new} =  \vec{v}^{\rm old} + \frac{q\delta t}{m}[\vec{E}
+\frac{1}{2}(\vec{v}^{\rm new} + \vec{v}^{\rm old})\times \vec{B}]
\label{n-lpupdate}
\end{eqnarray}

\begin{eqnarray}
\vec{r}^{\rm next} = \vec{r}^{\rm present} + \delta t \vec{v}^{\rm new}
\label{newposit}
\end{eqnarray}
 \rv{illustrating} that the position must be leap-frogged over velocities. A good physical interpretation of the steps in this explicit procedure is provided by Hartree and Boris \citep[e.g.,][]{toggweiler14,Winkel15}. Besides Hartree and Boris method, at least two different methods have been developed \citep[e.g.,][]{Vay08,Higuera17}. 
Many other particle pusher methods exist, each methods with pros \& cons, therefore choice depends on physics to be studied \citep[e.g.,][]{Ripperda18}.

\TRISTAN \rv{was originally developed} for relativistic plasma \rv{computations}, \rv{and, therefore,}
\rv{the code} solves the relativistic \rv{form of the} equation of motion
\begin{eqnarray}
\frac{d(m\vec{v})}{dt} = q(\vec{E}+\vec{v}\times \vec{B})
\label{relpupdate}
\end{eqnarray}
where $m = \gamma m_{0}$, with the rest mass $m_{0}$ and the Lorentz factor 
\begin{eqnarray}
\gamma=\frac{1}{\sqrt{1 -(\frac{v}{c})^{2}}}.
\label{loren}
\end{eqnarray}
We define $\vec{u} = \gamma \vec{v}$, or
\begin{eqnarray}
\vec{u}=\frac{c}{\sqrt{c^{2} -|\vec{v}|^{2}}}\vec{v}
\label{u}
\end{eqnarray}
which is equivalent to
\begin{eqnarray}
\vec{v}=\frac{c}{\sqrt{c^{2} +|\vec{u}|^{2}}}\vec{u}\equiv \vec{u}/\gamma_{\rm u},
\label{vel}
\end{eqnarray}
where $\gamma_{\rm u}=\sqrt{1+|\vec{u}|^{2}/c^{2}}$. Inserting (\ref{vel}) into (\ref{relpupdate}) gives
\begin{eqnarray}
\frac{d\vec{u}}{dt} = \frac{q}{m_{0}}(\vec{E}+\frac{c}{\sqrt{c^{2}
+|\vec{u}|^{2}}}\vec{u}\times \vec{B}).
\label{relpmod}
\end{eqnarray}
Defining the modified magnetic field
\begin{eqnarray}
\vec{B}_{\rm u} = \frac{c}{\sqrt{c^{2} +|\vec{u}|^{2}}}\vec{B}
\label{bu}
\end{eqnarray}
finally leads to
\begin{eqnarray}
\frac{d\vec{u}}{dt} = \frac{q}{m_{0}}(\vec{E}+\vec{u}\times \vec{B}_{\rm u})
\label{relpmodf}
\end{eqnarray}
which leads to the finite difference forms (\ref{n-lpupdate})
\begin{eqnarray}
\vec{u}^{\rm new} =  \vec{u}^{\rm old}  +\frac{q\delta t}{m_{0}}[\vec{E}
+\frac{1}{2}(\vec{u}^{\rm new} + \vec{u}^{\rm old})\times \vec{B}_{\rm u}]
\label{upupdate}
\end{eqnarray}
with $\vec{u}^{\rm new}=\vec{u}^{t+\delta t/2}$ and $\vec{u}^{\rm old}=
\vec{u}^{t-\delta t/2}$.

Based on the Boris method, the particle update \rv{is calculated} in several steps, as follows

\noindent
Step 1: calculate $\vec{u}^{\rm old}$ from $\vec{v}^{\rm old}$
\begin{eqnarray}
\vec{u}^{\rm old}=\frac{c}{\sqrt{c^{2} -|\vec{v}^{\rm old}|^{2}}}\vec{v}^{\rm old}
\label{uold}
\end{eqnarray}
\noindent
Step 2: update of $\vec{u}$ through half electric acceleration
\begin{eqnarray}
\vec{u}_{0}=\vec{u}^{\rm old}+\frac{q}{m_{0}}\frac{\delta t}{2}\vec{E}
\label{ehalf}
\end{eqnarray}
\noindent
Step 3: calculate $\vec{B}_{\rm u}$
\begin{eqnarray}
\vec{B}_{\rm u} = \frac{c}{\sqrt{c^{2} +|\vec{u}_{0}|^{2}}}\vec{B}^{\rm old}
\label{bu1}
\end{eqnarray}
\noindent
Step 4: cross with $\vec{B}_{\rm u}$
\begin{eqnarray}
\vec{u}_{1} = \vec{u}_{0}  +(\vec{u}_{1} + \vec{u}_{0})\times \vec{B}_{\rm
u}\frac{q\delta t}{2m_{0}}
\label{brotate}
\end{eqnarray}
The geometrical derivation of Eq.~(\ref{brotate}) is described in Fig.~3.1 in \cite{buneman93}.

\noindent
Step 5: another half electric acceleration
\begin{eqnarray}
\vec{u}^{\rm new}=\vec{u}_{1}+\frac{q}{m_{0}}\frac{\delta t}{2}\vec{E}
\label{ehalf1}
\end{eqnarray}

Eq.~(\ref{brotate}) \rv{which determines} $\vec{u}_{1}$ from $\vec{u}_{0}$ is still implicit, but its explicit form follows from (1) \rv{the} inner product of Eq.~(\ref{brotate}) with $(\vec{u}_{1} +\vec{u}_{0})$ to check that the magnetic 
field does not increase the \rv{particle} energy and that the magnitudes of 
$\vec{u}_{1}$ and $\vec{u}_{0}$ are the same, 
(2) \rv{the} inner product with $\vec{B}_{\rm u}$ to check that components along
$\vec{B}_{\rm u}$ are the same, (3) \rv{the} cross product with 
$\vec{b}_{0}=\frac{q\delta t}{2m_{0}}\vec{B}_{\rm u}$, and (4) substituting 
back, provides the explicit form of $\vec{u}_1$
\begin{eqnarray}
\vec{u}_{1} = \vec{u}_{0} +
\frac{2(\vec{u}_{0}+\vec{u}_{0}\times\vec{b}_{0})\times \vec{b}_{0}}{1+\vec{b}_{0}^{2}}.
\label{v1}
\end{eqnarray}

 \rv{Finally, the velocity} $\vec{v}$ and the new particle position \rv{are obtained} through
\begin{eqnarray}
\vec{v}=\vec{u}/\gamma_{\rm u}=\frac{c}{\sqrt{c^{2} +u_{\rm x}^{2}+u_{\rm
y}^{2} +u_{\rm z}^{2}}}\vec{u},
\label{vel1}\\
\vec{r}^{t+\delta t}=\vec{r}^{t}+\vec{v}^{t+\delta t/2}\delta
t=\vec{r}^{t}+\vec{u}^{t+\delta t/2}\delta t/\gamma_{\rm u}^{t +\delta t/2}
\label{move}.
\end{eqnarray}

A simple form of the Boris solver in PIC simulations is proposed by \cite{Zenitani18}. It employs an exact solution of the Lorentz-force equation\rv{, being}  equivalent to the Boris solver with a gyro-phase correction. As a favorable property for stable schemes, this form preserves a volume in the phase space. Numerical
tests of Boris solvers are conducted by test-particle and PIC simulations. The proposed form provides better
accuracy than the previous, popular forms \citep[e.g.,][]{buneman93},
while it requires \rv{only} insignificantly more computational time.

\begin{figure}[htb]
  \centering
\includegraphics[scale=0.5]{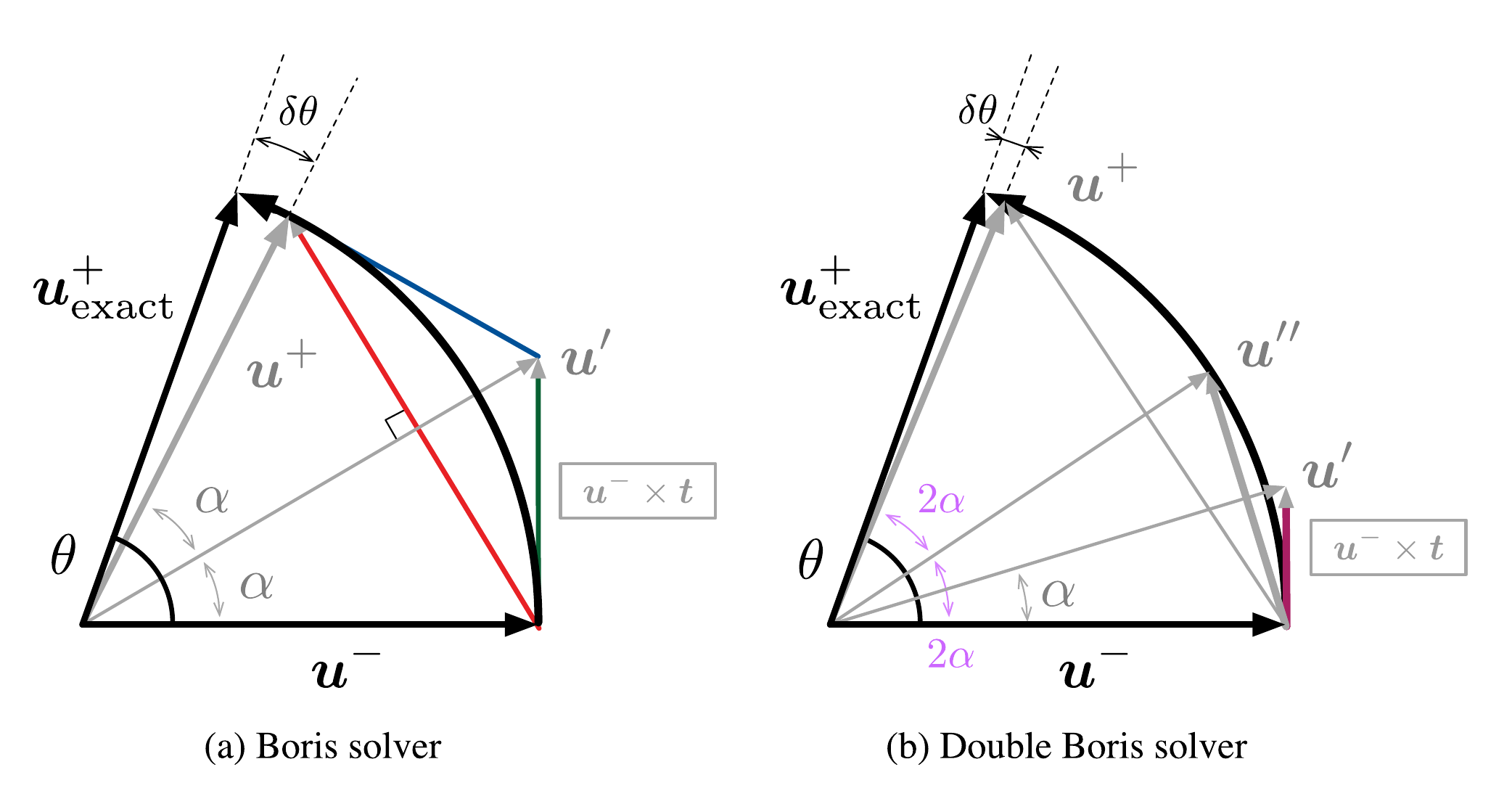}
\caption{(a) classical Boris solver and (b) double Boris solver. Adapted from Fig. 1 
in \citet{Zenitani2020}}
\label{Boris}
\end{figure}

Recently, \citet{Zenitani2020} developed a  more high performance, multiple Boris solver which combines the 2-step Boris procedure arbitrary $n$ times in the
Lorentz-force part. 
Figure~\ref{Boris} shows 
\rv{two schematic diagrams that illustrate the procedures performed for the classical Boris solver (a) and for 
the double Boris solver (b)}.
It should be noted that the 2-step Boris procedure as shown in Fig.~\ref{Boris}a is used by \cite{buneman93}. 
Using Chebyshev polynomials, a one-step form of the new solvers is provided. The  solvers \rv{by \citet{Zenitani2020}} give $n^{2}$ times smaller errors, allow larger time steps, and have a long-term stability as shown in Fig.~\ref{Boris}b. 
They presented numerical tests of the new solvers, in comparison with other particle integrator \citep{Ripperda18}.
We propose to implement these new methods if PIC simulations require more accurate particle trajectories. 

\subsection{Force interpolations}
\label{sec:3.5}

In Eq.~(\ref{upupdate}), $\vec{E}$ and $\vec{B}$ are interpolated from the grid points, i.e., the edge or center of a grid cell, to the particle positions on the Yee-lattice. Therefore, linear interpolation is employed to obtain a subgrid resolution which means that there is no stringent lower limit to the sizes of such quantities as the gyroradius or the Debye length. For quantities recorded on the integer mesh $x = i, y = j, z = k$, this means interpolating the eight nearest entries by applying weights, the  so-called ``volume" weights \citep{buneman93}. 

\begin{figure}[htb]
\centering
\includegraphics[scale=0.76]{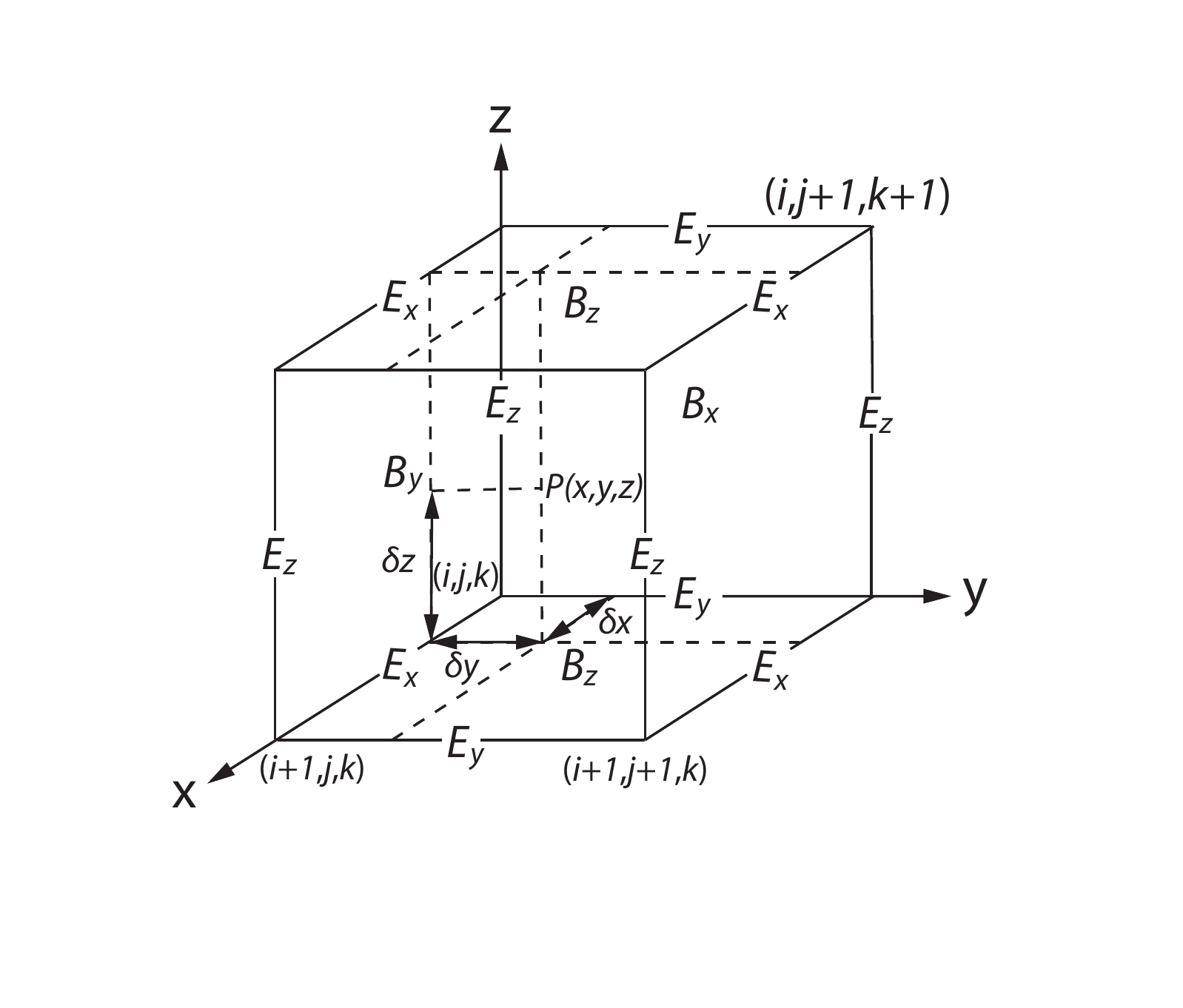}
\caption{The positions of field components on the Yee-lattice. A particle is located at the point $P$ with $x=i, y=j$ and $z=k$.}
\label{interp}
\end{figure}
For example, the ``volume'' weight is $(1 - \delta x)(1 -
\delta y)(1 - \delta z) = c_{\rm x} \cdot c_{\rm y} \cdot c_{\rm z}$ defining 
$c_{x,y,z}$ for $(i, j, k)$ and 
$\delta x \cdot \delta y \cdot \delta z$ (used as $dx, dy, dz$ in \TRISTAN) for 
$(i + 1, j + 1, k + 1)$. On the Yee-lattice as shown in Fig.~\ref{interp}, 
the interpolated force at $(x, j, k)$ exerted by the electric field components $e_{\rm x}$ 
is interpolated as 
%
\begin{eqnarray}
\vec{F}^{(x,j,k)}_{e_{\rm x}} = e_{\rm x}(i, j, k) + [e_{\rm x}(i + 1, j, k)
-e_{\rm x}(i, j, k)]\delta x,
\label{fex}
\end{eqnarray}
with
\begin{eqnarray}
e_{\rm x}(i, j, k) \Leftrightarrow\frac{1}{2}[e_{\rm x}(i, j, k) +e_{\rm x}(i-1, j, k)]
\nonumber\\
e_{\rm x}(i+1, j, k) \Leftrightarrow \frac{1}{2}[e_{\rm x}(i+1, j, k) +e_{\rm x}(i, j, k)].
\label{fex1}
\end{eqnarray}
Note that on the Yee-lattice, the electric (and magnetic) field is staggered, as sketched in Fig.~\ref{interp}. \rv{Therefore, the value of the electric field on the grid
needs to be averaged as Eqs.~(\ref{fex1})}. Thus, 
\rv{the interpolated force is}
\begin{eqnarray}
2\vec{F}^{(x,j,k)}_{e_{\rm x}} &=& e_{\rm x}(i, j, k) + e_{\rm x}(i -1, j, k)
\nonumber \\
& &+[e_{\rm x}(i+1, j, k) -e_{\rm x}(i -1, j, k)]\delta x.
\label{fex2}
\end{eqnarray}
The interpolated forces exerted by $e_{\rm x}$ at $(x, j +1, k), (x, j, k+1)$,
and $(x, j + 1, k + 1)$ are
\begin{eqnarray}
2\vec{F}^{(x,j+1,k)}_{e_{\rm x}} &=& e_{\rm x}(i, j+1, k) + e_{\rm x}(i -1,
j+1, k) \nonumber \\
&&+[e_{\rm x}(i+1, j+1, k)
-e_{\rm x}(i -1, j+1, k)]\delta x, \\
2\vec{F}^{(x,j,k+1)}_{e_{\rm x}} &=& e_{\rm x}(i, j, k+1) + e_{\rm x}(i -1, j,
k+1) \nonumber \\
&&+[e_{\rm x}(i+1, j, k+1)
-e_{\rm x}(i -1, j, k+1)]\delta x,
\label{fex3}
\end{eqnarray}
and
\begin{eqnarray}
2\vec{F}^{(x,j+1,k+1)}_{e_{\rm x}} &=& e_{\rm x}(i, j+1, k+1) + e_{\rm x}(
i -1, j+1, k+1) \nonumber \\
&&+[e_{\rm x}(i+1, j+1, k+1)
-e_{\rm x}(i -1, j+1, k+1)]\delta x,
\label{fex4}
\end{eqnarray}
respectively. Hence, the interpolated forces exerted by $e_{\rm x}$ at 
$(x, y, k),$ $(x, y, k +1)$, and $(x, y, z)$, are
\begin{eqnarray}
\vec{F}^{(x,y,k)}_{e_{\rm x}} &=& \vec{F}^{(x,j,k)}_{e_{\rm x}}
+[\vec{F}^{(x,j+1,k)}_{e_{\rm x}}-\vec{F}^{(x,j,k)}_{e_{\rm x}}]\delta y,  \\
\vec{F}^{(x,y,k+1)}_{e_{\rm x}} &=& \vec{F}^{(x,j,k+1)}_{e_{\rm x}}
+[\vec{F}^{(x,j+1,k+1)}_{e_{\rm x}}-\vec{F}^{(x,j,k+1)}_{e_{\rm x}}]\delta y,
\label{fex5}
\end{eqnarray}
and
\begin{eqnarray}
\vec{F}^{(x,y,z)}_{e_{\rm x}} = \vec{F}^{(x,y,k)}_{e_{\rm x}}
+[\vec{F}^{(x,y,k+1)}_{e_{\rm x}}-\vec{F}^{(x,y,k)}_{e_{\rm x}}]\delta z,
\label{fex6}
\end{eqnarray}
respectively. In the same way we can obtain the the interpolated components
$\vec{F}^{(x,y,z)}_{e_{\rm y}}$, $\vec{F}^{(x,y,z)}_{e_{\rm z}}$,
$\vec{F}^{(x,y,z)}_{b_{\rm x}}$,
$\vec{F}^{(x,y,z)}_{b_{\rm y}}$, and $\vec{F}^{(x,y,z)}_{b_{\rm z}}$
exerted by the electric field components $e_{\rm y}, e_{\rm z}$ and the magnetic field components $b_{\rm x}, b_{\rm y}$, and $b_{\rm z}$, respectively.

Note that the first step of calculating $\vec{F}^{(x,y,z)}_{b_{\rm x}}$ is as follows:
On the Yee-lattice, shown in Fig.~\ref{interp}, the interpolated force at 
$(x, j, k)$ exerted by the magnetic field components $b_{\rm x}$ is interpolated through
\begin{eqnarray}
\vec{F}^{(x,j,k)}_{b_{\rm x}} = b_{\rm x}(i, j, k) + [b_{\rm x}(i + 1, j, k)
-b_{\rm x}(i, j, k)]\delta x,
\label{ex}
\end{eqnarray}
where 
%
\begin{eqnarray}
b_{\rm x}(i, j, k) & = &\frac{1}{4}[b_{\rm x}(i, j, k)+b_{\rm x}(i, j-1, k) 
+b_{\rm x}(i, j, k-1) \nonumber \\
& & +b_{\rm x}(i, j-1, k-1)] \nonumber \\
b_{\rm x}(i+1, j, k)& = &\frac{1}{4}[b_{\rm x}(i+1, j, k)+b_{\rm x}(i+1, j-1,
k)  +b_{\rm x}(i+1, j, k-1) \nonumber \\
& &+b_{\rm x}(i+1, j-1, k-1)].
\label{ex1}
\end{eqnarray}

Subsequently, the staggering yields
\begin{eqnarray}
4\vec{F}^{(x,j,k)}_{b_{\rm x}} &=& b_{\rm x}(i, j, k)+b_{\rm x}(i, j-1, k) 
+b_{\rm x}(i, j, k-1) 
+b_{\rm x}(i, j-1, k-1) \nonumber \\
& &+[b_{\rm x}(i+1, j, k)+b_{\rm x}(i+1, j-1, k)  +b_{\rm x}(i+1, j, k-1)
\nonumber \\
& &+b_{\rm x}(i+1, j-1, k-1)  \nonumber
-\{b_{\rm x}(i, j, k)+b_{\rm x}(i, j-1, k)   \nonumber \\
& & +b_{\rm x}(i, j, k-1) +b_{\rm x}(i, j-1, k-1)\}]\delta x,
\label{ex2}
\end{eqnarray}
and analogously for $\vec{F}^{(x,y,z)}_{b_{\rm x}}$, $\vec{F}^{(x,y,z)}_{b_{\rm y}}$, 
and $\vec{F}^{(x,y,z)}_{b_{\rm z}}$. 

\subsection{Current deposit}
\label{sec:3.6}

\begin{figure}[htb]
\hspace*{1.2cm}
\includegraphics[scale=0.80]{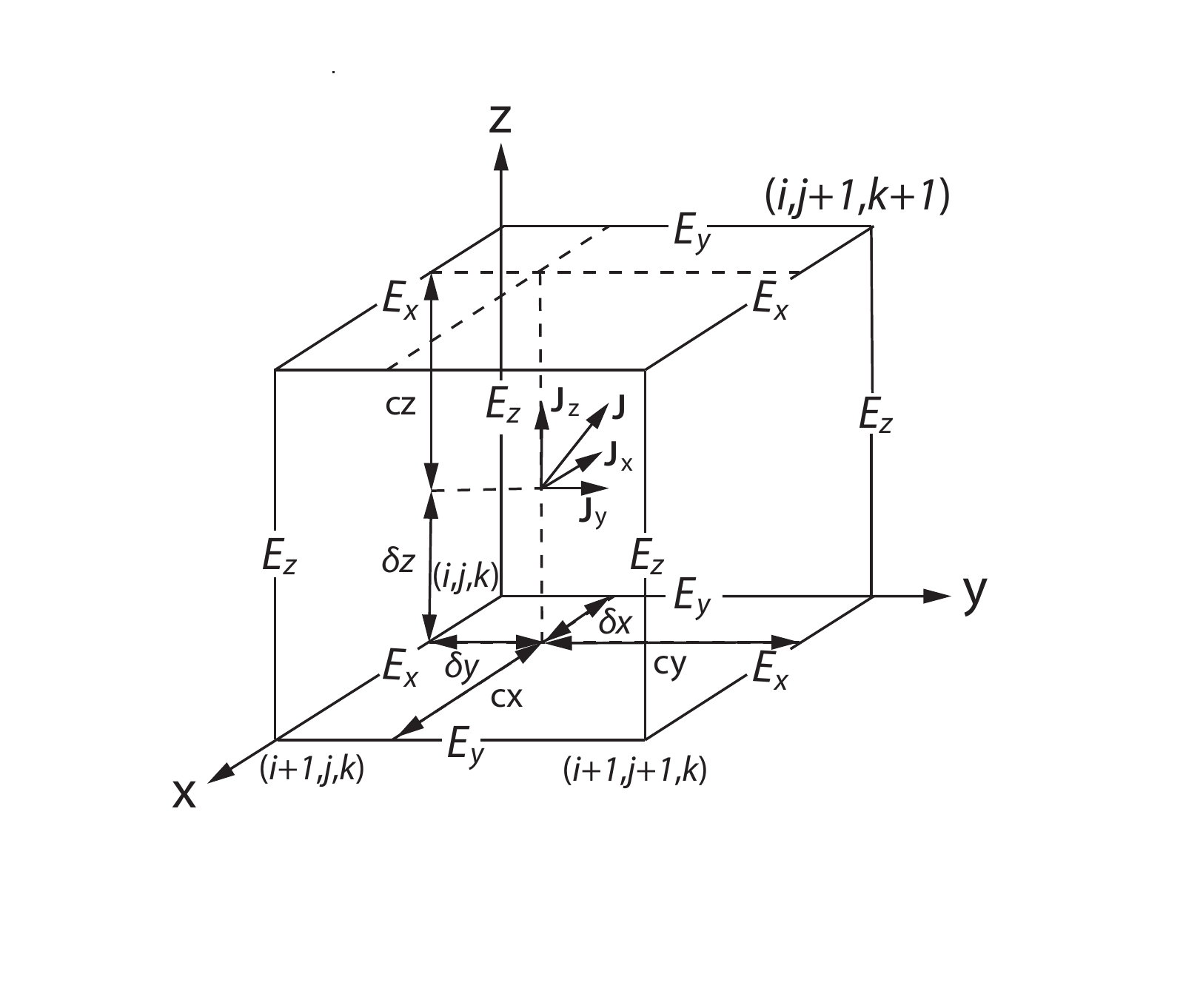}
\caption{The current components indicated by {\bf J} ($J_{\rm x}, J_{\rm y}, J_{\rm z}$)
recorded at the point $P(x,y,z)$ in the Yee-lattice.}  
\label{current}
\end{figure}

As we discussed in section \ref{sec:3.3.2}, the update of $\vec{E}$ through a cell surface (offset grid) is calculated in two steps: First, the curl of $\vec{B}$ around that surface is updated before the charge fluxes are  subsequently subtracted from $\nabla\times\vec{B}$.
Thus, only charge fluxes, i.e., the amount of charge flowing through the surfaces of Yee-lattice, are needed and, hence, \TRISTAN does not employ a charge density array. From the Maxwell equations, it follows that the Poisson equation will always be valid if the charge conservation condition
\begin{eqnarray}
\frac{\partial \rho}{\partial t}= - \nabla\cdot \vec{J}  
\label{pois}
\end{eqnarray}
is satisfied. Hence, if rigorous charge conservation is enforced numerically (see \cite{villasenor92,umeda03} for conservation methods), the electromagnetic field can be updated from the two curl equations (\ref{maxwB}) and (\ref{maxwE}) only. 
In the scheme described in detail by \citet{villasenor92}  and \citet{umeda03}, the current flux through every cell surface within a time step $\delta t$ is determined by counting the amount of charge carried through the Yee-lattice cell surfaces by particles as they move from $\vec{r}^{\rm n}$ to $\vec{r}^{\rm n+1}$ as shown in Fig.~\ref{current}. 
%

On Yee-lattice cell surfaces, the charge fluxes are
subtracted from each component of $\vec{E}$ field as follows 
\begin{eqnarray}
e_{\rm x}(i,j,k)& = & e_{\rm x}(i+0.5,j,k)    \nonumber \\
& = & e_{\rm x}(i,j,k) -J_{\rm x}*cy*cz \nonumber \\
e_{\rm x}(i,j+1,k) & = & e_{\rm x}(i+0.5,j+1,k)   \nonumber \\
& = & e_{\rm x}(i,j+1,k) -J_{\rm x}*\delta y*cz \nonumber \\
e_{\rm x}(i,j,k+1) & = & e_{\rm x}(i+0.5,j,k+1)   \nonumber \\
& = & e_{\rm x}(i,j,k+1) -J_{\rm x}*cy*\delta z \nonumber \\
e_{\rm x}(i,j+1,k+1)  & = & e_{\rm x}(i+0.5,j+1,k+1)   \nonumber \\
& = & e_{\rm x}(i,j+1,k+1) -J_{\rm x}*\delta y*\delta z \nonumber,
\label{excurrent}
\end{eqnarray}
\begin{eqnarray}
e_{\rm y}(i,j,k)     & = & e_{\rm y}(i,j+0.5,k)    \nonumber \\
& = & e_{\rm y}(i,j,k) -J_{\rm y}\cdot cx\cdot cz \nonumber \\
e_{\rm y}(i+1,j,k)   & = & e_{\rm y}(i+1,j+0.5,k)   \nonumber \\
& = & e_{\rm y}(i+1,j,k) -J_{\rm y}\cdot \delta x\cdot cz \nonumber
\\
e_{\rm y}(i,j,k+1)   & = & e_{\rm y}(i,j+0.5,k+1)   \nonumber \\
& = & e_{\rm y}(i,j,k+1) -J_{\rm y}\cdot cx\cdot \delta z \nonumber
\\
e_{\rm y}(i+1,j,k+1) & = & e_{\rm y}(i+1,j+0.5,k+1)   \nonumber \\
& = & e_{\rm y}(i+1,j,k+1) -J_{\rm y}\cdot \delta x\cdot \delta z 
\nonumber 
\label{eycurrent}
\end{eqnarray}
and
\begin{eqnarray}
e_{\rm z}(i,j,k)     & = & e_{\rm z}(i,j,k+0.5)    \nonumber \\
& = & e_{\rm z}(i,j,k) -J_{\rm z}\cdot cx\cdot cy \nonumber \\
e_{\rm z}(i+1,j,k)   & = & e_{\rm z}(i+1,j,k+0.5)   \nonumber \\
& = & e_{\rm z}(i+1,j,k) -J_{\rm z}\cdot \delta x\cdot cy \nonumber
\\
e_{\rm z}(i,j+1,k)   & = & e_{\rm z}(i,j,k+0.5)   \nonumber \\
& = & e_{\rm z}(i,j+1,k) -J_{\rm z}\cdot cx\cdot \delta y \nonumber
\\
e_{\rm z}(i+1,j+1,k) & = & e_{\rm z}(i+1,j,k+0.5)   \nonumber \\
& = & e_{\rm z}(i+1,j+1,k) -J_{\rm z}\cdot \delta x\cdot\delta y
\nonumber 
\label{ezcurrent}
\end{eqnarray}
with the $c_{x,y,z}$ as defined in section \ref{sec:3.4}.

In these equations each coordinate of $\vec{J}$ is proportionally distributed depending on the position of the particle and of the electric field components. For example, for the electric field 
$e_{\rm x}(i+0.5,j,k)$ the area $cy\cdot cz$ in the $y-z$ plane is factored with $J_{\rm x}$, which is the far side from the grid coordinates $(i+0.5,j,k)$.

During the calculation of the current density, a particle which does not leave its cell during the update will transport the charge through twelve surfaces, four \rv{surfaces} for each
orientation \rv{(treating a particle as a uniform cube)}. 
Assuming the particle move straight, we calculate how much charge is changed 
\citep[e.g.,][]{villasenor92,umeda03}. 
Within the code \TRISTAN, the current density is subtracted from the $\vec{E}$-components in the ``\DEPOSIT'' subroutine. 
If a particle moves into a neighbouring cell, one splits the move into two parts, one for each cells, and deposits the charge fluxes separately into each cells. 
This splitting is done in the three nested ``\SPLIT" subroutines before  the ``\DEPOSIT" is entered.

\subsection{Sorting and localization}
\label{sec:3.7}

\TRISTAN does not keep the particles sorted \citep{buneman93}. On some highly parallel computing facilities, it is, however, desirable to keep particles and fields in memory locations sorted by their physical locations. In order to perform simulations for large systems, \TRISTAN needs to be parallelized using \rv{MPI}
\cite[e.g.,][]{niemiec08,spit08a}. 
In this case the \SPLIT routine can be modified to re-index each particle that has left its cell.
Note that with such ordering all data traffic in \TRISTAN is strictly local: only nearest neighbor information is needed to update all quantities. Eliminating the Poisson solver was the most important step towards such localization: 
the solution of Poisson's equation anywhere depends on the charges everywhere.

\subsection{Smoothing}
\label{sec:3.8}

By distributing each particle onto a cubic grid, rather than treating it as a discrete delta function as Eq. (\ref{cur}), we already eliminate much of the unnatural noise created by coarse graining the (physically very fine-grained) population of particles. Further smoothing is achieved by \TRISTAN using a primitive form of filtering, which averages the particles over adjacent neighboring cells in all three dimensions (as it was explained in Sect.~\ref{sec:3.2}). 
The charge density $\vec{J}$ added in the \DEPOSIT subroutine as a source term in Maxwell's equations is divided into four contributions of which only two are kept in the cell, whilst the other two are deposited in the cells to each side. This is equivalent to ``convolving'' the flux array with the sequence $0.25$, $0.5$, and $0.25$ and is performed individually in each of the three dimensions.

Translated into Fourier space, this would mean to apply a low-pass filter which eliminates the unsatisfactorily aliased harmonics (the $\pi$-modes). In three dimensions each charge flux is in fact spread over $3\times 3\times 3$ neighboring cells whose weights are recorded once and for all in the smoothing array \texttt{sm(27)}
while \texttt{ms(27)} contains the corresponding index displacements accounting for the loop at the end of \DEPOSIT. 

In each time step, the \DEPOSIT subroutine takes most of the computer time since each particle references at least $12 \times 27 = 324$ field array data. 
The deposit procedure cannot be vectorized in the particles (unless the particles are carefully sorted or some extra storage arrays are used). Fortunately, the smoothing loop (of length $27$) can be vectorized on CRAY machines which offer the ``gather-scatter'' instruction.
Contrarily, the loop over the particles in the mover can be vectorized (without persuasion) by the CRAY compiler. Also note that in the \TRISTANMPI code developed by \cite{niemiec08}, these issues are modernized.

Whilst \TRISTAN does not keep a record of the charge density, $\mathrm{div}\,\vec{D}$ and $\mathrm{div}\,\vec{E}$ can always be calculated as the outflow of $\vec{D}$ or $\vec{E}$ from each unit cube through
\begin {eqnarray}
\nabla \cdot \vec{E} & = & e_{\rm x} (i, j , k ) - e_{\rm x}( i- 1, j , k ) 
+ e_{\rm y}( i ,j, k ) -
e_{\rm y}( i , j - 1, k ) \nonumber \\
& & +e_{\rm z}(i , j , k ) - c_{\rm z}( i, j , k - 1).
\end {eqnarray}
However, when post-processing simulations performed by \TRISTAN, one is usually interested in the densities of ions and electrons separately. To obtain these densities, one has to go through the ion or electron array and deposit charge according to the volume weighting rule shown above. 
For this purpose, we append a \texttt{DENSITY} subroutine \citep{buneman93} which will create each separate density array and apply the smoothing algorithm in conformity with \TRISTAN's smoothing of all sources of fields. 
In \TRISTANMPI \citep{niemiec08}, the subroutine 
\texttt{Vol\_Weighting} is used in order to obtain 3D data of physical quantities such as the densities, fluid velocities, kinetic energies, thermal velocities, drift velocities, and averaged electromagnetic energy. 

\subsection{Particle injection}
\label{sec:3.9}

For the injection of relativistic jet particles, sophisticated numerical algorithms to load relativistic Maxwell distributions in PIC and Monte-Carlo simulations are required. 
One distinguishes between stationary and relativistically shifted Maxwellian distributions.

The stationary relativistic Maxwell distributions known as the J\"uttner-Synge distribution 
\citep{Synge57,juettner11} is defined in the following form
\begin{equation}
f(u)d^{3}u  =  \frac{N}{4 \pi m^{2}cTK_{2}(mc^{2}/T)}
\exp\left(-\frac{\gamma mc^{2}}{T}\right)d^{3}u,
\label{Maxwell}
\end{equation}
where $\vec{u}=\gamma\vec{v}$ is spatial component of the 4-velocity, $\vec{v}$ is the the velocity, $\gamma=[1-(|\vec{v}|/c)^2]^{-1/2}$ is the Lorentz factor, $m$ is the rest mass, $c$ is the speed of light, $T$ is the temperature, and $K_{2}$ is the modified Bessel function of the second kind \citep{zenitani15}. 
For such a stationary relativistic Maxwellian, using spherical symmetry the absolute value
$u=|\vec{u}|$ is determined by the inverse transform method or the Sobol algorithm \citep{Pozdnyakov77,Pozdnyakov83,Zenitani18j}; 
subsequently the three components $(u_{\rm x}, u_{\rm y}, u_{\rm z})$ are calculated through 
\begin{eqnarray}
u_{\rm x}&=&u(2X_1-1) \\
u_{\rm y}&=&2u\sqrt{X_1(1-X_1)}\cos(2\pi X_2)\\
u_{\rm z}&=&2u\sqrt{X_1(1-X_1)}\sin(2\pi X_2)
\end{eqnarray}
with two uniform random numbers $X_{1,2}\in[0,1)$.


In order to inject particles with relativistic drift velocities (the relativistically shifted-Maxwellian \rv{distribution}) the Lorentz transformation of the stationary Maxwellian distribution is used. 
The general properties of this Lorentz transformation for particle distributions are described between two frames, $S$ and $S^{\prime}$ assuming that particles are stationary in the reference frame $S$, and that, without loss of generality, $S^{\prime}$ is the reference frame shifted with the 4-velocity $({\rm \Gamma},-{\rm \Gamma}\beta, 0, 0)$ 
such that particle velocities in $S^{\prime}$ are boosted by 
$({\rm \Gamma},+{\rm \Gamma}\beta, 0, 0)$. According to the previous nomenclature, in the following all quantities with a prime ($^{\prime}$) denote the quantities in $S^{\prime}$.

As the total particle number is conserved, the distribution function does not transform
\begin{equation}
f(\vec{x},\vec{u})dx^{3}du^{3}  =   f'(\vec{x}',\vec{u}')d^{3}x'd^{3}u'
\label{cons}
\end{equation}
where $d^{3}x =dx\ dy\ dz$ is the spatial volume element. Using the time element $dt$ in the same frame, we consider the 4-dimensional volume element of $dt$ and $d^{3}x$ that is moving at the 4-vector of $\vec{u}$.

After some calculation, one finds the explicit expression
\begin{eqnarray}
f(u) & = & f'(u') = \frac{N}{4 \pi m^{2}cTK_{2}(mc^{2}/T)} 
\exp\left(-\frac{\gamma mc^{2}}{T}\right), \nonumber \\
& =& \frac{N}{4 \pi m^{2}cTK_{2}(mc^{2}/T)}\exp\left(
-\frac{{\rm \Gamma}(\epsilon'-\beta mcu'_{\rm x})}{T}\right),
\label{ShifMaxwell}
\end{eqnarray}
for a relativistic shifted-Maxwellian \citep{zenitani15} where $\epsilon'=\gamma'mc^{2}$ is the particle energy, ${\rm \Gamma}$ and $\beta =v_{\rm jt}/c$ are bulk Lorentz factor and drift velocity. In PIC simulations of relativistic jets the thermal velocity of relativistic jets is small (${\rm \Gamma}=T/mc^{2}\ll 1$).

\citet{zenitani15} \rv{describes} various numerical algorithms  \rv{for loading} relativistic
Maxwellians in particle simulations. The inverse transform method and the Sobol method are useful to load the stationary Maxwellian as long as $T\gg 1$; if $T\ll 1$, one can switch to the Box-Muller method. 

For the relativistic shifted-Maxwellian, \citet{zenitani15} describes \rv{how} to adjust the particle number by a rejection method where a particle is accepted if the following condition is met:
\begin{equation}
\frac{1}{2{\rm \Gamma}}\left(\frac{\gamma'}{{\rm \gamma}}\right)=\frac{1}{2}
(1+\beta v_{\rm x}) > X_{\rm s}
\label{rej}
\end{equation}
with a uniform random number $X_{\rm S}\in[0,1)$. If the condition is not met, then the particle momentum is re-initialized.

Another method used for the particle injection is the flipping method (which is to be distinguished from the 
rejection method). If the following condition is met for a uniform random variable $X_{f}$,
\begin{equation}
\beta v_{\rm x} >X_{f}
\label{flip}
\end{equation}
then $u_{\rm x}$ is changed to $- u_{\rm x}$, before computing $u'_{\rm x}$.
Here, the two conditions of $-\beta v_{\rm x} < 0$ and 
$-\beta v_{\rm x}  >X_{f}$ are combined. Whilst the rejection method has an acceptance efficiency of 50\%, the flipping method has an efficiency of 100\%.

These rejection and the flipping methods are simple and physically transparent. They can be combined with arbitrary base algorithms which are useful in relativistic kinetic simulations in high-energy astrophysics.


\TRISTANMPI uses a similar method for the jet frame, as outlined in \cite{Pozdnyakov77,Pozdnyakov83,zenitani15}, but jet particles, initialized from relativistic 
Maxwellian distributions, are additionally transformed into the simulation frame. 

\subsection{Post-processing and other subjects}
\label{sec:3.10}

When running \TRISTAN one observes that space allocation is more problematic 
than time allocation where ``space'' here refers to both CPU (memory) space and disk space, 
not physical space. As with physical observations, data acquisition is 
only the first stage: data storage, data analysis, graphical display and
interpretation remain as post-simulation tasks
\citep{buneman93,niemiec08,nishikawa09,nishikawa16a,Nishikawa2020}.
This becomes an even more serious problem with recent large-scale simulations on the large 
high performance computing facilities such 
as Pleiades in the NASA Advanced Supercomputing facility at the NASA Ames Research
Center\footnote{\url{https://www.nas.nasa.gov/hecc/resources/pleiades.html}} and Blue Waters 
in the National Center for Supercomputing Applications at the University of Illinois at
Urbana-Champaign\footnote{\url{http://www.ncsa.illinois.edu/enabling/bluewaters}}. 
As a gradual development, some data is analyzed while main jobs are running, which 
lowers the burden of storing huge data. 

It was found expedient to keep control of the output data space while running
\TRISTANMPI by specifying the time step for the next full data dump and by
restarting the run from each of these dumps. 
This offers the opportunity of taking at least a superficial look at the
output before continuing. However, choosing each time step for outputting is quite 
impractical since this would take too much disc space and would slow down the simulations
because of the huge time impact of output operations. 
If one knows beforehand what information one would like to obtain of a simulation 
a run - typically orbits of some particular particles - the outputting could be performed 
at fixed intervals. However, first runs are often 
full of surprises and re-running with a new choice of output frequency 
and output material is then the most economical method.
Recently, it has become convenient to create movies (slide shows) from the outputted data 
for presentations or supplementary material which requires to output data more often 
for smoother movies.

Three-dimensional data of fluid quantities such as particle densities are generated by 
the \DENSITY subroutine or by post processes. Then two-dimensional plots are made in 
a few chosen slices as well as the projections of the magnetic field vectors 
in the middle of those slices have been found useful and reassures the credibility 
of the simulation. However, \TRISTANMPI output is extremely diverse for a variety 
of advanced graphics methods, especially because of the 
three-dimensionality of the output material.
Recently, more and more 3D visualization tools such as
\texttt{VisIt}\footnote{\url{https://wci.llnl.gov/simulation/computer-codes/visit/}},
\texttt{ParaView}\footnote{\url{https://www.paraview.org}}, and others, have become available.
These visualization tools assist in understanding complex 3D structures and their evolution 
allowing also for 
slicing 3D output in order to view inside the 3D structures. Several three-dimensional 
plots are shown in this review.

Our physics education has taught us to think in terms of waves, particles and their
interaction. We are familiar with dispersion, growth and decay, reflection 
and absorption which are features of linear wave systems. We try to describe non-linear
phenomena by wave-wave interaction, i.e., as small perturbations of linear  
behaviour. However, the wave analysis of data is expensive. \TRISTANMPI does not use
transformations since they are non-local and costly. Transforming in time is 
only possible as a post-processing operation and it requires a record  
for each time step which has been previously implemented into \TRISTAN and \TRISTANMPI.

If one acquires to investigate waves in the output of a simulation run, one first needs to
decide what variable(s) to subject for the wave analysis. One must allocate 
core space for performing the spatial transforms of the desired
variable(s) at each time step and output the necessary spatial harmonics after
each time step. Subsequently, it is possible to post process the data by transforming 
each spatial harmonics in time allowing to study the amplitudes and phases 
of waves for the quantities one is interested in. The necessary additions to 
\TRISTANMPI are possible; a more detailed overview of the extensive wave analysis for
one of best educational PIC codes is given in \citep{omura07}.

When a simulation code reproduces observed phenomena (such as the formation 
of our magnetosphere and magnetotail by the solar wind impinging on Earth's
dipole) one could carefully conclude that there is no longer any mystery in what we 
see, but whether we have ``explained'' or ``understood'' the observed phenomena (completely)
remains arguable \citep[e.g.,][]{buneman92}. 
The purpose of a simulation is not to ``explain'', rather but to offer a facility 
for controlling input parameters and making predictions, rather than having 
to wait for variations to occur naturally.  
Moreover, simulations offer the opportunity of going over the same data 
with a variety of different diagnostics, or, if needed, repeating a run 
with the same input, but recording different variables as continuous output. 
One example on how extensively a simulation code can be applied for 
one and the same topic is the very initial PIC simulations of solar 
wind-magnetosphere interaction \citep[e.g.,][]{nishikawa97}.

\subsection{Code test and performances of TRISTAN-MPI}
\label{sec:3.11}



The original version of \TRISTAN \citep[e.g.,][]{nishikawa06a} has been parallelized with 
MPI, optimized for speed, and improved to more efficiently handle small-scale numerical noise. 
%
%
This version, called \TRISTANMPI, is designed to be flexible and to perform well on different
computational platforms including user control over 
a number of optimization features. One example is the splitting of large loops into segments,
for which the relevant variable can be simultaneously held in the processor cache, 
thus avoiding memory access over the system bus. Another example is the periodic 
re-sorting of particles improving the speed of memory IO.

\TRISTANMPI \citep[see e.g.,][]{niemiec08} has been performed on several large high 
performance computing facilities such as Mercury, Abe, Kraken, Ranger, Ember which are 
located at the Computer Centers of 
the National Science Foundation (NSF) \citep{nishikawa09},
and on the Pleiades and Columbia systems of NASA Advanced Supercomputing (NAS). 
Validation tests performed on these platforms show that computation time scales linearly with 
the number of particles per cell and with the number of cells (volume) per single-processor 
domain. The code also has an excellent scalability, provided the memory 
load per core is kept above $\sim$100MB. Other tests have verified that the efficiency 
of the computational execution is indeed limited by the memory load per core, rather 
than by the number of cores. Therefore, running small simulations on a large number of 
cores is impractical, whilst large simulations can be efficiently executed on many cores, 
hence on an increasing number of cores with fixed domain size. 
Very large 
simulations of astrophysical jets have been performed using 556 nodes with 10,000 cores 
on Pleiades at NAS and reported in \citet{nishikawa16a} confirming that the code is 
well optimized up to 10,000 cores.  


\subsubsection{{Weak scaling law on Bridges and Frontera}}

The following scaling check is performed to show that using a large number of cores keeps 
the simulation time similar for the same number of time steps even when performing simulations 
for very large systems.
The scaling on the Bridges at the Pittsburgh Supercomputer Center 
has been checked by running simulations of relativistic cylindrical jets with different 
system sizes, one case with 324 and one with 2916 cores (processors). The conclusion is 
that \TRISTANMPI is well optimized using a large number of cores 
in order to satisfy the physical parameters in size of jet radius and other parameters 
starting with simulations for smaller system and finishing with simulations using the 
maximum capability of Bridges. 

\begin{figure}[htb]
\centering
\includegraphics[scale=0.66]{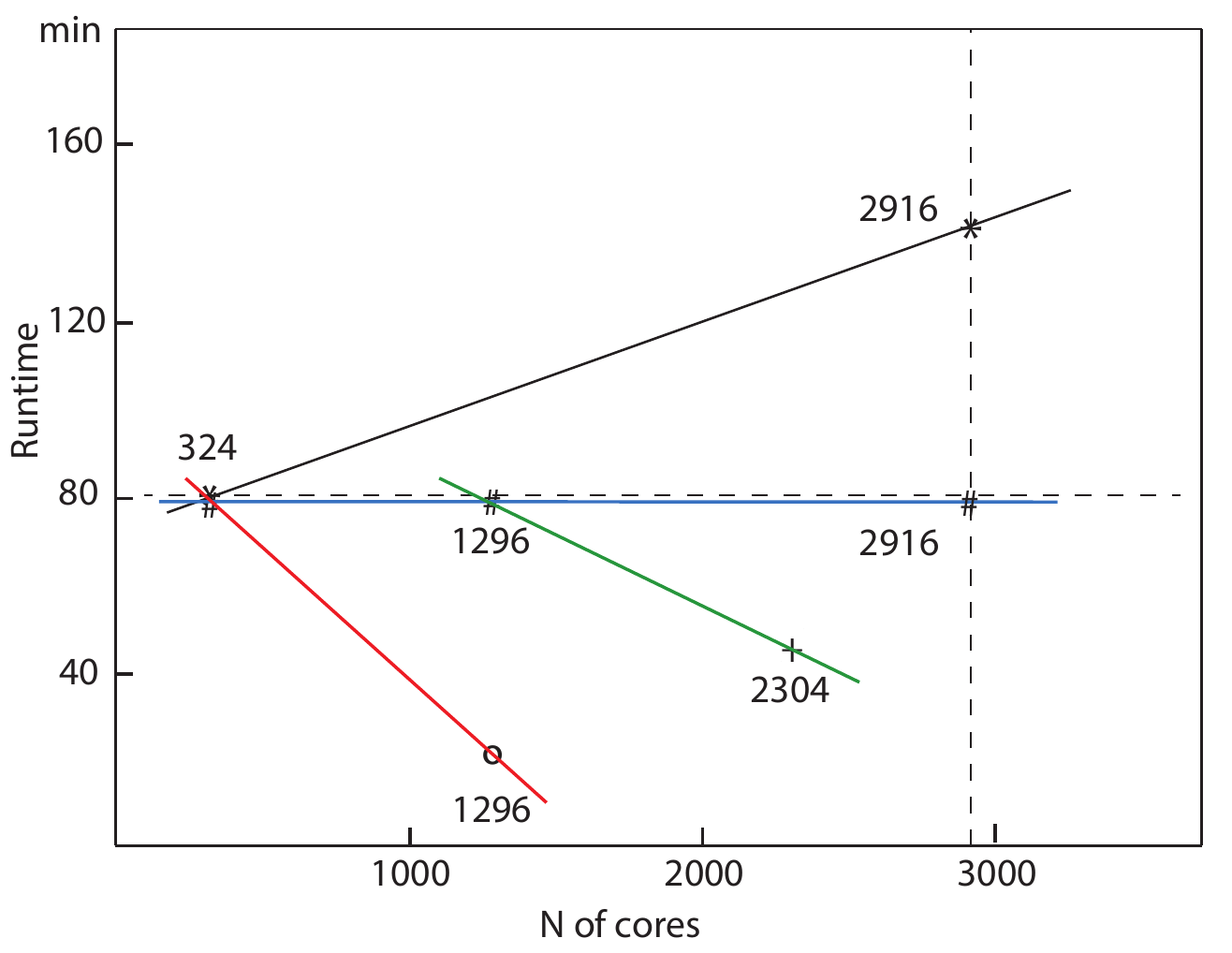}
\vspace*{-0.1cm}
\caption{The CPU run time as a function of the number $N$ of cores. The plot shows 
two results with 324 and 2916 cores indicated by ``$\ast$''. 
Runs 3, 4 and  5
are plotted with ``$\#$'' for the cases with 324, 1296 and  2916 cores.
Run 6 with 1296 cores is plotted with ``$\circ$'' (22 minutes), which $76/22=3.45$ faster 
than run 3 (theoretically 4.0).
Run 7 with 2304 cores is plotted with ``$+$'' (46 minutes), run4 runs in 78 minutes.} 
\label{scaling}
\end{figure}

In all the following comparisons, jet particles are traced for 5000 
time steps 
differences in these simulations are the size of jet radius and width. The system sizes are 
\begin{itemize}
\item run1: ($L_{x} \times L_{y} \times L_{z} = 645 \times 257 \times 257  \Delta^{3}$ $\Delta$: grid size)  
with jet radius $40 \Delta$ on 324 cores
\item run2: ($L_{x} \times L_{y} \times L_{z} = 645 \times 761 \times 761 \Delta^{3}$)  
with jet radius $120 \Delta$ on 2916 cores
\item run5: ($L_{x} \times L_{y} \times L_{z} = 645 \times 869   \times 869   \Delta^{3}$)  
with jet radius $40 \Delta$ on  2916 cores on Frontera
\end{itemize}
such that simulation run2 contains 9 time more jet and ambient particles than run1 as shown
by two ``$\ast$'' in Fig.~\ref{scaling}. The run time 
of both simulations is monitored by running with a different number of cores:

Although the particle number is increased by a factor of 9, the run time is only increase by 
a factor of 1.75. 
Considering some overall processes such as input 
and output, this shows that \TRISTANMPI is very well optimized.
It should be noted that run5 which is indicated by "$\sharp$" uses the same number of jet particles
runs in almost same time as shown by rushed blue line in Fig.~\ref{scaling}. 
This means that even more ambient particles in run5 than ran1, the 
run time does not increase.

\subsubsection{{Strong scaling law on Frontera}}

Strong scaling law test was performed on Frontera at the Texas Advanced Computing Center 
(TACC) for a more reasonable setting. The same jet radius $r_{\rm jt} =80\Delta$ used for 
the all calculations with 324, 1296 and 2304 cores as shown in Fig.~ref{scaling}. 
\begin{itemize}
\item run3: ($L_{x} \times L_{y} \times L_{z} = 645 \times 293  \times 293   \Delta^{3}$)  
with 324 cores
\item run6: ($L_{x} \times L_{y} \times L_{z} = 645 \times 293  \times 293   \Delta^{3}$)  
with  1296 cores
\item run4: ($L_{x} \times L_{y} \times L_{z} = 645 \times 581  \times 581  \Delta^{3}$)  
with 1296 cores
\item run7: ($L_{x} \times L_{y} \times L_{z} = 645 \times 581  \times 581   \Delta^{3}$)  
with  2304 cores
\end{itemize}
%
Run6 (indicated by ``$\circ$'') uses 4 times more cores than run3 for the same simulation 
size. Theoretically, it would run in 19 minutes. On Frontera it runs in 22 minutes and  $76/22=3.45$ faster than run3. It is 16\% slower than the theoretical prediction.
Run4 runs in 78 minutes and run7 (indicated by ``$+$'') runs 46 minutes, as shown by green line, run7 runs uses 1.8 times more cores, therefore run7 should run
in 43.9 minutes. Therefore, run7 runs in 4.8\% slower than the theoretical prediction. 
The Frontera is slightly faster than the Bridges.
This test shows that \TRISTANMPI is very well optimized on the Frontera.


\subsection{Numerical Cherenkov radiation}
\label{sec:3.12}

For one-dimensional electromagnetic particle simulation codes, linear dispersion relations are analyzed in order to determine numerical stability properties. It is found that fast particles may resonate with light waves of matching phase velocity to produce a severe numerical instability. 
A Courant condition for this instability is derived, and its restrictiveness amongst the various difference schemes is compared. At least two algorithms permitting reasonably large time steps for relativistic simulations are available \citep{GODFREY1974}.

\citet{NA2020} \rv{analyze} numerical Cherenkov radiation (NCR) effects produced by finite-element-based EM-PIC simulations involving relativistic plasma beams on different types of mesh. NCR is caused by spurious coupling between electromagnetic-field modes and multiple beam resonances. This coupling may result from the slow down of poorly-resolved waves due to numerical (grid) dispersion and from aliasing mechanisms. Complete dispersion diagrams over the first Brillouin zone were derived for periodic meshes with different element shapes and layouts. Analytical NCR predictions were compared against EM-PIC simulation results. Considering a relativistic plasma beam simulation, it \rv{is} observed  that the mesh element shape and mesh layout
have a marked influence on the ensuing NCR properties. In particular, it \rv{is} also observed that EM-PIC simulations on an unstructured mesh (with irregular triangular elements) does not support spatially coherent
NCR modes due to the aperiodic nature of the mesh layout. In this case, a diffusive-like behavior is observed for the NCR in the spatial domain. Importantly, it \rv{is} observed that the spurious energy produced by NCR on 
the unstructured mesh reaches saturation levels that are considerably lower than those on meshes based on periodic layout of (rectangular or triangular) elements. For simplicity, the analysis \rv{is} carried out in 2D but it is expected that similar conclusions should apply to 3D as well.
Results based on first order Whitney basis functions and without the use of high-order shape functions unveils the baseline NCR effect for a fair comparative analysis to be made across different meshes. It is expected that the use
of high-order shape functions should mitigate NCR effects. For that, the development of high-order shape functions on unstructured grids is an important line of future enquiry \citep[e.g.,][]{GREENWOOD2004}


\citet{Melzani2013} present the parallel \rv{PIC} code \texttt{Apar-T} and, more importantly, address the fundamental question of the relation amongst PIC models, the Vlasov--Maxwell theory, and real plasmas. 
They presented four validation tests: spectra from simulations of thermal plasmas, linear growth rates of the relativistic tearing and the filamentation instabilities, and nonlinear filamentation merging phase. Plasmas in nature contain millions to tens of billions of particles per Debye sphere, and relevant microphysical phenomena spread over numerous Debye lengths. It is impossible to track these particles one by one. Rather, the numerical particles represent numerous real particles, and are consequently called superparticles. A superparticle represents either $p$ real ions (having then a rest mass $m_{\rm sp} = p \times m_{\rm i}$ and a charge 
$q_{\rm sp} = p \times q_{\rm i}$), or $p$ real electrons (having then a rest mass $m_{\rm sp} = p \times m_{\rm e}$ and a charge $q_{\rm sp} = p \times q_{\rm e}$). 
The ratio of ion to electron charge is always $q_{\rm i}/q_{\rm e} = -1$, while that of their rest masses 
$m_{\rm i}/m_{\rm e}$ can be freely specified. 
Pair plasmas can thus be simulated. 
In \texttt{Apar-T} the number of real particles per superparticles $p$ is the same for all superparticles at all times.
For the filamentation instability they show that the effective growth rates measured on the total energy can differ by more than 50\% from the linear cold predictions and from the fastest modes of the simulation. 
They link these discrepancies to the superparticle number per cell and to the level of field fluctuations. 
A more subtle effect of coarse-graining is due to the loss of the dynamics of the $p$ particles represented by each superparticle.
They intuitively expect that it will lead to the overall loss of compressibility due to superparticle incompressibility, of the contribution to kinetic pressure of the particle velocity spreading within a superparticle, and of the multipole contribution to the electric and magnetic fields created by the distribution of particles within a superparticle. 
The relevance of these missing effects remains still unclear.

\cite{Melzani2013} \rv{descibe} a new method for initial loading of Maxwell-J\"uttner particle distributions with relativistic bulk velocity and relativistic temperature, and explain why the traditional method with individual particle boosting fails. For details see Sect.~\ref{sec:3.9}.
The formulation of the relativistic Harris equilibrium is generalized to arbitrary temperature and mass ratios. Both are required for the tearing instability setup. 

They \rv{turn} to the key point of \rv{their} paper and scrutinize the question of what description of (weakly coupled) physical plasmas is obtained by PIC models. 
These models rely on two building blocks: coarse-graining, i.e., grouping of the order of $p \sim  10^{10}$ real particles into a single computer superparticle, and field storage on a grid with its subsequent finite superparticle size. 
The size (shape) of superparticle is described in Sect.~\ref{sec:3.2}. They introduce the notion of coarse-graining dependent quantities, i.e., quantities depending on $p$. They derive from the PIC plasma parameter $\Lambda^{\rm PIC}$, which they show to behave as $1/\Lambda^{\rm PIC} \propto 1/p$. 
They explore two important implications. One is that PIC collision- and fluctuation-induced thermalization times are expected to scale with the number of superparticles per grid cell, and thus to be a factor $p \sim  10^{10}$ smaller than in real plasmas, a fact that they confirm 
with simulations. 
The other is that the level of electric field fluctuations scales as $1/\Lambda^{\rm PIC} \propto 1/p$. They \rv{provide} a corresponding exact expression, taking into account the finite superparticle size. 
They confirmed both expectations with simulations.

They \rv{compare} the Vlasov--Maxwell theory, often used for code benchmarking, to PIC simulations. 
The former describes a phase-space fluid with $\Lambda = +\infty$ and no correlations, while the PIC plasma features a small $\Lambda$ and a high level of correlations when compared to a real plasma. These differences have to be kept in mind when interpreting and validating PIC results against the Vlasov--Maxwell theory and when modeling real physical plasmas.

The dispersion relation of electromagnetic waves in vacuum is modified by the grid. This modification depends on the angle of propagation with respect to the grid, and waves can have a phase velocity smaller than $c$ \citep{GREENWOOD2004}. 
If superparticles with velocity close to c are present, they can overtake light waves and emit Cerenkov radiation. This results in the production of non-physical fields. The situation can be improved with a fourth order interpolation scheme for the fields \citep{GREENWOOD2004}.


\section{Short comparison between microscopic and macroscopic processes in plasma}
\label{sec:4}

\rv{In this section, we discuss the advantages and disadvantages of using microscopic and macroscopic simulations, or their coupling, when investigating the physics of plasmas.}

As mentioned in the previous sections, PIC simulations can investigate kinetic (microscopic) processes of plasma. Kinetic models describe the particle velocity distribution function at each point in the plasma and therefore do not need to assume a Maxwell-Boltzmann distribution. A kinetic description is often necessary for collisionless and collision dominated plasmas. 
The PIC technique includes the kinetic information by following the trajectories of a large number of individual particles. Such kinetic models are generally more computationally intensive than fluid models including 
solving the Vlasov equation or using a gyrokinetic approach.
The Vlasov equation may be used to describe the dynamics of a system of charged particles interacting with an electromagnetic field. In magnetized plasmas, a gyrokinetic approach can substantially reduce the computational expense of a fully kinetic simulation.

\rv{Vlasov equations are a type of hyperbolic partial differential equations describing the evolution of the phase-space distribution function $f_{\rm s}(\mathbf{r},\mathbf{v},t)$ of various species $s$ of charged particles. When coupled to the Maxwell equations, they can be applied to collisionless plasmas where the charged particles interact through electromagnetic 
fields. Let us consider test particles with charge $q_{\rm s}$, mass $m_{\rm s}$, and position $\mathbf{r}_{\rm s}$ in a plasma with stationary conditions. The particle $s$ will then feel the electric field $\mathbf{E}(\mathbf{r}_{\rm s},t)$ produced by all the other charges.  
The general form of the Vlasov-Maxwell equations describing the evolution of a single species plasma 
is given by}
\rv{\begin{equation}
    \frac{\partial{f}_{\rm s}}{\partial{t}} + \mathbf{v}\cdot\nabla_{\mathbf{r}}f_{\rm s} + 
     \frac{q_{\rm s}}{m_{\rm s}}(\mathbf{E}+\mathbf{v}\times\mathbf{B}) \cdot 
     \nabla_{\mathbf{v}}f_{\rm s} = 0 
    \label{eq:vlasov}
    \end{equation}
where the local fields are related only to local macroscopic quantities such as charge density and current through the Maxwell equations.}

\rv{The PIC method represents, in a way, an easier to handle approach to Vlasov solvers through re-graining the flow of the phase-space distribution functions via introducing macroparticles (superparticles) and providing the solution of differential equations of macroparticle motion (e.g., in 3D) instead of solving a system of hyperbolic partial differential equations in six dimensions.} 
In PIC simulations both electrons and ions are equally treated as particles. Therefore, the particle dynamics needs to be resolved in time and space. 
\rv{The scales in PIC simulation are set in such a way that the cell size $\Delta$ is much smaller than the length scale of the shortest physically relevant phenomenon and that the time step $\Delta t$ is short enough to resolve the fastest plasma fluctuations. 
For astrophysical plasma, for example in problems where the electrons are expected to be dynamically important, the electron plasma frequency $\omega_{\rm pe}$ and the skin depth $\lambda_{\rm e} \equiv c/\omega_{\rm pe}$ set the time and length scales such that the electron plasma oscillations are well resolved. 
Instead, if the ion dynamics is also of interest, the simulation should be large enough to include their skin depth and plasma oscillation period. For protons, approx. $\sqrt{m_{\rm p}/m_{\rm  e}} = 43$ more cell lengths 
and time steps, than what is needed for electrons, should be used. This factor is even larger for heavier ions. (In laser-plasma problems, the laser frequency is used instead of the electron plasma frequency.)} 
If, however, the ion dynamics is much more important and
electrons can be treated as an equalizing fluid, hybrid models can be used. 
\rv{In such simulations, the ions are treated as macroparticles for PIC calculations,} whereas the electron behavior is simulated as a fluid. This hybrid method is reviewed in, e.g., \citet{palmroth18}.

Astrophysical jets have extensively been investigated using hydrodynamic codes which are a particular class of fluid codes. In particular, since relativistic jets contain magnetic fields, RMHD codes need to be applied to study their propagation \citep[see, e.g.,][]{marti15}. More recently, jet formations have been investigated using GRMHD codes.

PIC simulations have become more feasible for global interaction between the solar wind and Earth magnetosphere \cite[e.g.,][]{nishikawa97,cai15}, which can be applied to study new aspects of, for example, astrophysical jet dynamics, in spite of the small size of the simulation domain. One of these aspects, which cannot be studied with the help of fluid equations, is the magnetic reconnection in the magnetotail generating burst flows of particles where electrons and ions are separated and flow in the opposite way due to the electric fields in the tail \citep{nishikawa97}. 

\rv{Recently, in order to avoid the weaknesses of PIC simulations for the
investigation of solar wind-magnetosphere interactions a Multiscale MHD-Kinetic 
PIC code has been developed \cite[e.g.,][]{Walker2019,Lapenta2020prl,Lapenta2020}.
For their multiscale approach they use a state from a global magnetohydrodynamics
(MHD) model to spawn a kinetic PIC simulation with an implicit PIC approach of 
a large portion of the tail. They directly investigate the energy fluxes resulting
from magnetic reconnection due to the substorm \citep{Lapenta2020}. 
Such a coupling provides information on additional processes absent in 
pure MHD simulations.}

It is important to understand the differences between kinetic (particle) and fluid methods. Since it is important to decide on a case-to-case basis which of the two approaches is the best method, we here now summarize their advantages and disadvantages including new hybrid methods between these two methods. 

 Since we describe the PIC simulations of relativistic jets in Sect. \ref{sec:5.4}, we  \rv{explain here} how PIC and RMHD simulations can handle relativistic jets \rv{from black holes}. \rv{(Those jets are supposed to be powered by the accreted matter around the black hole and/or by the rotational energy of the black hole itself. The matter orbiting a black hole and falling onto it is referred to as the accretion disk. A black hole surrounded by an accretion disk is one of the most effective machinery in the Universe for converting energy from the gravitational (disk) energy and rotational (black hole) energy into the kinetic energy of the jet particles.)} 
 The major differences between these two different methods are the units of simulation length and time step. In GRMHD simulations the unit of length is
\rv{usually set by the Schwarzschild radius of the black hole}: $r_{\rm S}=2GM/c^{2}$ ($G$:
gravitational constant, $M$: \rv{black hole} mass, and $c$: light speed) \rv{or by the gravitational
radius of the black hole, $r_{\rm g}$=$r_{\rm S}/2$}. 
Accordingly, the simulation time is measured \rv{in units of $r_{\rm S}/c$ or $r_{\rm g}/c$}.
Therefore, for example, \rv{the most extensive GRMHD simulations with $N_{\rm r} \times N_{\theta} \times N_{\phi} = 13440 \times 4608 \times 8092$ cells in whole simulation domain from inside the event horizon to $r = 10^5 r_{\rm g}$ 
and a run time exceeding $1.2$\id{$\times$} $10^6 r_{\rm g}/c$ provides a realistic size
for a relativistic jet from horizon to its global evolution in large scale \citep{Liska2019x,Liska2020L,Liska2020m}.}
The simulation parameters in RMHD simulations are scaled with jet radius, therefore it is very easy to accommodate the realistic size (length) based on the jet radius without including any kinetic processes. The simulation data can provide only phenomenological evolution of jets. 
In general, in these simulations current-driven kink instability is dominant if jet is highly magnetized. 
\citep[e.g.,][]{barniol17,Liska2019x,Davis2020,Dong20}.

On the contrary, \rv{as already mentioned,} in PIC simulations the length is scaled by the physical parameters such \rv{as the} electron Debye length:
$v_{th,e}/\omega_{\rm pe}$ and electron skin depth: $c/\omega_{\rm pe}$. 
Usually, the grid size $\Delta$ is similar or smaller than the Debye length.
If the real mass ratio ($m_{\rm proton}/m_{\rm e}= 1836$) is used, the ratio of these lengths between electrons and protons is $\sqrt{1836}\approx 43$, therefore it is difficult to include a large number of proton skin depth in the simulation system. Even for an electron-positron jet it is impossible to accommodate the 
realistic jet radius based on the electron skin depth. Therefore\rv{,} PIC simulations cannot 
provide any macroscopic scale phenomena as RMHD simulations can do phenomenologically. 
The assessment of physical parameters used in the simulation system to natural physical parameters has been done, and in the case of a jet with the radius $40\Delta$, the scaling implies that the jet axis of the PIC simulations (i.e., $240\Delta$ in length) corresponds to a physical size of $\sim$ 150 km.

However, RMHD simulations cannot provide any kinetic processes such as acceleration of non-thermal electrons which are responsible for high energy particles and observed gamma-ray flares.
Since one-fluid RMHD cannot distinguish between electron-positron and electron-ion (proton) jets.
\rv{While RMHD simulations capture well the large scales, they hardly capture the small scales which may be important, e.g., in turbulent cascades, anomalous resistivity, gyromotions, etc.}
Therefore, if the simulation size is large enough to accommodate a few modes of growing instabilities (MI, kKHI, and WI) along and transverse to jet propagating direction, we would be able to provide kinetic processes in relativistic jets described in Sect.~\ref{sec:5}. \rv{The recent PIC simulations show that two modes of MI are obtained in the jet radius ($100\Delta$) in the electron-positron jets \cite{Meli2021}.}
On the contrary for the electron-proton jet only one mode of MI is excited as show in \cite{Nishikawa2020}. The jet length in these simulations is large enough to accommodate a several modes of kKHI (WI) in the linear stage and at the end of simulations the nonlinear stage is established which provides an evolution of jets with kinetic processes.
In order to include ion kinetic processes the reduced mass ratio 
$m_{\rm i}/m_{\rm e}$ needs to be used. For example, if $m_{\rm i}/m_{\rm e}=9$ the ion skin depth becomes 3 times of the electron skin depth, therefore 
a few ion skin depths with jet radius ($300\Delta$ or larger) with 
the electron skin depth ($\lambda_{\rm e}= 10\Delta$) \rv{can be accommodated}. 
\rv{As the} evolution of kinetic instabilities 
depends on the mass ratio, 
even larger systems should be used to find out the required minimum mass ratio. Certainly, in order to minimize the weakness \rv{of the} PIC simulations, it is very important to develop an exa-scale PIC simulation which is in progress as described in Sect.~\ref{sec:3.12}. In order to \rv{mitigate} the weaknesses of PIC  simulations, the following new hybrid \rv{(coupling PIC and RMHD)} methods are developed.

As an example, the hybrid-Vlasov model \Vlasiator has recently been developed \citep[e.g.,][]{palmroth18} which models ions as a six-dimensional (6-D) space-velocity distribution and electrons are neglected apart from their charge-neutralizing behavior. The ion distribution function is propagated in time according to Vlasov's equation. The set of equations is then completed by Amp{\`e}re's law, Faraday's law, and a generalized Ohm's law such that the electric field is determined through
\begin{eqnarray}
\vec{E} = -\vec{V} \times \vec{B}+ \frac{1}{\rho_{\rm q}}\vec{j}\times
\vec{B}
\label{ohm}
\end{eqnarray}
where $\vec{V}$ is the ion bulk velocity, $\vec{B}$ is the magnetic
field, $\rho_{\rm q}$ is the ion charge density, $\vec{j}=\nabla 
\times \vec{B}/\mu_{0}$ is the current density,
and $\mu_{0} =4\pi \times 10^{-7} {\rm Hm}^{-1}$ is the vacuum permeability. The second term on the right-hand side of the equation
(\ref{ohm}) is the Hall term. Compared to resistive MHD, where the electric field is given by
\begin{eqnarray}
\vec{E} = -\vec{V} \times \vec{B} +\eta \vec{j}
\label{hall}
\end{eqnarray}
with resistivity $\eta$, including the Hall term produces
higher reconnection rates \citep{birn01}.
In \Vlasiator, there is no artificially added or enhanced resistivity and the magnetic reconnection is triggered by numerical diffusion. Any further
terms of the generalized Ohm's law, except those included in Eq.~(\ref{ohm}), are considered negligible at the current spatial
resolution of \Vlasiator which is approximately 300 km.

In summary, \Vlasiator includes the ion kinetics through a distribution function, but not the electron kinetics. This allows for the self-consistent global modeling of the
near-Earth plasma environment, including multi-temperature
non-Maxwellian ion populations that cannot be described by MHD realistically. Unlike \rv{PIC} approaches,
modeling ions as velocity distribution functions
produces solutions that are numerically noiseless, but also might
prevent the study of effects in the tail of the ion distribution.
Since the electron microscopic processes are not included in this method, any non-thermal distribution and radiations from electrons are not considered. 
\rv{For the devlopment} of astrophysical system\rv{s} the electron evolution is crucial, \rv{and} therefore this method is not appropriate for astrophysical research.

However, as mentioned beforehand, there are certain weaknesses of fluid
simulations that promote the use of PIC simulations instead.
Ideally, fluid RMHD simulations cannot distinguish kinetic effects of
electrons and ions and their skin depth (Debye length) is not accounted
for. These kinetic physical parameters are completely diminished in
RMHD simulations. Therefore, such models may miss important features.
For example, in the jet evolution kinetic processes may play a more dominant role, in
particular in the early linear stage. Certainly, only recently
performing large-scale PIC simulations has become feasible. So far RMHD
simulations of jets are considered as the ``gold standard'' of global jet simulations. 
At the present time, it is hard to investigate macroscopic phenomena using PIC models 
as compared to RMHD simulations. However, PIC simulations act complimentary to RMHD 
studies as computer power increases gradually \citep{moore_1965,mohseni_2017}.

\rv{\citet{MAKWANA2017}} describe a method for coupling an embedded domain in 
a MHD simulation with a PIC method. 
In this two-way coupling they follow the work of \rv{\citet{DALDORFF2014}} in which the PIC 
domain receives its initial and boundary conditions from MHD variables 
(MHD to PIC coupling) while the MHD simulation is updated based on the PIC variables 
(PIC to MHD coupling). This method can be useful for simulating large scale plasma systems, 
where kinetic effects captured by PIC simulations are localized but affect 
global dynamics. They \rv{describe} the numerical implementation of this coupling, its 
time-stepping algorithm, and its parallelization strategy, emphasizing the novel 
aspects of it. They tested the stability and energy/momentum conservation of this 
method by simulating a steady-state plasma. They performed the dynamics of this coupling 
by propagating plasma waves through the embedded PIC domain. Coupling with MHD shows 
satisfactory results for the fast magnetosonic wave, but significant distortion 
for the circularly polarized Alfv\'en wave. Coupling with Hall-MHD shows excellent 
coupling for the whistler wave. They also applied this methodology to simulate 
a Geospace Environmental Modeling (GEM) challenge type of reconnection with 
the diffusion region simulated by PIC coupled to larger scales with MHD and Hall-MHD. 
In both these cases they see the expected signatures of kinetic reconnection 
in the PIC domain, implying that this method can be used for reconnection studies.

Collisionless plasmas, mostly present in astrophysical and space environments, often
require a kinetic treatment as given by the Vlasov equation. Unfortunately, the
six-dimensional Vlasov equation can only be solved on very small parts of the 
considered spatial domain \citep{Lautenbach2018}. However, in some cases, e.g. 
magnetic reconnection, it is sufficient to solve the Vlasov equation in a localized 
domain and solve the remaining domain by appropriate fluid models. They describe 
a hierarchical treatment of collisionless plasmas in the following way. On the finest 
level of description, the Vlasov equation is solved both for ions and electrons. 
The next courser description treats electrons with a 10-moment fluid model
incorporating a simplified treatment of Landau damping. At the boundary between
the electron kinetic and fluid region, the central question is how the fluid
moments influence the electron distribution function. On the next coarser level
of description the ions are treated by an 10-moment fluid model as well. 
It may turn out that in some spatial regions far away from the reconnection 
zone the temperature tensor in the 10-moment description 
is nearly isotopic. In this case it is even possible to switch to a 5-moment
description. This change can be done separately for ions and electrons. 
To test this multi-physics approach, this full physics-adaptive simulations 
can be applied to the Geospace Environmental 
Modeling (GEM) challenge of magnetic reconnection. 

\cite{Rieke2015} \rv{present} a way to combine Vlasov and two-fluid codes for 
the simulation of a collisionless plasma in large domains while keeping full 
information on the velocity distribution in localised areas of interest. This is 
made possible by solving the full Vlasov equation in one region while the remaining 
area is treated by a 5-moment two-fluid code. In such a treatment, the main 
challenge of coupling kinetic and fluid descriptions is the interchange of physically 
correct boundary conditions between the different plasma models. In contrast to 
other treatments, they do not rely on any specific form of the distribution function, 
e.g. a Maxwellian type. Instead, they combine an extrapolation of the distribution 
function and a correction of the moments based on the fluid data. Thus, throughout 
the simulation both codes provide the necessary boundary conditions for each other. 
A speed-up factor of around 10 is achieved by using GPUs for the computationally 
expensive solution of the Vlasov equation. Additional major savings are obtained 
due to the coupling where the amount of savings roughly corresponds to the fraction 
of the domain where the kinetic equations are solved. The coupled codes were then 
tested on the propagation of whistler waves and on the GEM reconnection challenge.

{\cite{Bai2015} \rv{formulate} a \rv{MHD-PIC} method 
for describing the interaction between collisionless cosmic ray (CR) particles and 
a thermal plasma. The thermal plasma is treated as a fluid, obeying equations of 
ideal MHD, while CRs are treated as relativistic Lagrangian particles subject to 
the Lorentz force. Back reaction from CRs to the gas is included in the form of 
momentum and energy feedback. In addition, they include the electromagnetic feedback 
due to CR-induced Hall effect that becomes important when the electron-ion drift 
velocity of the background plasma induced by CRs approaches the Alfv\'en velocity. 
Their method is applicable on scales much larger than the ion inertial length, 
bypassing the microscopic scales that must be resolved in conventional PIC methods, 
while retaining the full kinetic nature of the CRs. They have implemented and tested 
this method in the Athena MHD code \cite{Stone_2008}, where the overall scheme is 
second-order accurate and fully conservative. As a first application, they describe 
a numerical experiment to study particle acceleration in non-relativistic shocks. 
Using a simplified prescription for ion injection. They \rv{reproduce} the shock structure 
and the CR energy spectra obtained with more self-consistent hybrid-PIC simulations, 
but at substantially reduced computational cost. They also \rv{show} that the CR-induced 
Hall effect reduces the growth rate of the Bell instability \citep[e.g.,][]{Reville2013} 
and affects the gas dynamics in the vicinity of the shock front. As a step forward, 
they are able to capture the transition of particle acceleration from 
non-relativistic to relativistic regimes, with momentum spectrum $f (p) \propto p^{-4}$ 
connecting smoothly through the transition, as expected from the theory of 
Fermi acceleration.}

\rv{\cite{Casse_2017} present simulations of magnetized astrophysical shocks taking
into account the interplay between the thermal plasma of the shock and supra-thermal
particles using particle-in-magnetohydrodynamics-cells. Such interaction is depicted 
by combining a grid-based magnetohydrodynamics
description of the thermal fluid with particle-in-cell techniques devoted
to the dynamics of supra-thermal particles. This approach, which incorporates the
use of adaptive mesh refinement, is potentially a key to simulate astrophysical
systems on spatial scales that are beyond the reach of pure PIC simulations.
They consider non-relativistic super-Alf\'{e}nic shocks with various magnetic field obliquities.
They recover all the features from previous studies when the magnetic field is parallel
to the shock front normal \citep{vanMarle2019}. In contrast to previous and hybrid
simulations, they find that particle acceleration and magnetic field amplification also
occur when the magnetic field is oblique to the shock front normal, but on larger
timescales than in the parallel case. They show that in their oblique shock simulations
the streaming of supra-thermal particles induces a corrugation of the shock front.
Such oscillations of both the shock front and the magnetic field then locally facilitate
the particles to enter the upstream region and to initiate a non-resonant streaming
instability finally inducing diffuse particle acceleration.}


Recently, \cite{drakefe19} \rv{have} developed a new computational model suitable
for exploring the self-consistent production of energetic electrons
during the magnetic reconnection in macroscale systems. These equations are
based on the recent discovery that parallel electric fields are
ineffective drivers of energetic particles during magnetic reconnection such that
the kinetic scales controlling the development of such fields can be
ordered out of the equations. The resulting equations consist of 
a MHD background with the energetic component
represented by macro-particles described by the guiding center
equations. Crucially, the energetic component feeds back on the MHD
equations, thus the total energy of the MHD fluid and of the energetic
particles is conserved. These equations correctly describe the firehose
instability, whose dynamics plays a key role in throttling magnetic reconnection
and in controlling the spectra of energetic particles. 

They describe a system with three distinct classes of particles: ions of
density $n$ and temperature $T_{\rm i}$, cold electrons with density
$n_{\rm c}$ and temperature $T_{\rm c}$ and energetic electrons with
density $n_{\rm h} = n - n_{\rm c}$ and temperature $T_{\rm eh}$. The hot electrons 
\rv{are} treated as macro-particles that evolve through the MHD grid by the guiding
center equations. Since this new method is an advanced two-fluid MHD
simulation, some of the fundamental properties of PIC method are
missing. This evidence of lacking PIC properties is found in the equation
\begin{eqnarray}
E_{\|} = -\frac{1}{ne}\left(\vec{b}\cdot \nabla P_{\rm ec} + \vec{B}\cdot \nabla
\frac{m_{\rm e}n_{\rm ec}v^{2}_{\rm ce\parallel}}{B}
+\vec{b}\cdot(\nabla \cdot {\mathbb T}_{\rm eh})\right),
\label{eparallel}
\end{eqnarray}
of the parallel electric field $E_{\|}$ where $P_{ec}$ is the pressure of the cold electrons. 
In their model the Debye length is ordered out, hence the system must
remain charge neutral. The ion density is calculated with a standard
continuity equation with a velocity given by the MHD momentum equation.
The energetic electron density is calculated by mapping the energetic
electrons onto the MHD grid with an appropriate interpolation scheme.
The cold electron density is then calculated by requiring that the sum
of the cold and hot electron densities matches that of the ions. The
physics leading to the charge neutrality is the strong parallel motion of
the cold electrons filling in for the hot electrons motion along 
the ambient magnetic field. 


\rv{In conclusion,} phenomena such as the electron
acceleration and associated radiation can be investigated with the use of PIC methods. 
In PIC methods, each particle is traced self-consistently using Maxwell's equations as described in Sect.~\ref{sec:3}. With this
method one can obtain all particle information and in principle also
macroscopic properties if the size of simulations is large
enough. However, in this method, one needs to resolve the spatial scale of the electron 
and ion Debye length, electron and ion skin depth and
electron cyclotron period. With the current computational constraints,
it is a daunting task to simulate a reasonable physical size in a 3D system for 
the direct comparison with fluid model. Fortunately, due to 
the great advance of computing power this limitation is becoming less serious. 


\section{Kinetic physics in astrophysical systems}
\label{sec:5}



\rv{In this section, we describe various applications of PIC codes for astrophysical plasmas, in particular for relativistic jets.}

\rv{Relativistic} jets are collimated plasma outflows associated with active
galactic nuclei (AGN), gamma-ray bursts (GRBs), and pulsars
\citep[e.g.,][]{hawley15,Cerutti2017,Blandford2019}. Amongst these astrophysical systems,
blazars and GRB jets produce the most luminous phenomena in the universe
\citep[e.g.,][]{peer14}. Despite extensive observational and theoretical
investigations, including simulation studies, our understanding of
their formations, interaction and evolution in an ambient plasma, and
consequently their observable properties, such as the time-dependent flux
and polarity \citep[e.g.,][]{nick18}, remain quite limited. 

The morphology of relativistic jets is very large and the macroscopic
views of jets are described well by RMHD simulations \citep[e.g.,][]{Marti2019}. 
However, these simulations cannot include the dynamics of particles, thus their 
acceleration in jets cannot be investigated. Therefore, PIC simulations
play an important role for the particle acceleration and the radiation from
accelerated particles in magnetic fields.

To date, models like shock-in-jet have failed to explain the extremely
rapid gamma-ray flares with a relatively harder spectrum (spectral
indices $<$1.5) \citep[e.g.,][]{sari99}. One of the key open questions in the study of 
relativistic jets is how they interact with the immediate plasma
environment on the microscopic scale. In response to this question,  
\citet{nishikawa16b,nishikawa19gal,Nishikawa2020} aim at examining
how relativistic jets containing helical
magnetic fields evolve under the influence of kinetic and MHD-like
instabilities that occur within and at the jet boundaries. They
conducted their examination with regards to consequences such as flares due
to magnetic reconnection besides other mechanisms such as recollimation shocks
and moving internal shocks \citep[e.g.,][]{zaman14}. 
Therefore, magnetic reconnection which takes place in a short time and
accelerates particles, could be a viable option.

Jet outflows are commonly thought to be dynamically hot (relativistic)
magnetized plasma flows that are launched, accelerated, and collimated
in regions where the Poynting flux dominates over the particle (matter) 
flux \citep[e.g.,][]{blandford1977,aloy00,oliber17}. This scenario involves
a helical large-scale magnetic field structure in some AGN jets
providing a unique signature in the form of observed asymmetries across
the jet width, particularly in the polarization
\citep[e.g.,][]{liang81,aloy00,clausen11,sasha15}.


Large-scale, ordered magnetic fields have been invoked to explain the
launching, acceleration, and collimation of relativistic jets from the
central nuclear region of an active galaxy \citep[e.g.,][]{meier08},
and coalescing and merging stars (neutron star and black hole)
\citep[e.g.,][]{piran05}. The magnetic field structure and particle
composition of the jets are still not well constrained observationally.

The circular polarization (CP; measured as Stokes parameter V) in the radio
continuum emission from AGN jets provides a powerful diagnostic for
deducing magnetic structure and particle composition because, 
unlike linear polarization (LP), CP is expected to remain almost
completely unmodified by external screens  
\citep[e.g.,][]{O'Sullivan13,nick18}. 


The differences in the global jet evolution among different particle species, as
described later, may be compared with observations of jet images in
order to test the jet composition. PIC studies of the jet evolution with
helical magnetic fields using a large simulation system may provide morphological
structures of jet evolution and some possible identification of species of jets. 

In this review, we \rv{also} discuss the progresses made in developing state-of-the-art PIC simulations 
of relativistic jets used to investigate the generation of rapid flares of very high energy 
(in particular, in $\gamma$- and X-ray bands), the composition of jet plasma, and
the polarized radiation observed in blazars and GRB jets, accounting
for microscopic and macroscopic processes with helical magnetic
fields. In particular, formation of jets-in-jet due to magnetic reconnection may provide 
a possible mechanism for production of flares and their rapid variability, with timescales 
as short as hours or even minutes \cite[e.g.,][]{Aharonian06,Aharonian07}. 

\subsection{PIC simulations of beam-induced instabilities}
\label{sec:5.1}

PIC simulations can shed light on the microphysics within relativistic shocks. \rv{Such} simulations have shown that particle acceleration occurs within downstream jets
\citep[e.g.,][]{silva03,nishikawa03,frederiksen04,heledal04,hededakk05,nishikawa05,jaroschek05,nishikawa06a,nishikawa06,nishikawa08,spit08a,spit08b,chang08,dieckmann08,nishikawa09,martins09,nishikawa11ad}.
In general, these simulations confirm that a relativistic shock in a weakly or non-magnetized plasma is dominated by {the Weibel instability
\citep{Fried1959,weibel59,Bret2009}.} The associated current filaments and magnetic fields \citep[e.g.,][]{medvedev99} accelerate electrons \citep[e.g.,][]{nishikawa06} and cosmic-rays, subsequently affecting the pre-shock medium \citep{medvedev09}. 

\subsubsection{Weibel instabilities with various flavours}
\label{sec:5.1.1}

The Weibel instability has been investigated \rv{with}
different flavours, such as counter-streaming \citep[e.g.,][]{Takamoto2019,kumar2020} 
and bump-in-tail instabilities \cite[e.g.,][]{Ziebell2012,Thurgood2015}.

\citet{kumar2020} have performed 2D PIC simulations of counter-streaming, unmagnetized electron-positron plasmas. They discovered that, if the electron-positron plasma is distributed homogeneously, the magnetic field, produced by a Weibel instability, amplifies exponentially in the linear regime and rapidly decays after saturation. However, for inhomogeneous
electron-positron plasmas, as they might originate from supernova remnants and GRBs, the magnetic field amplifies again after saturation where such amplification depends on the degree of inhomogeneity. This re-amplification is caused by the thermal anisotropy created by the inhomogeneous plasma distribution. 

Similarly, \citet{Takamoto2019} performed 3D simulations of counter-streaming electron and ions beams leading to a relativistic ion-electron Weibel instability. In their simulations, they compare the growth of the Weibel instability with the evolution of a MHD kink instability. They found that the growth time of the kink instability is larger than the evolution time of the Weibel instability. Hence, the Weibel instability is capable of producing magnetic fields, possibly seeding large-scale MHD dynamos.

\citet{Thurgood2015} have performed the bump-in-tail instability in order to understand the generation mechanism of Type III solar burst for plasma radio emission.
In first case using a more tenuous and fast beam, they found evidence consistent with all stages of the three-wave based fundamental and harmonic emission mechanisms, which including the beam-mode to Langmuir mode coupling, the growth of a population of counter-propagating Langmuir/electrostatic waves via back scattering and decay processes, the action of fundamental emission, and the coalescence of the counter-propagating population to produce harmonic emission. 
Using a fully kinetic and electromagnetic PIC simulation, they confirm the role of all such stages whilst taking care to distinguish signals above the inherent noise levels associated with particle methods.
Thus, first case is arguably the first unambiguous confirmation of the three-wave based emission processes resulting from a single electron beam in the literature using the fully kinetic PIC approach.

For second case using a slower and denser beam, they demonstrated the sensitivity of the parameter space by considering a more dense beam with the density of beam/density of ambient plasma, $n_{\rm b}/n_0$, a similar density to past works, and found that the processes are significantly suppressed due to the resulting non-Langmuir characteristics of the beam mode. 
Whilst a full parameter study may prove useful, but is beyond available computational resources, they hope that second case demonstrates plainly that caution must be applied when attempting to simulate astrophysical beam-plasma systems using unrealistically dense beams. 
Whilst a larger density ratio reduces relaxation times (i.e. computer time), the results are unlikely to be physically representative of the intended system due to the sensitivity to beam parameters.

Recently, as for an astrophysical system, \citet{Vafin2018} have investigated the electrostatic instability for blazar-induced pair beams propagating through the intergalactic medium (IGM) using linear analysis and PIC simulations. 
To study the nonlinear beam relaxation, they performed PIC simulations that take into account a realistic wide-energy distribution of beam particles. They investigated the growth rate of the instability for arbitrary wave vectors considering, in particular, the effect of the transverse beam temperature (realistic beams have $\rm{rms}(p_{\perp}) \approx m_{\rm e}c/2$). The parameters of the simulated beam plasma system provide an adequate physical picture that can be extrapolated to realistic blazar-induced pairs. 
In their simulations, the beam looses only 1\% of its energy, and they analytically estimate that the beam would lose its total energy over about 100 simulation times. 
If the beam had no angular spread, then the growth rate would reach its maximum at wave vectors perpendicular to the beam. However, for a realistic finite angular spread, the growth rate is the largest at wave vectors quasi-parallel to the beam direction, and the maximum growth rate is reduced, in agreement with \citet{Miniati_2013}, but it is still by more than a factor of 10 larger than the peak growth rate of the strictly parallel electrostatic mode studied by \citet{Schlickeiser_2013}.
An analytical scaling is then used to extrapolate the parameters of realistic blazar-induced pair beams. They found that they can dissipate their energy slightly faster by the electrostatic instability than through inverse-Compton scattering. 
The uncertainties arising from, e.g., details of the primary gamma-ray spectrum are too large to make firm statements for individual blazars, and an analysis based on their specific properties is required.

They studied the impact of the realistic distribution function of pairs resulting from the interaction of high-energy gamma-rays with the extragalactic background light. 
They have presented analytical and numerical calculations of the linear growth rate of the instability for the arbitrary orientation of wave vectors. Their results explicitly
demonstrate that the finite angular spread of the beam dramatically affects the growth rate of the waves, leading to the fastest growth for wave vectors quasi-parallel to the beam direction and a growth rate at oblique directions that is only a factor of 2-4 smaller compared to the maximum.


In all these simulations no shock has been found comparing with reflected and injected jet simulations. More advanced kinetic instabilities are found in magnetorotational instability 
(MRI) \cite[e.g.,][]{Hoshino2015PhRvL,Kunz2016,Hirabayashi2017,Kimura2019}.

\cite{Hirabayashi2017} show a series of stratified-shearing-box simulations of collisionless accretion disks in the recently developed framework of kinetic MHD, which can handle finite non-gyrotropy of a pressure tensor. Although a fully kinetic simulation predicted a more efficient angular-momentum transport in collisionless disks than in the standard MHD regime, the enhanced transport has not been observed in past kinetic-MHD approaches to gyrotropic pressure anisotropy. For the purpose of investigating this missing link between the fully kinetic and MHD treatments, they have explored the role of non-gyrotropic pressure and makes the first attempt to incorporate certain collisionless effects into disk-scale, stratified disk simulations. 
When the timescale of gyrotropization was longer than, or comparable to, the disk-rotation frequency of the orbit, they found that the finite non-gyrotropy selectively remaining in the vicinity of current sheets contributes to suppressing magnetic reconnection in the shearing-box system. 
This leads to increase both in the saturated amplitude of the MHD turbulence driven by MRI and in the resultant efficiency of angular-momentum transport.
Their results seem to favor the fast advection of magnetic fields toward the rotation axis of a central object, which is required to launch an ultra-relativistic jet from a black hole accretion system in, for example, a magnetically arrested disk state.

The theory and modelling of the dynamo effect in turbulent fluids and plasmas have been reviewed by \citet{Rincon2019}. 
The primary focus is on the physical and mathematical concepts underlying different (turbulent) branches of dynamo theory, with some astrophysical, geophysical and experimental contexts disseminated throughout the document. These sections are complemented by an overview of a selection of current active research topics in the field, including the numerical modelling of the geo- and solar dynamos, shear dynamos driven by turbulence with zero net helicity and MHD-instability-driven
dynamos such as the magnetorotational dynamo. Turbulence in collisionless plasma is described in Sect.\ref{sec:5.4.2}.

For the thin-shell-instability \citep[e.g.,][]{Dieckmann2017}, we refer the reader to section \ref{sec:5.2.1}.
%
%
The PIC simulations of kink instability is described separately at the end of  Sect.~\ref{sec:5.4} \id{\citep[e.g.,][]{Davelaar2020}}.

In recent work, in order to generate a shock, a relativistic plasma stream is injected from one end of the computational domain and reflected from a rigid wall at the opposite end, for instance as in a 1D simulation by \cite{hoshino02,amano07}, in 2D simulations by \cite{amano09,spit08a,spit08b,martins09}, or in a 3D simulation by \cite{guo14}. 
This method resembles the collision of two identical counter-streaming beams and reduces the number of calculations by one half, which is used to investigate the particle acceleration at termination shocks historically. 
This approach simulates only a forward moving shock (FS). In these settings the backward (reverse) shock is degenerated to an FS. Recently, \citet{Vanthieghem2020} reviewed physics and phenomenology of weakly magnetized, relativistic astrophysical shock waves in the reflected scheme.

In the work by \citet{nishikawa09} a particle jet is injected into an ambient plasma without the reflection at the opposite boundary, such that a double shock structure (forward and reverse shock) is fully captured. 
With this setup the jet-to-ambient density ratio can be changed and, by assuming a density ratio greater than one, the shock formation process can be properly handled by smaller scale (temporal and spatial) simulations. 
In such a scenario, the deceleration of the jet flow by the ambient plasma results in a contact discontinuity (CD) being the location where the electromagnetic field and the velocity of the jet and ambient plasmas are similar and the density changes. 
Two shocks propagate away from the CD into the jet and ambient upstreams in the CD frame \citep{nishikawa09,choi14,ardaneh15,ardaneh16}. 
\citet{ardaneh15} have illustrated that forward and reverse shocks and one CD split up the jet and ambient plasma, 
and the structures of these shocks depend on the density ratio of the jet and ambient densities.

Therefore, this injection scheme is more universal and used with a more advanced setting as described later in order to investigate the propagation of jets including the instabilities generated in the velocity-shear regions. 

Next, we present the three-dimensional simulation results for
an electron-positron jet injected into an electron-positron plasma using a long simulation grid \citep{nishikawa09}. A leading and trailing shock system develops with strong electromagnetic fields accompanying the trailing shock.

\subsubsection{Injection scheme}
\label{sec:5.1.2}

\rv{To setup a slab model for PIC simulations of relativistic jets, \citet{nishikawa09} use \TRISTANMPI, an MPI-based parallel version of the relativistic electromagnetic particle code \TRISTAN \citep{buneman93,nishikawa03,niemiec08}.} \rv{Simulations} have been performed using a grid with $(L_{\rm x},L_{\rm y},L_{\rm z}) = (4005, 131, 131)$ cells and a total of $\sim 1$  billion particles (12 particles/cell/species for the ambient plasma) in the active grid. The electron skin depth, 
$\lambda_{\rm s} = c/\omega_{\rm pe} = 10.0\Delta$,
where $\omega_{\rm pe} = (e^{2}n_{\rm a}/\epsilon_{0}m_{\rm e})^{1/2}$ is the electron plasma frequency, 
and the electron Debye length $\lambda_{\rm D}$ is half of the cell size $\Delta$, where $\Delta = 1$. 
This computational domain is six times longer than in their previous simulations \citep{nishikawa06,ramirez07}. 
The jet-electron number density in the simulation reference frame is 0.676$n_{\rm a}$, where $n_{\rm a}$ is the ambient electron density, and the jet Lorentz factor is $\gamma_{\rm j} = 15$. The jet-electron/positron thermal velocity is $v_{\rm j,th} = 0.014 c$ in the jet reference frame. 
The electron/positron thermal velocity in the ambient plasma is $v_{\rm a,th} = 0.05 c$. Similar to the previous work done by \citet{nishikawa06}, the jet is injected in a plane across the
computational grid located at $x = 25\Delta$ in order to eliminate effects associated with the boundary at $x = x_{\min}$. Radiating boundary conditions are used on the planes at $x = x_{\min}$ and $x = x_{\max}$ and periodic boundary conditions on all transverse boundaries \citep{buneman93}. 
The jet makes contact with the ambient plasma at a two-dimensional interface spanning the computational domain.
Here the formation and dynamics of a small portion of a much larger shock are studied in a spatial and temporal way that includes the spatial development of nonlinear saturation and the dissipation from the injection point to the jet front defined by the fastest moving jet particles.

\begin{figure}[htb]
\centering
\includegraphics[width=0.8\textwidth]{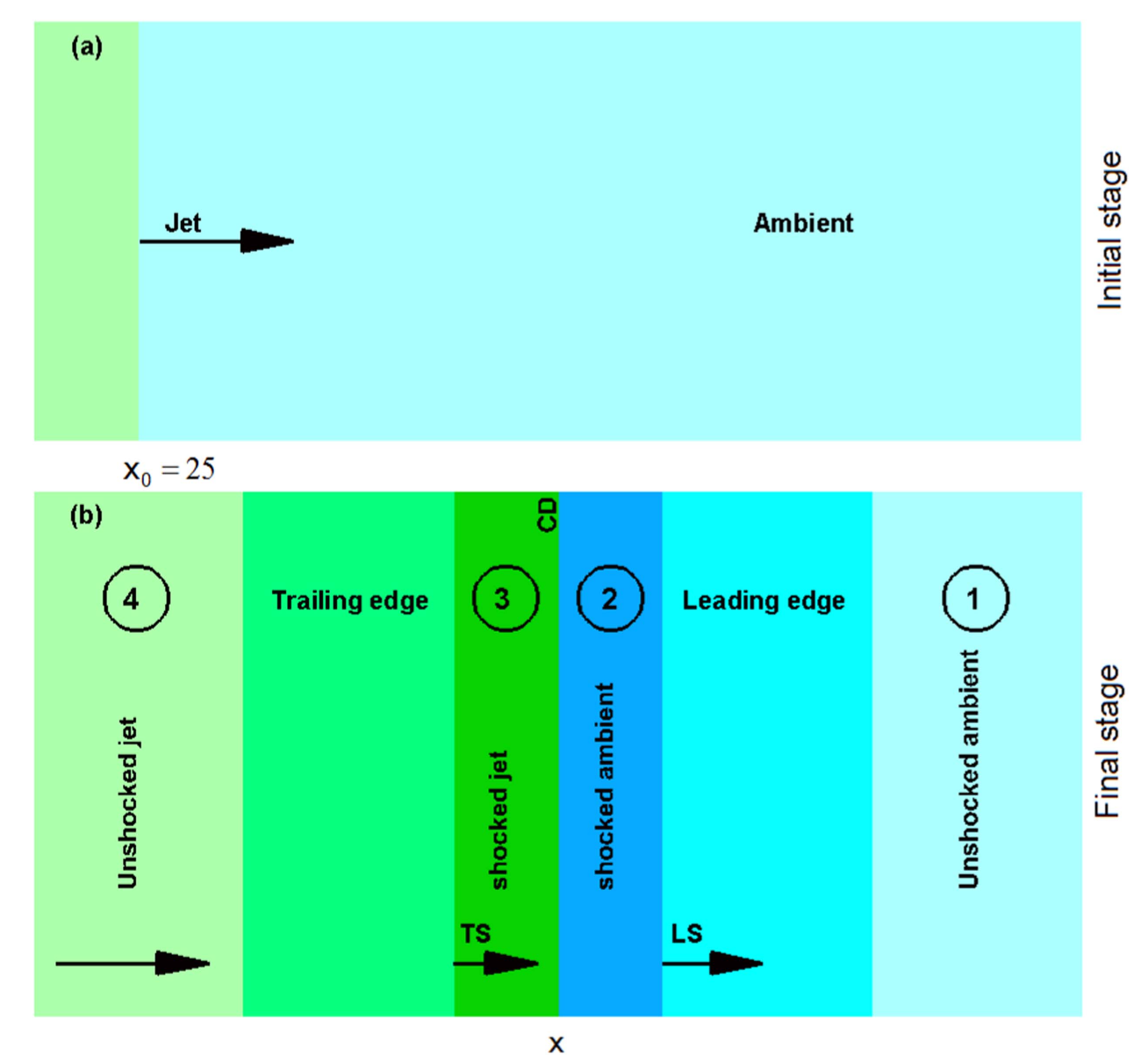} 
\caption{Illustration of the jet--ambient interaction showing (a) a particle jet being injected into an ambient plasma and (b) the resulting double shock structure. The shocks are named according to \citet{nishikawa09}. Image reproduced with permission from \citet{ardaneh16}, copyright by AAS.}
\label{ardaneh16fig1}
\end{figure}

In this simulation a  relativistic jet is injected into an ambient plasma as shown in Fig.~\ref{ardaneh16fig1}a, therefore the shock structures are different from those that are obtained when using numerical schemes with reflecting boundary conditions \citep[e.g.,][]{Vanthieghem2020}.

Figure \ref{ardaneh16fig1} shows (a) the initial and (b) the final stages of plasma evolution, where a double shock structure is formed resembling what is schematically illustrated. 
The deceleration of the jet stream by magnetic fluctuations (excited in the beam--plasma interactions) results in a CD and two shock waves that divide the jet and ambient plasmas into four regions: (1) unshocked ambient, (2) shocked ambient, (3) shocked jet, and (4) unshocked jet. 
These subscripts are not used further on.

\begin{figure}[htb]
\includegraphics[scale=0.76]{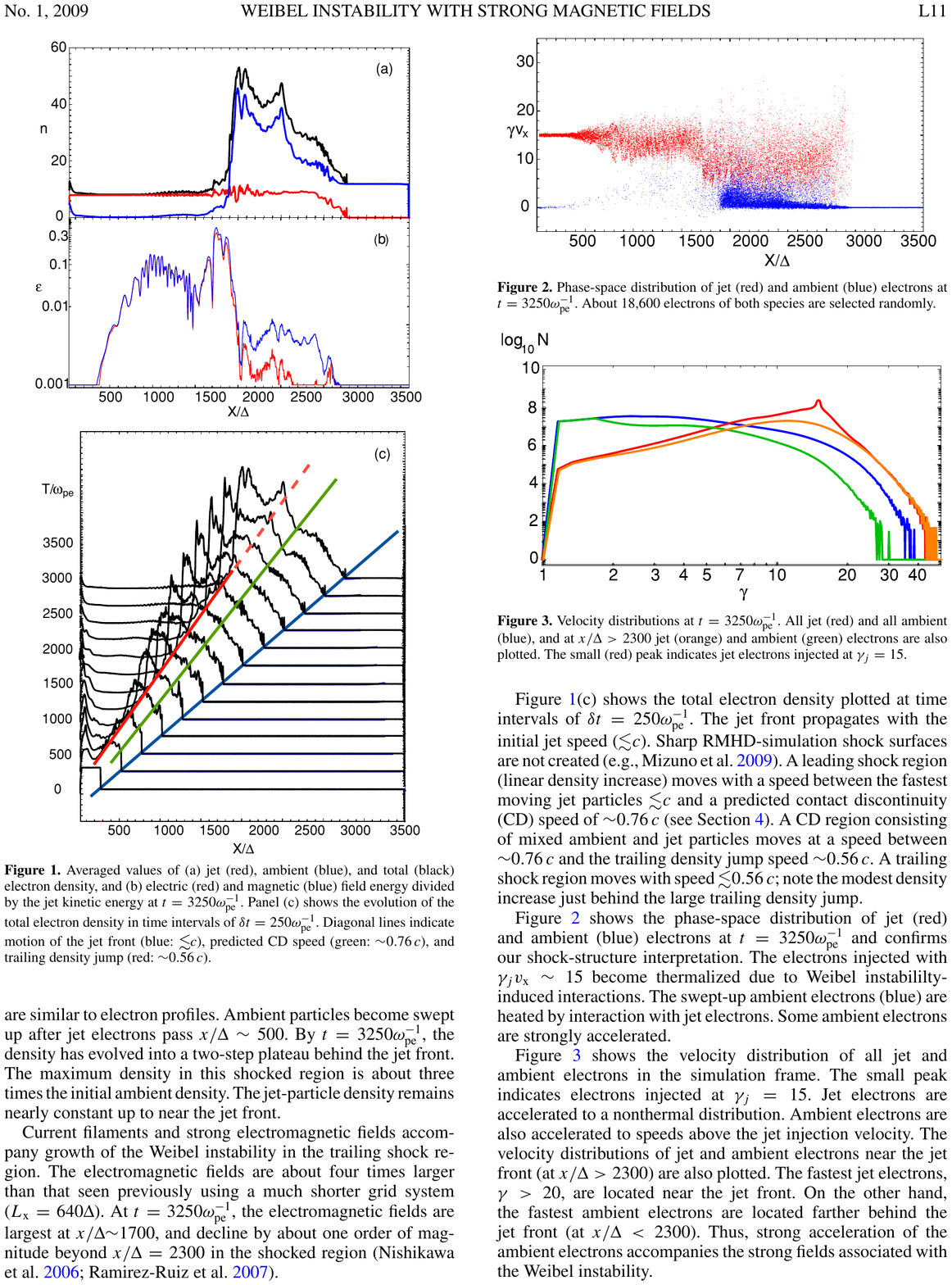}
\includegraphics[scale=0.66]{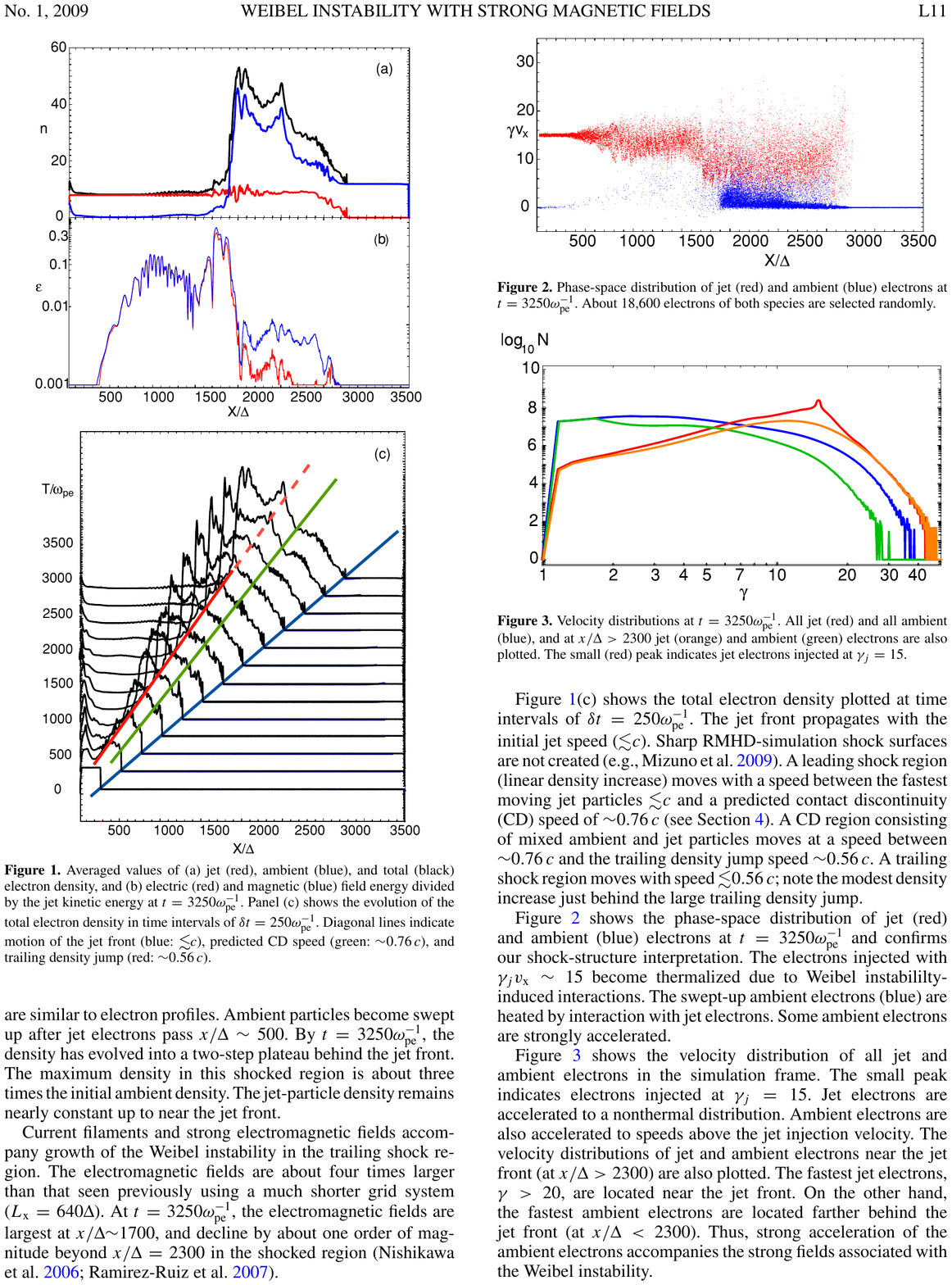}
\caption{Averaged values of (a) jet (red), ambient (blue), and total (black) electron density, and (b) electric (red) and magnetic (blue) field energy divided by the jet kinetic energy at 
$t = 3250\,\omega^{-1}_{pe}$. Panel (c) shows the evolution of the electrons in time intervals of $\delta t = 250\omega^{-1}_{\rm pe}$. Adapted from Fig.~1 in \citet{nishikawa09}.}
\label{nishi09fig1}
\end{figure}

Figure~\ref{nishi09fig1} shows the averaged (a) jet (red), ambient (blue), and total 
(black) electron density and (b) the electromagnetic field energy divided by the total jet kinetic energy ($E^{\rm j}_{\rm t} = \sum_{\rm i =e,p}m_{\rm i}c^{2}(\gamma_{\rm j} - 1))$ in the $y-z$ plane after $t = 3250\,\omega^{-1}_{pe}$
where ``e'' and ``p'' denote electrons and positrons. Positron density profiles are similar to the ones for the electron density. Ambient particles become swept
up after jet electrons pass by $x/\Delta \sim 500$. By $t = 3250\,\omega^{-1}_{pe}$, the density has evolved into a two-step plateau behind the jet front. The maximum density in the shocked region is about three times higher than the initial ambient density. The jet-particle density remains nearly constant up to near the jet front.

Current filaments and strong electromagnetic fields accompany the growth of the Weibel instability in the trailing shock region.
The electromagnetic fields are about four times larger than that seen previously using a much shorter grid system
($L_{\rm x} = 640\Delta$). At $t = 3250\,\omega^{-1}_{pe}$, the electromagnetic fields are largest at $x/\Delta \sim 1700$, and decline by about one order of magnitude beyond $x/\Delta = 2300$ in the shocked
region \citep{nishikawa06,ramirez07}.

Figure~\ref{nishi09fig1}c shows the time evolution of the electron plotted at time intervals of $\delta t = 250\omega^{-1}_{\rm pe}$. The jet front propagates with the initial jet speed ($\le c$). Sharp shock surfaces seen in RMHD simulations are not created\citep[e.g.,][]{mizuno09}. A leading shock region (for which the linear density increase) moves with a speed between the fastest
moving jet particles $\le c$ and a predicted contact discontinuity (CD) speed of $\sim 0.76 c$ \citep{nishikawa09}. 
The CD region consisting of mixed ambient and jet particles moves at a speed between
$\sim 0.76 c$ and the trailing density jump speed $\sim 0.56 c$. 
A trailing shock region moves with $\sim 0.56 c$; note the modest density increase just behind the large trailing density jump.

\begin{figure}[htb]
\centering
\includegraphics[scale=0.88]{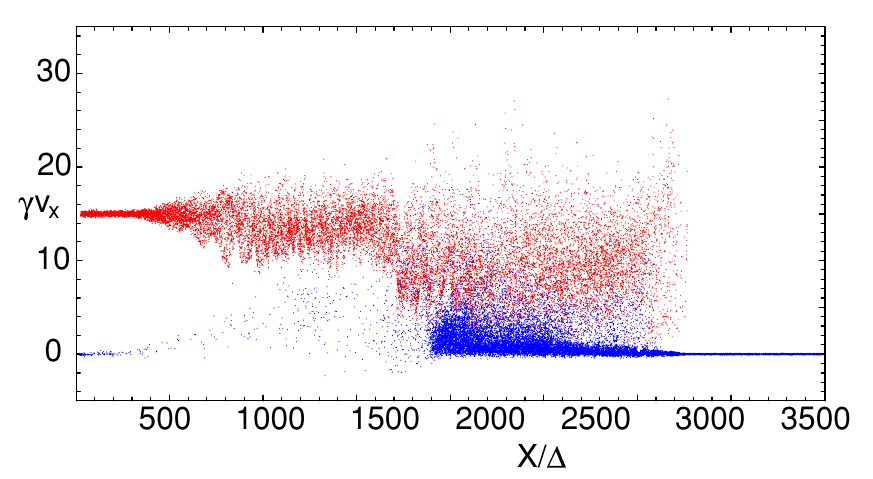}
\caption{Phase-space distribution of jet (red) and ambient (blue) electrons at $t = 3250\,\omega^{-1}_{pe}$. About 18,600 electrons of both media are selected randomly {\citep{nishikawa09}}.}
\label{nishi09fig2}
\end{figure}

Figure \ref{nishi09fig2} shows the phase-space distribution of jet (red) and ambient (blue) electrons at $t = 3250\,\omega^{-1}_{pe}$ and confirms the shock-structure interpretation exposed in the paragraph above. Some jet electrons injected with $\gamma_{\rm } v_{\rm x} \sim 15$ become thermalized due to the Weibel instability induced interactions whilst some swept-up ambient electrons (blue) are strongly accelerated and heated by the interaction with jet electrons. 

\subsubsection{3D quasi-perpendicular shock simulations}
\label{sec:5.1.3}

\begin{figure}[htb]
  \centering
\includegraphics[width=0.8\textwidth]{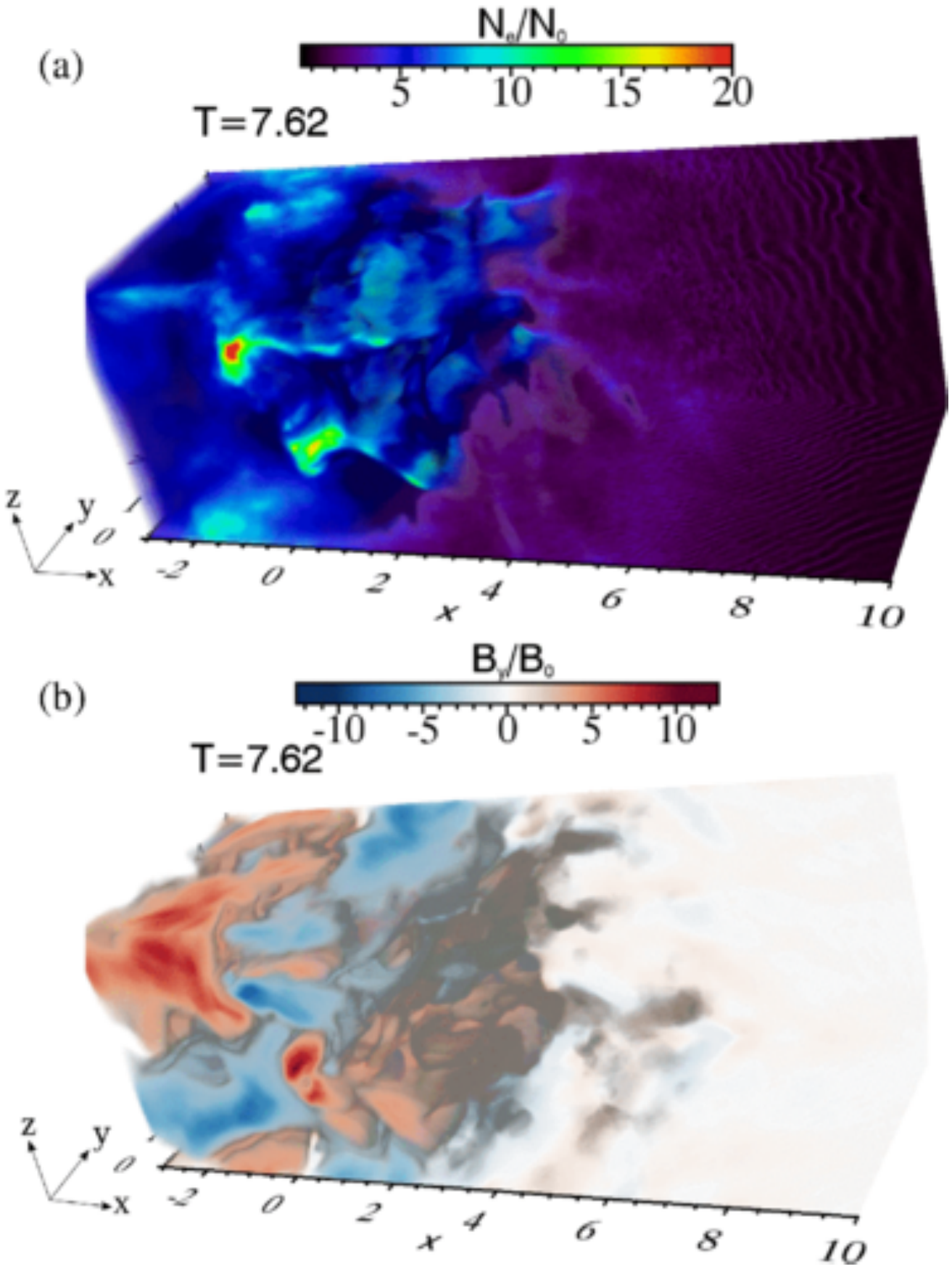} 
\caption{The 3D structure around the shock front ($x=0$) obtained at time $T=7.62\Omega_{\rm gi}^{-1}$ from the initiation of the experiment
($\Omega_{\rm gi}$: ion gyrofrquency).
The structures of (a) the electron density and (b) the $y$-component of the magnetic field are visualized by a volume rendering technique with cross-sectional profiles in the $x-y$ ($z=0$) and $x-z$ ($y=L_{\rm y}$) planes. The quantities and the spatial scale were normalized to the upstream values and upstream ion inertia length, respectively. Videos of time evolution corresponding to (a) and (b) are provided as supplemental material at
\url{http://link.aps.org/supplemental/10.1103/PhysRevLett.119.105101}. Image reproduced with permission from \citet{matsumoto17}, copyright by APS.).
}
\label{matsumoto17fig1}
\end{figure}
3D PIC simulations of a high Mach-number quasi-perpendicular shock have been performed by \citet{matsumoto17}. They found an extended electron acceleration through the investigation of the electron motion in a fully self-consistent fashion that included all of the essential ingredients: shock drift acceleration (SDA), shock surfing acceleration (SSA), and a strong Weibel magnetic turbulence.

Coherent electrostatic Buneman waves and ion-Weibel magnetic turbulence coexist in a strong-shock structure whereby particles gain energy during shock surfing and subsequent stochastic drift accelerations \citep[see][]{nishikawa09}. 
Energetic electrons that initially experienced the surfing acceleration undergo pitch-angle diffusion by interacting with magnetic turbulence and continuous acceleration during the confinement in the shock transition region. The ion-Weibel turbulence is the key to the efficient non-thermal electron acceleration.

Figure~\ref{matsumoto17fig1} shows a 3D shock structure in a fully developed stage after its initiation. Electron-scale coherent structures were found to persist during the simulation, as can be seen from the stripes of electron density at the leading edge of the shock ($8 < x < 10$) in Fig.~\ref{matsumoto17fig1}a. 
The shock transition (foot) region ($0 < x < 6$) is dominated by the ion-Weibel instability because of the interaction between the upstream and reflected ions, resulting in rib structures
(Fig.~\ref{matsumoto17fig1}a) and strong magnetic turbulence (Fig.~\ref{matsumoto17fig1}b). 
The $y$-component of the magnetic field is a component newly generated by the instability and is further amplified up to 20 times the upstream value by the shock compression, similar to the other $x$- and $z$-components.

These acceleration mechanisms through SDA and SSA in strong magnetic turbulence are considered to be important in relativistic jets as will be
discussed later in Sect.~\ref{sec:5.4.1}.


\subsection{PIC simulations of velocity-share instabilities}
\label{sec:5.2}

\subsubsection{With slab model setup}
\label{sec:5.2.1}

In addition to producing shocks, outflow interactions with an ambient medium include velocity shears. In particular, the Kelvin--Helmholtz instability (KHI) has been investigated on the macroscopic scale as a mean to generate magnetic fields in the presence of strong
relativistic velocity shears in AGNs and GRB jets
\citep[e.g.,][]{d'amgelo65,gruzino08,mizuno07,perucho08,zhang09}. 
Recently, PIC simulations have been employed to study the magnetic field generation and the particle acceleration in velocity shears at the
microscopic scale using counter-streaming setups. Here, the shear interactions are associated with the kinetic Kelvin--Helmholtz instability (kKHI), also referred to as the electron-scale Kelvin--Helmholtz instability \citep[ESKHI; e.g.,][]{alves12,nishikawa13,liang13a,grysmayer13a,grysmayer13b,liang13b,alves14,liang17,liang18}. 

\citet{alves12} found that alternating current, hereafter AC, and magnetic field modulations seen in the non-relativistic regime are less noticeable in the relativistic regime because they are masked by a strong and relatively steady direct current, hereafter DC, and an associated magnetic field. 
As the amplitude of the kKHI modulations grows, the electrons from one flow cross the shear surface and enter the counter-streaming flow. In their simulations, the heavier protons remain unperturbed and DC sheets pointing in the direction of the proton velocity form
around the shear surface. These DC sheets induce a DC component in the magnetic field which is dominant in the relativistic scenario because a higher direct current is set up by the deceleration of electrons relative to the protons and also because the growth rate of the AC dynamics is lowered by $\gamma^{3/2}_{0}$
compared to the non-relativistic case. 
It is important to note here, that the DC magnetic field is not captured in MHD simulations
\citep[e.g.,][]{zhang09} or in fluid theories because it results from intrinsically kinetic phenomena whilst such currents are not seen in single fluid MHD.

To date, kKHI and mushroom instability (MI) simulations have been performed in slab
\citep{alves12,nishikawa13,alves14,nishikawa14b,alves15,liang13a,liang13b} and in cylindrical geometries \citep{alves10,nishikawa14c},
but on periodic grids with no shock system. 
Full scale shock simulations have not incorporated velocity shear interaction with the ambient plasma (interstellar medium)
\citep[e.g.,][]{nishikawa09,choi14,ardaneh15,ardaneh16},
and to date only global simulations using a very small simulation system have been performed
\citep{nishikawa03,nishikawa06,nishikawa14c,ng06}. 
Simulations by \citet{ng06} pointed to the growth of kKHI, but a much larger simulation system is required to study properly both shocks and 
velocity shear instabilities (kKHI and MI)
\citep[e.g.,][]{nishikawa16a}. 

\cite{alves14} have extended the theoretical analysis and the simulations of the ESKHI to electron-ion plasmas with various density ratios  with a velocity shear gradient across the shear surface, and to warm as well as cold shear flows. 
For counter-streaming flows, they found that unequal densities lead to ``drift when the density symmetry is broken'', the most rapid growth occurs for equal densities, that a velocity shear gradient slows the growth rate. 
\citet{grysmayer13a,grysmayer13b} found a persistent equipartition of DC saturation magnetic field that ``persists longer than the proton timescale.'' 
A saturation electric field with
$E_{\rm sat} \sim \sqrt{\gamma_{0}}cm_{c}\omega_{\rm pe}/e$ 
(where $\omega_{\rm pe} \equiv \sqrt{n_{\rm e}e^{2}/\epsilon_{0}m_{\rm e}}$) results in a maximum electron energy gain of $\Delta E_{\max}
\sim E_{\rm sat}/(k_{\max}\Delta v) \propto \gamma_{0}^{4} m_{\rm e}c^{2}$, where $\Delta v = v_{\rm e} - v_{0}$ is the difference
between the accelerated electron speed and the flow speed and
$1/k_{\max} = \sqrt{8/3}\gamma^{3/2}_{0}c/\omega_{\rm pe}$.
In the next section, we will discuss 3D PIC simulations that were performed to investigate the cold kKHI using a relativistic jet core.

\subsubsection{With core-sheath jet setup}
\label{sec:5.2.2}

\begin{figure}[htb]
\includegraphics[width=\textwidth]{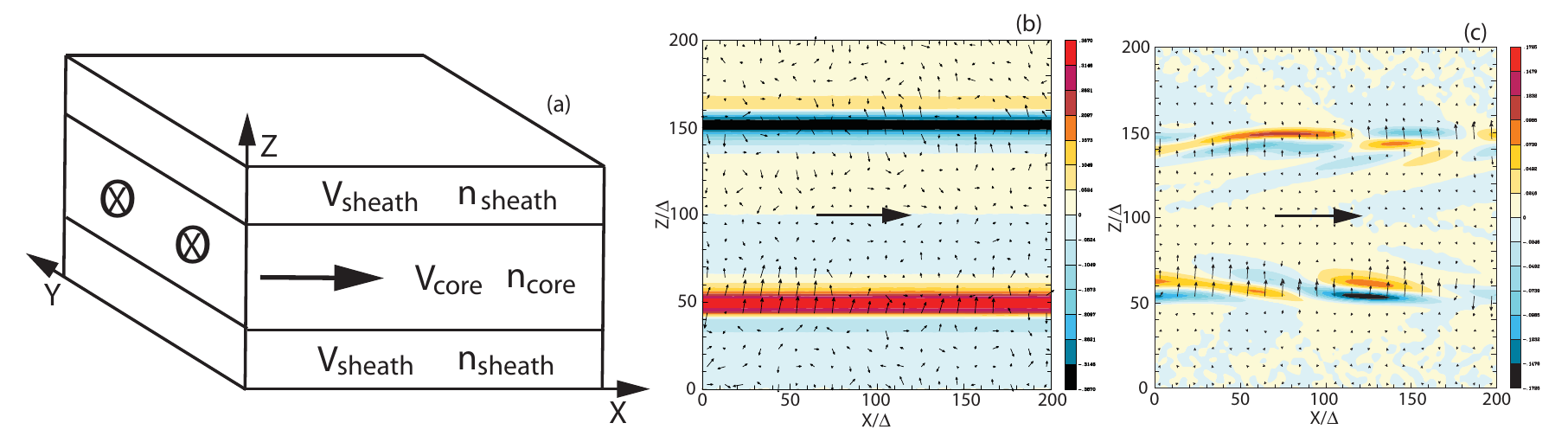}
\caption{Panel (a) shows the 3D simulation setup. Panels (b) and (c) show the magnetic field component $B{\rm y} > 0$ (red) and $B{\rm y} < 0$ (blue) plotted in the $x-z$ plane (jet flow indicated by large arrows) at the center of the simulation box, $y = 100\Delta$ at $t = 300 \omega_{\rm pe}^{-1}$, (b) for the $e^{-} - p^{+}$ case and (c) for the $e^{\pm}$ case, both with $\gamma_{\rm core} = 15$.
The smaller arrows indicate the magnetic field direction in the plane. Panels (b) and (c) cover one-fifth of the simulation system length in the $x$ direction. The maximum and minimum magnetic field strengths are $B_{\rm y} \sim$ (b) $\pm$0.367, and (c) $\pm$0.173. Adapted from \citet{nishikawa14b}.}
\label{kkhi}
\end{figure}

In the simulation study of \citet{nishikawa14b} a core-sheath plasma jet structure was employed instead of the counter-streaming plasma setup used in previous simulations \citep{alves12,alves14,grysmayer13a,grysmayer13b,liang13a,liang13b}. 
The basic setup and illustrative results are shown in Figure \ref{kkhi}.
In the setup, a jet core with velocity $\gamma_{\rm core}$ in the positive $x$ direction resides in the middle of the computational box. The upper and lower quarters of the numerical grid contain a sheath plasma that can be stationary or moving with velocity $v_{sheath}$ in the positive $x$ direction \citep{nishikawa13,nishikawa14b}. 
This model is similar to the setup used in the work by \citet{mizuno07} for RMHD simulations with a cylindrical jet core. 

However, \citet{nishikawa14b} represented the jet core and sheath as plasma slabs, where the system is initially charged and the current is neutral (jets). 
The simulations have been performed using a numerical grid with $(L_{\rm x},L_{\rm y},L_{\rm z}) = (1005\Delta, 205\Delta, 205\Delta)$ (simulation cell size:
$\Delta = 1$) and periodic boundary conditions in all dimensions.
The jet and sheath (electron) plasma number densities measured in the simulation frame are $n_{\rm jt} = n_{\rm am} = 8$. 
The electron skin depth, $\lambda_{\rm s} = c/\omega_{\rm pe} = 12.2\Delta$, where $\omega_{\rm pe} = (e^{2}n_{\rm am}/\epsilon_{0}m_{\rm e})^{1/2}$ is the electron plasma frequency and the electron Debye length for the ambient electrons $\lambda_{\rm D}$ is 1.2$\Delta$. 
The jet-electron thermal velocity is $v_{\rm jt,th,e} = 0.014c$ in the jet reference frame, where $c$ is the speed of light. 
The electron thermal velocity in the ambient plasma is $v_{\rm am,th,e} = 0.03c$, and ion thermal velocities is scaled down by a factor
$(m_{\rm i}/m_{\rm e})^{1/2}$. Simulations were performed using an electron-positron ($e^{\pm}$) plasma or an electron-proton $e^{-} - p^{+}$ plasma for jet Lorentz factors of 1.5, 5.0, and 15.0 with the sheath plasma at rest ($v_{\rm sheath} = 0$).

Figures~\ref{kkhi}b and \ref{kkhi}c show an illustration of the development of the velocity shear surfaces for $e^{-} - p^{+}$ and $e^{\pm}$ plasmas with $v_{\rm core} = 0.9978 c$ 
($\gamma_{\rm core} = 15$) at $y=100\Delta$ after $t=300 \omega_{pe}^{-1}$. 
For the $e^{-} - p^{+}$ case, a nearly DC magnetic field is generated at the shear surfaces perpendicular to the jets which shows the growing MI.
The $B_{\rm y}$ magnetic field component is generated with negative values (blue) at $z = 150\Delta$ and positive values (red) at 
$z = 50\Delta$. Additionally, a $B_{\rm z}$ (and $B_{\rm x}$) magnetic field component, shown by the small arrows in Figures~\ref{kkhi}b and \ref{kkhi}c, is generated at the shear surfaces by current filaments. 
Additionally, for the $e^{\pm}$ case, a relatively long wavelength ($\sim 100\Delta$) AC magnetic field is generated at the shear surfaces. Note the alternating $B_{\rm y} > 0$ (red) and $B_{\rm y} < 0$ (blue) in Figure~\ref{kkhi}c along the flow direction. Similar results have been obtained by \citet{liang13a,liang13b}, though the evolution of plasma shows some different patterns as their simulations model with counter-streaming flows in 2D.

The dominantly growing modes depend on the composition of the plasma and the jet Lorentz factor. In the $e^{-} - p^{+}$ setup
a DC magnetic field is generated in the shear plane, perpendicular to the relative velocity ($B_{\rm y}$ with $E_{\rm z}$), which show the MI 
is a dominant growing mode.
On the contrary, the $e^{\pm}$ cases generate AC electric and magnetic fields which shows that the kKHI is growing. In the $e^{\pm}$ cases, current filaments are generated similar to those found associated with the filamentation (Weibel-like) instability. 
In the simulations, the initial growth appears in the $E_{\rm z}$ electric field component perpendicular to the velocity shear surface. 
This growth is followed by the growth of the magnetic field component $B_{\rm y}$ in the velocity shear plane transverse to the flow direction in the $e^{-} - p^{+}$ case. 
For the $e^{\pm}$ case, the growth is seen in both the magnetic field components $B_{\rm y}$ and 
$B_{\rm z}$ as current filaments dominate the structure near the velocity shear surface.
In all cases, fluctuation structures along the jet are much longer than transverse fluctuation structures. In the $e^{-} - p^{+}$ plasma setup,
interactions and magnetic fields do not extend far from the initial velocity shear surface whilst interactions and magnetic fields extend farther from the initial velocity shear surface for $e^{\pm}$ jets, although they extend mostly into the jet side for higher jet Lorentz factor.

\begin{figure}[htb]
\includegraphics[width=\textwidth]{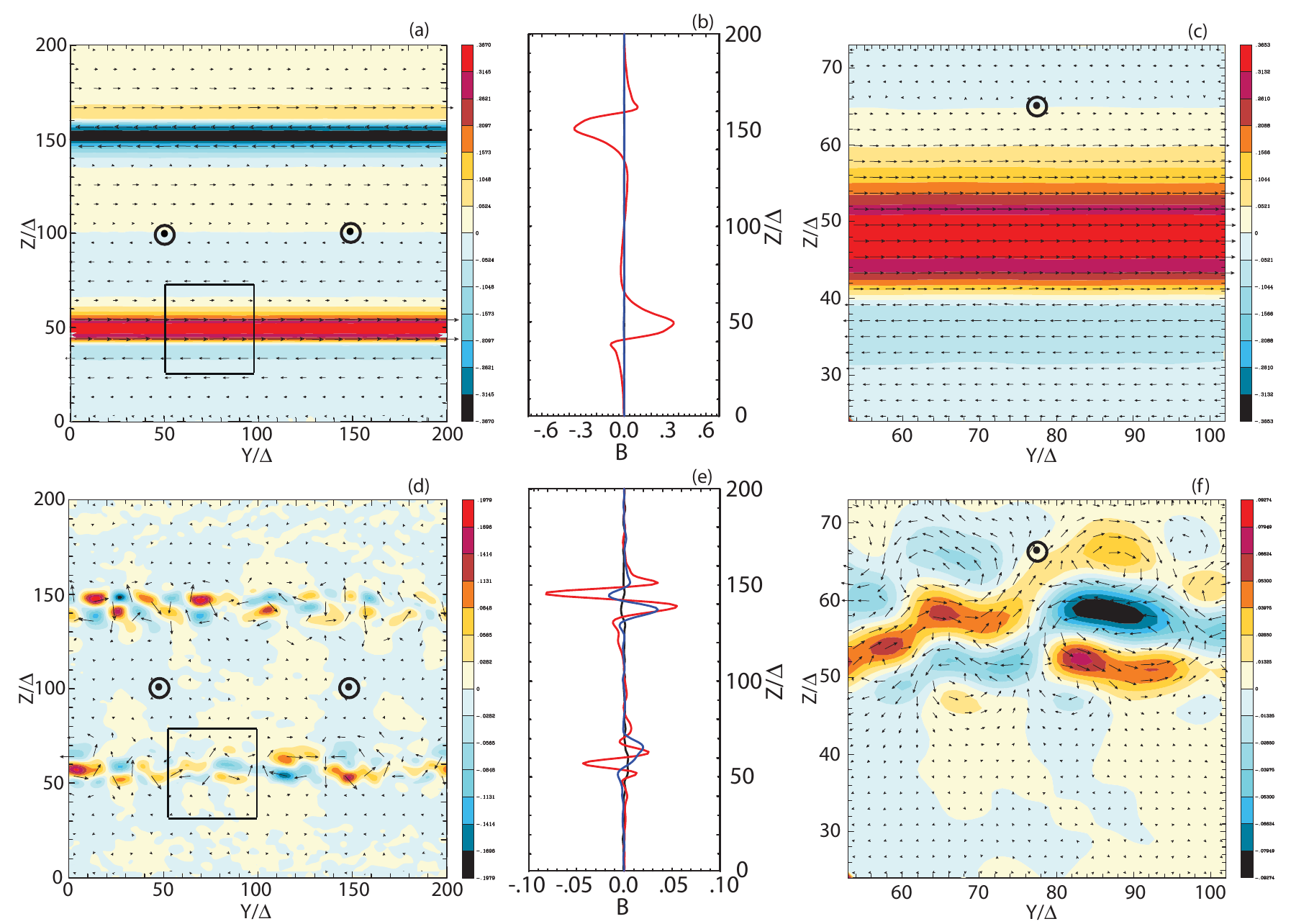}
\caption{a,d) Magnetic field structure $B_{\rm y}$ transverse to the flow direction for $\gamma_{\rm jt} = 15 $ is shown in the $y-z$ plane (jet flows out of the page) at the center of the simulation box, $x = 500\Delta$ for the $e^{-} - p^{+}$ case (upper row) and the $e^{\pm}$ case (lower row) after $t = 300 \omega_{\rm pe}^{-1}$. The small arrows show the magnetic field direction in the transverse plane (the arrow length is not scaled to the magnetic field strength). b,e) 1D cuts along the $z$ axis of magnetic field components $B_{\rm x}$ (black), $B_{\rm y}$ (red), and $B_{\rm z}$ (blue) are plotted at $x = 500\Delta$ and $y = 100\Delta$ for the $e^{-} - p^{+}$ case (b) and the $e^{\pm}$ case (e). Note that the magnetic field strength scales in panels (a) ($\pm 0.367$) and (d) ($\pm 0.198$) are different. c,f) Enlargement of the shear surface structure in the regions indicated by the squares in the left panels (a,d). Adapted from \citep{nishikawa14b}.}
\label{Nishikawa14Fig6}
\end{figure}

Figure~\ref{Nishikawa14Fig6} shows the structure of the $B_{\rm y}$ component of the magnetic field in the $y-z$ plane (jet flows out of the page) at the midpoint of the simulation box, $x = 500\Delta$, and 1D cuts along the $z$ axis showing the magnitude and direction of all three magnetic field components at the midpoint of the simulation box, $x = 500\Delta$ and $y = 100\Delta$ for the $e^{-} - p^{+}$ case (upper row) and the $e^{\pm}$ case (lower row) at simulation time $t = 300 \omega_{\rm pe}^{-1}$ with $\gamma_{\rm jt}=15$ in both cases.
The comparison of the transverse structures in the $y$ direction at the velocity shear surfaces (a, d) with the parallel structures in the $x$ direction (Figs.~\ref{kkhi}b and \ref{kkhi}c)  shows that the fluctuations transverse to the jet in the $y$ direction are much more rapid than fluctuations along the jet in the $x$ direction. 

In the $e^{-} - p^{+}$ case, magnetic fields appear relatively uniform at the velocity shear surfaces along the transverse $y$ direction just as we have seen at the velocity shear surfaces along the parallel $x$ direction, with almost no transverse fluctuations visible in the magnetic field (small fluctuations in the $y$ direction over distances on the order of $\sim 10\Delta$ are visible in the currents, whereas small longitudinal mode fluctuations in the $x$ direction occur over distances $\sim 100\Delta$). 
It should be noted that if the mass ratio is small, for example, $m_{\rm i}/m_{\rm e} =100$, 
the wavy structure of MI would be seen in Fig.~\ref{Nishikawa14Fig6}a.

For the $e^{\pm}$ case, the magnetic field alternates in both the $y$ and $z$ directions and these transverse fluctuations occur over distances on the order of $\sim 10\Delta$, whereas longitudinal mode fluctuations in the $x$ direction occur over distances $\sim 100\Delta$.
This shows that the MI is also growing with a shorter wave length that that of the MI. 
The 1D cuts (b, e) show that the $B_\mathrm{y}$ field component dominates in the $e^{-} - p^{+}$ case whilst in the $e^{\pm}$ case, it is about an order of magnitude smaller than that in the $e^{-} - p^{+}$ case and the $B_{\rm z}$ component is more significant.
Panels~(b) and (e) also show that there is a magnetic field sign reversal on either side of the maximum that is relatively small for the $e^{-} - p^{+}$ case, but rather significant for the $e^{\pm}$ case, which is also apparent in Figure~ \ref{Nishikawa14Fig6}d. 
More details are revealed by the enlargement of the region indicated by the squares in the left panels. 
On the one hand, for the $e^{-} - p^{+}$ case, the generated relatively uniform DC magnetic field is symmetric about the velocity shear surface, e.g., note that $B_{\rm y} > 0$ immediately around the shear surface and $B_{\rm y} < 0$ in the jet and ambient plasmas at larger distances from the shear surface. 
On the other hand, for the $e^{\pm}$ jet, the generated AC magnetic field resides largely on the jet side of the velocity shear surface.

In very-high-resolution radio images, the appearance of a jet is dependent on the structure of the magnetic field, where the latter is determined by the type of the jet plasma. In a cylindrical jet consisting of $e^{-} - p^{+}$ plasma, the magnetic fields would be predominantly generated in the toroidal direction at the velocity shear surface and, as a consequence, the magnetic field would be quasi-parallel to the line of sight at the limbs of the jet for typical aspect jet angles $\theta \approx \gamma^{-1}$. 
In contrast, a $e^{\pm}$ plasma jet would generate sizable radial field components that are only about a factor of two weaker than the toroidal field. 



The strong electric and magnetic fields in the velocity shear zone would also facilitate the particle acceleration. 
The PIC simulations presented above 
are, however, too short for definitive statements on the efficiency of electromagnetic fields on the particle acceleration and the resulting spectra. The interpretation of the emission spectra will be complicated by the organization of the field in compact regions. 
A spatially resolved treatment of particle acceleration and transport would be mandatory for a realistic assessment.
Relativistic electrons, for example, will suffer little synchrotron energy loss outside of the thin layer of strong magnetic fields.
Thus, the synchrotron emissivity will be dominated by the shear layer and in general, the emissivity will depend on how efficiently electrons can flow in and out of the shear layer and be accelerated
in the regions of strong magnetic fields. 
An immediate consequence for radiation modeling is that the energy loss time of electrons cannot be calculated with the same mean magnetic field that is used to compute emission spectra because the former includes the volume filling factor of the strong-field regions.

These structures of electric and magnetic fields can be useful to determine the polarity in jets. However, since such structures are generated in the slab model, they cannot be applied to cylindrical jets \rv{yet}. 

\subsubsection{With other setups}
\label{sec:5.2.3}

As a variation of kKHI instability, \cite{Dieckmann2017} have examined the collision of two clouds of electrons and protons at a speed that exceeded the ion acoustic speed by a factor of 3.5. 
Their initial contact boundary was sinusoidally displaced along the collision direction. 
The displacement of the contact boundary is formed by the interpenetrating plasma clouds and this corrugation seeds the non-linear thin-shell instability (NTSI). 
They have confirmed that a wavelength of the seed perturbation, which exceeded that used in the previous simulation study \citep{Dieckmann_2014}
by a factor of 4, is unstable. A wide range of wavenumbers of the seed perturbation is thus subjected to the NTSI.

They have identified the proton--proton beam instability as the process that limits the lifetime of the thin shell. 
This instability is known to destroy planar double layers and electrostatic shocks 
\citep{Karimabadi1991,Kato2010,Dieckmann2015}
and it also affects the non-planar ones.

The amplitude of the shell's spatial oscillation grew because the NTSI introduces a spatially varying velocity of the thin shell in the reference frame that moves with the mean speed of the shell.
The modulus of the velocity peaked at the extrema of the shell's oscillation and the velocity at these positions reached 70 per cent of the ion acoustic speed. The amplitude of the thin shell's spatial displacement grew during the simulation to almost three times its initial value before the shell was destroyed by the proton--proton beam instability.

\begin{figure}[htb]
\hspace{0.5cm}
\includegraphics[width=0.8\textwidth]{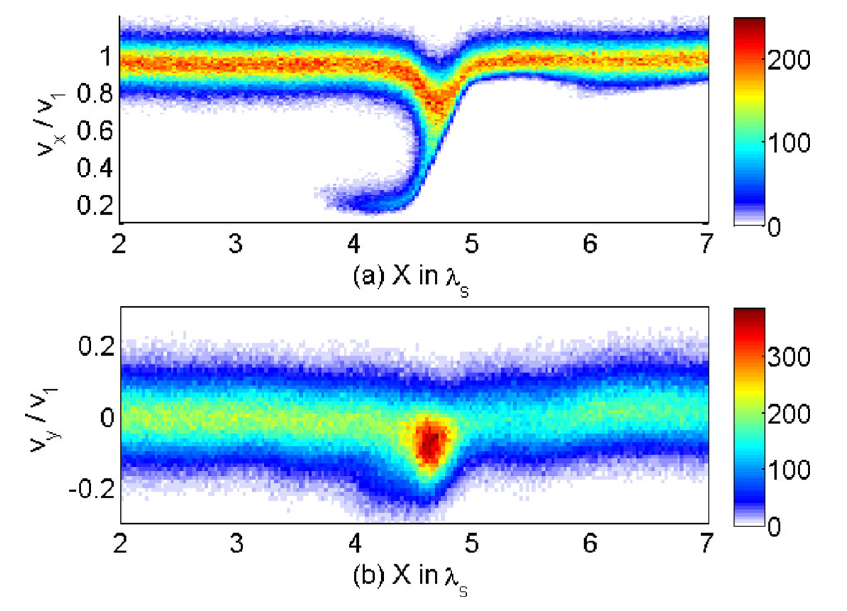}
\caption{Projections of the phase-space density distribution of the protons of cloud 1 on to the $x, v_{\rm x}$ plane (a) and on to the $x, v_{\rm y}$ plane (b) at $t = t_{4}= 1.6 \times 10^{3}$. Here $\lambda_{\rm s}=c/\omega_{\rm pe}$. 
The colour scale is linear. {Image reproduced with permission from \cite{Dieckmann2017}, copyright by the authors.}
}
\label{Dieckmann17fig5}
\end{figure}

Before the shell is destroyed, Fig.~\ref{Dieckmann17fig5} shows the phase-space 
density distribution of the protons of cloud 1 averaged over the $y$-interval, which is delimited by the thin region (vertical lines in Fig.~3). The thin shell is located at $x \approx 4.7$. 
The protons that enter the thin shell reach their lowest mean speed $vx \approx 0.75v_{1}$ at $x \approx x_{4}$ in Fig.~\ref{Dieckmann17fig5}a 
and their mean speed along the $y$-direction reaches $v_{\rm y} \approx -0.1v_{1}$ in Fig.~\ref{Dieckmann17fig5}b. 
Figure~\ref{Dieckmann17fig5}a shows that the protons are reaccelerated by the double layer at $x \approx 4.9$ when they leave the thin shell and move into the upstream region at $x \approx 5$. 
Some of the protons are reflected by the electrostatic shock at $x \approx 4.5$ and they form the beam at $x \approx 4$ and $v_{\rm x} \approx 0.2$. 
These protons fall behind the thin shell, which moves at the mean speed $v_{1}/2$ and they thus constitute a shock-reflected proton beam. 
The proton beam in Fig.~\ref{Dieckmann17fig5}a in the interval $x > 5$ is not spatially uniform. The density is lower for $5 < x < 5.8$ than for $x > 5.8$. 
The growth in time of the thin shell's potential relative to the upstream results in an increasing proton compression within the shell, which reduces temporally the number of protons that exit the thin shell via the double layer.

Their simulation data hint at a possible coupling of the NTSI with ion acoustic oscillations along the thin shell. They have explained the change of the NTSI's growth rate at late times in terms of these oscillations, which would make the collisionless NTSI overstable.
It should be noted that \cite{Vishniac1983} has shown that thin shells are linearly stable, but non-linearly unstable in the hydrodynamical regime.

The ambipolar electric field at the boundaries of the thin shell deflected the inflowing upstream electrons and protons into different directions. The relative drift of the electrons and the protons resulted in a net current and, thus, in the growth of magnetic fields.


\subsection{Global PIC simulations of cylindrical jets}
\label{sec:5.3}
\subsubsection{Unmagnetized jets}
\label{sec:5.3.1}

In order to investigate the possible structures of electromagnetic fields in jets including
instabilities which are generated due to the velocity-shear, simulations with cylindrical jets 
need to be performed because relativistic jets and internal filamentary structures are more 
suitably modeled as intrinsically cylindrical shape \citep{nishikawa14c,nishikawa16a}. 

\citet{nishikawa16a} have performed 3D PIC simulations involving a cylindrical jet injection into an ambient plasma without a magnetic field in order to investigate shocks 
(Weibel instability) and velocity shear instabilities (kKHI and MI) simultaneously .
Previously, these two processes (Weibel instability and velocity shear instabilities) have been studied separately. For example, kKHI and MI have been examined for sharp velocity shear slab and cylindrical geometries extending across the computational grid
\citep[e.g.,][]{nishikawa14b,nishikawa14c} 
and the injection across the entire inlet grid plane has been used for the study of collisionless shock (Weibel) instabilities \citep[e.g.,][]{nishikawa09}. 


\begin{figure}[htb]
\centering
\includegraphics[scale=0.26]{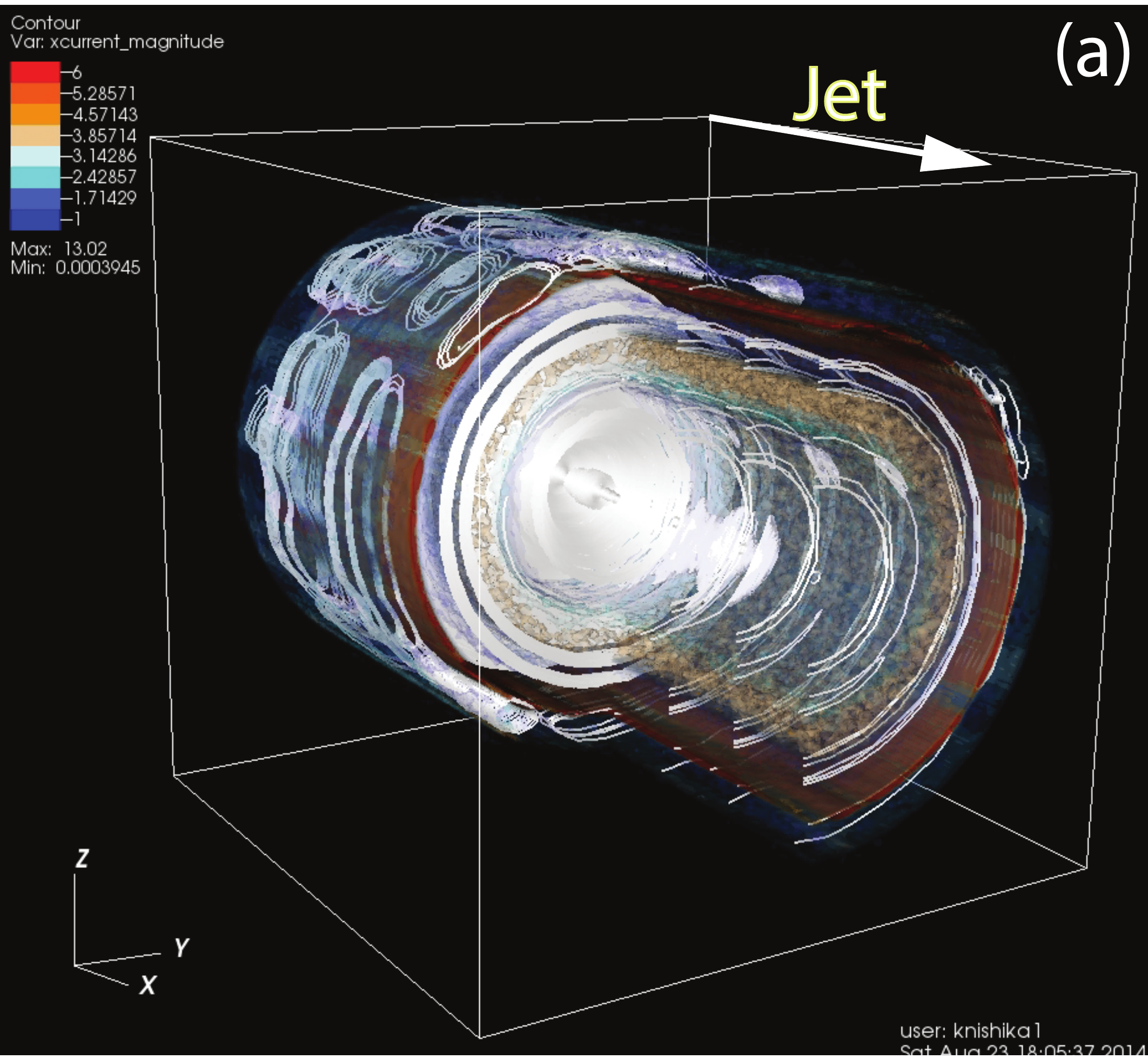}
\includegraphics[scale=0.26]{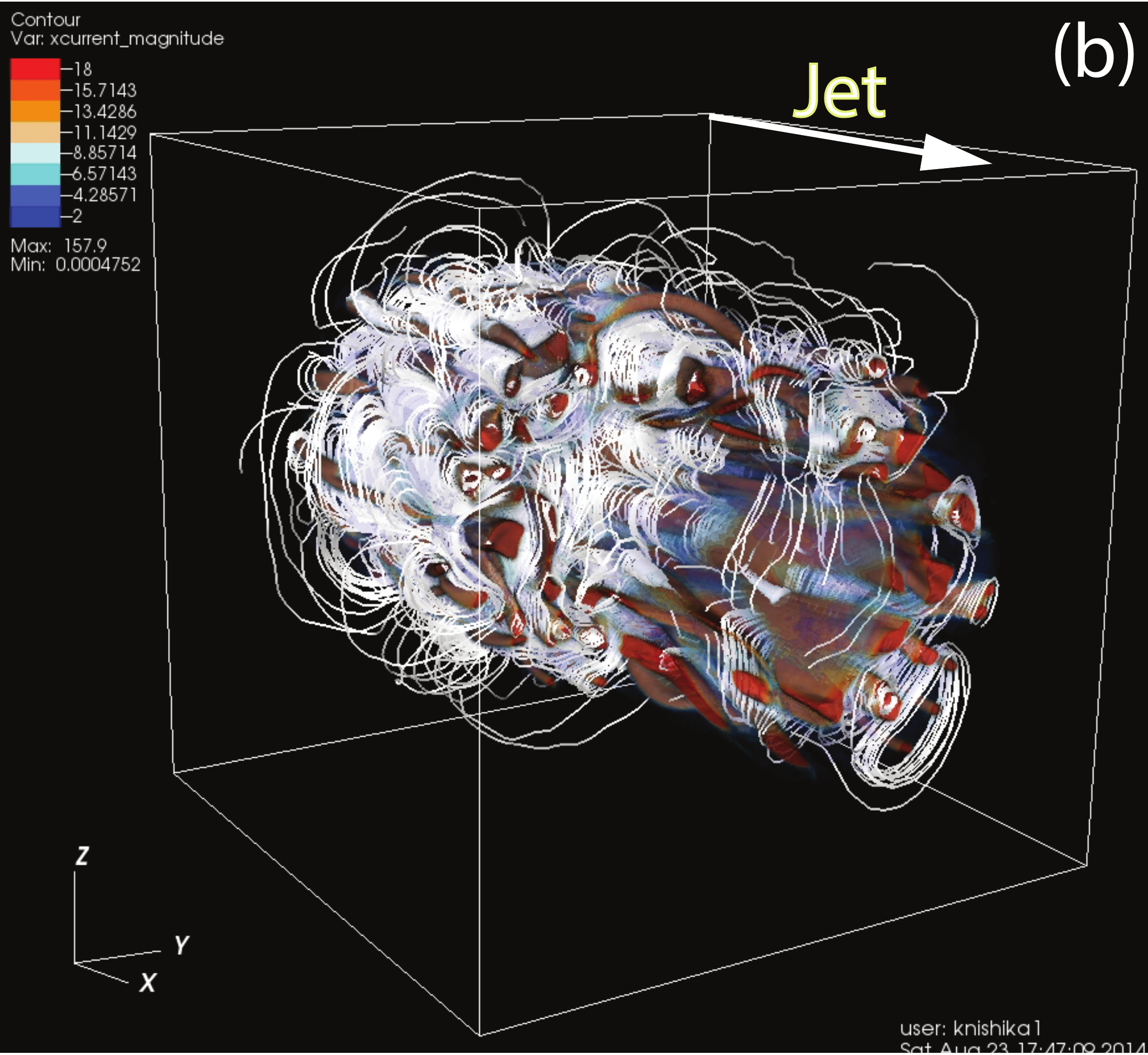}
\caption{Isocontour plots of current density component $J_{\rm x}$ with magnetic field lines (one fifth of the jet size) for (a) an $e^{-} - p^{+}$ and (b) an $e^{\pm}$ jet after $t = 300 \omega_{\rm pe}^{-1}$. The 3D displays are clipped along the jet and perpendicular to the jet in order to view the interior.
\citep{nishikawa14c}.}
\label{cylind}
\end{figure}

Isocontour images of the $x$ component $J_{\rm x}$ of the current density along the magnetic field lines generated by the velocity-shear instabilities are shown in Figure~\ref{cylind}. 
These images indicate that in the $e^{-} - p^{+}$ jet currents are generated in sheet like layers and the magnetic fields are wrapped around the jet, which is generated by MI. 
On the contrary, for $e^{\pm}$ jet plasmas, many distinct current filaments are generated near the velocity shear and the individual current filaments are wrapped by the magnetic field. 
The clear difference in the magnetic field structure between these two cases may make it possible to distinguish different jet compositions via differences in circular and linear polarization.

\begin{figure}[htb]
\includegraphics[width=\textwidth]{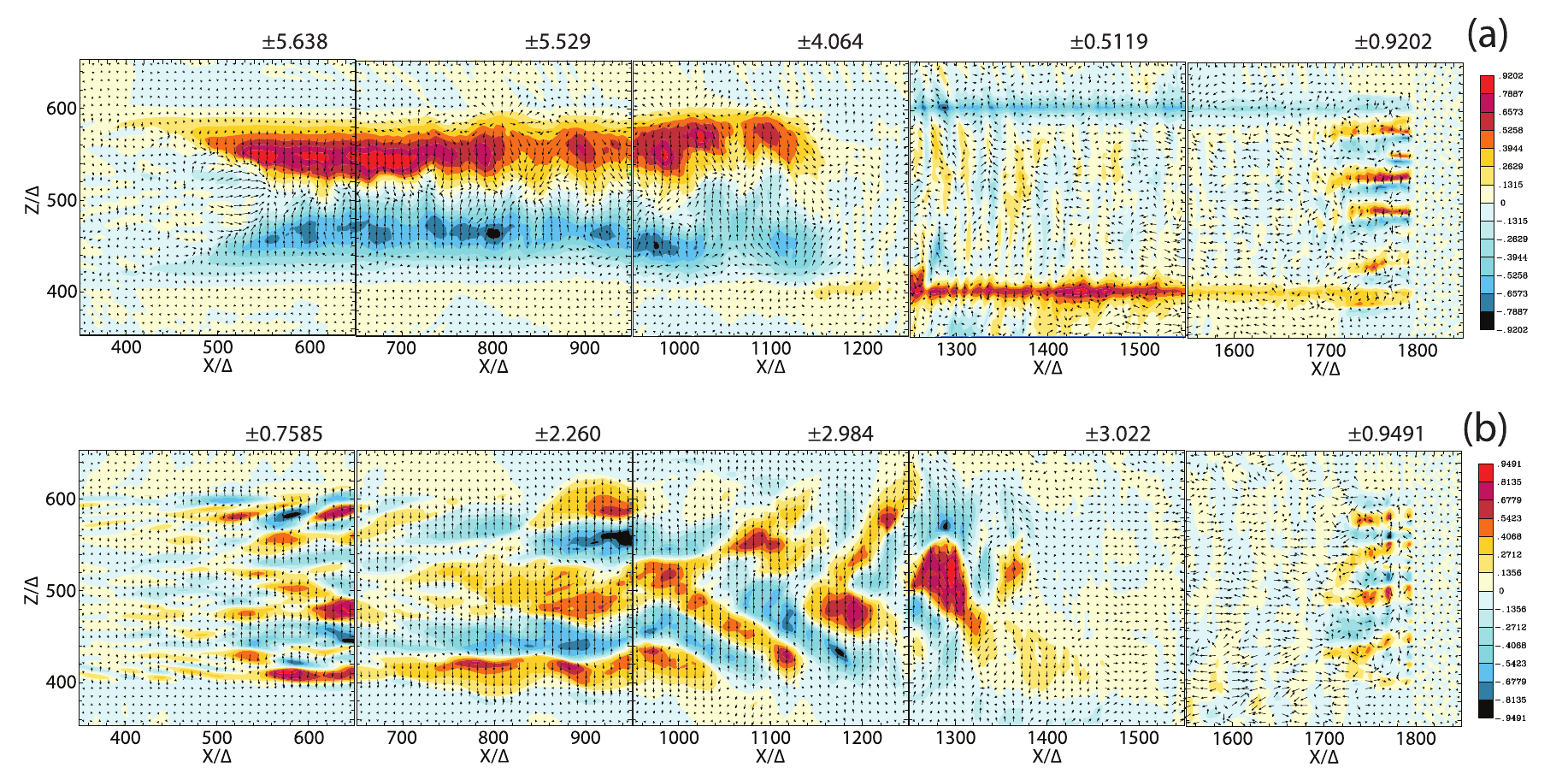}
\caption{2D slices of $B_{\rm y}$ for (a) an $e^{-}-p^{+}$ and (b) an $e^{\pm}$ jet at time $t = 1700 \omega_{\rm pe}^{-1}$.
Arrows show $E_{\rm x,z}$. The color scale at the right is for only the rightmost panel (region $1550 < x/\Delta <1850$). The maximum and minimum values in each of the five jet regions are indicated at the upper right. Adapted from Fig.~11 in \citet{nishikawa16a}.}
\label{3djetwohm}
\end{figure}

Figure~\ref{3djetwohm} shows 2D slices of $B_{\rm y}$ at the middle plane for the $e^{-} - p^{+}$ (upper) and $e^{\pm}$ (lower) jet plasma after $t = 1700 \omega_{\rm pe}^{-1}$. 
In the  $e^{-} - p^{+}$ jet, strong toroidal magnetic fields collimating the jet are generated by the mushroom instability. This strong toroidal magnetic field is produced by the strong current $-J_{\rm x}$ by collimated jet (ambient) electrons as shown in Figure~\ref{3djetwohm}a. 
At the nonlinear stage the collimation relaxes around $x/\Delta\simeq 1150$, the polarity of the toroidal magnetic fields switches from clockwise in the rightmost panels to counter-clockwise in the leftmost panels as viewed from the jet front. The counter-clockwise magnetic fields are created by the current layer ($+J_{\rm x}$) at the jet boundary. 
Owing to the perpendicularly accelerated jet electrons in the collimated region, the electrons are expelled from the collimated region and are
slowed down. 
Consequently, the MI is saturated and the strong toroidal magnetic field disappears and releases the collimation. 
At the same time, the concentrated electron current near the center of the jet decelerates and jet electrons expand outward. 
Heavy jet protons maintain the original jet border line, and as a result of the decrease of electrons near the jet boundary, the current $+J_{\rm x}$ 
is produced, which generates the counter-clockwise magnetic field. 
Furthermore, the patterns of strong toroidal magnetic fields are seen near the jet boundary indicating the excitation of MI. Comparing with simulations including helical magnetic fields, see Fig.~\ref{JxB}a, there are two differences: the MI is a dominate growing instability and at the nonlinear stage at the edge of the jet the polarity of the toroidal magnetic field is reversed.
However, including helical magnetic fields, the kKHI also grows; hence at the nonlinear stage a few clusters of magnetic field islands are generated near the edge of the jet as shown in Figs.~\ref{3DBV}b and \ref{3DBV}c.

For the $e^{\pm}$ plasmas, filaments of alternating $B_{\rm y}$ are initially generated along the jet by the current filaments, and at the
nonlinear stage they move out of the jet \rv{as shown in Fig.~\ref{3djetwohm}b}. 

For both cases at the jet front the strong striped $B_\mathrm{y}$ components are created by the current filaments, which have been observed in previous simulations of the Weibel instability \citep{nishikawa09,ardaneh15,Nishikawa2020}. 


\subsubsection{Magnetized jets}
\label{sec:5.3.2}

Motivated by understanding how (mildly) relativistic jets are formed in the vicinity of X-ray binaries, \citet{Dieckmann2019} used the PIC code \texttt{EPOCH} to study the interface layer of an electron-positron cloud, formed by photons in the ambient corona of this binary, in the ambient electron-ion plasma. 
In this set-up the initial magnetic field is aligned with the mean initial velocity of the electron-positron cloud. 
They consider a two-dimensional simulation domain $L_{\rm x}\times L_{\rm y}=24.6\times 12.3$ with 14000 and 7000 grid cells in $x-$ and $y-$direction respectively. 
In their set-up, the initial energy of the ambient plasma is $T_{\rm 0}=$ 2 keV, of the electron-positron cloud is 400 keV, the initial ambient plasma density of each species is $n_0=14$ particles per cell and the initial density of the cloud electrons and positrons is $n_{\rm pc}(x,y) = 5 - (45x^{2} + 5y^{2})W^{2}_{\rm jet}$ if $n_{\rm pc}(x, y) \ge 0$ and zero otherwise. 
The half-width of the jet is $W_{\rm jet} = 1.7$. 

\begin{figure}[htb]
\includegraphics[width=\textwidth]{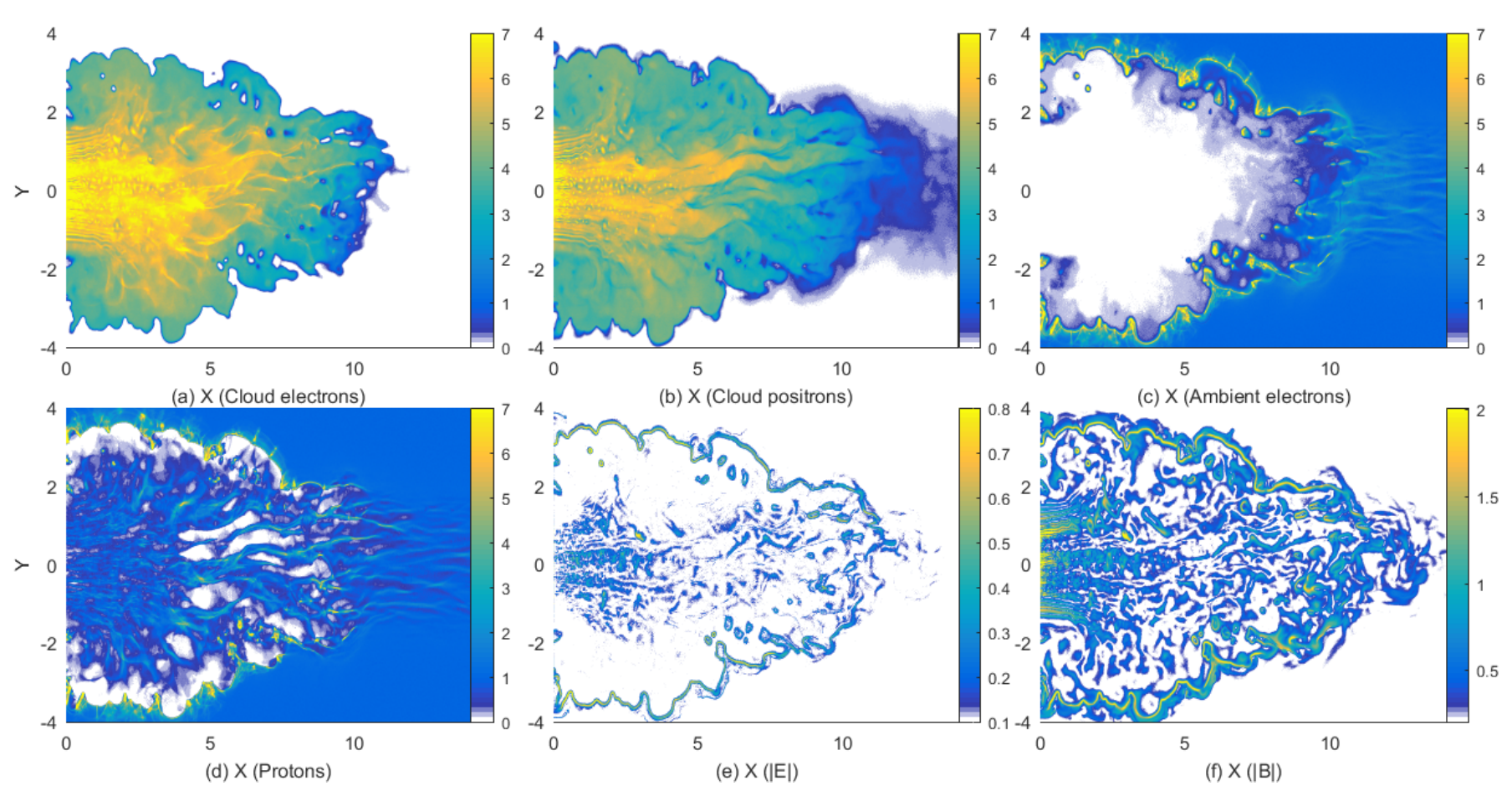}
\caption{Plasma and field distributions at the interface layer of an electron-positron cloud in an ambient electron-proton plasma after $t_{\rm sim} = 34$. (a), (b) spatial distribution of the cloud electrons and positrons; (c), (d): ambient plasma electrons and protons; (e), (f) electric and magnetic field strengths $|E|$ and $|B|$. The electric and magnetic strengths are truncated at values of 0.1 and 0.2 to remove noise. Image reproduced with permission from \cite{Dieckmann2019}, copyright by ESO.}
\label{dieckmann.fig}
\end {figure}

Figs.~\ref{dieckmann.fig}a and \ref{dieckmann.fig}b show the distributions of the cloud electrons and positrons. The electrons and protons of the ambient plasma are shown in panels (c) and (d) at $t_{\rm sim} = 34 \omega_{\rm pp}^{-1}$
where $\omega_{\rm pp}=\sqrt{(n_0 e_0^2/(m_{\rm p} \epsilon_0)}$ is the plasma frequency of protons. 
It shows that the cloud leptons tend to move outwards into the ambient electron-proton plasma expelling the ambient electrons outwards leading to a (collisionless) contact discontinuity. 
Such discontinuity also appears for the protons; however, there are still residual protons in the cloud region through a filamentation instability between the ambient plasma and the fast-moving pair cloud from earlier time steps. 
This electromagnetic piston separates the jet plasma from the ambient plasma and its role is that of the contact discontinuity in hydrodynamic jets.
The important point is that this structure is coherent on scales which are much larger than particle inertial scales and may be thus a small step to better understand the transition from the micro-scale to the macro-scale processes.

The motion and distribution of charged particles lead to the modulation of the magnetic and electric fields depicted in Figs.~\ref{dieckmann.fig}e and \ref{dieckmann.fig}f at the same time. 
Whilst the motion of ambient electrons is blocked by the magnetic field, it creates an electrostatic field strong enough to accelerate the protons to high non-relativistic velocities which subsequently drive a shock in the ambient material forming the border of the jet's outer cocoon.
Finally, the interaction of the pair cloud with the strong magnetic field at the border might contribute to the radio emission of X-ray binaries.   

\subsubsection{Magnetized jets with a helical magnetic field structure}
\label{sec:5.3.3}

Large scale magnetic fields have been invoked to explain the launch, acceleration, and collimation of relativistic jets from the central nuclear region of AGNs \citet[e.g.,][]{meier08}, and from coalescing and merging compact objects (neutron stars and black holes) \citep[e.g.,][]{piran05,rezzolla11,Ruiz2020c}. 
The magnetic field structure and particle composition of such jets have not been well constrained observationally yet. It is natural to inject jets with helical magnetic fields \citep[e.g.,][]{Meier01}.

\medskip
\noindent
{\bf{Helical magnetic field structure}}
\medskip

Since in PIC simulations helical magnetic fields are not generated with jets powered by a self-consistent rotating black hole, a force-free helical magnetic field is implemented at the jet orifice.
For initial simulations, \cite{nishikawa16b,nishikawa17} used the force-free helical magnetic field whose poloidal ($B_x$) and toroidal ($B_{\phi}$) coordinates in the laboratory frame are \citep{mizuno14}
\begin{eqnarray}
B_{x} &=& \frac{B_{0}}{[1 + (r/a)^2]^{\alpha}},  \label{hmfwp1} \\
B_{\phi} &=& \frac{B_0}{(r/a)[1 + (r/a)^2]^{\alpha}}\sqrt{\frac{[1 + (r/a)^2]^{2\alpha} -1 -2\alpha (r/a)^{2}}{2\alpha -1}}, \label{hmfwp2}
\end{eqnarray}
where $r$ is the radial position in cylindrical coordinates, $B_0$ is the magnetic field (the toroidal field component is a maximum at $a$ for constant magnetic pitch), and $\alpha$ is a pitch profile parameter.



PIC simulations are performed in Cartesian coordinates.
Since $\alpha =1$, Eqs.~(\ref{hmfwp1} and \ref{hmfwp2}) 
reduce to:
\begin{eqnarray}
  B_{x} = \frac{B_{0}}{[1 + (r/a)^2]}, \qquad
  B_{\phi} = \frac{(r/a)B_{0}}{[1 + (r/a)^2]}.
\label{hmfcp}
\end{eqnarray}

The toroidal magnetic field is created by a current $+J_{x}(y, z)$ in the positive $x$-direction, such that the $y$- and $z$-coordinate of the magnetic field in Cartesian coordinates become:
\begin{eqnarray}
B_{y}(y, z) =  \frac{((z-z_{\rm jc})/a)B_{0}}{[1 + (r/a)^2]}, \qquad
B_{z}(y, z) = -\frac{((y-y_{\rm jc})/a)B_{0}}{[1 + (r/a)^2]}.
\label{hmfcar}
\end{eqnarray}
Here, $a$ is the characteristic length-scale of the helical magnetic field, $(y_{\rm jc},\, z_{\rm jc})$ is the center of the jet, and  $r = \sqrt{(y-y_{\rm jc})^2+(z-z_{\rm jc})^2}$. 
The chosen helicity is defined through Eq.~(\ref{hmfcar}), which has a left-handed polarity with positive $B_{0}$. At the jet
orifice, the helical magnetic field is implemented without the motional electric fields. This corresponds to a toroidal magnetic field generated self-consistently by jet particles moving along the $+x$-direction.

\rv{It should be noted that the scheme for injecting jet particles with helical magnetic fields was adjusted. For such simulations, we found for $a=4$ that the narrow current generating the toroidal magnetic field excites a kinetic current-driven instability before the growth of MI and kKHI. Therefore, this kinetic current-driven instability is included into the evolution of jets with MI and kKHI \citep[e.g.,][]{Nishikawa2020}. However, by using small $a$ MI and kKHI grow faster than the current-driven instability. Therefore, we have revisited and modified our injection scheme; new simulations with $a= 2.0$ with a better injection scheme will be reported in \citep{Meli2021}}.

\begin{figure}[htb]
\hspace{4.cm} (a) 
\hspace{5.8cm} (b) 
\vspace*{-.05cm}
\includegraphics[scale=0.38]{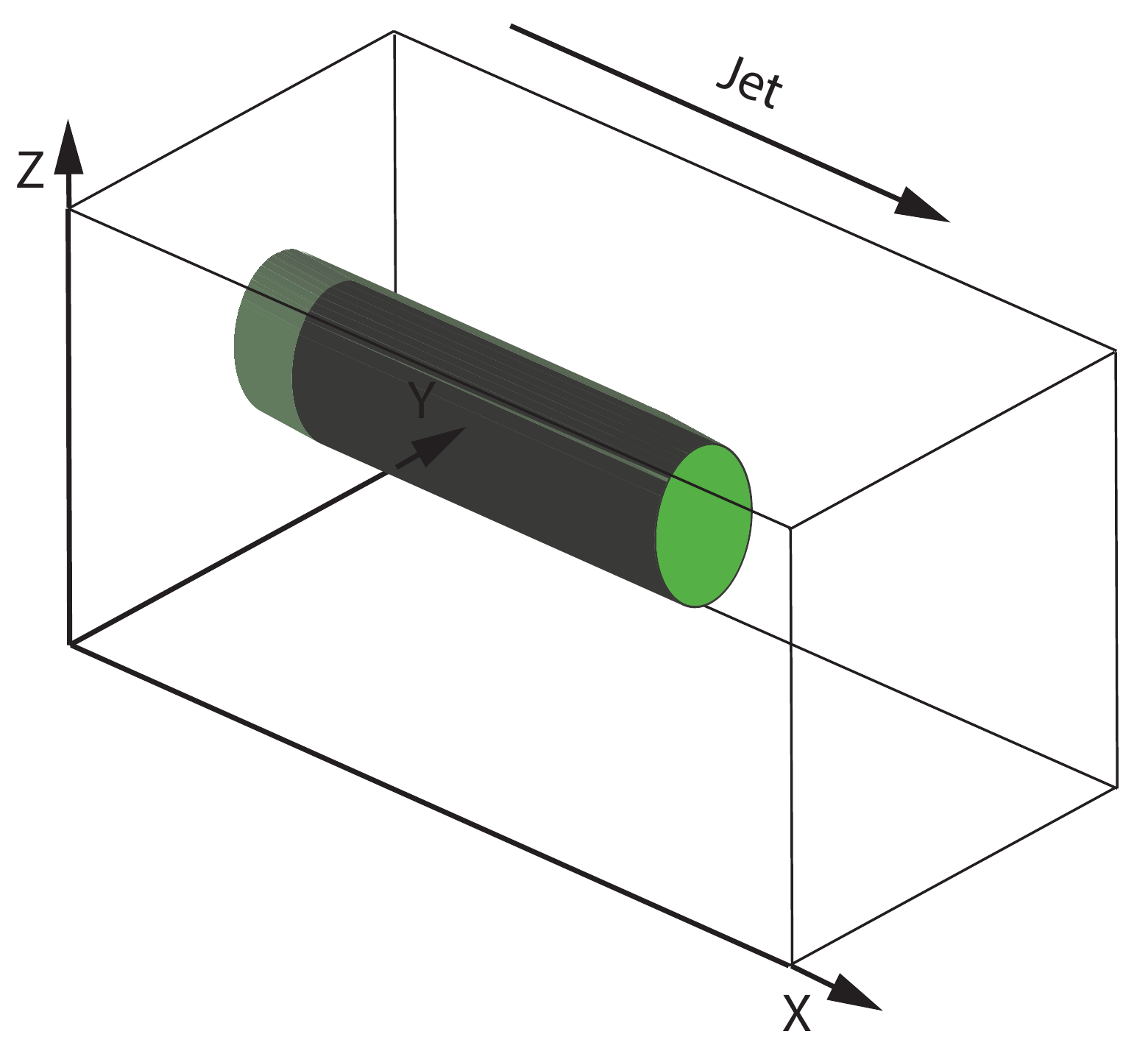}
\hspace*{-0.cm}
\includegraphics[scale=0.4]{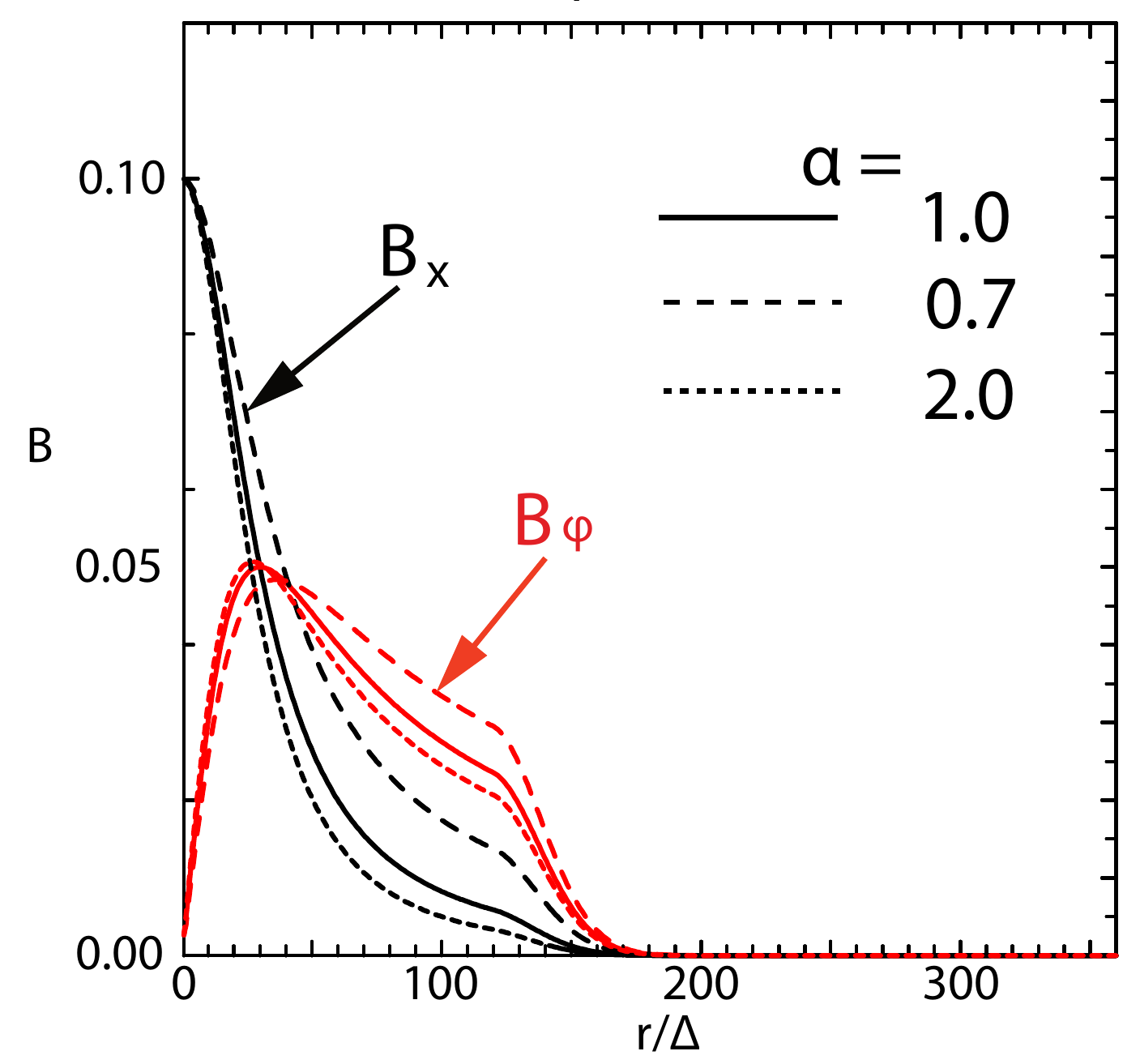}
\vspace{-.0cm}
\caption{
Panel (a) shows the schematic simulation setup of a global jet. The jet is injected at $x = 100\Delta$ with a jet radius of $r_{\rm jt}$ at 
the center of the $(y,z)$ plane (not scaled). Panel (b) shows the helical magnetic fields, $B_{x} ({\rm black}), B_{\phi} ({\rm red})$ for
the pitch profile $\alpha=$ 0.7 (dashed), 1.0 (solid), and 2.0 (dotted) with damping functions outside the jet with $b = 800.0$. 
The jet boundary is located at $r_{\rm jt}=120\Delta$ \citep{nishikawa17}.
}
\label{hmfpitch}
\end{figure}

The jet particles are injected at $x=100\Delta$ and jet particles propagate as shown by the cylindrical shape in Fig.~\ref{hmfpitch}a.
Fig.~\ref{hmfpitch}b presents the poloidal ($B_x$: black) and toroidal ($B_{\phi}$: red) components of helical magnetic fields with different pitch profiles. 
The toroidal magnetic fields become zero at the center of the jet as seen in the red lines. 
They have checked the structure of the helical magnetic field around constant pitch 
($\alpha =1$) (solid lines). If the pitch profile parameter $0.5 < \alpha < 1$, the magnetic 
pitch increases with radius. If $\alpha  > 1$, the magnetic pitch decreases. For comparison, Fig.~\ref{hmfpitch}b shows the pitch profile $\alpha =0.7$ (increasing: dashed) and $\alpha =2.0$ (decreasing: dotted). 
PIC simulations have been performed with a constant pitch ($\alpha = 1$) and with $b=200\Delta$ (damping factor outside the jet boundary: multiply by $\exp(-r/b)$ for $r > r_{\rm jt}$) using $r_{\rm jt}=20, 40, 80,$ and $120\Delta$ \citep{nishikawa16b,nishikawa17}. 

The structure of the jet formation region is more complicated than what they have implemented in their PIC simulations at the present time
\citep[e.g.,][]{broderick09,moscibrodzka17}.

The initial profile of helical magnetic fields,
jet densities and velocities can be refined based on GRMHD simulations \citep[e.g.,][]{oliber17}. 

Initial global jet simulations with helical magnetic fields show the effects of helical magnetic fields on kKHI, MI, and the Weibel instabilities \citep{nishikawa16b,nishikawa17}, which will be further described in the next section. 


\medskip
\noindent
{\bf{Helical magnetic fields in relativistic jets with larger jet radius}}
\medskip

As an initial step, \citet{nishikawa17}  examined how the helical magnetic field modifies the jet evolution using a small simulation domain before performing larger-scale simulations; for this first step a similar simulation setup and particle injection is used as in their previous work \citep{nishikawa16b}.
In these small system simulations, a numerical grid with 
$(L_{x}, L_{y}, L_{z}) = (645\Delta, 131\Delta, 131\Delta)$
(simulation cell size: $\Delta = 1$) and periodic boundary conditions in transverse directions is used, where the jet radius is $r_{\rm jt} = 20\Delta$. The jet and ambient (electron) plasma number density measured in the simulation frame is $n_{\rm jt}= 8$ and  $n_{\rm am} = 12$,  respectively. 
This set of density of jet and ambient plasmas is used in previous simulations \citep{nishikawa16a,nishikawa16b,dutan17}

In these simulations, the electron skin depth is $\lambda_{\rm s} = c/\omega_{\rm pe} = 10.0\Delta$ and the electron Debye length for the ambient electrons is $\lambda_{\rm D}=0.5\Delta$ where $c$ is the speed of light and $\omega_{\rm pe} = (e^{2}n_{\rm am}/\epsilon_0 m_{\rm e})^{1/2}$ is the electron plasma frequency. The jet-electron 
thermal velocity is $v_{\rm jt,th,e} = 0.014c$ in the jet reference frame whereas the electron thermal velocity in the ambient plasma is $v_{\rm am,th,e} = 0.03c$, and the ion thermal velocities are smaller by a factor of $(m_{\rm i}/m_{\rm e})^{1/2}$. Simulations were performed using an $e^{\pm}$ plasma or an $e^{-}-p^{+}$ and with the ambient plasma at rest ($v_{\rm am}= 0$).

Other plasma parameters used in the simulation are: the initial magnetic field amplitude parameter $B_{0}=0.1c$, ($c=1$),
the magnetization parameter
~\mbox{{(}$\sigma = B^2/n_{\rm e}m_{\rm e}\gamma_{\rm jt}c^{2}=2.8\times 10^{-3}$ {)}}, and the characteristic radius $a =  
0.25r_{\rm jt}$. The helical field structure inside the jet is defined by Eq.~(\ref{hmfcar}). 
For the external magnetic fields,  a damping function
$\exp{[-(r-r_{\rm jt})^{2}/b]}$ 
$(r \ge r_{\rm jt})$ is used that multiplies Eq. (\ref{hmfcar}) with the 
tapering parameter $b=200$. The final profiles of the helical 
magnetic field components are similar to that in the case where 
jet radius $r_{\rm jt} = 20\Delta$, the only difference is 
$a= 0.25*r_{\rm jt}$ \citep{nishikawa16a}. 

\begin{figure}[htb]
\hspace{5.20cm} (a) \hspace{5.2cm} (b)
\vspace{-0.cm}
\hspace{0.cm}
\includegraphics[scale=0.33]{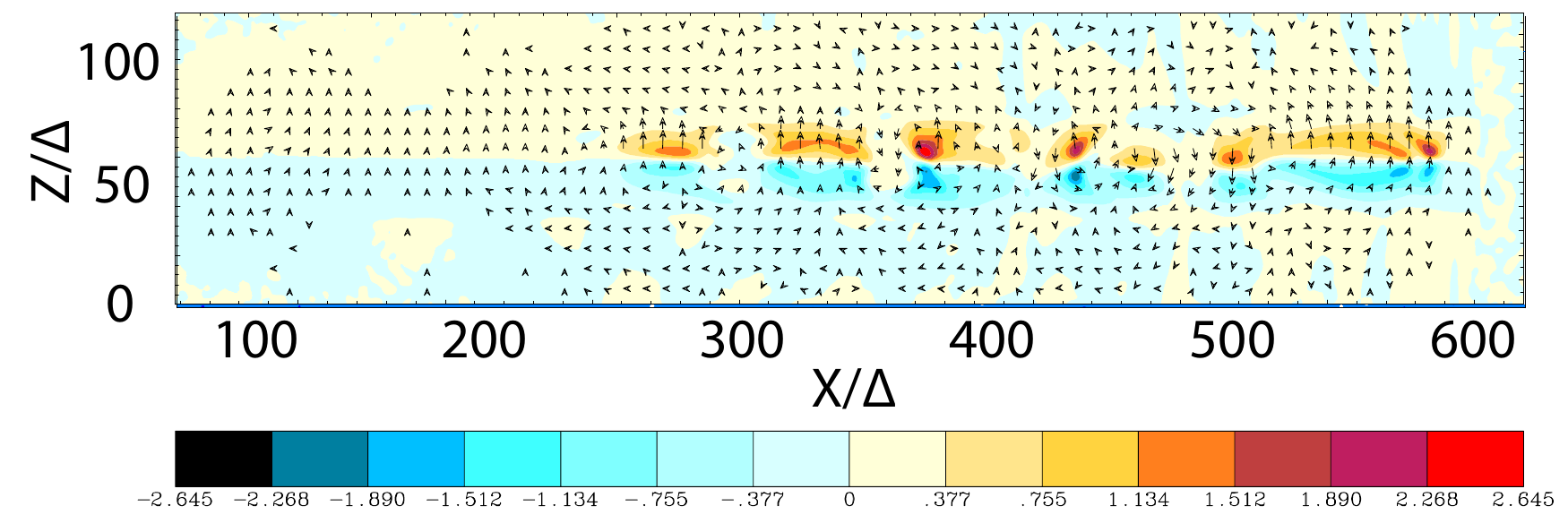}
\hspace*{-0.3cm}
\includegraphics[scale=0.33]{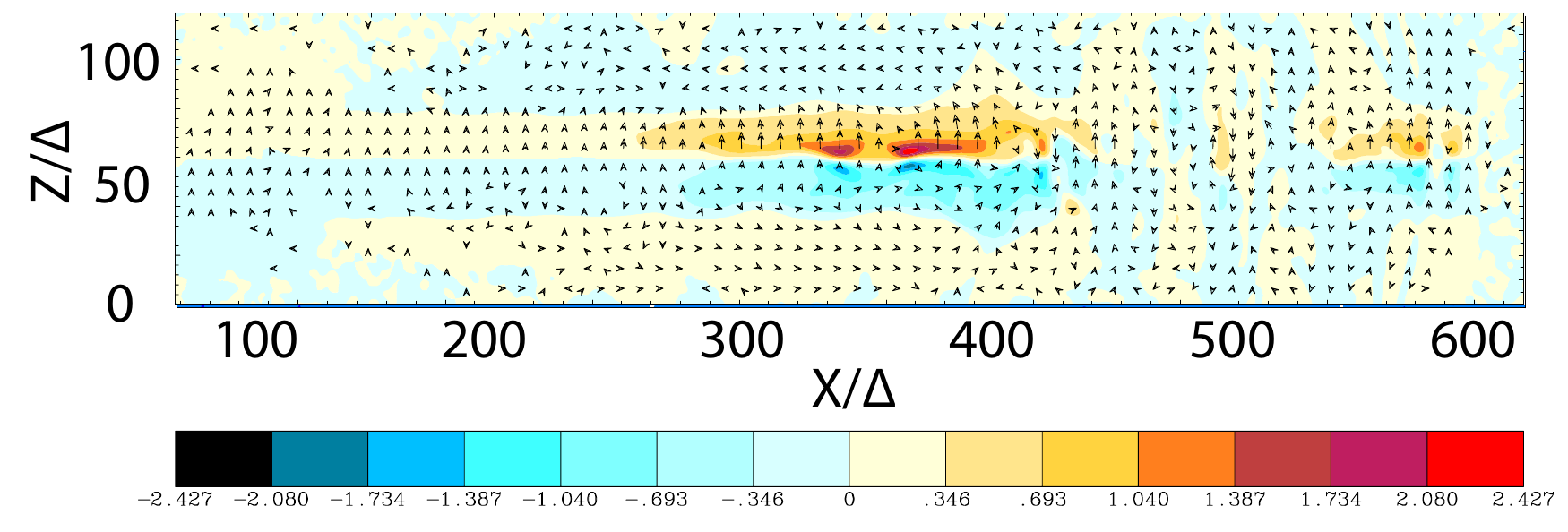}
\vspace*{-0.cm}
\hspace{5.20cm} (c) \hspace{5.2cm} (d)
\hspace*{0.cm}
\includegraphics[scale=0.33]{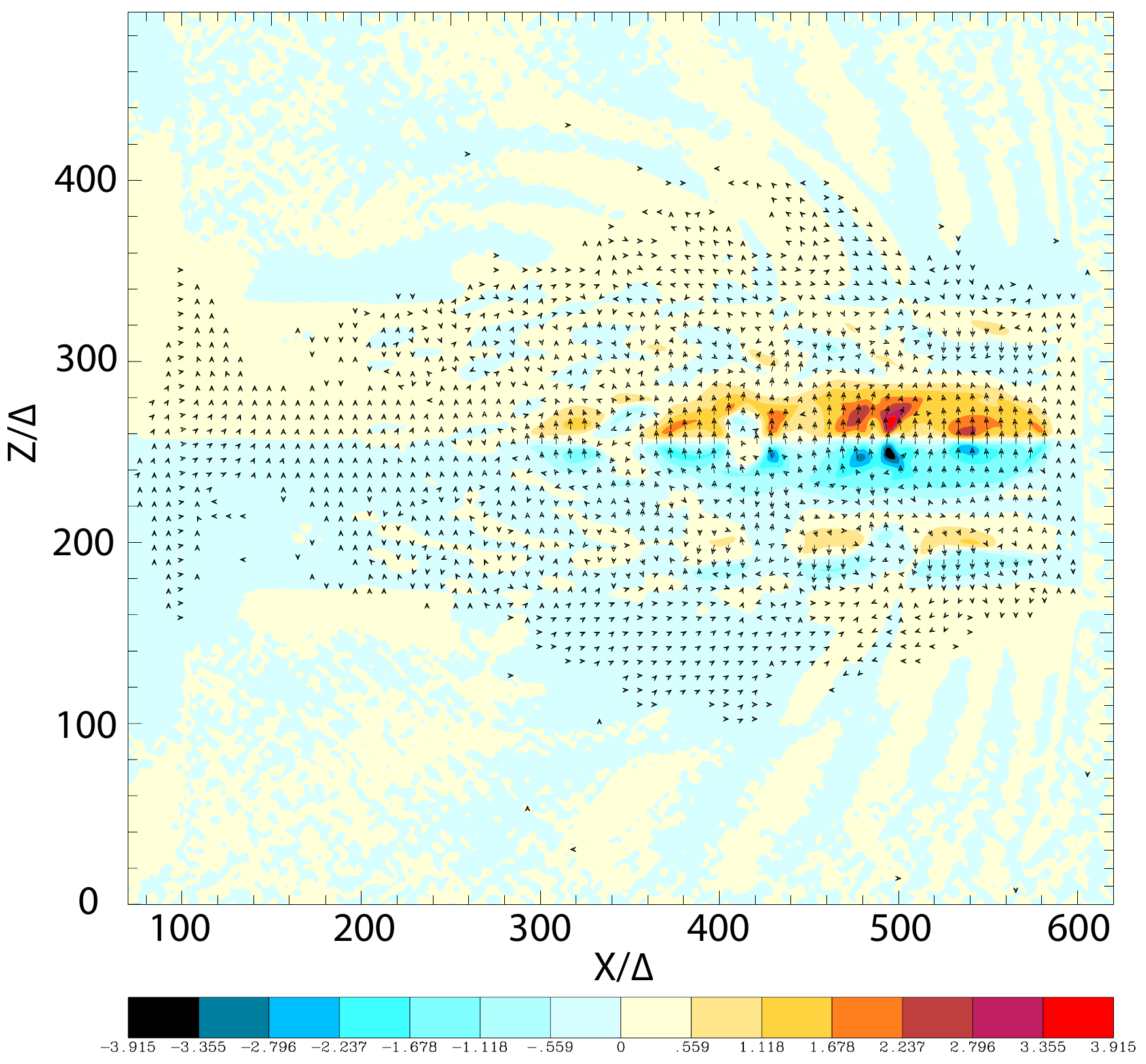}
\hspace*{-0.3cm}
\includegraphics[scale=0.33]{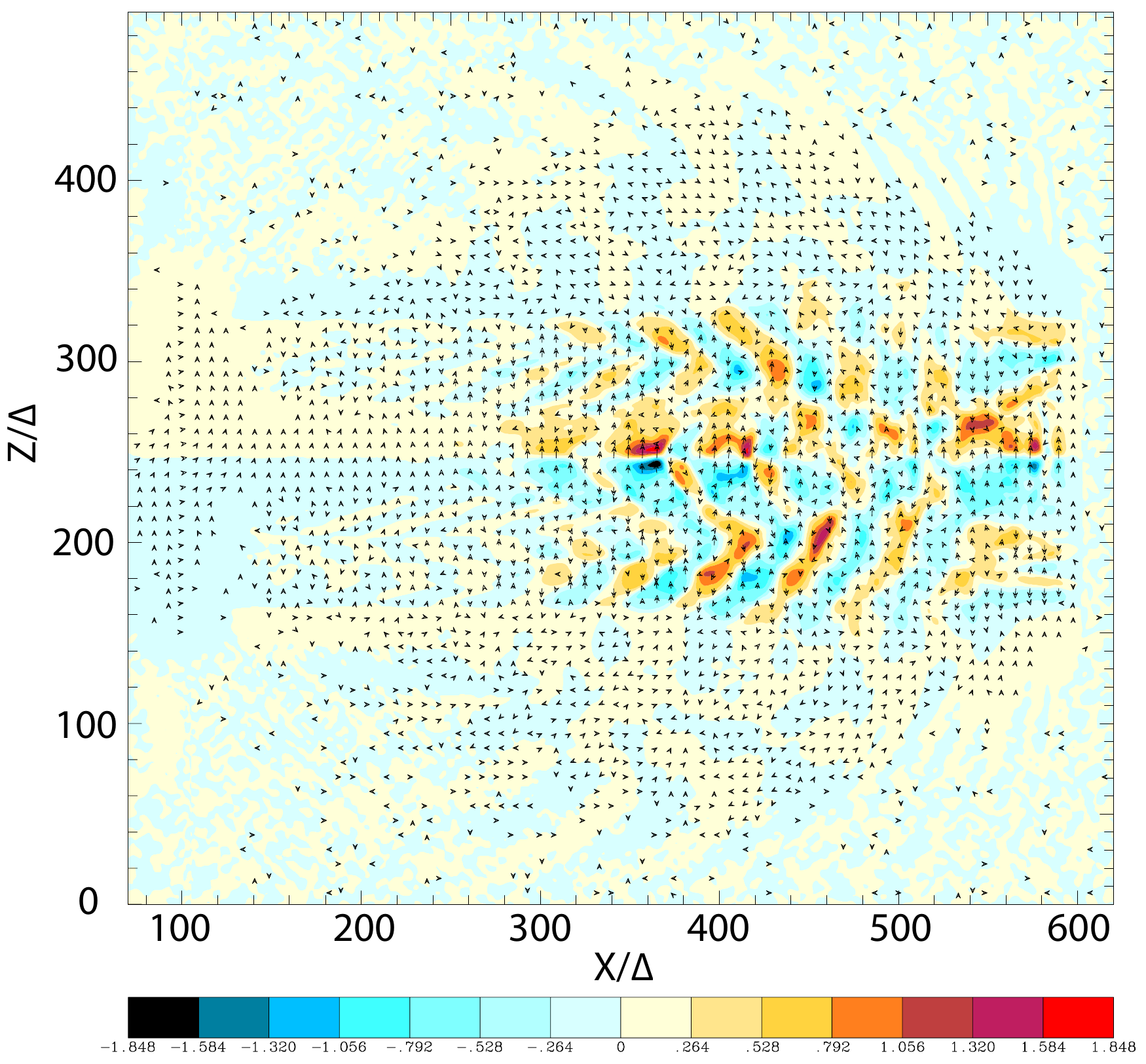}
\caption{Isocontour plots of the azimuthal component of magnetic field $B_y$ intensity at the center of the jets  for  $e^{-}-p^{+}$  ((a) and (c))  $e^{\pm}$  ((b) and (d)) jets;  with $r_{\rm jt}=20\Delta$ ((a) and (b)) $r_{\rm jt}=80\Delta$ ((c) and (d)) at time $t =  500\omega_{\rm pe}^{-1}$. The disruption of helical magnetic fields are caused by instabilities and/or reconnection. 
The max/min numbers of  panels are (a) $\pm$2.645, (b) $\pm$ 2.427, (c) $\pm$  3.915, (d) $\pm$1.848. Adapted from \cite{nishikawa17}.}
\label{ByBxz}
\end{figure}

All simulation parameters are maintained as described above except the jet radius and the simulation size, the latter being adjusted based on the jet radius. \citep{nishikawa17} performed simulations with larger jet radii $r_{\rm jt} =40\Delta,\, 80\Delta,\, {\rm and}\, 120\Delta$. 
The cylindrical jet with jet radius $r_{\rm jt} = 120\Delta$ is  injected in the middle of {the} $y-z$~plane ($(y_{\rm jc}, z_{\rm jc}) =  (381\Delta, 381\Delta)$) into the system with $(L_{\rm x}, L_{\rm y}, L_{\rm z})=(645\Delta, 761\Delta, 761\Delta)$) at $x= 100\Delta$.
The largest value of the jet radius with $r_{\rm jt} =120\Delta$ is larger than that used previously in \cite{nishikawa16a} ($r_{\rm jt} =100\Delta$), but the simulation length is much shorter with $x = 2005\Delta$.

Figure~\ref{ByBxz} shows the $y$ component of the magnetic field ($B_{\rm y}$) in the jet radius with $r_{\rm jt}=20\Delta\, {\rm and}\,  80\Delta$.
The initial helical magnetic field (left-handed; clockwise viewed from the jet front) is enhanced and disrupted due to the instabilities for both cases. 

Even though the simulation system is short, the growing instabilities are affected by the helical magnetic fields. These complicated patterns of $B_{\rm y}$ are produced by the currents which are generated by instabilities in jets. The larger jet radius contributes more modes of instabilities to grow in the jets, which make the jet structures more complicated. 
The simple recollimation shock generated in the small jet radius is shown in Figs.~\ref{ByBxz}a and \ref{ByBxz}b.  
Longer simulations are needed for the investigation of full development of instabilities and jets with helical magnetic fields. 

\vspace*{-0.cm}
\begin{figure}[htb]
\hspace{5.3cm}(a) \hspace{5.4cm}(b)
\vspace*{-0.1cm} 
\includegraphics[scale=0.3,angle=-90]{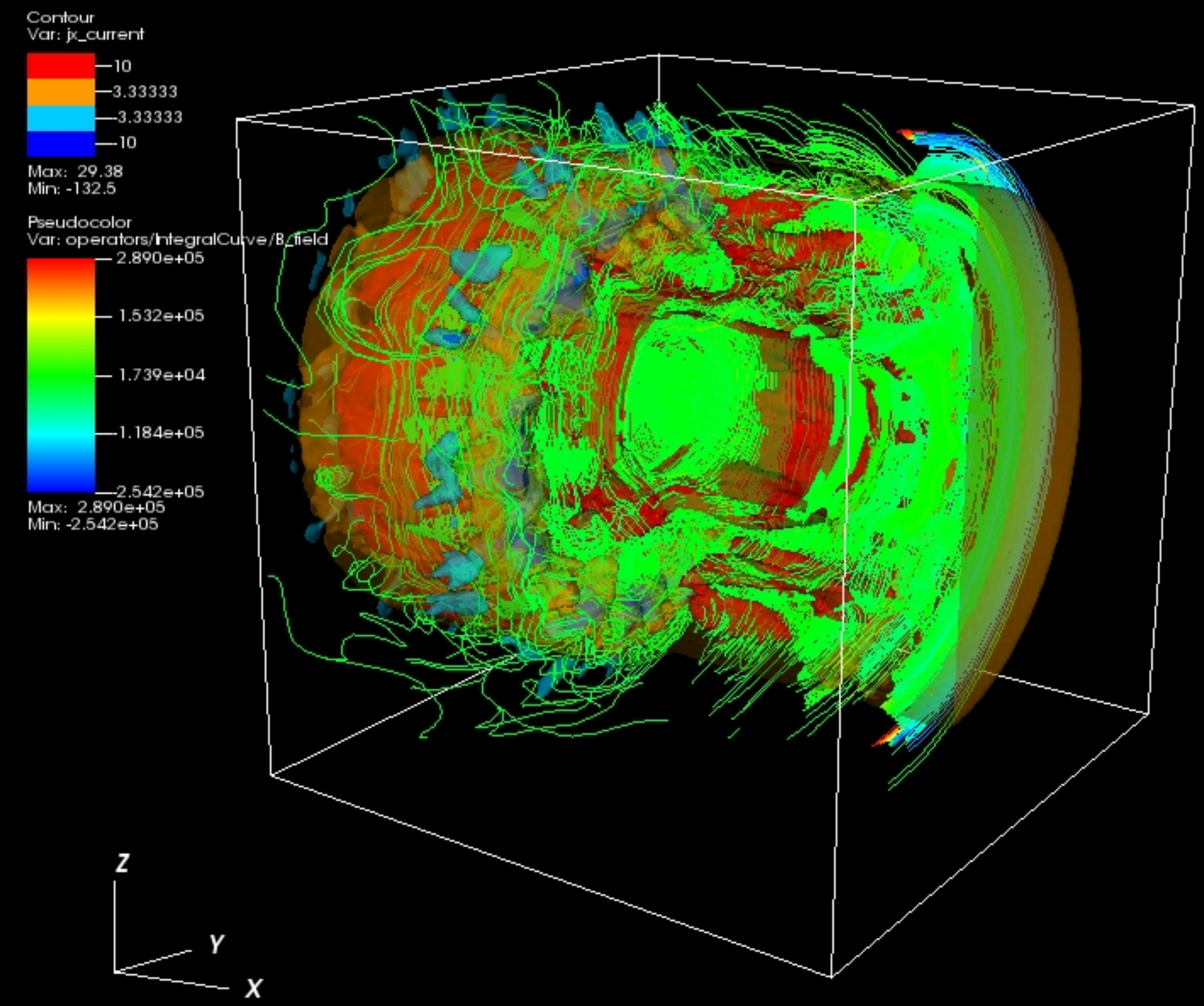}
\hspace*{-0.1cm}
\includegraphics[,scale=0.3,angle=-90]{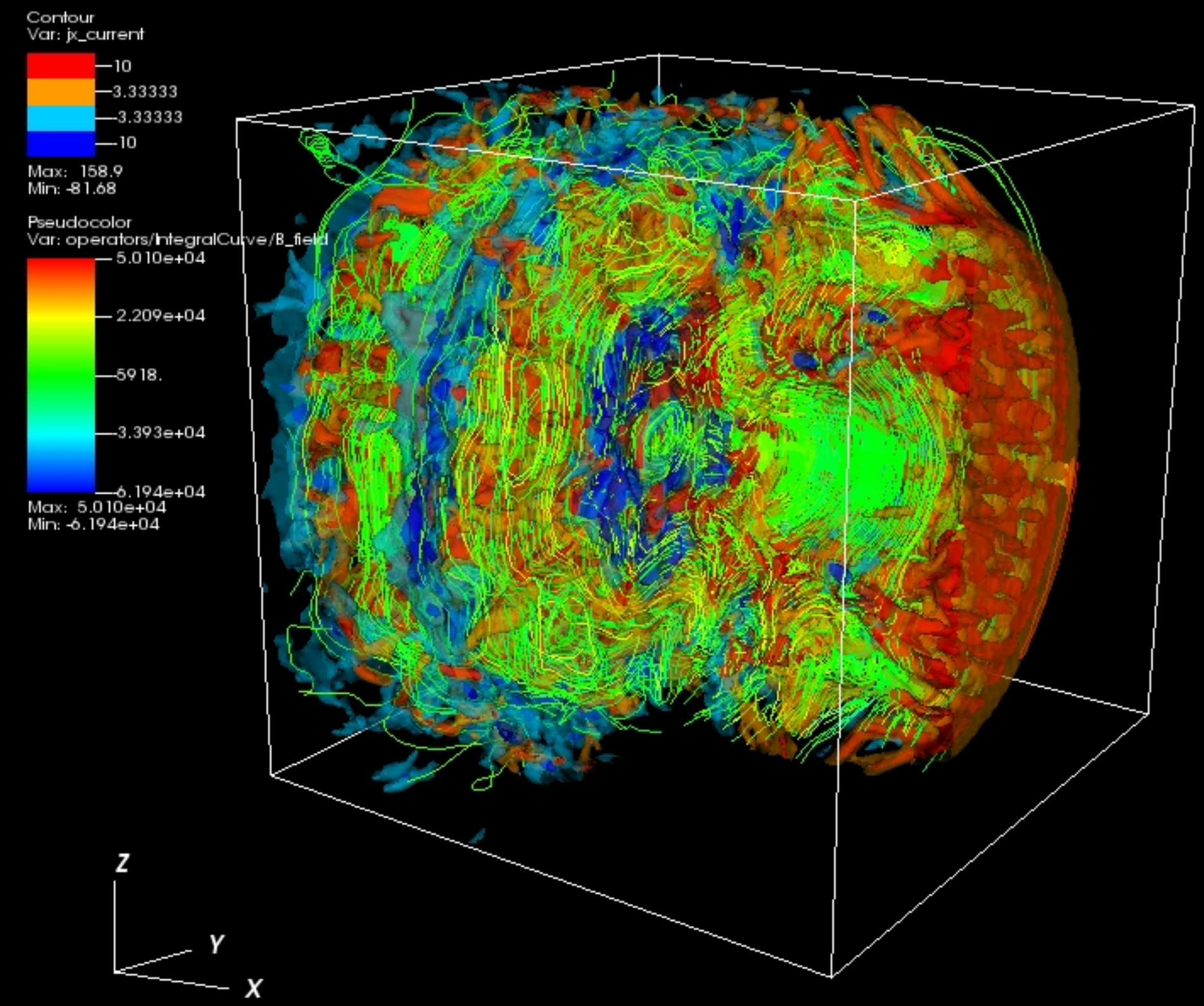}
\vspace*{-0.cm}
\caption{Panels show 3D iso-surface plots of  the current ($J_{\rm x}$) of jet electrons for $e^{-}$ {--}$p^{+}$ (a) and $e^{\pm}$ (b) ~jet with  $r_{\rm jt}=80\Delta$  at time  $t =  500 \omega_{\rm pe}^{-1}$. The lines show the magnetic field stream lines in the quadrant of the front part of jets.
The color scales for contour (upper left): red 10; orange  3.33; right blue $-3.33$. blue $-10$. The color scales of streaming lines (a) (2.89, 1.53, 0.174, $-1.2$, $-2.54$) $\times 10^{5}$; (b) (5.01, 2.21, $-0.592$, $-3.39$, $-6.19$) $\times 10^{4}$. Adapted from \cite{nishikawa17}.}
\label{3dJxBS}
\end{figure}

Furthermore, the 3D structures of averaged jet electron current ($J_{\rm x}$) are investigated  in the front
($420 \le x/\Delta \le 620$,  $152 \le y,z/\Delta \le 352$). 

Figure~\ref{3dJxBS} presents the current ($J_{\rm x}$) of $e^{-}$ {--}$p^{+}$ (a) and $e^{\pm}$ (b) ~jet. The cross sections at $x/\Delta = 520, y/\Delta=252$ and surfaces of jets show complicated patterns, which are generated by instabilities with the magnetic field lines. 

\begin{figure}[ht]
\hspace{4.3cm}(a) \hspace{5.5cm}(b)
\hspace*{0.cm}
\includegraphics[scale=0.35]{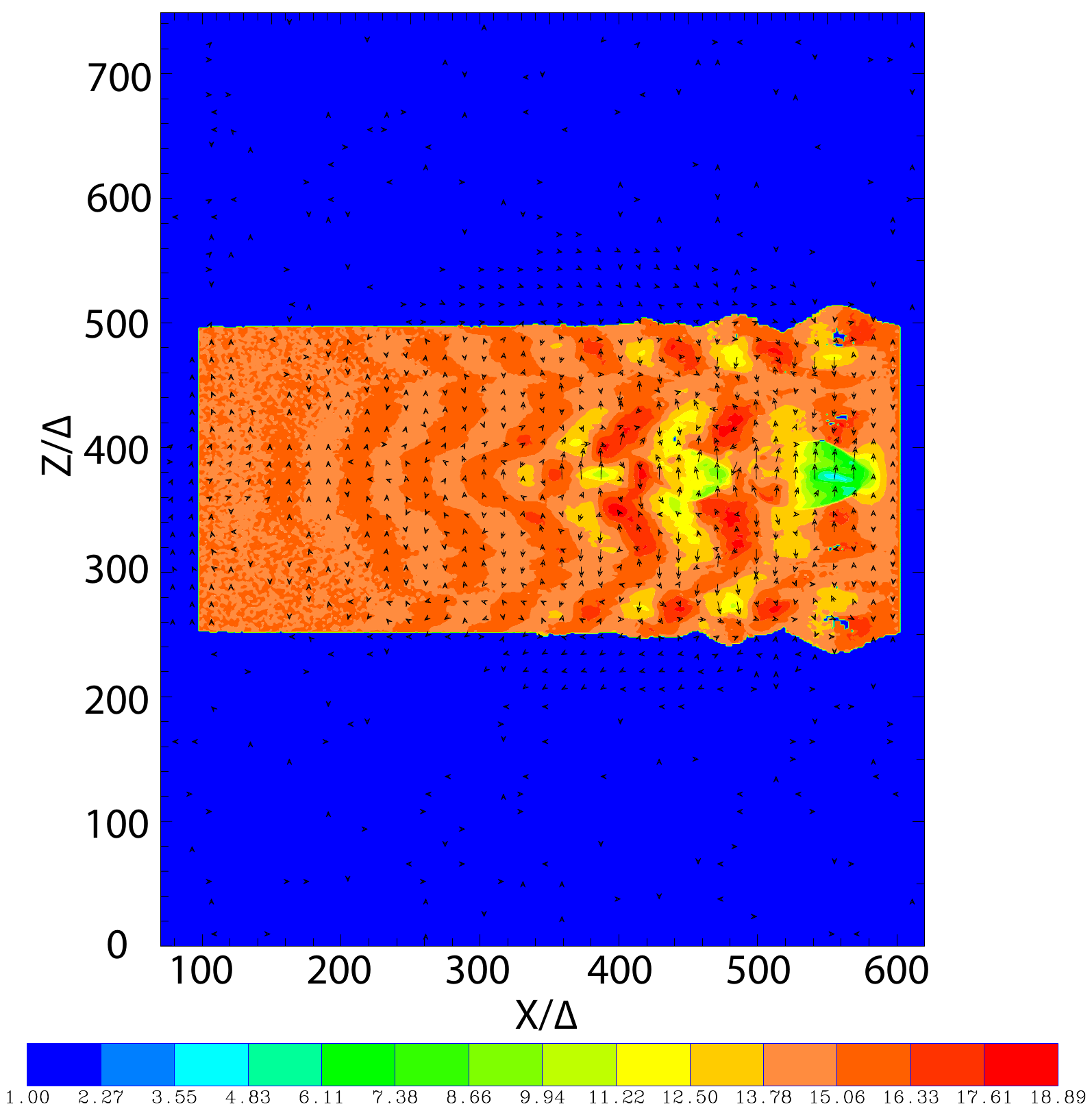}
\hspace*{0.1cm}
\includegraphics[,scale=0.35]{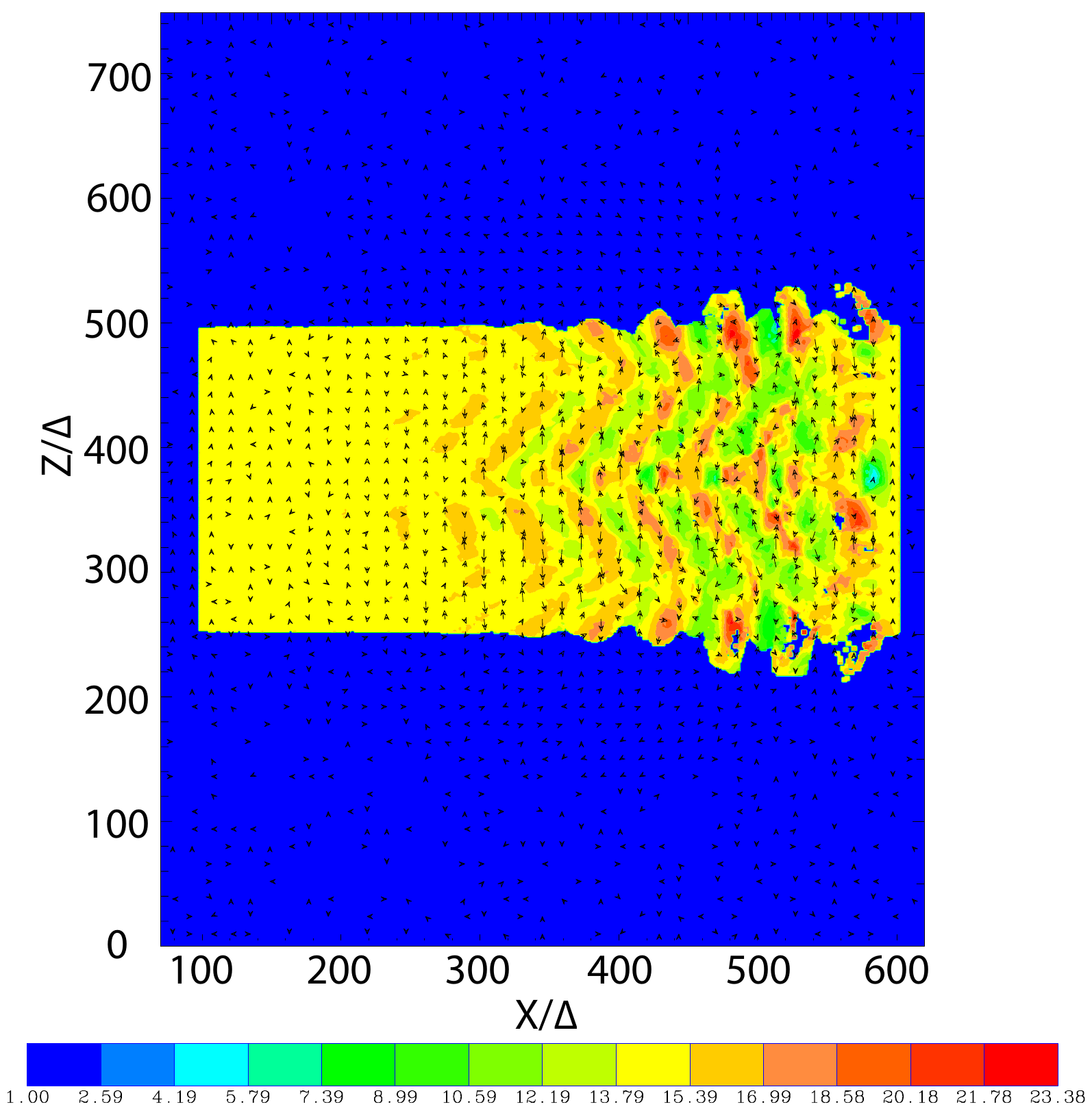}
\caption{Panels (a) and (b) show 2D plot of  the Lorentz factor of jet electrons for $e^{-}$ {--}$p^{+}$ (a) and $e^{\pm}$ (b)~jet with  $r_{\rm jt}=120\Delta$ at time  $t =  500 \omega_{\rm pe}^{-1}$. 
The arrows (black spots) show the magnetic fields in the $x-z$plane. 
Adapted from \cite{nishikawa17}.}
\label{lor2d}
\end{figure}

In order to determine particle acceleration the Lorentz factor of jet electrons in the cases with
$r_{\rm jt}=120\Delta$ is also calculated, as shown in Fig.~\ref{lor2d}. These patterns of Lorentz factor coincide with the changing directions of local magnetic fields that are generated by instabilities.
The directions of magnetic fields are indicated by the arrows (black spots), which can be seen with magnification. The directions of magnetic fields are determined by the generated instabilities.
The structures at the edge of jets are produced by the kKHI. 
The plots of Lorentz factor in the $y-z$ plane show the MI in the circular edge of the jets.

\begin{figure}[htb]
\hspace{5.cm}(a) \hspace{5.1cm}(b)  
\vspace*{-0.1cm} 
\hspace*{0.3cm}
\includegraphics[scale=0.25,angle=-90]{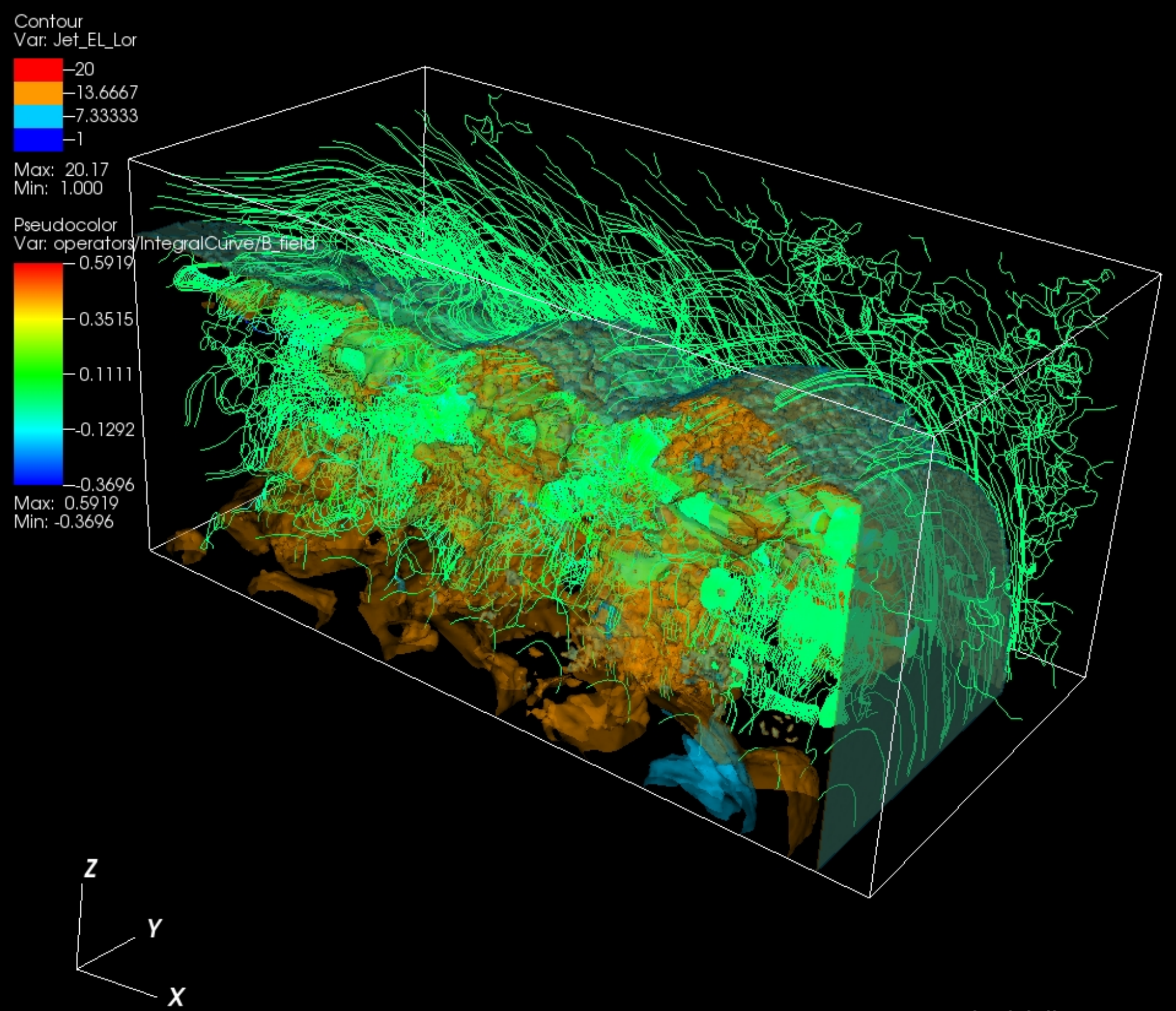}
\hspace*{-0.1cm}
\includegraphics[,scale=0.25,angle=-90]{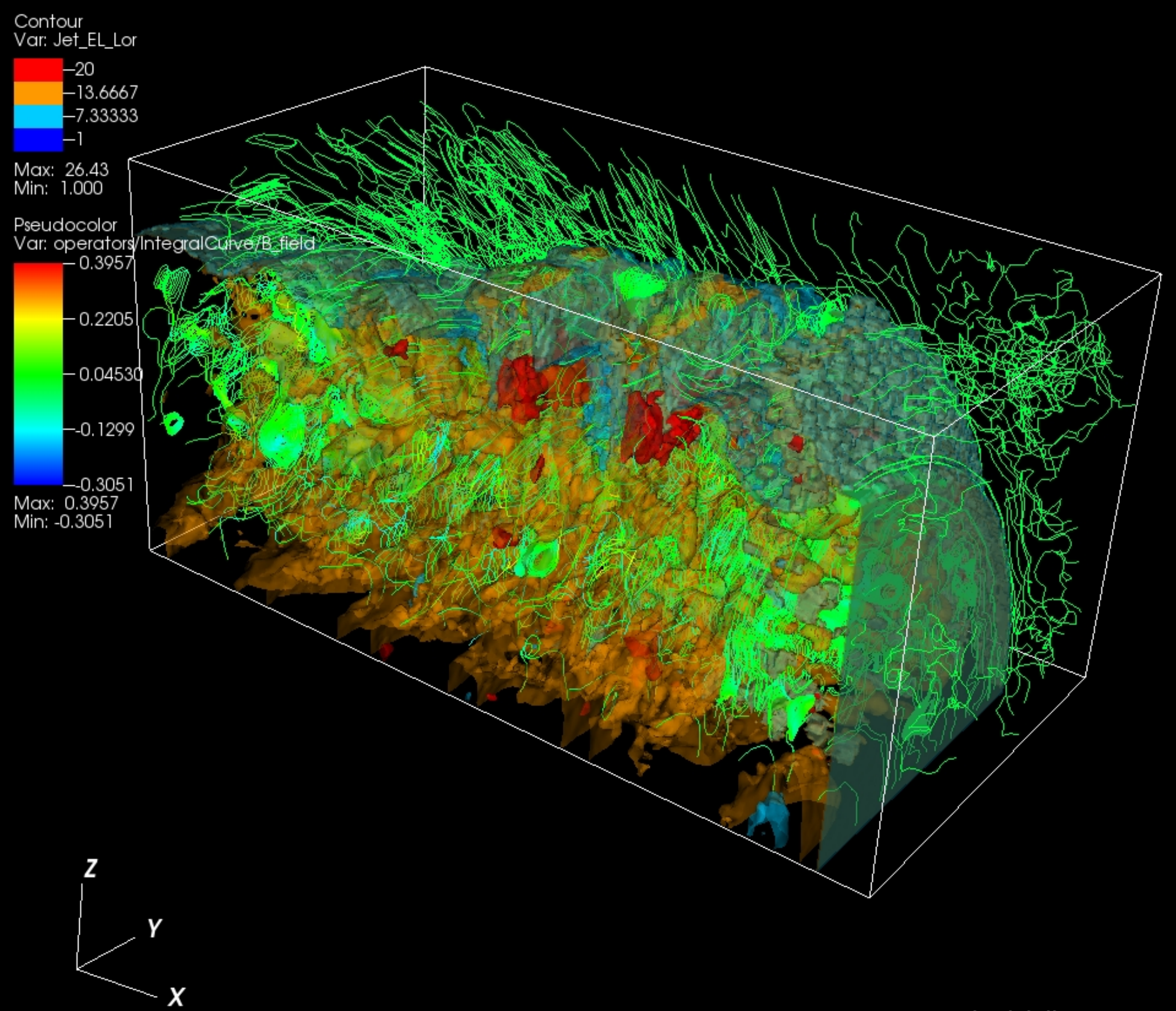}
\caption{Panels show 3D iso-surface plots of  the Lorentz factor of jet electrons for $e^{-} {--}p^{+}$ (a) and $e^{\pm}$ (b) ~jet with  $r_{\rm jt}=120\Delta$  at time  $t =  500 \omega_{\rm pe}^{-1}$. The lines show the magnetic field stream lines in the quadrant of the front part of jets. Adapted from \cite{nishikawa17}.}
\label{3dlorBS}
\end{figure}

The 3D structures of the averaged jet electron Lorentz factor is also investigated by plotting iso-surfaces of this factor in a quadrant of the jet front ($320 \le x/\Delta \le 620$,  $381 \le y,z/\Delta \le 531$),
%
for $e^{-}$ {--}$p^{+}$ (a) and $e^{\pm}$ (b) ~jet (Fig.~\ref{3dlorBS}). The cross sections and surfaces of jets show complicated patterns that are formed by instabilities with the magnetic field lines. 

In both plasma type cases where the jet radii is larger than $r_{\rm jt}=80\Delta$, kKHI and MI are generated at the jet surfaces, whereas inside the jets the Weibel instability is created with kink-like instability, in particular in the $e^{-}-p^{+}$ jet.
However, further investigations using different
plasma parameters (the magnetization factor), including the characteristic radius $a$, which determines the structure of helical magnetic fields in Eq.~(\ref{hmfcar}).  

One of the key questions is how the helical magnetic fields affect the growth of the kKHI, MI, and WI, and how and where in the jet structure particles are accelerated in the nonlinear stage. In the latter respect, of special interest is the role of magnetic reconnection, and its ability to aid in the rapid merging and breaking of the helical magnetic fields carried by relativistic jets.  
RMHD simulations demonstrate that jets with helical fields develop kink instabilities (KI) \cite[e.g.,][]{mizuno14,singh16,barniol17}, and similar structures were found in PIC simulations \cite[see, e.g.,][]{nishikawa19gal}.
PIC simulations of a single flux rope modeling the jet that undergoes internal KI showed signatures of secondary magnetic reconnection \citep{Markidis_2014}. 
Recently, it has also been demonstrated that the development of the KI in relativistic strongly magnetized jets with helical magnetic fields lead to the formation of highly tangled magnetic fields and a large-scale inductive electric field promoting the rapid energization of particles \citep{alves18eff,Alves19ion,Davelaar2020}. 

In particular, \citet{Davelaar2020} has studied the KI in non-rotating force-free jets using PIC simulations.
The magnetic field profile consists of a strong vertical field, $B_{\rm z}$, dominated core surrounded by a region dominated by a toroidal field component, $B_{\phi}$. 
The relative strength of the two components is set by the value of the magnetic pitch, $P = rB_{\rm z}/B_{\phi}$, on the axis.  
The radial profile of the pitch is important for the global evolution of the instability. In the case where the pitch is increasing with the cylindrical radius (IP), resonant surfaces confine the instability to the kink unstable core, while in the case of a decreasing pitch profile (DP) the instability becomes disruptive. 
They also consider a force-free setup, which has a non-monotonic pitch profile and a strong confining vertical magnetic field outside of the kink unstable core. 
They termed this profile as embedded pitch (EP).  In what follows, the initial cylindrical radius of the jet's core is defined as $r_{\rm core}$.

\begin{figure}[htb]
\includegraphics[width=\textwidth]{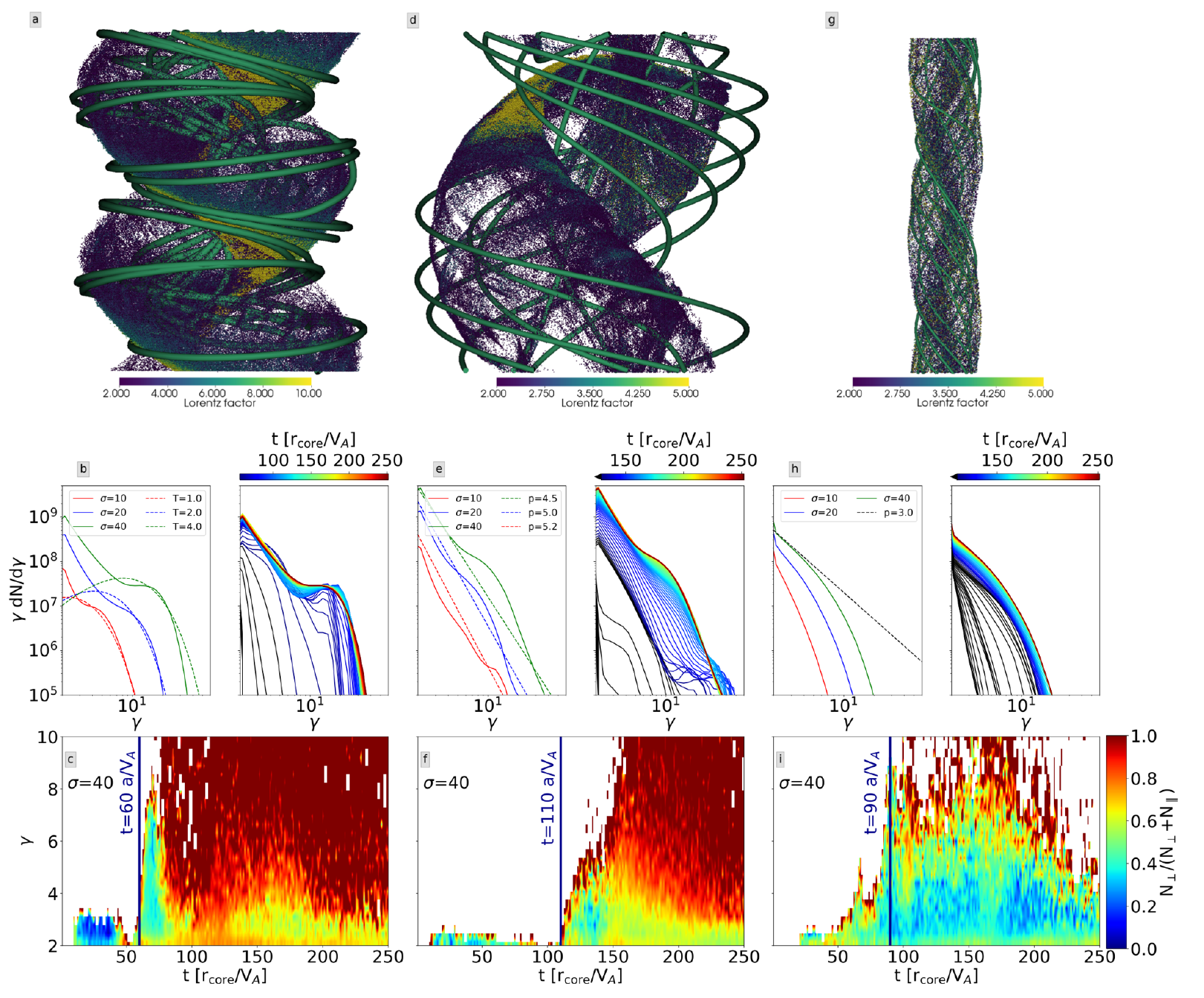}
\caption{From left to right: decreasing pitch (DP), increasing pitch (IP), and embedded pitch (EP) cases. In the top row thick green lines show magnetic field lines. 
Subsampled distribution of energetic particles is visualized as dots color-coded by their Lorentz factors. Plots are computed at $t =$ 60; 110; 90 $r_{\rm core}/V_{\rm A}$ correspondingly. Middle row shows DFs for all three setups, each set of two plots shows DFs at the end of the simulation on the left for all three $\sigma =$ 10; 20; 40 values, and the time evolution of the spectrum of the $\sigma =$ 40 run on the right. 
Panel b also includes Maxwellians fitted to the DFs, panel e and h show power-laws fitted to the DFs. Bottom row shows statistics of the acceleration events as a function of simulation time and particle energy. 
For a given particle at a particular energy they classify the acceleration episode based on if parallel or perpendicular electric field dominates particle energization. 
$N_{\parallel}$ and $N_{\perp}$ are the numbers of parallel and perpendicular acceleration events, respectively.
Adapted from \citet{Davelaar2020}.}
\label{Davelaar20xFig1}
\end{figure}

They have performed simulations in the frame comoving with the jet with three different magnetic field profiles (IP, DP and EP), thus the plasma is initially at rest. 
This means that no bulk flow in the simulations. Figure~\ref{Davelaar20xFig1} shows magnetic field lines, sub-sampled distribution of energetic particles, distribution functions (DFs), and time evolution of accelerated particles. 

They obtained similar overall evolution of the instability as found in MHD simulations \citep{mizuno09}. 
Magnetic reconnection and turbulence in collisionless plasma were studied so far in idealized periodic boxes. Their study shows how KI can be self-consistently excited and energize particles in the growth of KI in highly magnetized jets. 
\rv{They found that the acceleration in current sheets dominates for low particle energies and occurs due to strong non-ideal electric fields present at the magnetic reconnection sites. It leads to the formation of steep power-laws in the distribution function.}
While they observed plasmoid formation, their limited scale separation did not allow the formation of a full plasmoid chain, and to study the Fermi-like process of particle acceleration in
plasmoids. Later evolution shows heating of the plasma which is driven by weak turbulence
induced by the KI. These two processes energize particles due to a combination of ideal and non-ideal electric fields.

However, as an initial condition these simulations assumed helical magnetic fields supported by symmetrically streaming electrons and positrons (ions), corresponding to zero bulk flow of the pair plasma, across the simulation domain, therefore, the development of shocks at the head of jet and their propagation have not been simultaneously investigated. 
Furthermore, since no bulk flow is set as an initial setup in their simulation, no velocity-shear instability such as kKHI and MI will be excited in their simulations.
Therefore, only KI is excited in their simulations with periodic conditions.

\subsection{PIC simulations of magnetic reconnection}
\label{sec:5.4}

Magnetic reconnection is ubiquitous in solar and magnetosphere plasmas \citep[e.g.,][]{Lazarian2020,Marcowith2020}. It is proposed that it provides an important additional particle acceleration mechanism for AGN and GRB jets
\citep[e.g.,][]{giannios09,komissarov09,giannios10,Zhang_2010,Uzdensky11,granot11,granot12,mckinney12,komissarov12,sironi14,sironi15,sironi16,barniol16,petropoulou16,Werner18,Giannios19}


In order to study the importance of magnetic reconnection, researchers have performed different types of simulations including i) PIC simulations in the slab model \citep[e.g.,][]{Zenitani01,zenitani05prl,Zenitani07,zenitani08,oka08,Daughton11,Sironi_2011,kagan13,sironi14,karimabadi14,Guo_2015,Guo_2016,Guo_2019}, ii) resistive RMHD simulations \citep[e.g.,][]{komissarov07,zenitani10,takahashi11b,mizuno13,Baty13}, 
and iii) two-fluid RMHD simulations \citep[e.g.,][]{2009ApJ...696.1385Z,2009ApJ...705..907Z}. 

Two-fluid RMHD simulations are considered two fluid components, electron fluid and ion fluid 
and interaction between them. 
\cite{2009ApJ...696.1385Z,2009ApJ...705..907Z} have used two-fluid RMHD codes with generalized Ohm's law in order to avoid the limit of electron Debye length and skin depth which generally encounter the limitation in PIC simulations.  \citep{arnold19} have developed a new computational model, \texttt{kglobal}, to explore energetic electron production via magnetic reconnection in macro-scale systems. 
To convert this initial setup to a PIC simulation, they had to make sure that the smallest length scale in \texttt{kglobal} was much larger than the Debye length since this spatial scale is not resolved in \texttt{kglobal}. Thus they equated 
the transition width between the two regions of hot and cold electrons to 30 times the Debye length.

Investigations of magnetic reconnection have also been summarized in books \citep[e.g.,][]{Birn2007} and reviews \citep[e.g.,][]{lazarian12r,Kagan2015}.

An important point is that in spite of the extensive research on magnetic reconnection, most of all PIC simulations have been performed with the Harris sheet \citep{harris62}, where the unperturbed magnetic fields ${\bf B}$ are anti-parallel (${\bf B} = -\tanh(x){\bf e}_{\rm y}$) and are generated by a current sheet. 
For example, in the 3D slab model, \citet{Guo_2015} have demonstrated that relativistic magnetic reconnection is highly efficient at accelerating particles through a first-order Fermi process accomplished by the curvature drift of particles along the electric field induced by the relativistic flows. 
They initiate a plasma of electrons and positrons in a current layer with $\vec{B} \sim \tanh(z)\vec{e}_x + \textnormal{sech}(z)\vec{e}_y$ and perform PIC simulations of the acceleration of these leptons during magnetic reconnection in different domain sizes and with different electron densities, i.e. plasma frequencies. 

\begin{figure}[htb]
\includegraphics[width=\textwidth]{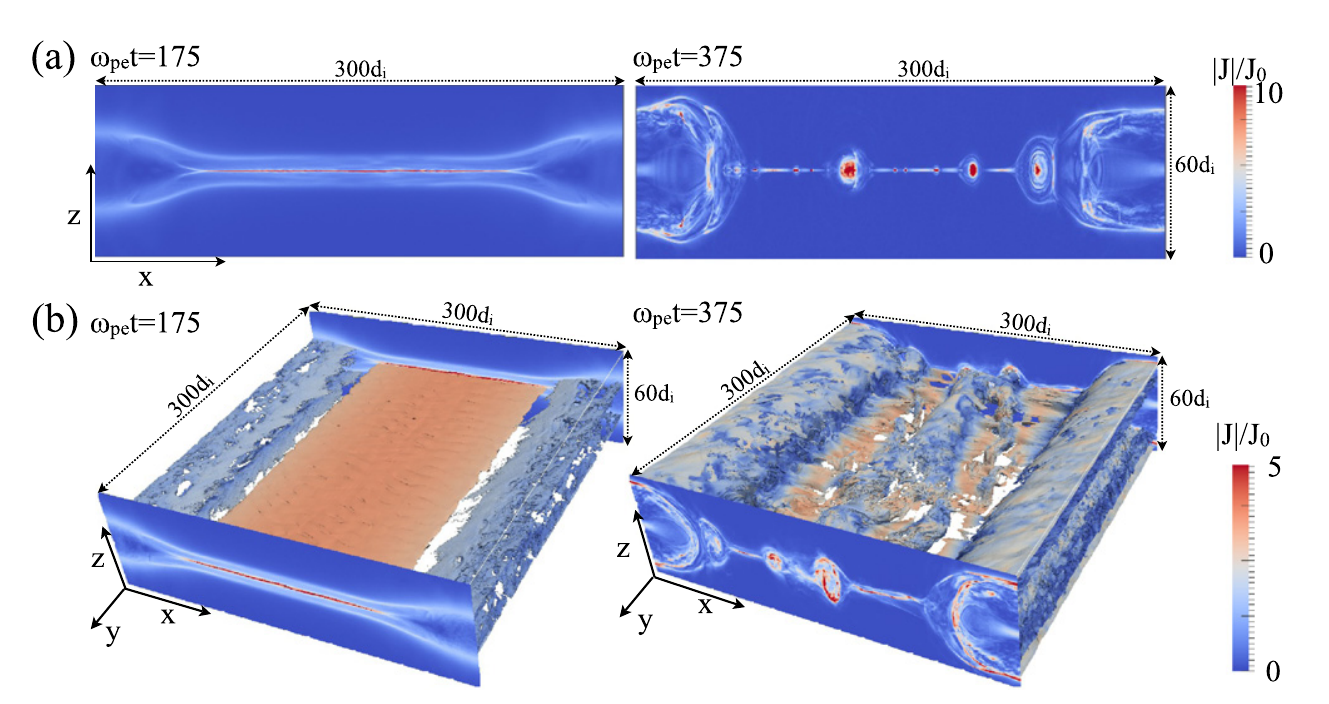}
\caption{Evolution of 2D and 3D simulations of an electron-positron plasma 
field $\vec{B} \sim \tanh({\rm z})\vec{e}_{\rm x} + {\rm sech}({\rm z})\vec{e}_{\rm y}$
with $\sigma = 100$ and domain size $L_{\rm x}\times L_{\rm z}= 300d_{\rm i} \times 194d_{\rm i}$\, ($L_{\rm y}=300d_{\rm i}$ 
for 3D): (a) 2D: color-coded current density from at $\omega_{\rm pe}t = 175$ and $\omega_{\rm pe}t = 375$; (b) 3D: current density and a 3D isosurface of the plasma density colored by the current density at $\omega_{\rm pe} t = 175$ and $\omega_{\rm pe}t = 375$. Image reproduced with permission from \cite{Guo_2015}, copyright by AAS.}
\label{3drecon}
\end{figure}

As an example, Fig.~\ref{3drecon} gives an overview of the current layer evolution where  $\sigma = 100$ and domain size $L_{\rm x}\times L_{\rm z}= 300d_{\rm i}\times 194d_{\rm i}$ for 2D and $L_{\rm x}\times L_{\rm z}= 300d_{\rm i}\times 300d_{\rm i}$ for 3D simulations.
For 2D, panel (a) shows the color-coded current density, and for 3D, panel (b) shows a 2D slice of the current density and a 3D isosurface of plasma density colored by the current density at $\omega_{\rm pe}t = 175$ and $\omega_{\rm pe}t = 375$, respectively. Starting from an initial perturbation, the current sheet becomes smaller as the current density is concentrated in the middle of the domain. In 2D, due to subsequent tearing instabilities, the extended thin current sheet breaks into a number of fast moving plasmoids ($\omega_{\rm pe}t \approx 225$) which finally merge into a single island at the edge of the simulation domain. However, in 3D, the KI develops and interacts with the tearing mode \citep{daughton99}, such that a turbulent evolution is initiated  \citep{yin08}. 
However, although strong 3D effects modify the current layer, small-scale structures develop as a result of secondary tearing instabilities.
Therefore, the energization of particles in 2D and 3D simulations moving during magnetic reconnection is very similar and sufficiently strong to create sources of non-thermal energetic particles. 

The importance of magnetic reconnection in jets has been proposed 
\citep[e.g.,][]{giannios09}. 

\begin{figure}[htb]
\begin{minipage}{65.0mm}
\hspace{0.cm}
\includegraphics[scale=0.68,angle=0]{{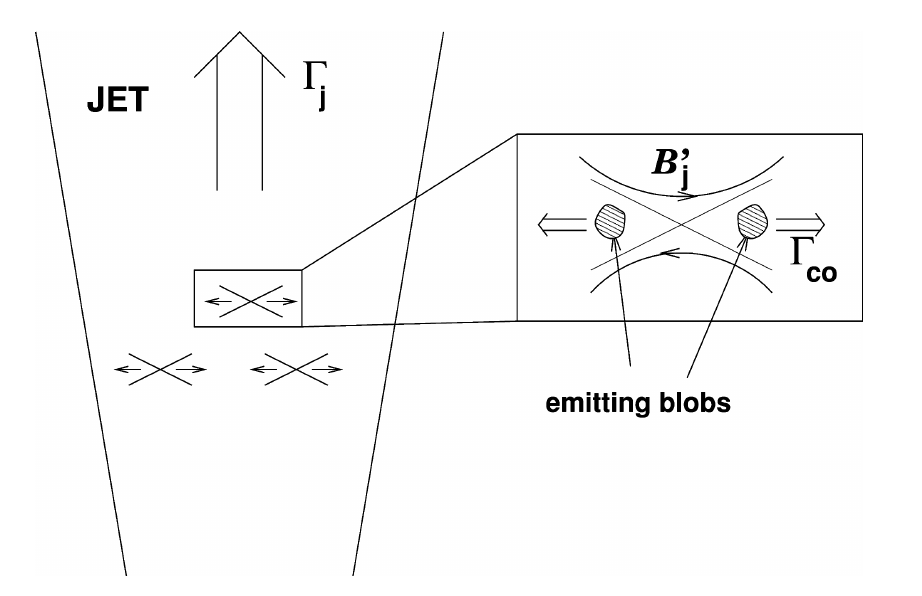}}
\end{minipage}
\begin{minipage}{48.0mm}
\caption{
Schematic representation of the geometry of the `jets in a jet' {\it shown in a frame co-moving with the jet}. Right: the reconnection region enlarged. Plasma heated and compressed by magnetic reconnection leaves the reconnection region at relativistic speed ${\rm \Gamma}_{\rm co} \gg1$ within the jet in the form of blobs. Each blob emits efficiently through synchrotron-self-Compton in a narrow beam within the jet emission cone, powering a fast evolving soft X-ray and TeV flare. 
Image reproduced with permission from \cite{giannios09}, copyright by the authors.} 
\label{giaf1}
\end{minipage}
\end{figure}
%
Based on the jets-in-jet scenario (see Fig.~\ref{giaf1}), further investigations have been done \cite[e.g.,][]{barniol16}. 
\rv{Such reconnection and plasmoid models based on 2D PIC simulations have been used calculate the $\gamma$-ray radiation in relativistic jets
\cite[e.g.,][]{Giannios2013,sironi15,petropoulou16,Petropoulou2018,christie19}.}
These investigations encourage them to simulate relativistic jets with helical magnetic fields, which may excite (kinetic) KI and, consequently, 
magnetic reconnection occurs. 



It should be noted that in jets the magnetic reconnection does not start in the Harris model 
in the slab geometry. Therefore the embedded figure is somehow misleading. One need to 
investigate how the magnetic reconnection occurs in helical magnetic fields (not simple 
anti-parallel magnetic fields in the slab model).

\subsubsection{Reconnection in jets as a possible mechanism for high energy flares} 
\label{sec:5.4.1}

Since the particle acceleration in shocks generated by the Weibel instability and other kinetic instabilities such as kKHI and MI is 
not enough to accelerate particles to very high energies \citep[e.g.,][]{nishikawa09,nishikawa16a}, 
the importance of reconnection in jets has been proposed  
\citep[e.g.,][]{giannios09}. 
Extensive simulation studies have been performed in a slab model with the Harris configuration \citep{harris62}.  
Since the Poynting-flux is stored in helical magnetic fields with currents parallel to the jet direction, the particle acceleration performed in the Harris configuration with sheet currents cannot be applied to the particle acceleration in global jets. 
Furthermore, no kinetic simulations of global jets with helical magnetic fields have been performed previously except for their own simulations \citep{nishikawa16b,nishikawa17}. 

The simulated jet has a radius $r_{\rm jt}$  and is assumed to propagate in an initially unmagnetized ambient medium. For the magnetic fields external to the jet they use a damping function, $\exp{[-(r-r_{\rm jt})^2/b]} \, (r\geq r_{\rm jt})$, that multiplies the expressions in Equation~(\ref{hmfcar}) with the tapering parameter $b=200\Delta$, where $\Delta $ is the grid scale. 
They further assume a characteristic radius $a =  0.25*r_{\rm jt}$. 
The profiles of the resulting helical magnetic field components are shown in Figure~6b in \cite{nishikawa19gal}. The toroidal magnetic field vanishes at the center of the jet (red line in  Fig.~6b of \cite{nishikawa19gal}). 
Note, that the simulation setup adopted in this work has been used in their preliminary studies of helical jets \citep{nishikawa16b,nishikawa17,nishikawa19gal} with the modifications $B_0 = 0.1$ and $r_{\rm jt} =100$.

The jet head assumed here has a flat-density top-hat shape which is a strong simplification of the true structure of the jet-formation region \cite[e.g.,][]{broderick09,moscibrodzka17}. 
A more realistic jet structure such as a Gaussian shape will be implemented in future studies.


\cite{Nishikawa2020} have performed new PIC simulations with larger a numerical grid of size \mbox{$(L_{x}, L_{y}, L_{z}) = (1285\Delta, 789\Delta, 789\Delta)$} in Cartesian coordinates.
They use open boundaries at the $x/\Delta=0$ and $x/\Delta=1285$ and impose periodic boundary conditions in the transverse directions. 
Since the jet is located at the center of the numerical box far from the boundaries and the simulation time is still shorter than the propagation time to tangential direction of jet, there are no visible effect of the periodic boundary conditions on the system. 

The jet has a radius with $r_{\rm jt} = 100\Delta$ and is injected at $x = 100\Delta$ in the center of the $y-z$ plane. The large computational box allows to follow the jet evolution long enough to permit the investigation of the strongly nonlinear stage. The longitudinal box size, $L_{x}$, and the simulation time, $t_{\rm max}=1000\omega_{\rm pe}^{-1}$, are a factor of two larger than in their previous studies \citep{nishikawa16b,nishikawa17,nishikawa19gal}, in which jets with radii $r_{\rm jt} = 20, 40, 80, 120\Delta$ were investigated.

Jets of different plasma composition exhibit distinct dynamical behavior that manifests itself in the morphology of the jet and its emission characteristics \citep{nishikawa16a,nishikawa16b,nishikawa17,nishikawa19gal}.
However, \cite{Nishikawa2020} have performed only electron-proton plasma in both the jet and the ambient medium. The initial particle number per cell is $n_{\rm jt}= 8$ and $n_{\rm am} = 12$, respectively, for the jet and ambient plasma. 
The electron skin depth is 
$\lambda_{\rm se} =  c/\omega_{\rm pe} = 10.0\Delta$, $\omega_{\rm pe} = (e^{2}n_{\rm
am}/\epsilon_0 m_{\rm e})^{1/2}$ is the electron plasma frequency;
the electron Debye length of ambient electrons is $\lambda_{\rm D}=0.5\Delta$. 
The thermal speed of jet electrons is $v_{\rm jt,th,e} = 0.014c$ in the jet reference frame; 
in the ambient plasma it is $v_{\rm am,th,e} = 0.03c$. Temperature equilibrium is assumed, 
thus the thermal speed of ions is smaller than that of electrons by the factor 
$(m_{\rm p}/m_{\rm e})^{1/2}$. In the simulation, the realistic proton-to-electron mass ratio is used $m_{\rm p}/m_{\rm e}=1836$. 
At the initial state, The jet plasma is weakly magnetized, and the magnetic field amplitude parameter assumed, $B_{0}=0.1c$, corresponds to plasma magnetization 
\mbox{$\sigma = B^2/n_{\rm e}m_{\rm e}\gamma_{\rm jt}c^{2} =2.8\times 10^{-3}$}. 
The jet Lorentz factor is $\gamma_{\rm jt}=15$.

\begin{figure*}[htb]
\includegraphics[width=\textwidth]{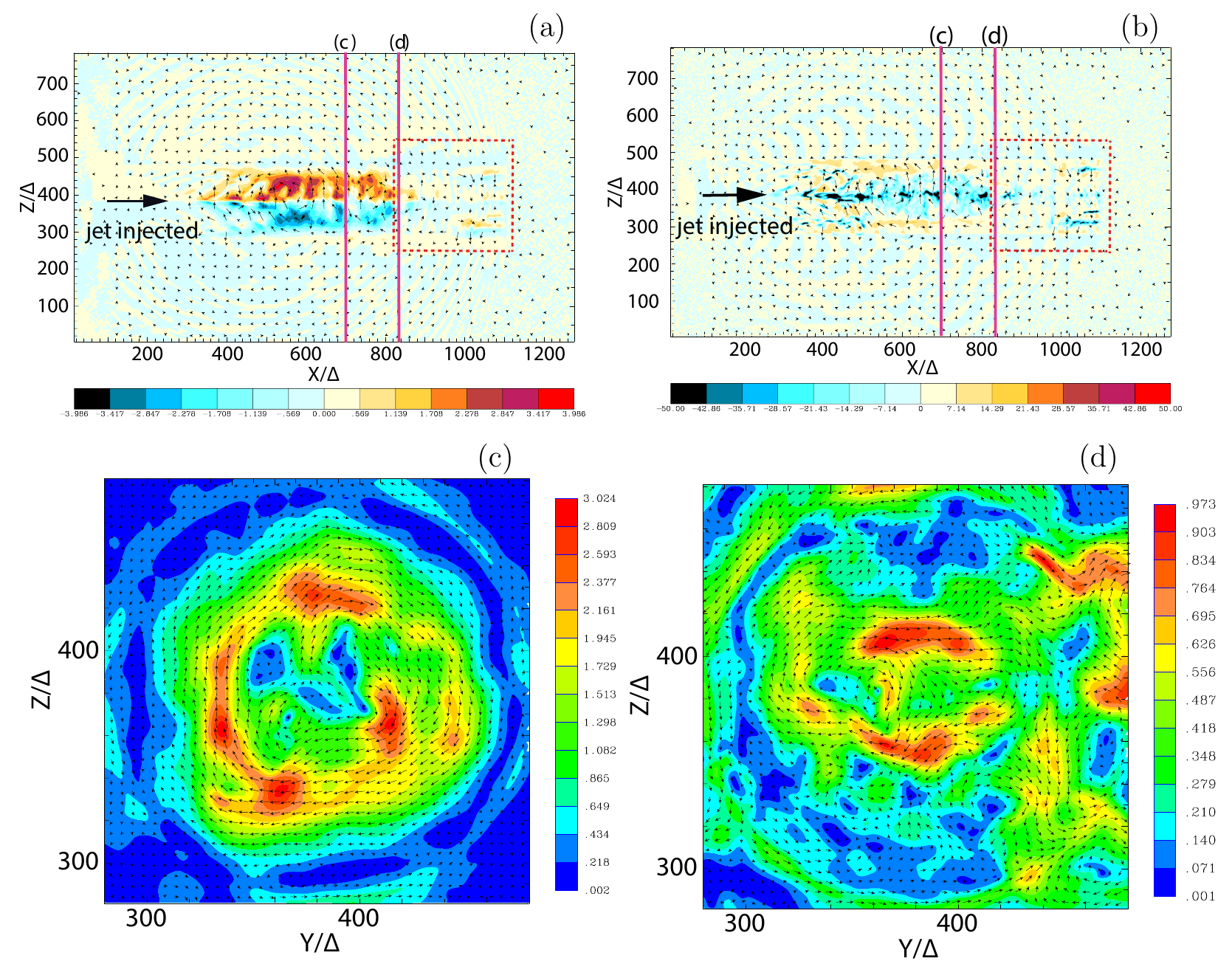}
\caption{Upper panels: (a) the $y$-component of the magnetic field, $B_{\rm y}$, with $x$-$z$ electric field depicted by arrows, and (b) the $x$-component of the electron current density, $J_{\rm x}$, with the $x$-$z$ magnetic field 
as arrows, both in the $x-z$ plane at $t =1000\, \omega_{\rm pe}^{-1} $. 
The lower panels show the total magnetic-field strength in the $y-z$ plane at $x/\Delta=700$ (c) and $x/\Delta=835$ (d). The arrows indicate the magnetic field ($B_{\rm y}, B_{\rm z}$). Adapted from \cite{Nishikawa2020}.}
\label{JxB}
\end{figure*}

Since simulation domain in PIC simulations is large enough to capture macroscopic instabilities, a kink-like instability was found in the pair and 
$e^{-} - p^{+}$ jet cases with jet radius $r_{\rm jt} =120 \Delta$
\citep[e.g.,][]{nishikawa17,nishikawa19gal}.

Figure~\ref{JxB} shows cross-sections through the center of the jet at time $t =1000\, \omega_{\rm pe}^{-1}$ with (a) the $y$-component of the magnetic field, $B_{\rm y}$, with the $x$-$z$ electric field as arrows, and (b), the $x$-component of the electron current density, $J_{\rm x}$, with the $x$-$z$ magnetic field depicted as arrows. The jet propagates from the left to right. 
Very strong helical magnetic field in the jet is evident at $400\lesssim x/\Delta \lesssim 830$. The amplitude is with $B/B_0\approx 40$ (Fig.~\ref{JxB}a) much larger than that of the initial field. 
As in the unmagnetized case \citep{nishikawa16a}, this field results from MI and kKHI. However, in the presence of the initial helical magnetic field the growth rate of the transverse MI mode is reduced, and the field structure is strongly modulated by longitudinal kKHI wave modes as shown by the bunched  $B_{\rm y}$ field (Fig.~\ref{JxB}a). 
This causes multiple collimation along the jet by pinching of the jet electrons toward the center of the jet, as visible in the electron current density (Fig.~\ref{JxB}b). 
It should be noted that in Fig~\ref{JxB}b the color scale for $J_{\rm x}$ does not extend beyond $J_{\rm x} = -50$, as they intend to show the weak positive (return) current. 
The collimation becomes weaker further along the jet and eventually disappear at  $x/\Delta\gtrsim 830$. 
At this point $B_{\rm y}$ is weak. This demonstrates that the nonlinear saturation of the MI leads to the dissipation of the helical magnetic field.

They selected possible magnetic reconnection sites in Figures~\ref{JxB}a and \ref{JxB}b, indicated by the two red lines at $x/\Delta=700$ and $x/\Delta=835$, and show the magnetic-field structure in the $y-z$ plane in Figures~\ref{JxB}c and \ref{JxB}d, respectively. 
At $x/\Delta=700$ clockwise-circular magnetic field is split near the jet into a number of magnetic structures, which demonstrate the growth of MI. 
They are surrounded by field of opposite polarity that is produced by the proton current framing the jet boundary \citep[see][]{nishikawa16a}. The magnetic field at $x=835\Delta$ is strongly turbulent; its helical structure is distorted and reorganised into multiple magnetic islands, which reflect the nonlinear stage of MI and kKHI. 
The  magnetic islands interact with each other, providing conditions for magnetic reconnection. In their 3D geometry, magnetic reconnection does not occur at a simple $X$-point as in 2D slab geometry. 
Instead, reconnection sites can be identified with regions of weak magnetic field surrounded by oppositely directed magnetic field lines. 
An example of a possible location of magnetic reconnection can be found at $(y/\Delta, z/\Delta)=(380, 340)$, where the total magnetic field becomes minimal (the null point,
Fig.~\ref{JxB}d. The evolution of the magnetic field at different locations in the jet ($600 < x/\Delta < 1100$) is shown in the supplemental movie. 
Note that the filamentary structure at the jet head (Figs.~\ref{JxB}a and \ref{JxB}b) is formed 
by the electron WI. One can see in the movie that nonlinear evolution of the filaments also leads to the appearance of the magnetic structures that are prone to magnetic reconnection.
\rv{The complex structures of 3D reconnections have been investigated
in \cite[e.g.,][]{Lalescu2015,Lazarian2020, Borissov2020}. In order to determine the reconnection 
locations analytically, we would need to investigate the eigenvalues of Jacobian matrix, which is beyond the scope of this work, for more details, see [e.g.,][]{cai07}.} 

Recently \cite{Dong20} show complimentary RHMD simulations of jets compared with PIC
simulations.
Relativistic jets are launched with a significant amount of magnetic energy, which can
be dissipated to accelerate non-thermal particles and give rise to electromagnetic radiation at larger scales. 
KI can be an efficient mechanism to trigger dissipation of jet magnetic energy. 

In \cite{Dong20}, the simplified central engine is assumed to be a perfectly conducting sphere of radius $r_0 = l_0$, where $l_0$ is the code length unit, threaded by radial magnetic field lines.
The jet is highly magnetized at its base, where the initial magnetization factor $\sigma$, which is the magnetic energy density
over enthalpy, is $\sigma_0 = 25$. The sphere is rotating at a constant angular frequency of $\Omega =0.8c/r_0$. This is mimicking a supermassive black hole and its accretion disk rotation.
The rotation then coils up the radial magnetic field lines into helical structure and launch two oppositely directed magnetized jets into the surrounding medium. Initially, the surrounding
medium is cold and static. In the neighborhood of the central engine, the surrounding medium is expected to be dominated by winds from the accretion disk, resulting in a steep density profile, but at larger distances the interaction
between jet/wind and the interstellar medium can become important resulting in a possibly flatter density profile. 
This motivates their assumed broken power-law density profile for the gas, where the break point is at radius $r_{\rm B} = 100l_0$, so that
\begin{eqnarray}
\rho =
\begin{cases}
\rho_0(\frac{r}{r_0})^{-3} & , r < r_{\rm B} \nonumber \\
\rho_0(\frac{r_{\rm B}}{r_0})^{-3}(\frac{r}{r_{\rm B}})^{-1} & , r \ge r_{\rm B}.
\end{cases} \label{rho}
\end{eqnarray}

\vspace*{-0.cm}
\begin{figure}[htb]
\vspace*{-0.1cm} 
\begin{minipage}[t]{68mm}
\hspace*{0.0cm}
\includegraphics[scale=0.77,angle=0]{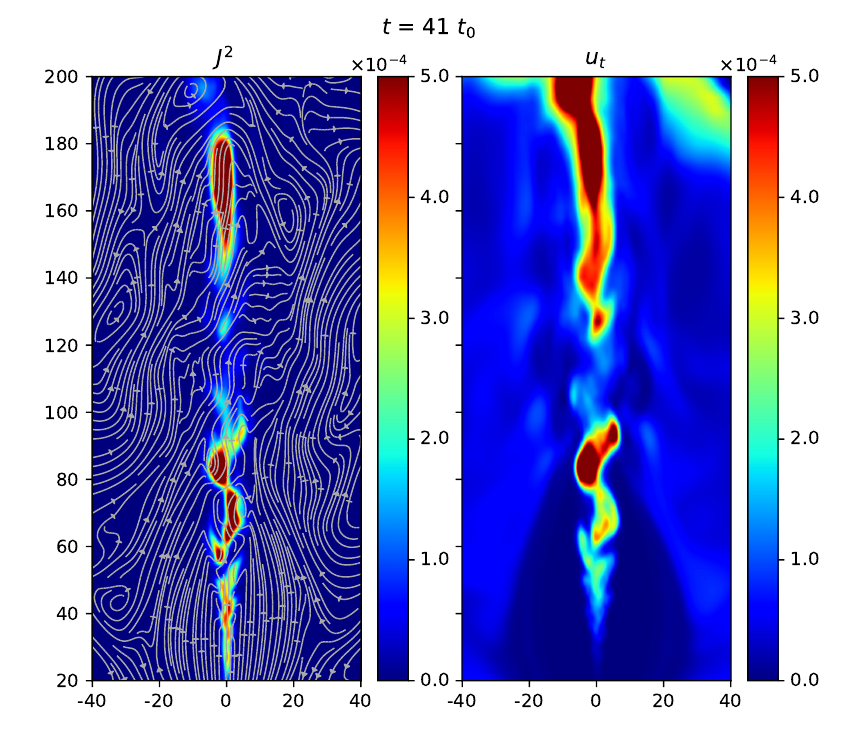}
\end{minipage}
\begin{minipage}[t]{48mm}
\vspace*{-5.0cm}
\caption{A snapshot of the jet at $t = 41t_{0}$ from $r = 20 l_{0}$ to $r = 200 l_{0}$ in jet propagation direction, with a radius of 40$l_{0}$. At this snapshot, the jet head has mostly moved out of $r = 200 l_0$.
Left Panel: the current density with magnetic fields lines projected on the same plane. The twisted structures indicate the location of kink instabilities. Right Panel: thermal energy density. Adapted form Figure 1 in \citet{Dong20}}
\label{Dong20fig1kink}
\end{minipage}
\end{figure}

They aimed to self-consistently study radiation and polarization signatures from KI in relativistic jets. They combine RMHD simulations with polarized radiation transfer of a magnetized jet, which emerges from the central engine and propagates through the surrounding medium. They observed that a localized region at the central spine of the jet exhibits the strongest KI, which identified as the jet emission region as shown in Fig.~\ref{Dong20fig1kink}. 
Very interestingly, quasi-periodic oscillation (QPO) signatures are seen in the light curve from the emission region. Additionally, the polarization degree appears to be anti-correlated to ares in the light curves. 
Their analyses show that these QPO signatures are intrinsically driven by KI, where the period of the QPOs is associated to the kink growth time scale. 
The latter corresponds to weeks to months QPOs in blazars. The polarization signatures offer
unique diagnostics for QPOs driven by KI.

It should be noted that in RMHD simulations kinetic instabilities are not included, therefore the development of the KI in relativistic strongly magnetized jets with helical magnetic field
leads to the formation of highly tangled magnetic field as shown in Fig.~\ref{Dong20fig1kink}.
Since kinetic KHI and MI grow faster than current-driven kink instabilities due to strong velocity shear and moderate magnetization.
PIC simulation\rv{s} \rv{of relativistic jets} do not show a kink-like instability as seen in the simulations of \citet{Dong20}. In PIC simulation the particles are accelerated by the turbulent magnetic reconnection that is initiated
by the growth of kinetic instabilities such as kKHI and MI\citep{Nishikawa2020}.


\begin{figure}[htb]
\begin{minipage}{65.0mm}
\hspace{0.2cm}
\includegraphics[scale=0.3,angle=0]{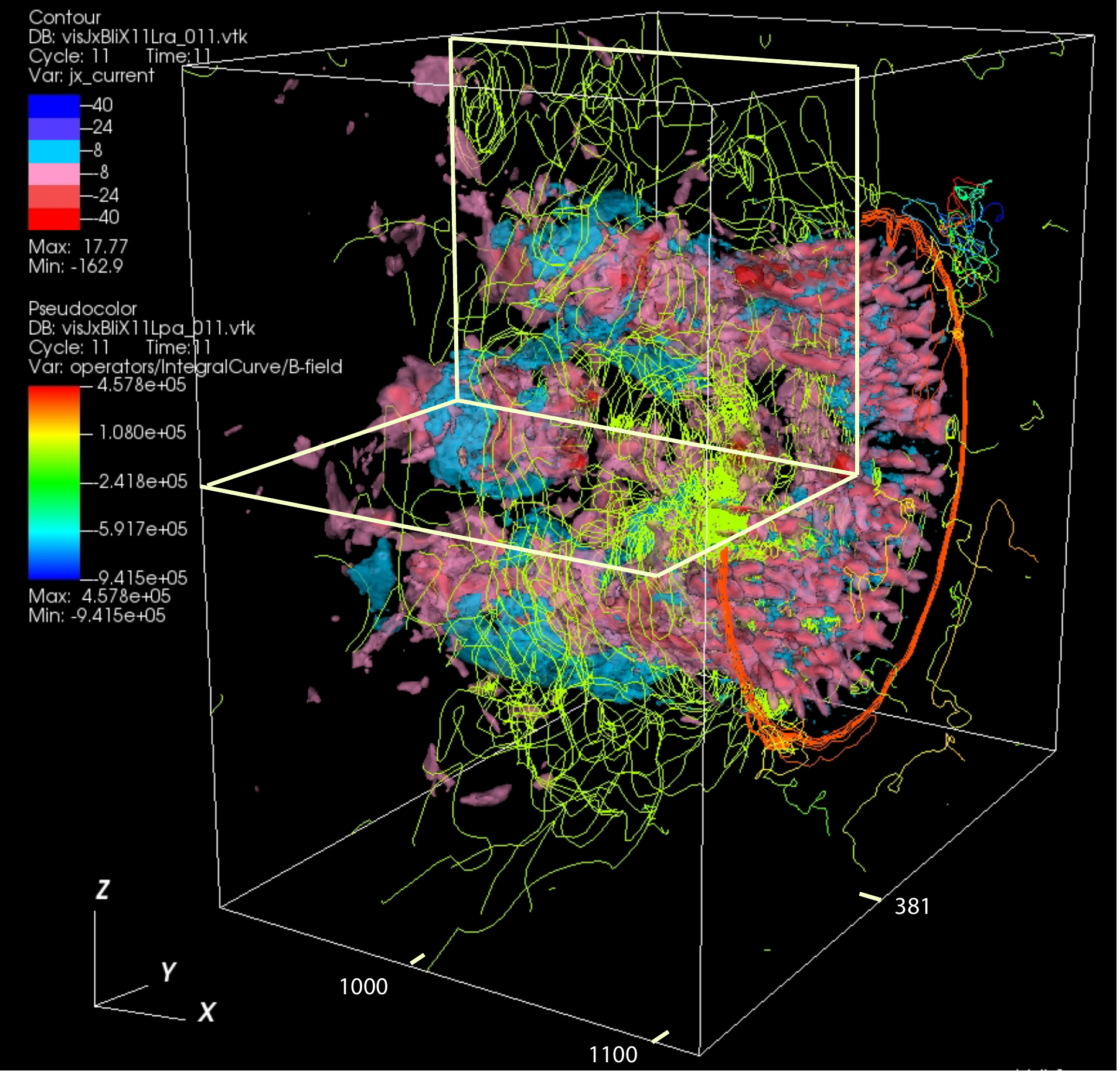}
\vspace{0.1cm}
\end{minipage}
\begin{minipage}{48.0mm}
\caption{Isosurface of the $x$-component of the electron current density, $J_{\rm x}$, with  the magnetic field lines (yellow) in a rectangular section of the simulation grid ($920 < x/\Delta < 1120; 231 < y/\Delta, z/\Delta < 531$)  
at time $t =1000\, \omega_{\rm pe}^{-1} $. To illustrate the distribution of $J_{\rm x}$ inside the jet, a quadrant of the regions is clipped at the center of the jet in the $x-z$ plane ($381 < z/\Delta < 531$) and in the $x-y$ plane ($231 < y/\Delta < 381$). The jet front is located at $x/\Delta=1100$. Adapted from {\citet{Nishikawa2020}}.}
\label{3DJX}
\end{minipage}
\end{figure}

\begin{figure}
\begin{center}
\hspace{3.0cm} (a) \hspace{3.30cm} (b) 
\includegraphics[scale=0.26]{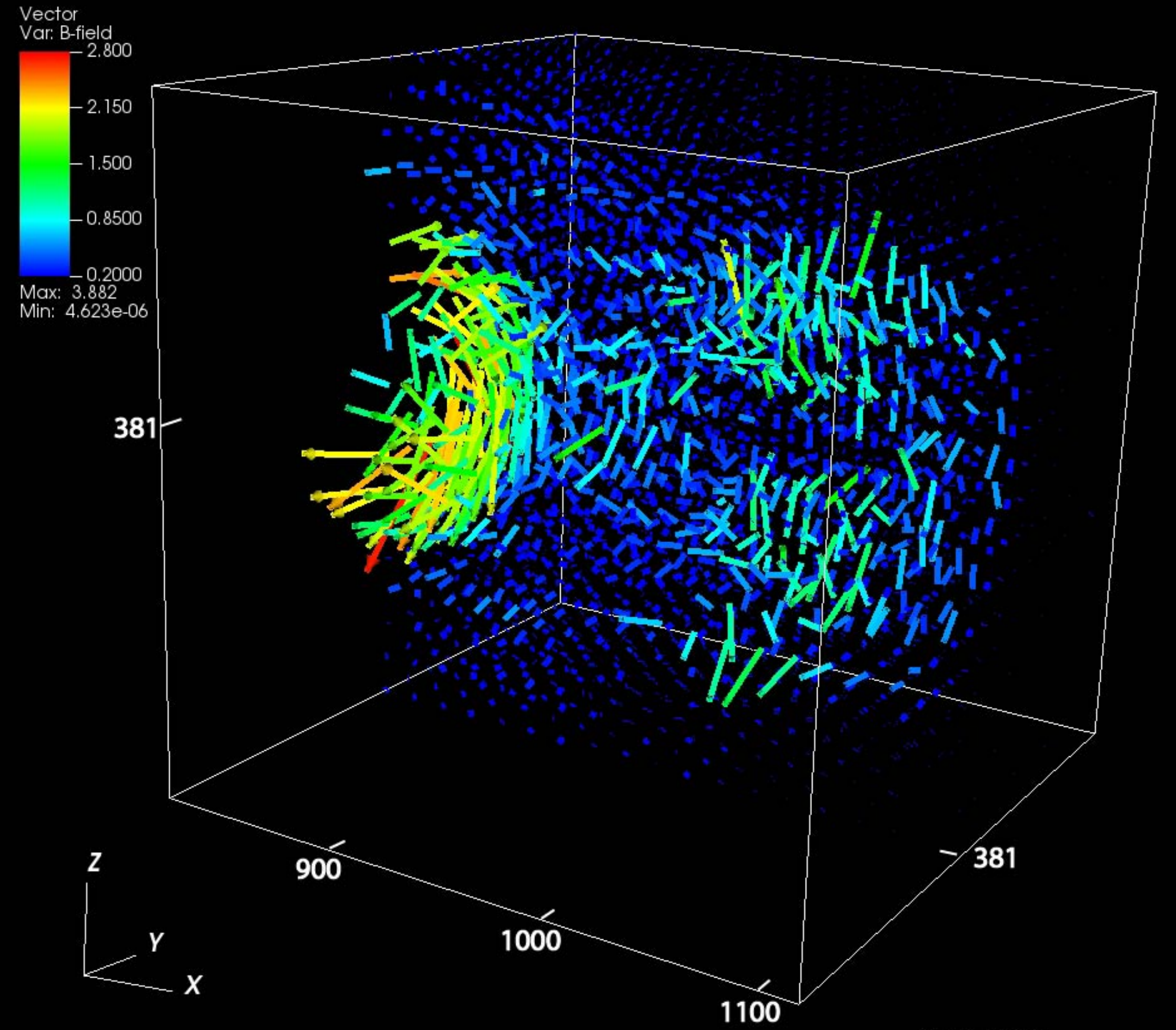}
\hspace{0.1cm}
\includegraphics[scale=0.26]{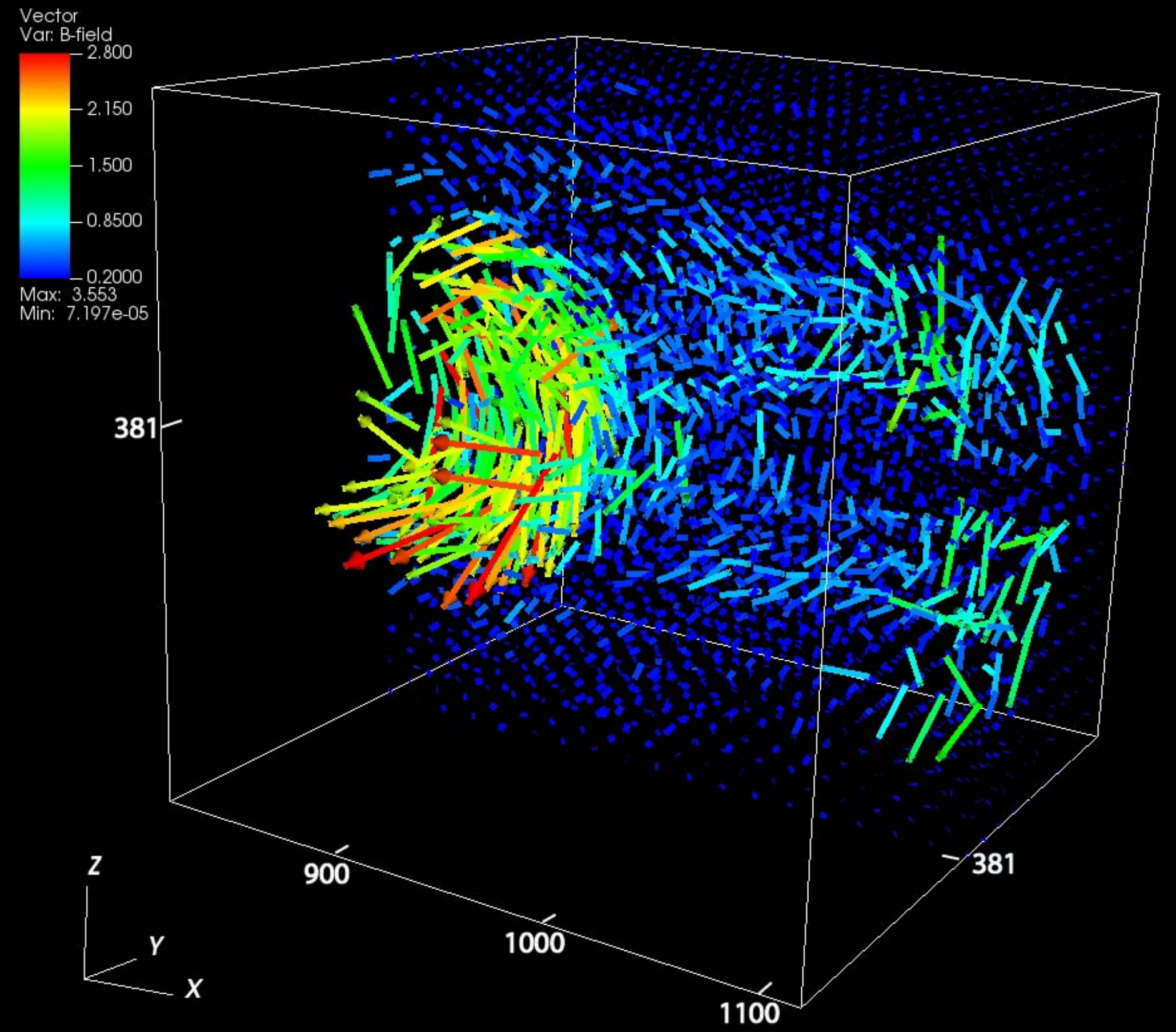}
\end{center}
\caption{Magnetic-field vectors within a cubic section of the simulation grid ($820 < x/\Delta < 1120; 231 < y/\Delta, z/\Delta < 531$)  
at time $t =900\, \omega_{\rm pe}^{-1} $ (a) and at $t =1000\, \omega_{\rm pe}^{-1}$ (b). 
To illustrate the magnetic field inside the jet, the plots show the rear half of the 
jet with cut} in the $x-z$ plane ($381 < y/\Delta < 531$). Adapted from  {\citet{Nishikawa2020}}.
\label{3DBV}
\end{figure}

In Fig.~\ref{3DJX} 3D isosurfaces of the $x$-component of the electron current density at the jet head are shown together with magnetic field lines plotted in yellow obtained by 3D PIC simulations of propagating jet with helical magnetic field \citep{Nishikawa2020}. 
Two rectangles indicate the visible area in the jet. 
Near the jet head, the current filaments generated by the WI are evident. 
This result is similar to that obtained by \cite{ardaneh16}  (their Fig.~3), who 
investigate the structure of the head of an electron-ion jet in slab geometry with $m_{\rm i}/m_{\rm e}=16$. 
They demonstrated the acceleration of jet electrons in the linear stage of the instability. Figure~\ref{3DJX} shows the merging of current components (the front of the nonlinear region) behind the current filaments, where some jet electrons are slightly accelerated. 
More complicated structures are seen near the jet boundary, that they shall investigate in more detail below. 

The 3D morphology of the jet's magnetic field is presented in Figure~\ref{3DBV} 
where they plot magnetic-field vectors at $t =900\, \omega_{\rm pe}^{-1}$  (Fig.~\ref{3DBV}a) 
and  $t=1000\, \omega_{\rm pe}^{-1} $ (Figs.~\ref{3DBV}a 
and \ref{3DBV}b). The regions displayed ($820 < x/\Delta < 1120; 231 < y/\Delta, 
z/\Delta < 531$) are indicated by the red dashed rectangle in Figure~\ref{JxB}b. 
The plot is clipped at the center of the jet at $y/\Delta =381$. 
One can see that the edge of the helical magnetic field in the jet moves from 
$x/\Delta =780$  at $t =900\, \omega_{\rm pe}^{-1}$ to $x/\Delta = 830$ at  
$t =1000\,\omega_{\rm pe}^{-1}$, which is much slower than the jet speed
(if moving with the jet velocity, the front at $t =1000\, \omega_{\rm pe}^{-1}$ should 
be located at $x/\Delta\simeq 880$).
This seems to indicate that the front edge of the helical magnetic field is peeled off as 
the jet propagates. This may indicate that the helical magnetic field is braided by kinetic 
instabilities and subsequently becomes untangled as discussed in \cite{Blandford2017}. 
The untangling of helical magnetic field results from magnetic-reconnection-like phenomena, 
that push the helical magnetic field away from the center of the jet at the forward position.
Two smaller magnetic islands can be identified in the jet seen 
in Figure~\ref{3DBV} (half of the jet is shown).
The supplemental movie\footnote{dBtotByz11MF\_011.mp4: the total magnetic fields in the $y-z$-plane; $280 <y/\Delta<480, 280<z/\Delta<480$} shows how the helical magnetic field is untangled and magnetic islands evolve along the jet at $t =1000,
\omega_{\rm pe}^{-1}$. 
For example at $x/\Delta =950$ the centers of magnetic islands are located around $(y/\Delta, z/\Delta)=$(435, 420) (upper), (440, 318) (lower) 
(visible in Fig.~\ref{3DBV}b), (320,350), and (280, 400) (not visible).


\subsubsection{Particle acceleration in forced magnetic field turbulence}
\label{sec:5.4.2}

In order to simulate the energies of electrons and ions in turbulent plasmas, \citet{Zhdankin19} have used a PIC simulation. 
In their simulations, ultra-relativistic electrons and sub-relativistic ions are initiated as they might occur in a variety of astrophysical systems \citep{shapiro_1976,quataert_1999,ryan_2018}. In a collisionless plasma the turbulence leads to a more significant growth of the ion energy compared to the electron energy. 
They found that the ratio of the energy growth $\Delta E_{e}$ and $\Delta E_{i}$ can be approximated through $\Delta E_{e}/\Delta E_{i}\sim \left(\varrho_{e}/\varrho_{i}\right)^{2/3}$ where $\varrho_{e,i}$ are the gyro radii of electrons and ions. 


\begin{figure}[htb]
\centering
\includegraphics[scale=0.88,angle=0]{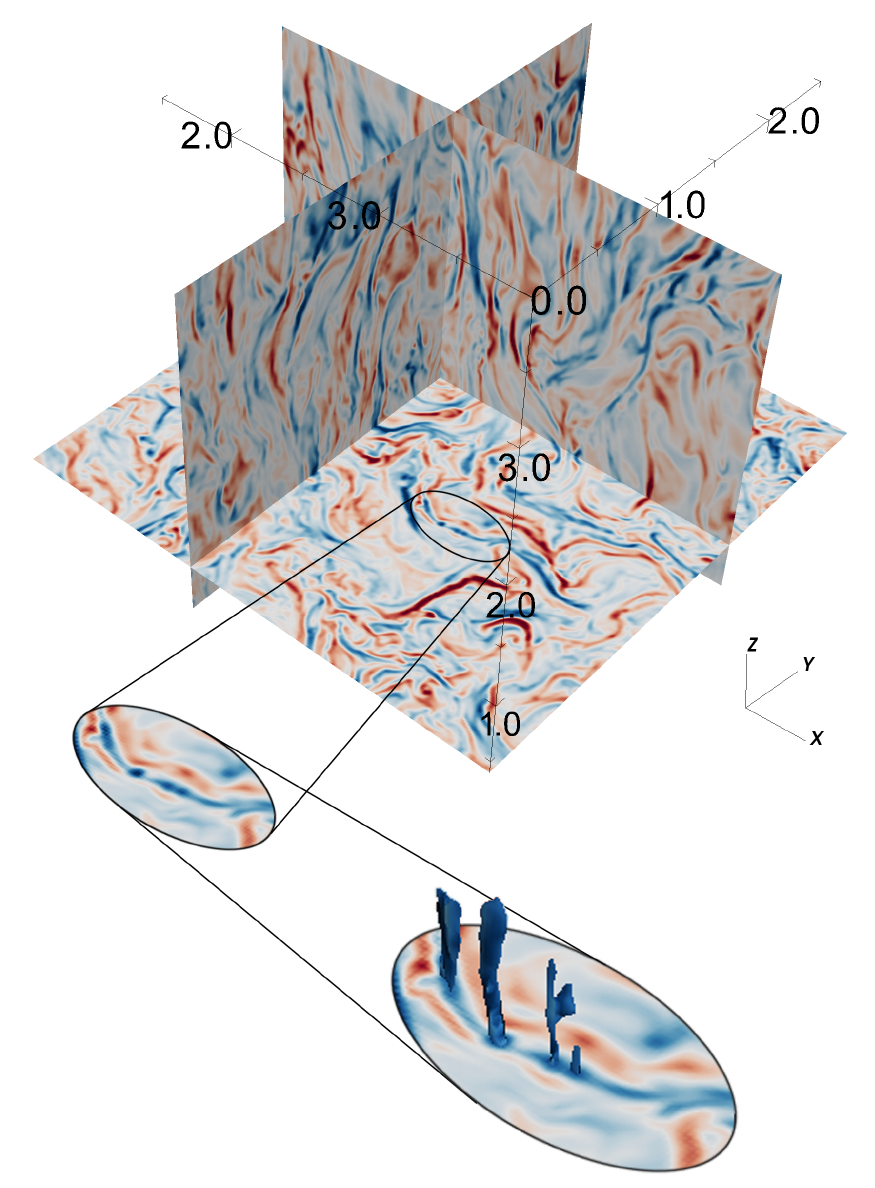}
\caption{Current densities formed in a reconnecting current sheet developing in 3D turbulence (with $\sigma_{0} = 10$, $\delta B_{\rm rms0}/B_{0}= 1$, and $L=d_{e0} = 820$). 
In the zoomed region, the isosurfaces (in blue color) show the current density $J_{\rm z}$, highlighting four 3D plasmoids as flux ropes along $\vec{e}_z$, parallel to the mean magnetic field. In all panels, blue isocontours indicate regions with $J_{\rm z} < 0$, while red indicates regions with $J_{\rm z} > 0$. 
Adapted from {\citet{Comisso19}}}.
\label{3dtub}
\end{figure}

\cite{Comisso19} have investigated the interplay between turbulence and magnetic reconnection in generating non-thermal particles in magnetically-dominated pair plasmas. 
They found that the turbulence evolution leads to the generation of a non-thermal particle distribution  with a power-law tail. The slope is harder for larger magnetizations and stronger turbulence fluctuations, and  it can be as hard as $p \lesssim 2$. 
The further particle injection is controlled by plasmoid-mediated reconnection as a self-consistent phenomenon in the turbulent plasma. Fig.~\ref{3dtub} shows an example of reconnecting current sheets for $\sigma_{0} = 10$, $\delta B_{\rm rms0}/B_{0}= 1$, and $L=d_{e0} = 820$ where the magnetic field is parallel to the $z$-direction. 
It illustrates that these current sheets randomly host regions with current densities smaller than zero (red contour lines) and larger than zero (blue contour lines). 
The close-up regions shows the emergence of four 3D plasmoids parallel to the magnetic field. It is the typical signature of fast plasmoid-mediated reconnection which controls the further particle injection.

Once particles are injected, they are further accelerated by the stochastic scattering off turbulent fluctuations. The electric fields in the reconnection layers contribute in two different ways: Whilst the parallel electric fields contribute the initial energy increase of the system, the perpendicular electric fields of turbulent fluctuations dominate the energization of high-energy particles.


Their simulations suggest that hard radio spectrum of the Crab Nebula \citep{lyutikov_2019} can be explained by large plasma magnetizations and strong turbulent fluctuations and that, in general, magnetically dominated plasma turbulence lead to particle acceleration which might be relevant for extreme astrophysical systems \citep{takahashi_2009}.

It should be noted that electrons can be further accelerated to higher Lorentz factors on account of turbulent acceleration, as in kinetic simulations of driven magnetized turbulence \citep[e.g.,][]{Comisso19,Zhdankin19}.
In these simulations turbulent magnetic fluctuations are externally forced in the simulation system, and so the energy source for turbulence is not self-consistent. In contrast
to the driven turbulence, in our simulations turbulent magnetic field (multiple magnetic field islands) as shown in Fig.~\ref{3DBV} are self-consistently generated in the relativistic jet through the untangling of the helical magnetic field \citep{Nishikawa2020}. Particles can be directly accelerated in the magnetic reconnection regions and also through interactions with the magnetic islands that are visible in e.g. Fig.~\ref{JxB}. 

\subsection{Radiation spectra in PIC simulations of relativistic jets}
\label{sec:5.5}

To determine a synthetic radiation spectrum from PIC simulations is quite challenging because of computer memory constraints. Nevertheless, using different approaches on the numerical calculation of the radiation a good approximation to the observational spectra can be reached. 

There two key methods for calculating the synthetic spectra:
\begin{itemize}
\item directly by integrating the expression of the radiated power, derived from the Li\'enard--Wiechart potentials for a large number of representative particles in the PIC representation of the plasma, e.g., using a test-like particle approach \citep{Hededal2005PhDT},
self-consistently in injected jets into an ambient medium
\citep{nishikawa08,nishikawa09,Nishikawa11AIPC,Nishikawa2012grb,Nishikawa2013EPJ}, 
in reflecting wall generated shocks with test-particles
\citep{Sironi_2009syn}, and self-consistently in counter-streaming flows
\citep{Frederiksen2010ApJ} 
\item post-processing the large output data of the PIC simulations that list the position and velocity of the simulation particles and the values of the electromagnetic fields at each time step \citep[e.g.,][]{Reville2010ApJ,Kagan2016},
\end{itemize}
where both methods have their advantages and disadvantages.

One disadvantage of calculating the synthetic spectra directly from the PIC simulations using 
the Li\'enard--Wiechart potentials \citep[e.g.,][]{Jackson1999} 
is the fact that the method is very 
expensive in terms of computing resources.

\subsubsection{Self-consistent calculations of synthetic spectra}
\label{sec:5.5.1}

A self-consistent (or in situ) synthesis of spectra from 3D PIC simulations was developed by \citet{nishikawa08,nishikawa09} for both $e^{\pm}$ and $e^{-}-p^{+}$ plasma jets. Here, a brief summary of the main points of their procedure is included. 

\begin{figure*}[h]
\hspace{2.5cm}
\includegraphics[scale=.68]{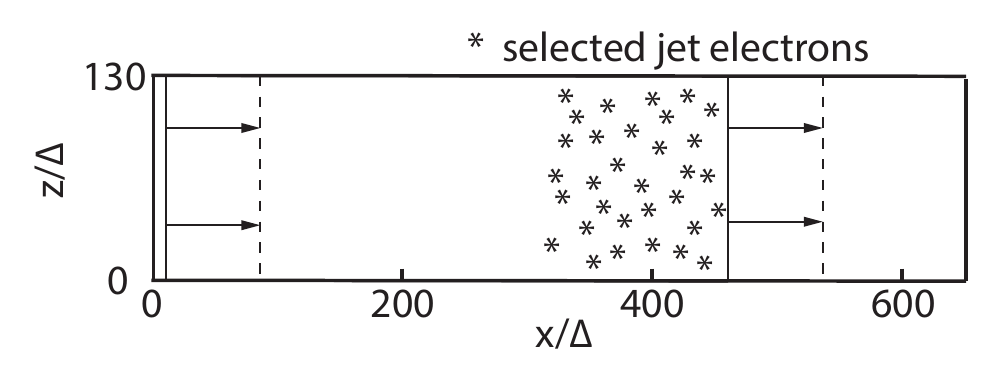}
\caption{Schematic representation of the frontal third of the jet volume, from where the electrons can be randomly selected when performing calculations of the radiation spectra 
(Eq.~\ref{spect}), after tracing the particles over a sampling time  $t_{\rm s}$ with a high temporal resolution.}
\label{scheme_syn}
\end{figure*}

Let a particle be at position ${\rm r}_{0}(t)$ at time $t$. At the same time, the electric field is observed from the particle at position {\bf r}. However, because of the finite velocity of light, the particle at an earlier position 
${\rm r}_{0}(t^{'})$ is observed where it was at the retarded time $t^{'} = t -\delta t^{'} =t -{\bf R} (t^{'} )/c$. 
Here ${\bf R} (t^{'}) = |{\bf r}- {\bf r}_{0} (t^{'})|$ is the distance from the charge (at the retarded time $t^{'}$) to the observer. After some calculation and simplifying assumptions, the total energy $W$ radiated per unit solid angle per unit frequency from a charged particle moving with instantaneous velocity $\beta$ under acceleration $\dot{\beta}$ can be expressed as: 

\begin{equation}
\frac{d^{2}W}{d\Omega d\omega} = \frac{\mu_{0}cq^{2}}{16\pi^{3}} \bigg\vert \int^{\infty}_{-\infty}\frac{{\bf n}\times[({\bf n}-\mathbf{\beta})\times\dot{{\bf \beta}}]}{(1-{\bf \beta}\cdot{\bf n})^{2}}e^{\frac{i\omega(t^{'}-{\bf n}\cdot{\bf r}_{0}(t^{'})}{c}}dt^{'}\bigg\vert^{2}.
\label{spect}           
\end{equation}
Here, ${\bf n}\equiv {\bf R}(t^{'})/|{\bf R}(t^{'})|$ is a unit vector that points from the retarded position of the particle towards the observer and $q$ is the  unit charge. 

Since extremely-large computing resources would be needed to compute the spectra of the radiation emitted by electrons over the whole simulation grid, a moment of time $t$ is selected that corresponds to a region where the particles show increased acceleration (which can be determined, for example, from running a diagnostic on the phase-space distribution of the jet electrons). Thus, to obtain the radiation spectrum, the particles are traced over a time interval around $t$, with a higher temporal resolution than that used for the whole PIC simulation (Fig.~\ref{scheme_syn}). Once the positions, the velocities, and the accelerations of the particles are sampled, the radiation spectra can be calculated by integration in Eq.~(\ref{spect}). 

The method described above was further developed by \citet{Nishikawa2012grb}, being applied to uniform relativistic jets using a slab model, where the jet electrons are accelerated by the Weibel instability. In the simulations performed by \citet{Nishikawa2012grb} for an $e^{\pm}$ plasma jet, the grid size is $(L_x,L_y,L_z)=(645\Delta,131\Delta,131\Delta$) 
(with $\Delta =1$: grid size) and the total number of particles is $\sim$ half of a billion (or 12 particles/cell/species for the ambient plasma). The jet front is located at about
$x=480\Delta$. By randomly selecting 16,200 electrons near the jet front, they calculated 
the emission during the sampling time $t_s=t_2-t_1=75\omega_{\rm pe}^{-1}$, with the 
Nyquist frequency $\omega_{\rm N}=1/2\Delta t=200\omega_{\rm pe}$, the simulation time step 
$\Delta t=0.005\omega_{\rm pe}^{-1}$, and the frequency resolution
$\Delta\omega=1/t_s=0.0133\omega_{\rm pe}^{-1}$. The mass ratio is $m_{\rm e}/m_{\rm p}=1$.

\begin{figure*}[htb]
\centering
\includegraphics[scale=.68]{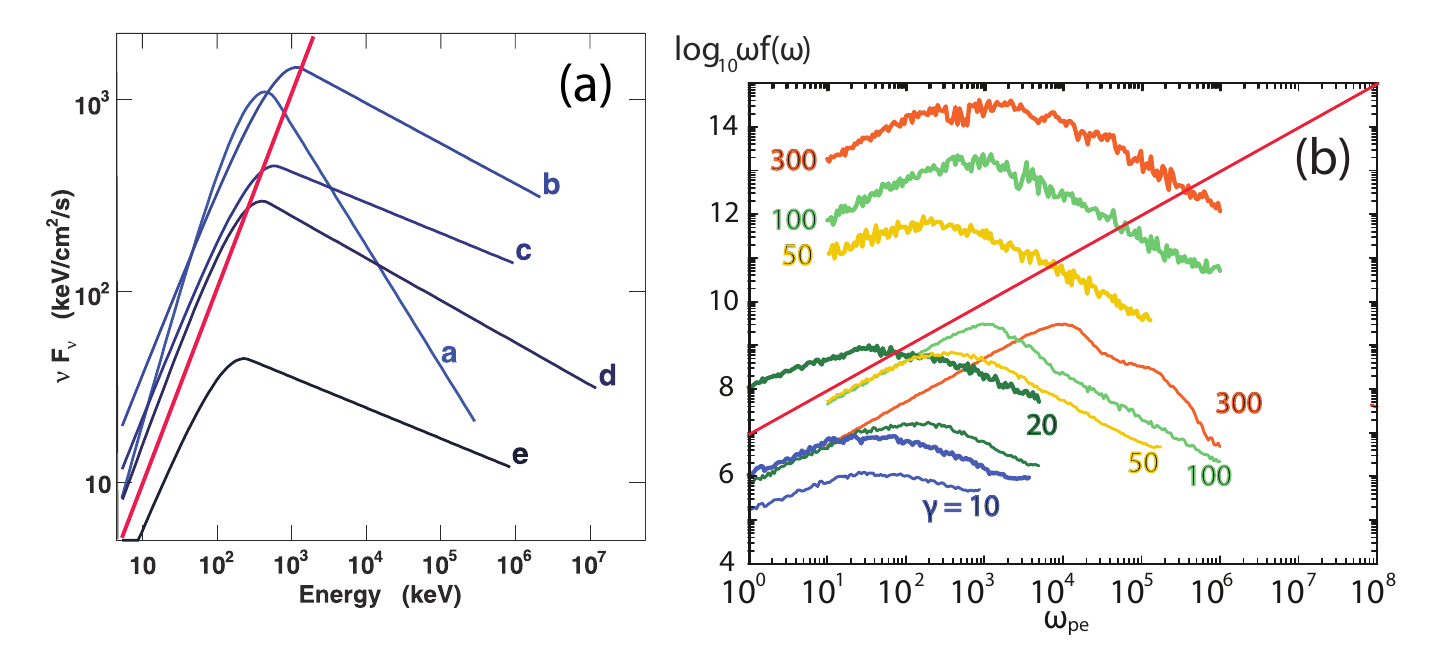}
\caption{Comparison of synthetic spectra (b) with Fermi observations for GRB 080916C (a). 
Panel (a) shows modeled Fermi spectra in $\nu F(\nu)$  units at five different times. 
A flat spectrum would indicate equal energy per decade in photon energy. The changing shapes show the evolution of the spectrum over time.
Panel (b) shows PIC generated spectra for jet Lorentz factors of $\gamma = 10, 20, 50,100$, and $300$. For each case two simulations are represented by the same color for cold (thin, lower lines) and warm (thick, upper lines) jet electrons. The low frequency slope is approximately 1. 
Adapted from \citet{Nishikawa2012grb}}
\label{KenSpectra}
\end{figure*}

\citet{Nishikawa2012grb} have also compared the shape of the calculated spectra with that of the spectra obtained from the observation taken by the Fermi \citep{Abdo09Sci}. 
Figure~\ref{KenSpectra} shows how the synthetic spectrum matches with spectra obtained from 
Fermi observations. Figure~\ref{KenSpectra}a presents the observed spectra in $\nu F(\nu)$ 
as modeled by \citep{Abdo09Sci} at five different times. The red lines in Fig.~\ref{KenSpectra}a and
\ref{KenSpectra}(b) indicate a slope of one, and except for the Fermi spectrum at time ``a'' the low frequency slopes are all approximately one. As seen in Fig.~\ref{KenSpectra}a, prompt emission spectra of GRB 080916C exhibit a remarkably constant shape (low-energy and high-energy slopes are different but do not evolve significantly). On the other hand the peaks of the SED vary significantly with time. Based on the PIC simulations, this behavior can be interpreted 
in terms of declining bulk Lorentz factor.

Behind the reverse shock electrons are accelerated and strong magnetic fields are generated
\cite[e.g.,][]{nishikawa09}. 
Therefore, this region would seem more likely to produce the emission that is observed. 
In the future we will examine PIC computed spectral changes over time varying plasma conditions such as plasma composition and ambient magnetic fields along with jet Lorentz factors and jet thermal temperatures.

Many results on synthesizing spectra from PIC simulations are presented in the literature \citep[e.g.,][]{Hededal2005PhDT,Sironi_2009syn,Medvedev_2011}, 
however their physical setups are entirely different. The resulted spectra are tested against
different mechanism of producing radiation (e.g., synchrotron, jitter, Bremsstrahlung, or 
undulator radiation), and their properties are studied based on the simulation setups. 

Unlike synthetic spectra which are calculated self-consistently (i.e., they are collected in situ), several works \citep[e.g.,][]{Hededal2005PhDT,Sironi_2009syn,Medvedev_2011} employ 
a static (or frozen) electric field, although this field was self-consistently obtained in the PIC simulations. Thus, the particles are traced around a chosen time step while the electric field is kept frozen. Therefore, the particles can be seen as test particles. 
The calculations of the radiation power were performed based on the Li\'enard--Wiechart potentials (similar to Eq.~(\ref{spect})).

As a final remark, it should be noted that a self-consistent calculation 
of radiation spectra directly from PIC simulations by tracing particles 
using Eq.~(\ref{spect}), without making assumptions about the magnetic 
field, particle orbit, and so forth, plays an important role when
interpreting astrophysical observed spectra. 
Including realistic physical
description of the plasma, including plasma jets that contain helical
magnetic fields is crucial. 
Extension of calculating the spectra with radiative particle cooling,
synchrotron self-absorption, and inverse Compton radiation is also desirable.
This effort may also be accompanied by transcription of the codes for 
running on graphical processing units to reduce the computing resources
needed.
\subsection{PIC simulations of pulsars}
\label{sec:5.6}

A pulsar is a highly magnetized rotating neutron star that emit\rv{s} beams of electromagnetic radiation from its magnetosphere \citep[e.g.,][]{Cerutti2017r}. 

\rv{On November 28, 1967 Jocelyn Bell Burnell and Antony Hewish observed the first pulsar. 
The discovery of the Crab pulsar using Arecibo Observatory confirmed the rotating neutron star model of pulsars. In 1974, Antony Hewish and Martin Ryle developed revolutionary radio telescopes, and became the first astronomers to be awarded the Nobel Prize in Physics, with the Royal Swedish Academy of Sciences noting that Hewish played a "decisive role in the discovery of pulsars".}

\rv{In 1974, Joseph Hooton Taylor, Jr. and Russell Hulse discovered for the first time a pulsar in a binary system, PSR B1913+16. Observations of the pulsar confirmed the orbital energy due to the gravitational radiation predicted by Einstein's theory of general relativity.
In 1993, the Nobel Prize in Physics was awarded to Taylor and Hulse for the discovery of this pulsar.}

Neutron stars are very 
dense, and have short, regular
rotational periods. Pulsars contain pair plasmas including pair production and due to the rotation and pair discharge the dynamics of pulsars is very active. Furthermore, 
relativistic magnetic reconnection occurs at the equatorial current sheet, which is supposed to be one of the candidates for the source of ultra-high-energy cosmic rays. 
Based on these fundamental phenomena with pulsars, PIC simulations have been playing 
an essential role in investigation of pulsars. 

\citet{sironi17} have used the PIC approach to study the origin of particle emission in the Crab Nebulae, as a prototype of Pulsar Wind Nebulae (PWNe). 
They assume that the termination shock of the pulsar wind is responsible for the particle acceleration from PWNe. However, in such a scenario, the individual particle properties cannot be retrieved from MHD simulations which leads to the necessity of PIC simulations. Indeed, they found that in addition to the Fermi process, magnetic reconnection in PWNe powers the particle acceleration. 

Global PIC simulations of pulsar magnetosphere have been performed by 
\cite{Philippov_2018}, including pair production, ion extraction from the surface, frame-dragging corrections, and high-energy photon emission and propagation. 
In the case of oblique rotators, the effects of general relativity increase the fraction of the open field lines that support active pair discharge. 
They found that the plasma density and particle energy flux in the pulsar wind are 
highly non-uniform with latitude. 
A significant fraction of the outgoing particle energy flux is carried by energetic ions, which are extracted from the stellar surface. Their energies may extend up to a large fraction of the open field line voltage, making them interesting candidates for ultra-high-energy cosmic rays. They have shown that pulsar gamma-ray radiation is dominated by synchrotron emission, produced by particles that are energized by relativistic magnetic reconnection close to the $\gamma$-point and in the equatorial current sheet. In most cases, the calculated light curves contain two strong peaks, which is in general agreement with Fermi observations. The radiative efficiency decreases with increasing pulsar inclination and increasing efficiency of pair production in the current sheet, which explains the observed scatter in $L_{\gamma}$ versus $\dot{E}$.
The high-frequency cutoff in the spectra is regulated by the pair-loading of the current sheet. Their findings lay the foundation for quantitative interpretation of Fermi observations of gamma-ray pulsars.

\begin{figure}[htb]
\centering
\includegraphics[scale=.55]{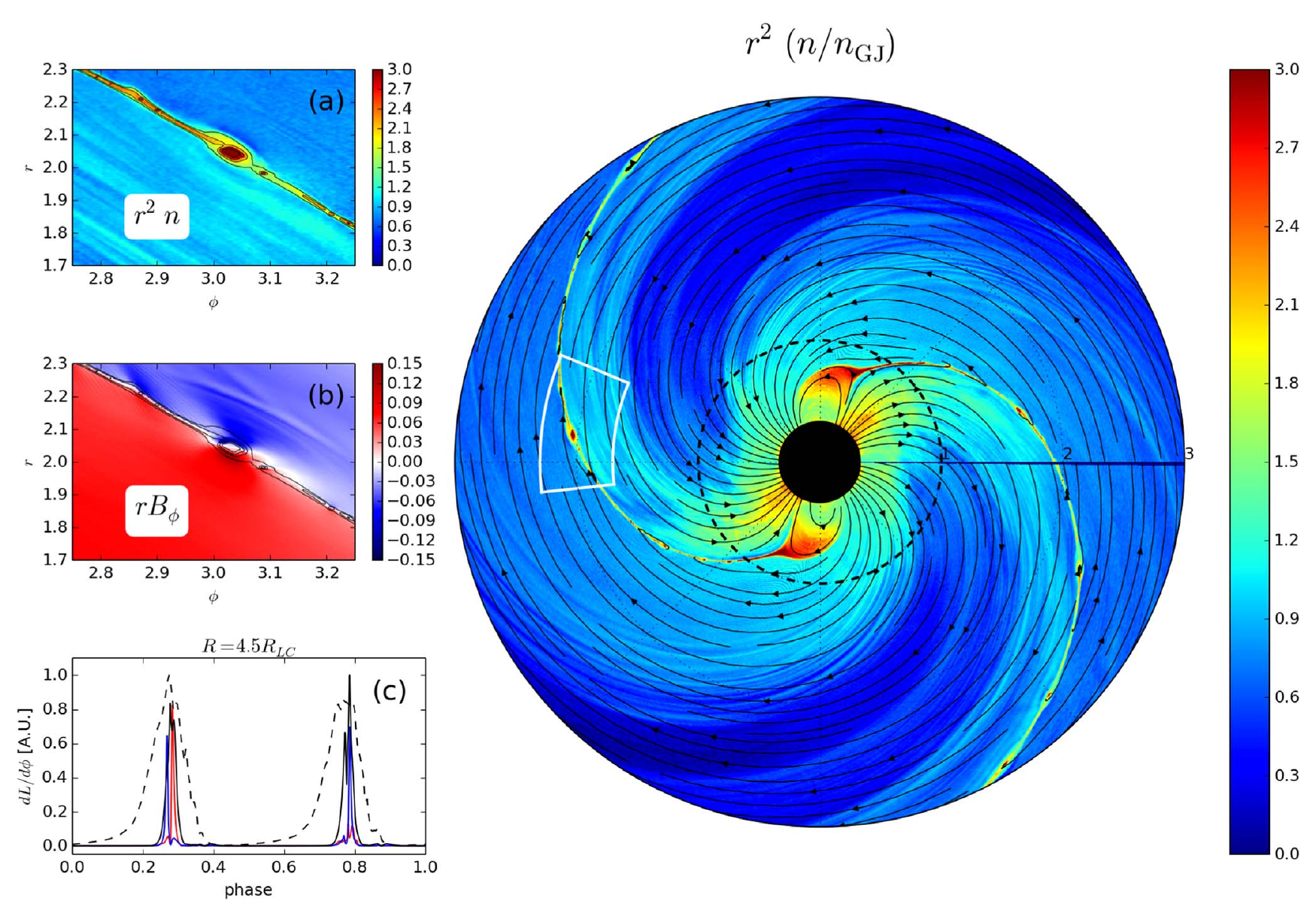}
\caption{Plasma density in the 2D global simulation of the striped wind in the equatorial plane, for plasma multiplicity at the stellar surface. The current sheet beyond the light cylinder is unstable to tearing instabilities and is divided into plasmoids. The black dashed circle shows the pulsar's light cylinder. Panels a)--c) zoom into the plasmoid in the region highlighted with the white rectangle and present (a) the plasma density, and (b) the toroidal magnetic field component. Panel (c) shows the light curve of two radio pulses (blue and red lines), the radio light curve averaged over two rotational periods (black solid line) and the gamma-ray profile (black dashed line). 
Adapted from \cite{Philippov_2019}}
\label{2dpul}
\end{figure}

More recently, \citet{Philippov_2019} simulated the magnetic reconnection in the magnetospheric current sheet beyond the light cylinder (LC) of different pulsars. 
Their simulations have shown that the magnetic reconnection leads to an efficient plasmoid-dominated regime. 
Simulations quickly settle to a solution with closed and open field lines, and a current sheet beyond the LC, separating
regions of oppositely directed magnetic field. The typical structure of the plasma density is shown in Figure \ref{2dpul}.

Figure~\ref{2dpul} shows the typical structure of the plasma density in the simulations by \citet{Philippov_2018}; here, for a simulation of $\kappa = 100$ and $\kappa/ R_{\rm LC} \approx 0.0028$, it shows that a vigorous plasmoid chain develops whose plasma density is partly shown in Figure~\ref{2dpul}a. 
%
%
Furthermore, Figures~\ref{2dpul}b and \ref{2dpul}c show the toroidal magnetic field component, the light curves of two radio pulses, and the associated gamma-ray profile.
They found that the number of plasmoids within $2\,R_{\rm LC}$ scales approximately as $R_{\rm LC}/\delta$ where $\delta$ is the current sheath width. The merging plasmoids 
are in turn producing fast plasma waves which become apparent in the toroidal magnetic field component. Such plasmoids subsequently collide with each other and emit electromagnetic waves which become apparent as radio emissions from pulsars (panel c) where they have observed common features such as the coincidence with high-energy emissions, nano-second duration and the extreme instantaneous brightness of individual pulses concluding that pulsar activities are associated to the magnetic reconnection in the magnetospheric current sheath.

\begin{figure}[htb]
\centering
\includegraphics[width=0.8\textwidth]{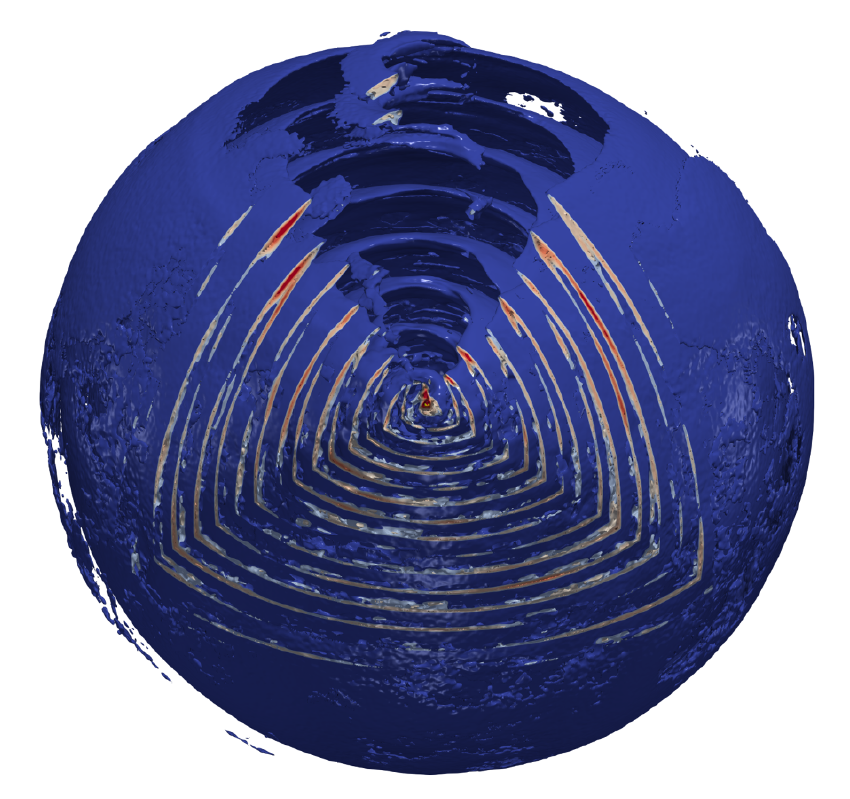}
\caption{3D rendering of plasma density isosurfaces for $\chi= 60^{\circ}$ from the star surface (yellow dot at the center) up to $50\,R_{\rm LC}$. A spherical wedge has been removed to show the internal structure of the striped wind.
Image reproduced with permission from \cite{Cerutti2020A&A}, copyright by the authors.}
\label{Cerutti3d}
\end{figure}

Recently, \cite{Cerutti2020A&A} performed large 3D PIC simulations of pulsar winds with a focus on magnetic dissipation and particle acceleration within the striped region. 

Shortly after the onset of the simulations, the plasma blown from the star establishes a force-free split-monopole-like configuration as it propagates out as shown in Fig.~\ref{Cerutti3d}. It is characterized by a toroidal-dominated magnetic structure whose polarity reverses across the current sheet. The shape of this sheet is consistent with the prediction of \citep{Bogovalov1999}: it is a 3D spherical
Archimedean spiral of wavelength $2\pi R_{\rm LC}$ defined within the spherical wedge $\pi/2 - \chi <\theta < \pi/2 + \chi$ (Fig.~\ref{Cerutti3d}). This region, which includes both the sheet and the plasma in between, is the striped wind \citep{Michel1971,Coroniti1990,Kirk2009}. 

The global structure of the striped wind is consistent with the split-monopole prediction \citep{Michel1973,Bogovalov1999}. 
The current sheet is prone to the plasmoid instability shortly after its launching at the light cylinder. This instability leads to an efficient fragmentation of the sheet into a network of interconnected flux ropes separated by secondary thin current layers where the field reconnects, a structure reminiscent of 3D plane-parallel reconnection studies. This chain
is highly dynamical and flux ropes form and merge to form bigger structures, which may result in bright, short bursts of radio emission \citep{Philippov_2019,Lyubarsky2019} aligned with the incoherent gamma-ray-pulsed emission from the sheet \citep{Cerutti2016}.
The sheet is also kink unstable, which leads to a significant widening of the striped wind region at low magnetic obliquity. Reconnection in this environment proceeds in a highly asymmetric way, in a qualitatively similar manner as the day side of the Earth's magnetopause. More efforts are needed to better understand asymmetric reconnection in the relativistic regime using local studies.

\rv{Since PIC simulations of pulsars provide essential processes which cannot be investigated by RMHD simulations, in the future further simulations with additional physical processes will be performed in order to interpret observations such as fast radio bursts.}


\subsection{General relativistic PIC simulations of black hole and neutron star}
\label{sec:5.7}

An algorithm is presented that incorporates the Kerr--Schild metric of rotating black 
holes in PIC codes \citep{watson10}. The algorithm and implementation for the simulation of charged particles are described in the region of a spinning black hole. 
They have tested the overall model by using a `toy' black hole and accretion disk system in a uniform magnetic field which has shown the production of bipolar jets. 


\begin{figure}
\begin{minipage}{7.5cm}
	\includegraphics[scale=0.70]{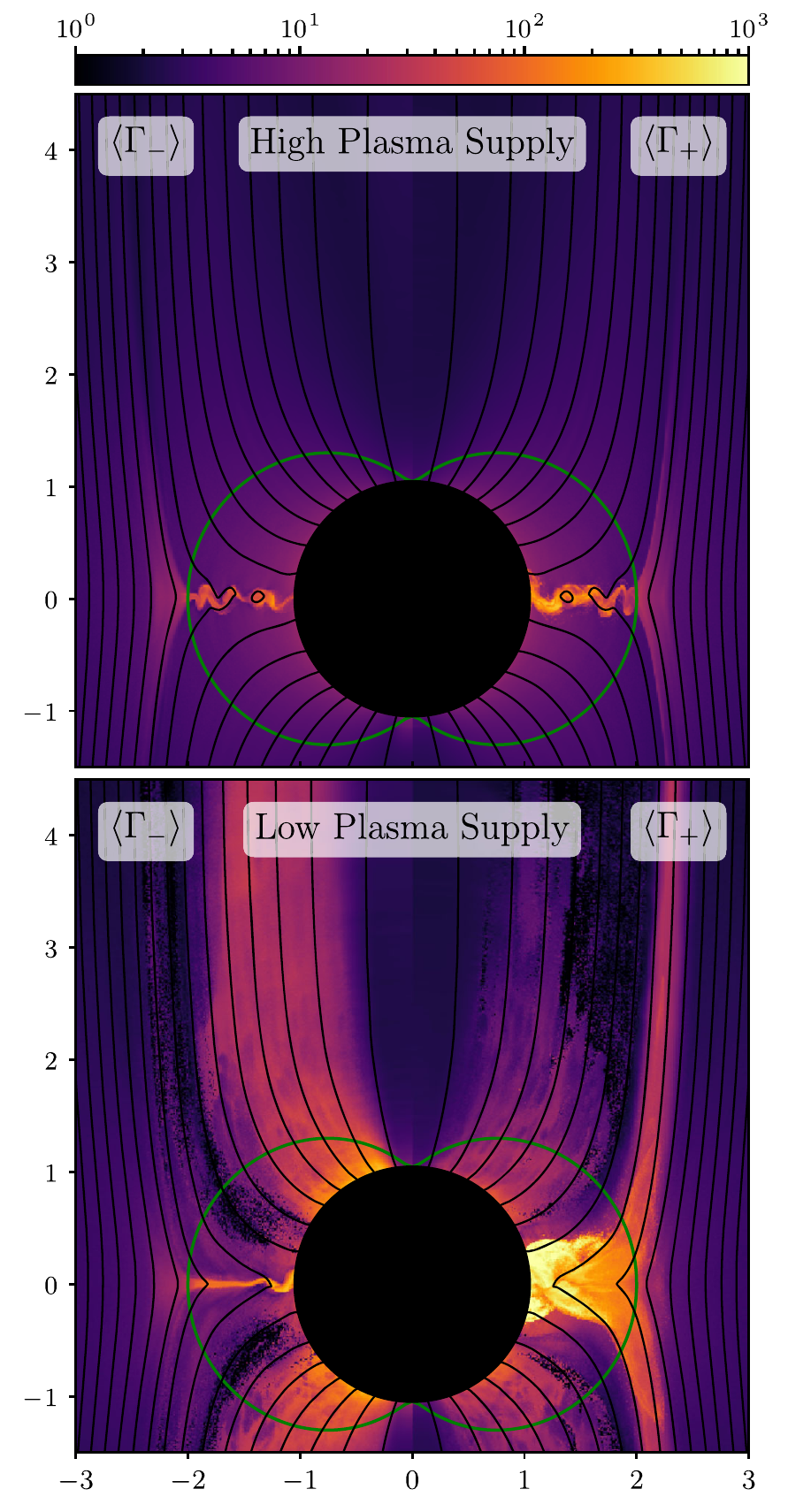}
\end{minipage}
\begin{minipage}{3.5cm}
\vspace*{-3.0cm}
\caption{Average FIDO (the orthonormal basis of the local fiducial observer)-measured Lorentz factors in the two steady states (top) High plasma supply and (bottom) Low plasma supply; the full potential corresponds to ${\rm \Gamma}_{\max}= 10^{3}$.
Image reproduced with permisison from \citet{Parfrey19}, copyright by APS.}
\label{Parfrey19Fig3}
\end{minipage}
\end{figure}

\cite{Parfrey19} have shown that black holes drive powerful plasma jets to relativistic
velocities. This plasma should be collisionless, and self-consistently supplied by pair creation near the horizon. They present general-relativistic collisionless plasma simulations of Kerr black hole magnetosphere which begin from vacuum, inject $e^{\pm}$ pairs based on local unscreened electric fields, and reach steady states with electromagnetically powered Blandford--Znajek jets and persistent current sheets. 
Particles with negative energy at infinity are a general feature, and can contribute significantly to black-hole rotational-energy extraction in a variant of the Penrose process. The generated plasma distribution depends on the pair-creation environment, and Fig.~\ref{Parfrey19Fig3} shows two distinct realizations of the force-free electrodynamic solution. This sensitivity suggests that plasma 
kinetics will be useful in interpreting future horizon-resolving sub-millimeter and infrared observations \citep[e.g.,][]{EHT2019_6,Gravity17}.

\begin{figure}[htb]
\begin{minipage}{6.5cm}
\includegraphics[scale=0.72]{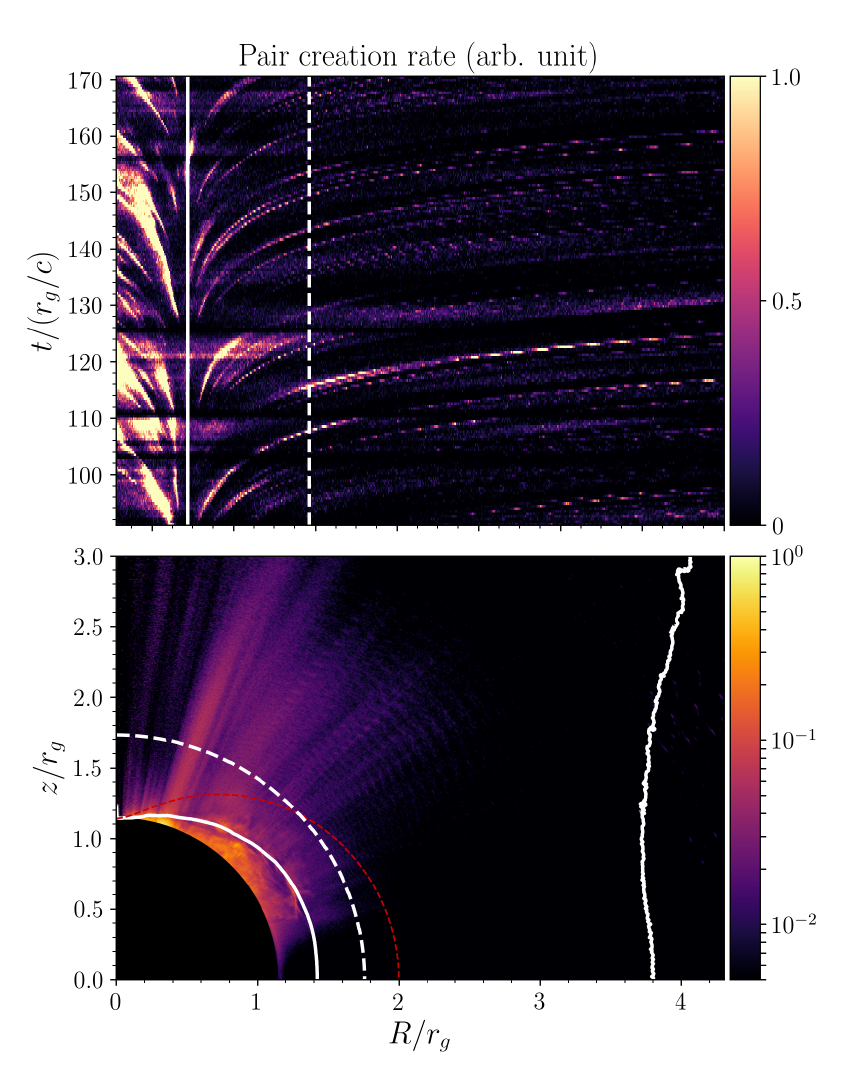}
\end{minipage}
\begin{minipage}{5.0cm}
\caption{Top panel: spacetime diagram of the pair creation rate at $\theta = \pi/4$ for the high opacity simulation, in arbitrary units. The white solid (resp. dashed) line marks the location of the inner light surface (resp. the null surface) at $\theta = \pi/4$.
Although pair creation is continuous in time in the simulations, trajectories look discretized because of down sampling. Bottom panel: 2D map of the time-averaged pair creation rate. The white solid (resp. dashed) line marks the time-averaged reconstructed light surfaces (resp. analytical null surface).
Image reproduced with permission from \citet{Crinquand20x}, copyright by APS.}
\label{Crinquand20xFig2}
\end{minipage}
\end{figure}

Recently, \cite{Crinquand20x} have shown that black holes can launch powerful relativistic jets and emit highly variable $\gamma$-ray radiation though GRPIC simulations. 
How these jets are loaded with plasma remains poorly understood. Spark gaps are thought to drive particle acceleration and pair creation in the black-hole magnetosphere. They have performed 2D axisymmetric GRPIC simulations of a monopole black-hole magnetosphere with a realistic treatment of inverse Compton scattering and pair production. They found that the magnetosphere can self-consistently fill itself with plasma and activate the Blandford-Znajek mechanism. 

A highly time-dependent spark gap opens near the inner light surface which injects pair plasma into the magnetosphere as shown in Fig.~\ref{Crinquand20xFig2}. 
These results may account for the high-energy activity observed in AGNs and explain the origin of plasma at the base of the jet.

\rv{Recently, \cite{Hirotani2021} have investigated the temporal evolution of an axisymmetric magnetosphere around a rapidly rotating stellar-mass black hole by using a 2D PIC simulation. Adopting homogeneous pair production and assuming that the mass accretion rate is much less than the Eddington limit, they find that the black hole's rotational energy is preferentially extracted from the middle latitudes and that this outward energy flux exhibits an enhancement lasting approximately 160 dynamical timescales. 
It is demonstrated that the MHD approximation cannot be justified in such a magnetically dominated magnetosphere because Ohm's law completely breaks down and the charge-separated electron-positron plasma is highly non-neutral.}

\begin{figure}[htb]
\begin{minipage}{6.8cm}
\includegraphics[scale=0.72]{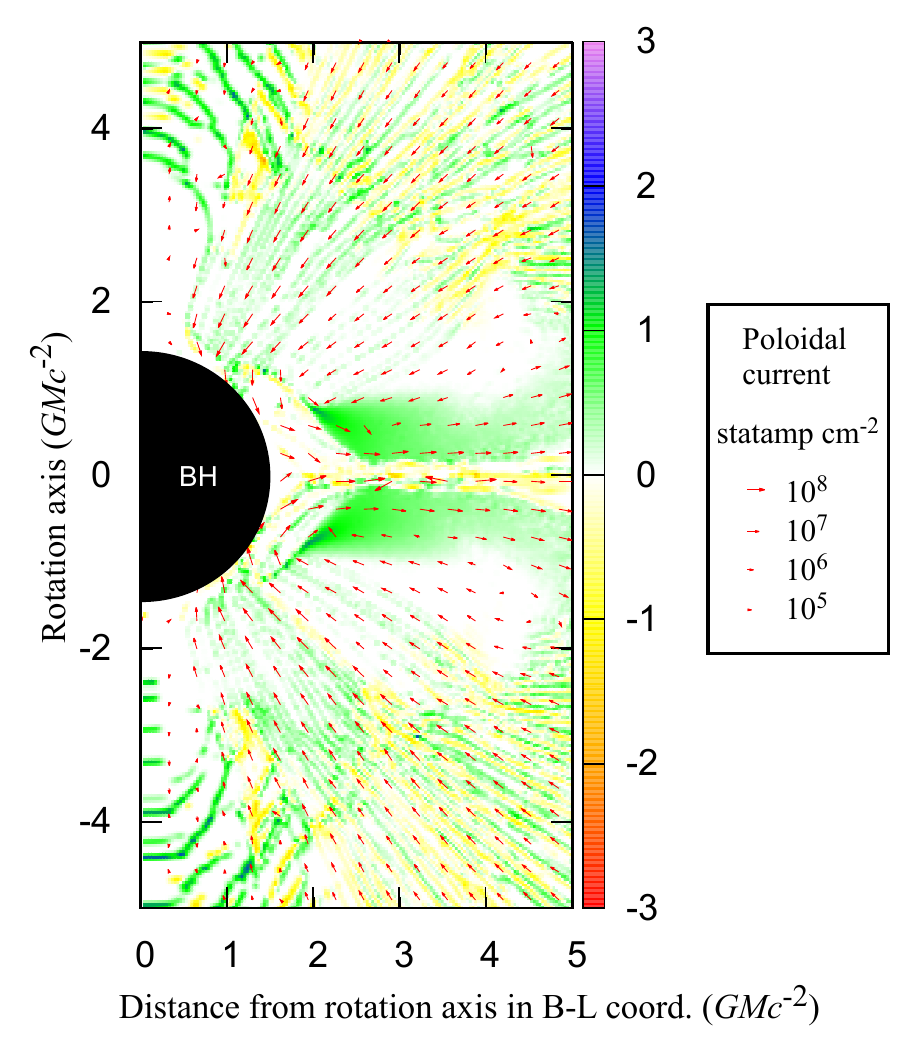}
\end{minipage}
\begin{minipage}{4.6cm}
\caption{\rv{Dimensionless charge density $(n_{+} - n_{-})/n_{\rm GJ}$ (colorbar) and poloidal electric current (red arrows). The ordinate represents the distance along the rotational axis of the BH, while the abscissa represents the distance $r \sim q$ from the rotation axis. Both axes show lengths in the Boyer-Lindquist (BL) coordinates normalized by the gravitational radius, $r_{\rm g} = GMc^{-2}$.
The equatorial plane corresponds to the ordinate of zero. The black filled circle shows the BH. The charge density is plotted in linear scale. The green (or yellow) regions show positive (or negative) dimensionless charge densities. The right panel shows four example arrow lengths corresponding to the indicated strengths of the poloidal current densities in statampere ${\rm cm}^{-2}$.
Image reproduced with permission from \citet{Hirotani2021}, copyright by AAS.}}
\label{Hiriotani2021Fig9}
\end{minipage}
\end{figure}

\rv{Figure~\ref{Hiriotani2021Fig9} shows the current distribution with the direction and strength of the poloidal currents as red arrows. 
The current density is averaged over the area in which the authors compute the direction and length of each arrow. 
These averaged currents mostly flow in the middle and lower latitudes; thus, the low-density regions in the polar funnels do not essentially affect the entire structure of the magnetosphere. This figure also plots the charge density, $(n_{+} - n_{-})/n_{\rm GJ}$ (color-coded) where $n_{+}$ and $n_{-}$ refer to the number densities of positrons and electrons and $n_{\rm GJ}$ denote the rotationally induced Goldreich-Julian number density.
To plot the values, they take the decadic logarithm of the absolute value of a quantity, then put the same sign as the quantity
($F = {\rm sign}\{\lg[\max(|f|, 1)], f\}$).
It follows that the real charge density becomes even greater than the GJ value, which indicates that the electron-positron plasmas become highly non-neutral near the BH.}

\rv{It is noteworthy that equatorward return currents appear inside the ergosphere in the lower-middle latitudes. They enable the magnetosphere to electromagnetically extract the BH's rotational energy via the Blandford-Znajek (BZ) process \citep{blandford1977}.
Henceforth, the problem with the BZ process in relation to the causality and the plasma's inertia \citep{Punsly1996},
which assumed a stationary and axisymmetric magnetosphere, can be overcome by the formation of these strong (but finite) meridional return currents near (but outside) the event horizon, when a magnetically-dominated BH magnetosphere becomes highly time-dependent due to plasma oscillations (see Figure 4 of \cite{Hirotani2021}).}

In the future as the full GRMHD simulations with the Einstein equations full GRPIC simulation codes will be developed.
These GRPIC codes will provide kinetic processes to compliment to GRMHD simulations.

\subsection{Mergers of neutron stars and black holes and associated jets}
\label{sec:5.8}

The mergers of BNS, BBH, and NS-BH have been extensively investigated using GRMHD simulations \citep[e.g.,][]{etienne17,baiotti17,ruiz16,ruiz17,ruiz18,kelly17} 
and provide a prime source for current and future interferometric gravitational wave observatories and promising candidates for coincident electromagnetic counterparts \citep[e.g.,][]{ruiz17,kelly17,sun17}. These systems are thought to be progenitors of strong electromagnetic emissions, including short gamma-ray bursts (sGRBs) in the case of NS-NS and NS-BH mergers.

\begin{figure}[htb]
\hspace{6.20cm} (a) \hspace{3.80cm} (b)
\vspace*{-.cm}
\hspace*{-0.cm}
\includegraphics[scale=1.52]{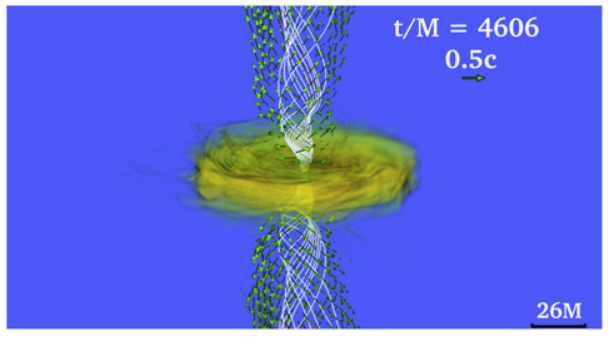}
\includegraphics[scale=0.38]{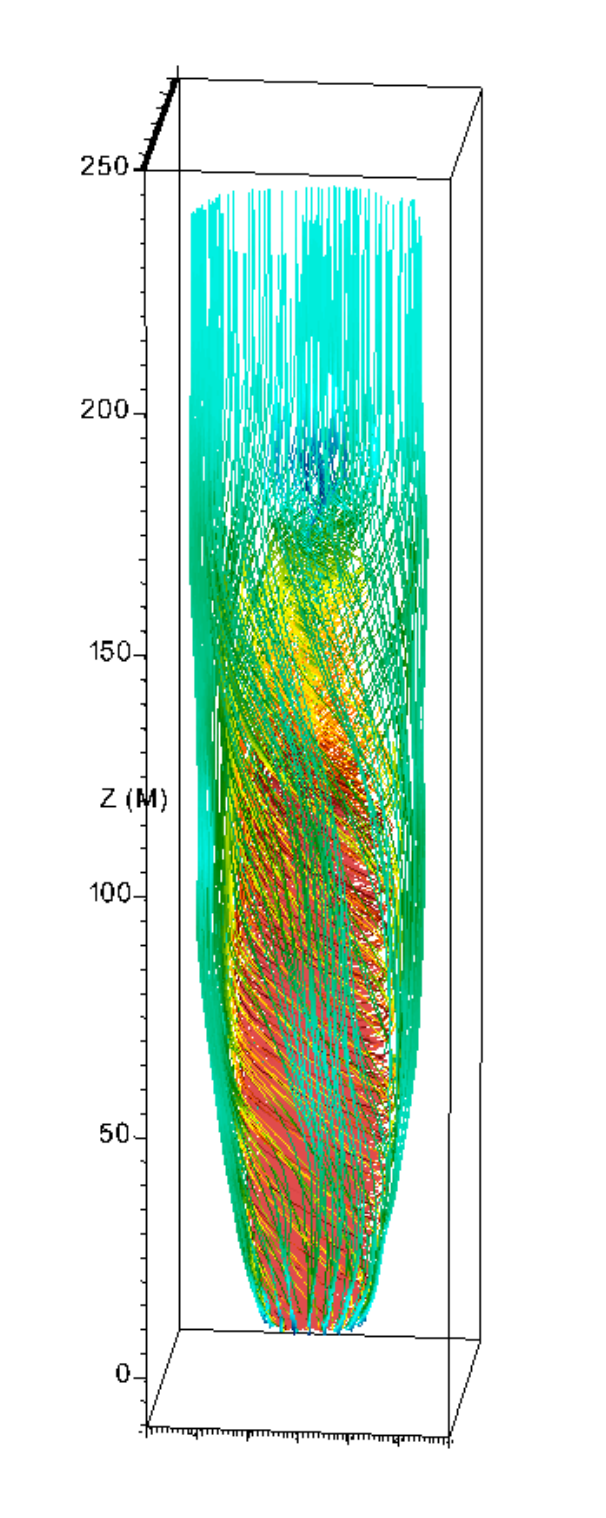}
\caption{a) Snapshot of the rest-mass density, normalized to its initial maximum value at $t/M = 4606$.
The arrows indicate plasma velocities, and the white lines show the $B$-field structure. This panel highlights the system after an incipient jet is launched
(adapted from Fig.~1 in \cite{ruiz16}). b) Magnetic field streamlines in the polar region, $t/M \sim 1100$ after merging
\citep{kelly17}.}
\label{hmfjet}
\end{figure}

\cite{ruiz16} have performed GRMHD simulations of quasi-circular, equal-mass, merging binary neutron stars. They explored 
two types of magnetic-field geometries where each star was endowed with a dipole magnetic field extending from the interior to the exterior. 
The outcome of this merging was a hypermassive neutron star that, with a small delay, further collapsed into a black hole (spin parameter $a/M_{\rm BH} \sim 0.74$) immersed in a magnetized accretion disk. About $4000M \sim 60(M_{\rm NS}/1.625\,M_{\odot}$) ms following this merging, the region above the black hole poles became strongly magnetized, and 
a collimated, mildly relativistic outflow -- an incipient jet -- was launched as shown in Fig.~\ref{hmfjet}a. 
The lifetime of this accretion disk, which likely equals the lifetime of the jet, is in the order of $\Delta t \sim 0.1 (M_{\rm NS}/1.625\,M_{\odot}$) s. 
In contrast to BH-NS mergers, they found that incipient jets are launched even when the initial magnetic field is confined to the interior of the stars. 


Recently, full GRMHD simulations of BNS 
mergers have been performed undergoing a prompt collapse to explore the possibility of jet formation from black hole-light accretion disk remnants \citep{ruiz18}. 
These simulations show that after $t - t_{\rm BH} \sim 26(M_{\rm NS}/1.8M_{\odot}$) ms (with the ADM mass $M_{\rm NS}$) following the prompt black hole formation, there is no evidence of mass outflow or magnetic field collimation. 
The rapid formation of the black hole following the merging prevents the magnetic energy from approaching force-free values above the magnetic poles, which is required to launch a jet by the usual Blandford-Znajek mechanism \citep{blandford1977}. 
The detection of gravitational waves in coincidence with sGRBs, may provide constraints on the nuclear equation of state (EOS): the fate of a BNS merger with a delayed or prompt collapse to a black hole, and hence the appearance or nonappearance of an sGRB, depends on the critical value of the total mass of the binary which is sensitive to the EOS. 

The relativistic jet formation is also observed in the merger of supermassive BBH \cite[e.g.,][]{Palenzuela10,Giacomazzo12,d'Ascoli18}
Black hole-neutron star (BHNS) mergers \citep[e.g.,][]{Ruiz2020c}, and magnetar \citep[e.g.,][]{Moesta2020ApJ} which makes it interesting to study the possible formation of a gamma-ray emission under such circumstances \citep[e.g.,][]{Nakar2020PhR,Biscoveanu2020ApJ}.

Massive BBH mergers may often take place in plasma-rich environments, leading to the exciting possibility of a concurrent electromagnetic signal observable by traditional astronomical facilities. 
\cite{kelly17} explored mechanisms that may drive electromagnetic counterparts with MHD simulations treating a range of scenarios involving equal-mass black-hole binaries immersed in an initially homogeneous fluid with uniform, orbitally aligned magnetic fields. 
They found that the time development of the Poynting luminosity, which may drive jet-like emissions, is relatively insensitive to aspects of the initial configuration. 

Figure~\ref{hmfjet}b shows magnetic field streamlines in the polar region of a BBH merger around $1100M$ \citep{kelly17}. 
The field lines are twisted into a helical pattern, concentrated at the origin. This helical structure propagates outward at the ambient Alfv\'{e}n speed $v_{\rm Alf} = 0.07433c$ ($c=1$), replacing the initially
vertical magnetic fields (still visible at large $z$).  

Since these GRMHD simulations of merging BNS and BBH systems provide conditions to generate jets, they intended to perform PIC simulations based on these jet conditions. Up to date, global jet simulations with helical magnetic fields based on force-free magnetic fields \citep{mizuno15} with top-hat density structure have been performed \citep[e.g.,][]{Nishikawa2020}. In the future, these GRMHD simulations will provide initial conditions for PIC simulations studying jet structures with more realistic magnetic fields in order to investigate multi-frequency generation associated with GW observations.

\rv{In the future, new models and more computational power will allow merge simulations with GRPIC codes  with the Einstein equations (fully dynamic metrics) similar to GRMHD simulations.}

%
\subsection{Future PIC simulations of electromagnetic radiation from relativistic jets generated by binary mergers}
\label{sec:5.9}
The first direct observation of gravitational waves was made on 14 September 2015 and was announced by the LIGO and Virgo collaborations on 11 February 2016
\cite[e.g.,][]
{Abbott2016PhRvL.116f1102A,AbbottPhysRevLett.116.241102}. 
The waveform, detected by both LIGO observatories, matched the predictions of general relativity 
for a gravitational wave emanating from the inward spiral and merge of a pair of black holes of around 36 and 29 solar masses and the subsequent ``ringdown'' of the single resulting black hole. 
This was the first observation of a binary black hole merger, demonstrating both the existence of binary stellar-mass black hole systems and the fact that such mergers could occur within the current age of the universe.

One year later, a gravitational wave (GW170817) signal was observed by the LIGO and Virgo detectors on August, 17th 2017 which was produced during the last minutes of two neutron stars spiralling closer to each other and finally merging, and is the first GW observation which has been confirmed by the accompanying EM waves \cite[e.g.,][]{abbott17a,abbott17b}.
Unlike the five GW detections prior to this event, which were caused by merging black holes not expected to produce a detectable electromagnetic signal, the signal of this merger was also seen by 70 observatories on several continents and in space, across the electromagnetic spectrum, marking a significant breakthrough for multi-messenger astronomy \cite[e.g.,][]{abbott17a,abbott17b,abbott17c}. 
Recently, \cite{Abbott2020LRR} present the current best estimate of the plausible observing scenarios for the Advanced LIGO, Advanced Virgo and KAGRA gravitational-wave detectors over the next several years, with the intention of providing information to facilitate planning for multi-messenger astronomy with gravitational waves.

\begin{figure}[h]
\hspace{4.80cm} (a) \hspace{5.30cm} (b)
\vspace*{-0.1cm}
\hspace*{-0.cm}
\includegraphics[scale=0.23]{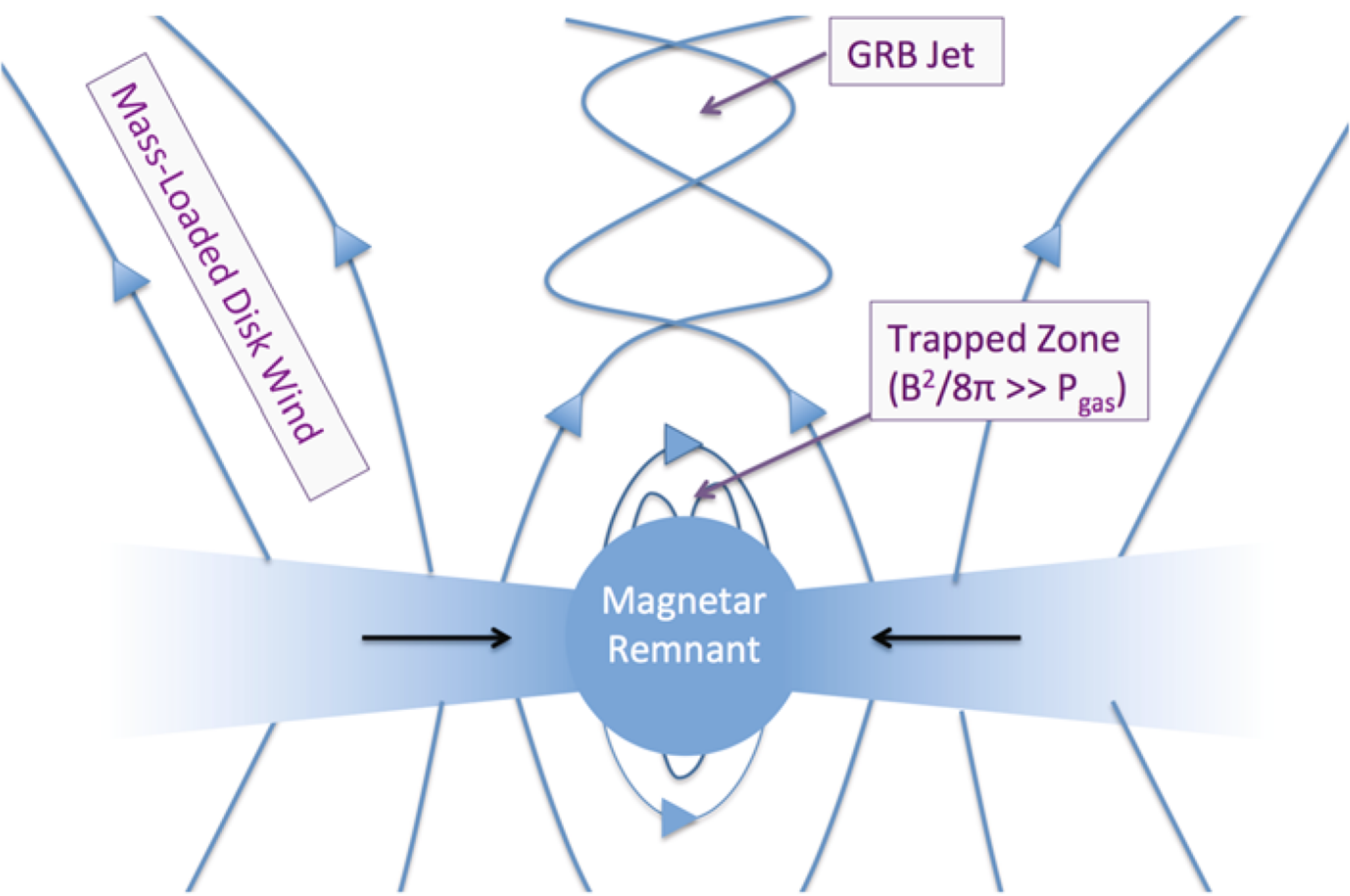}
\includegraphics[scale=0.61]{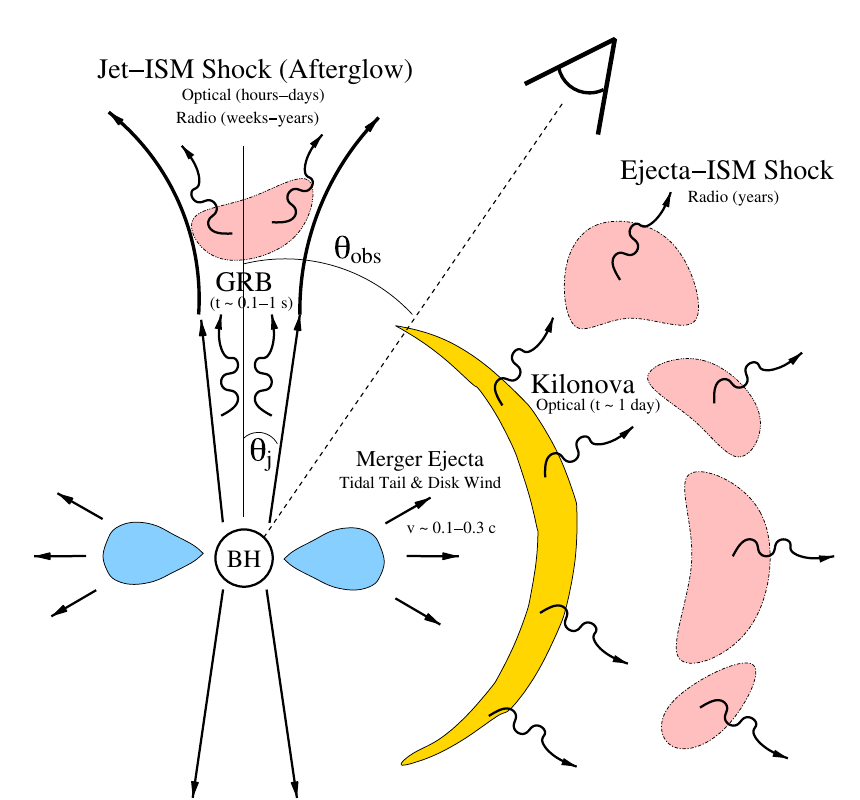}
\caption{a) Schematic illustration of a possible scenario by which the accretion onto the magnetar remnant of a BNS merger could power an ultra-relativistic short GRB jet \cite{Metzger2017r}.
b) Summary of potential EM counterparts of BNS/BBH mergers \cite{Metzger12}. 
} \label{kilonova}
\end{figure}

Figure~\ref{kilonova}a shows the schematic illustration of a possible scenario by which the accretion onto the magnetar remnant of a binary neutron star merger could power an ultra-relativistic short GRB jet \citep{nagakura14,metzger17x,Metzger2017r,shapiro17,ciolfi17}.
Strong magnetic fields in the polar region confine the hot atmosphere of the proto-NS \citep{thompson03}, preventing the formation of a steady neutrino-driven wind in this region. 
Open magnetic field lines, which thread the accretion disk or shear boundary layer, carry the Poynting flux powering the GRB jet \citep[e.g.,][]{rezzolla11}. These field lines are relatively devoid of baryonic matter due to the large centrifugal barrier, enabling the outflow to accelerate which attains high asymptotic Lorentz factors. At larger radii within the disk, outflows will be more heavily mass-loaded and form a potential collimating agent for the jet. 
\rv{Recently, \cite{Ruiz2020a,Ruiz2020b,Ruiz2020c,Ruiz2021Multi} have shown that black hole-neutron star (BHNS) and neutron star-neutron star (NSs) mergers are thought to be sources of gravitational waves (GWs) with coincident electromagnetic (EM) counterparts. To further probe whether these systems are viable progenitors of short gamma-ray bursts (sGRBs) and kilonovae, and how one may use (the lack of) EM counterparts associated with LIGO/Virgo candidate BHNS GW events to sharpen parameter estimation, they study the impact of neutron star spins in BHNS mergers.}

The observed short GRB 170817A \citep{abbott17a,abbott17b,abbott17c,goldstein17,savchenko17,granot18,shibata17,Lazzati18} constrains the properties of a jet associated with GW170817 \citep{ioka18}. 
The emission from the jet is beamed into a narrow (half-)angle $\sim 1/{\rm \Gamma}$  where ${\rm \Gamma}$ is the Lorentz factor of the jet, while the de-beamed off-axis emission is also inevitable outside $\sim 1/{\rm \Gamma}$ as a consequence of relativistic effects (see Fig.~\ref{kilonova}b). 
\cite{abbott17c} proposed three potential jet viewing geometries and jet profiles that could explain the observed properties of GRB 170817A. 
To begin with, they have considered the most simple uniform top-hat jet with uniform brightness and a sharp edge. 
It would be very important to simulate structured jets with jet densities and velocities as described in Fig.~5 of \cite{abbott17c}. The third potential jet viewing geometry is a uniform jet and cocoon. In the future, these potential viewing geometries will be examined with PIC simulations of global jets in a similar 
way as performed by \cite{nishikawa16b,nishikawa17,nishikawa19gal,Nishikawa2020}. 

\cite{abdalla17} provided the time line of observations following the detection of GW170817 with a focus on the high-energy, non-thermal domain. A complete picture of the multi-wavelength and multi-messenger campaign is given by \cite{abbott17a}. 
Future PIC simulations of global jets and associated radiation will provide a possible insight of the timeline of multi-messenger
observations. 

Further investigation needs to be performed on the short gamma-ray bursts generated by 
merging binary neutron stars (BNS) \citep{ruiz18,east16,kawamura16}, 
by binary black holes (BBH) \citep{kelly17}, or a neutron star and black hole \citep{paschalidis17,paschalidisST2017}. 
GRMHD simulations of mergers will provide GRB jet conditions as seen in Fig.~\ref{kilonova}a. 
Figure~\ref{kilonova}b shows detailed structures of the jet, the ejecta and their interactions, generating various sequential radiation as described by \cite{abbott17a} and \cite{abdalla17}.
The investigation of the kinetic jet effects subsequently requires PIC simulations based on the prior results generated by GRMHD simulations of mergers. 

In a future research project, GRMHD codes will be used to simulate the generation of 
gravitational waves (GW) and jets;  afterwards the radiation with multi-frequency observed in short GRB jets will be addressed through PIC simulations. 
This approach accounts for both microscopic and macroscopic processes with helical magnetic fields and their complicated structures in the interaction zone between jets and ejecta generated with mergers as shown in Fig.~\ref{kilonova}b. 
In particular, jets-in-jet due to magnetic reconnection may provide a possible mechanism for rapid flares and rapid variability.

\subsection{PIC simulations of laser-plasmas physics}
\label{sec:5.10}

In addition to studying astrophysical plasmas, the recent development of computational power allows the PIC simulation study of the interaction of plasmas and laser beams, such as the high-speed acceleration of electrons through the laser wakefield acceleration (LWFA) where an intense laser pulse is fired into a plasma creating  a density wake whose electric field pushes charged particles like electrons \citep[e.g.,][]{Esarey09,Palastro2020}.
Furthermore, these new PIC simulations of laser-plasma Physics will be applied to studies of astrophysics. 

\citet{Esarey09} reviewed laser-driven plasma-based accelerators, which are capable of supporting fields in excess of 100 GV$/$m, including the laser wakefield accelerator, the plasma beat wave accelerator, the self-modulated laser wakefield accelerator, plasma waves driven by multiple laser pulses, and highly nonlinear regimes. 
They have discussed the properties of linear and nonlinear plasma waves as well as the electron acceleration in plasma waves. Additionally, they have summarized methods for injecting and trapping plasma electrons in plasma waves as well as limits to the electron energy, including laser pulse diffraction, electron dephasing, laser pulse energy depletion, and beam loading limitations. 
They have also described the basic physics of laser pulse evolution in under-dense plasmas including the propagation, self-focusing, and guiding of laser pulses in uniform plasmas and with preformed density channels. 
Instabilities relevant to intense short-pulse laser-plasma interactions, such as Raman, self-modulation, and hose instabilities, have been discussed. 
Experiments demonstrating key physics, such as the production of high-quality electron bunches at energies of 0.1--1 GeV, are summarized.

\begin{figure}
\begin{minipage}{7.5cm}
\includegraphics[scale=1.1]{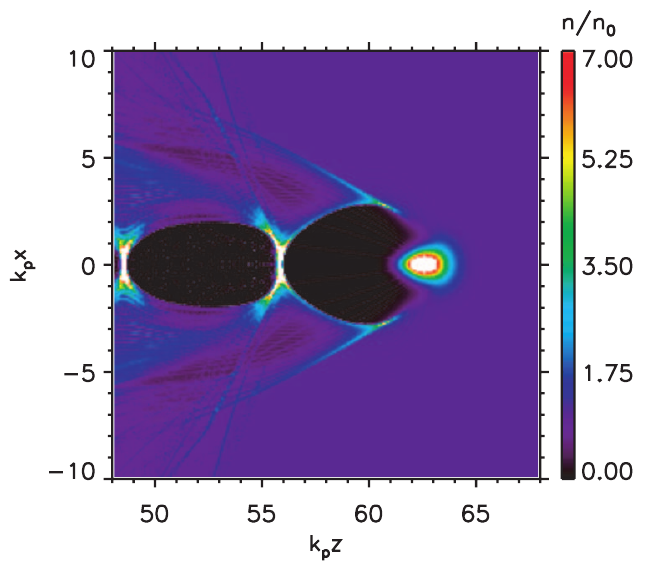}
\end{minipage}
\begin{minipage}{3.8cm}
\caption{(Color) Electron density wake from an electron beam with energy of 0.5 GeV, peak density $n_{\rm b} = 5n_{0}$, $n_{0}=5\times 10^{17}$cm$^{-3}$, and rms beam sizes $k_{\rm p} \sigma_{\rm x}=k_{\rm p} \sigma_{\rm y}=k_{\rm p} \sigma_{\rm y}=1/\sqrt{2}$ (in Gaussian profiles). The electron beam is moving toward the right with its center located at $k_{\rm p}z=62.5$.
In each cell, there are four particles and the cell size (in all directions) is $0.7\mu {\rm m}$. This figure is adapted from Fig.~17 in \citet{Esarey09}.}
\label{Esarey09F17}
\end{minipage}		
\end{figure}

\citet{Esarey09} showed the 3D PIC simulation of a nonlinear plasma wakefield accelerator (PWFA) in the bubble, blow-out and cavitation regime where the focusing force in the latter two regions can be very large. 
For example, the radial space charge field of a long ion channel is $E_{\rm r}=E_{0}(k_{\rm p}r/2)$. At the edge of an electron beam with radius $\sigma_{\rm r}$, this can be written in convenient units as
\begin{eqnarray}
E_{\rm r}({\rm MV}/{\rm m}) & \simeq & 9.06\times10^{-15}n({\rm cm}^{-3})
\sigma_{\rm r}(\mu {\rm m}). 
\end{eqnarray}

This radial force will cause a relativistic electron with $\gamma \gg 1$ to perform betatron oscillations about the axis with a betatron wavelength
$\lambda_{\beta}=(2\gamma)^{1/2}\lambda_{\rm p}$. Here the plasma wavelength is defined through $\lambda_{\rm p}=2\pi c/\omega_{\rm p}=2\pi/k_{\rm p}$ with the plasma frequency $\omega_{\rm p}=(4\pi n_{0}e^{2}/m_{\rm e})^{1/2}$.

One consequence of the betatron motion of a relativistic electron in the plasma focusing fields is the emission of betatron (i.e., synchrotron) radiation which has been observed in the blow-out regime for both the electron beam-driven and laser-driven wakes.
This radiation is characterized by the betatron strength parameter $\alpha_{\beta}=\gamma k_{\beta} r_{\beta}$, which is analogous to the undulator strength parameter in conventional synchrotron, where $r_{\beta}$ is the betatron orbit amplitude and $k_{\beta}=2\pi/\lambda_{\beta}= 2\pi/\sqrt{2\gamma}\lambda{\rm p} =k_p/\sqrt{2\gamma}$ the betatron's wave number. 
As the beam radiates, the mean energy decreases and the normalized energy spread can increase. For plasma accelerators, $\alpha_{\beta}$ can be large 
and the radiation can extend into the hard $x$-ray regime. 

An example of an electron beam-driven wake in the blow-out regime is shown in Fig.~\ref{Esarey09F17}, illustrating the spatial plasma density response to an electron beam with energy of 0.5 GeV, density $n_{\rm b}=5n_{0}$, $n_{0}=5\times 10^{17}$cm$^{-3}$, and rms longitudinal and transverse beam sizes $k_{\rm p} \sigma_{\rm x}=k_{\rm p} \sigma_{\rm y}=k_{\rm p} \sigma_{\rm y}=1/\sqrt{2}$ (in Gaussian profiles). 
It shows that the electron beam, propagating initially in a uniform plasma, moves towards the right.



Recently, \cite{Yano19} have analized the propagation of short and ultra-intense laser pulses in a semi-infinite space of over-dense hydrogen plasma via fully-relativistic, real geometry PIC simulations including radiation friction. 
The relativistic transparency and hole-boring regimes are found to be sensitive to the transverse plasma field, backward light reflection, and laser pulse filamentation. 

\citet{Popruzhenko_2019} have shown that in the interaction of laser pulses of extreme intensity ($>10^{23}\mathrm{\,W\,cm}^{-2}$) with high-density and thick plasma targets, significant radiation friction losses in simulations, in contrast to thin targets for which such losses are negligible. 
They have presented an analytical calculation, based on classical radiation friction modeling, of the conversion efficiency of the laser energy into incoherent radiation in the case when a circularly polarized pulse interacts with a thick plasma slab of overcritical initial density. 
By accounting for three effects including the influence of radiation losses on the single electron trajectory, the global `hole boring' motion of the laser-plasma interaction region under the action of radiation pressure, and the inhomogeneity of the laser field in both longitudinal and transverse direction, they found a good agreement with the results of 3D PIC simulations. 
Overall, the collective effects greatly reduce radiation losses with respect to electrons driven by the same laser pulse in vacuum, which also shift the reliability of classical calculations up to higher intensities.

In the regime of interest here, an ultra-intense laser pulse of frequency $\omega$ and dimensionless field amplitude $a_{0} = eE_{\rm L}/m_{\rm e}\omega c$ (with $E_{\rm L}$ the electric field amplitude) interacts with a strongly over-dense (electron density $n_{\rm e} \gg n_{\rm c}=m_{\rm e}\omega^{2}/4\pi e^{2}$, the cut-off density) plasma target which remains opaque to the laser light. 
The radiation pressure of the laser light is high enough to produce `hole boring' (HB) in the target, i.e., the plasma surface is driven at an average velocity 
\begin{eqnarray}
\frac{v_{\rm HB}}{c}=\frac{\sqrt{\Xi}}{1+\sqrt{\Xi}}, \, \, \Xi=\frac{I_{\rm L}}{\rho
c^{3}}=\frac{Zn_{\rm c}m_{\rm e}}{An_{\rm e}m_{\rm p}}a_{0}^{2}
\label{HB}
\end{eqnarray}
where $I_{\rm L} = cE_{\rm L}^{2}/4\pi= m_{\rm e} c^{3}n_{\rm c} a^{2}_{0}$
is the laser intensity. 
Equation (\ref{HB}) can be obtained by balancing the mass and momentum flows at the surface \citep{Robinson_2009} and is valid for total reflection of the laser light \rv{in the frame co-moving with the surface}, i.e., in the absence of dissipative effects. If a fraction $\eta$ of the laser intensity is dissipated, for example due to RF losses, Eq.~(\ref{HB}) may be modified by replacing $I_{\rm L}$ with $I_{\rm L} (1 - \eta/2)$. 
In the case of their simulations this would lead at most to $a \simeq 5\%$ decrease in $v_{\rm HB}$ at the highest intensity considered ($a_{0}=800$).

\begin{figure}
	\begin{minipage}{7.5cm}
		\includegraphics[scale=0.65]{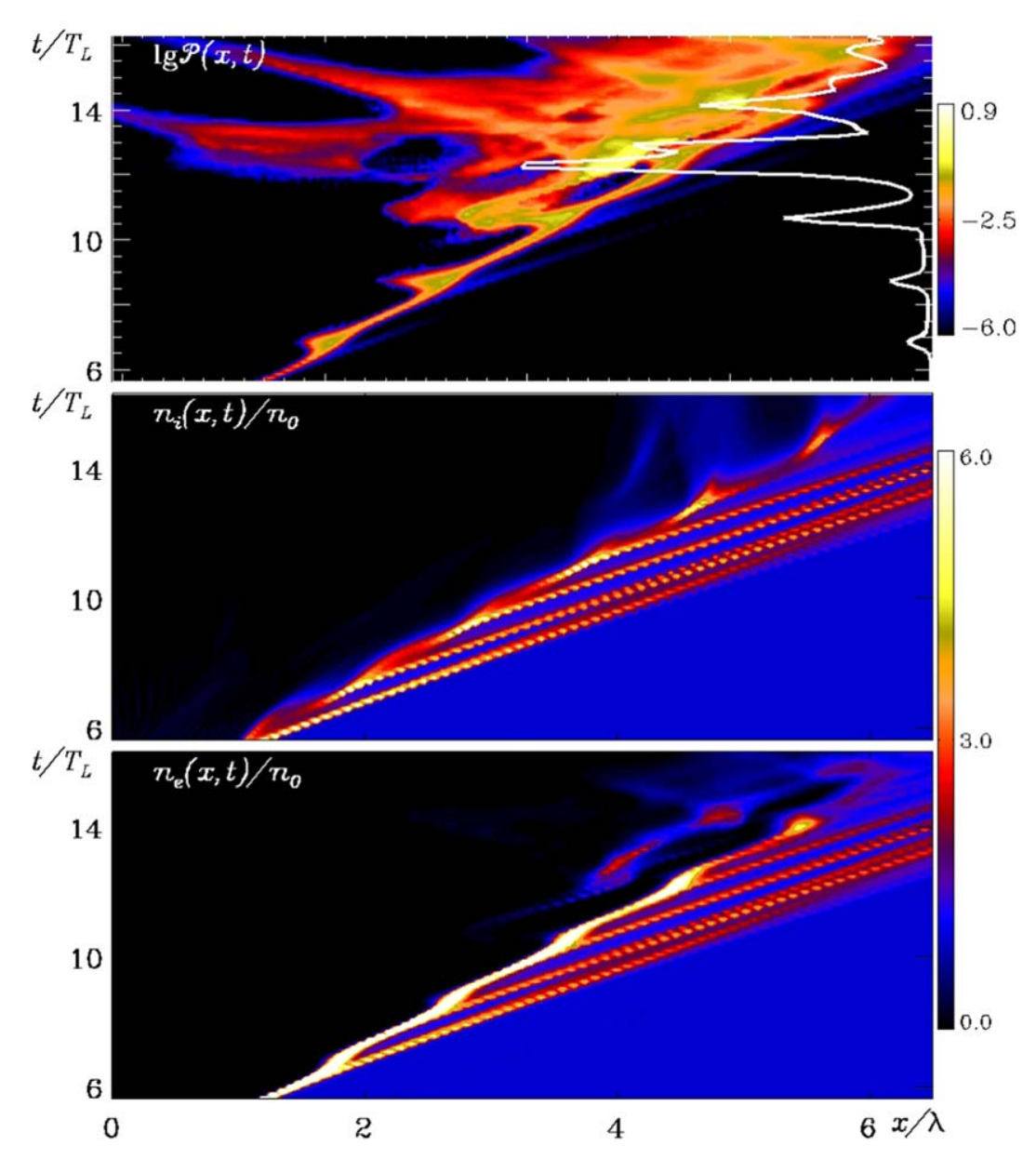}
	\end{minipage}
\begin{minipage}{3.5cm}
\caption{Space--time plots of the radiation power density $\mathcal{P}(x, t)$ (top, logarithmic scale, arbitrary units), ion density $n_{\rm i}(x, t)$ (middle) and electron density $n_{\rm e}(x, t)$ (bottom) all evaluated at $r=1\lambda$ distance from the axis. 
A white curve on the upper panel shows the radiation power integrated over $x$. Adapted from Fig.~2 in \citep{Popruzhenko_2019}}
\label{Popruzhenko_2019F2}
\end{minipage}
\end{figure}

An analysis of the 3D distribution functions of the radiation power density $\mathcal{P}(x, r, v_{\rm x})$ (calculated as $\mathcal{P} = -n_{\rm e} \vec{v}\cdot \vec{F}_{\rm rad}$ for the radiation reaction (or friction) force 
$\vec{F}_{\rm rad}$) and of the electron and ion density $n_{\rm e, i}(x, r, v_{\rm x})$ extracted from the PIC simulation shows that most of the emitted radiation comes from electrons having velocities $v_{\rm x} >0$, and located close to the receding front of the ion density. 
This is illustrated for the $a_{0}=500$ case in Fig.~\ref{Popruzhenko_2019F2} where space--time plots in the ($x, t$) plane are shown for the radiation power and the particle densities at $r=1\lambda$, where the former has its radial maximum. 

The density fronts move in the forward direction with average velocity $\simeq 0.41c$, in fair agreement with the value 
$\nu_{\rm HB} = 0.47c$ given by Eq.~(\ref{HB}). 
Small oscillations in the front position are visible in correspondence of the generation of plasma bunches in the forward direction. 
The power density plot shows that most of the emission originates close to the HB front. Emission due to returning electrons with velocity $\simeq c$ is visible after $t = 11T_{\rm L}$, but its contribution to the total emitted power is small, presumably because of the low density in the returning jets (as seen on the $n_{\rm e}(x, t)$ plot).

As clearly seen from the plot of the $x$-integrated radiation power shown on the upper panel, spikes of radiation occur in correspondence of the generation of plasma bunches. Such spikes may be explained by the enhanced penetration of the laser field into the plasma at these time instants. 
Since the spikes remain close to the HB front, no strong modification of $v_{\rm x}$ is correlated with them. Consistently with these observations, they assume that on the average the radiating electrons move with velocity 
$v_{\rm x} = v_{\rm HB}$ given by Eq.~(\ref{HB}). 
The account of the longitudinal motion improves the agreement.


\citet{Arber_2015} have explained the conceptual design of the PIC package of 
the \texttt{EPOCH} code. \texttt{EPOCH} is one of the famous packages in the
laser-plasma research community.
There are several approaches on radiation processes, however, with increasing 
computer power, the combination of PIC codes with a Monte-Carlo method including 
quantum electrodynamics (QED) cross-sections or probability rates becomes more and 
more feasible.

Modern PIC codes tend to add to high-order shape functions describing the spatial distribution of individual, real particles in a computational super-particle, Poisson preserving field updates, collisions, ionisation, a hybrid scheme for solid density and high-field QED effects. 
In addition to these physics packages, the increase in computing power now allows simulations with real mass ratios, full 3D dynamics and multi-speckle interaction.
\citet{Arber_2015} have presented a review of the core algorithms used in current laser-plasma specific PIC codes. 
They also report estimates of self-heating rates, convergence of collisional routines and test of ionisation models which are not readily available elsewhere. 
Having reviewed the status of PIC algorithms they have presented a summary of recent applications of such codes in laser-plasma physics, concentrating on stimulated Raman scattering, short-pulse laser-solid interactions, fast-electron transport, and QED effects.

\citet{Gonsalves19} have reviewed common extensions of PIC schemes which account for strong field phenomena in laser-plasma interactions. 
After describing the physical processes of interest and their numerical implementation, they provided solutions for several associated methodological and algorithmic problems. 
They proposed a modified generator that precisely models the entire spectrum of incoherent particle emission without any low-energy cutoff, and which imposes close to the weakest possible demands on the numerical time step. 
Based on this, they also developed an adaptive event generator that subdivides the time step for locally resolving QED events, allowing for efficient simulation of cascades. Further, they presented a new and unified technical interface for including the processes of interest in different PIC
implementations. Two PIC codes which support this interface, \texttt{PICADOR} 
and \texttt{ELMIS}, are also briefly reviewed.

\begin{figure}
\begin{minipage}{7.5cm}
	\includegraphics[scale=0.80]{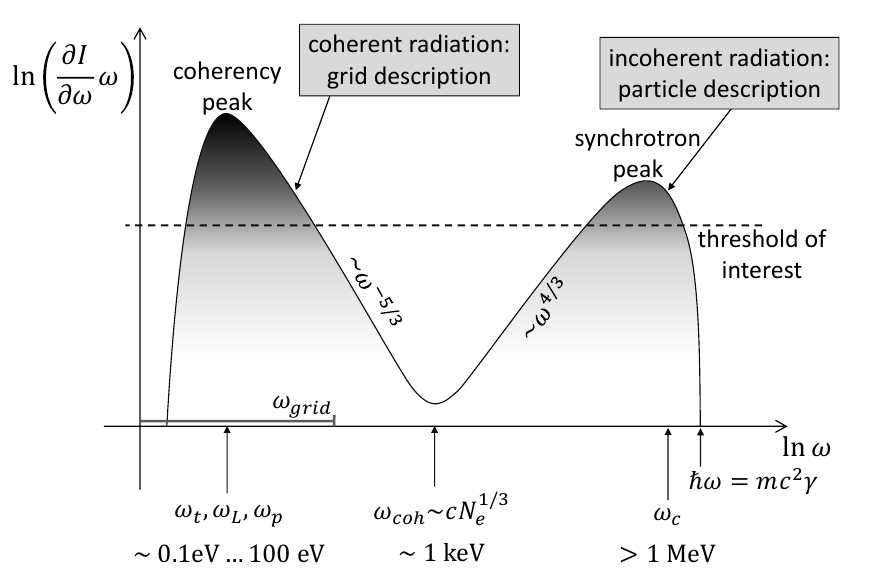}
\end{minipage}
\begin{minipage}{3.5cm}
	\caption{Schematic representation of the typical EM-energy deposition as a function of laser frequency for a typical lasma-plasma interaction. 
Adapted from Fig.~1 in \citep{Gonoskov15}}
\label{Gonoskov15Fig1}
\end{minipage}
\end{figure}

\citet{Gonoskov15} have discussed the electromagnetic energy distribution in a laser-plasma interaction using the function  $\omega \partial I/\partial \omega$, the integral of which (w.r.t.\ $\ln \omega$) gives the emitted energy shown in Fig.~\ref{Gonoskov15Fig1}. 
It illustrates the separation of two regions of electromagnetic energy deposition, which is the basis of the dual treatment of the EM-field: the radiation that can be resolved by the numerical grid (coherent peak) for low frequency emissions and the interaction of the electromagnetic field with charged particles (incoherent peak) for high-frequency emissions. 
The incoherent peak is not necessarily separated for low intensities, but then it will contain a negligible part of the emitted energy only. 
However, if the intensity becomes sufficiently high converting a significant part of the particle energy into the incoherently emitted radiation, the incoherent peak will appear in a well which makes the separated spectral region and the dual description of the electromagnetic field possible.

\begin{figure}[htb]

\includegraphics[width=\textwidth]{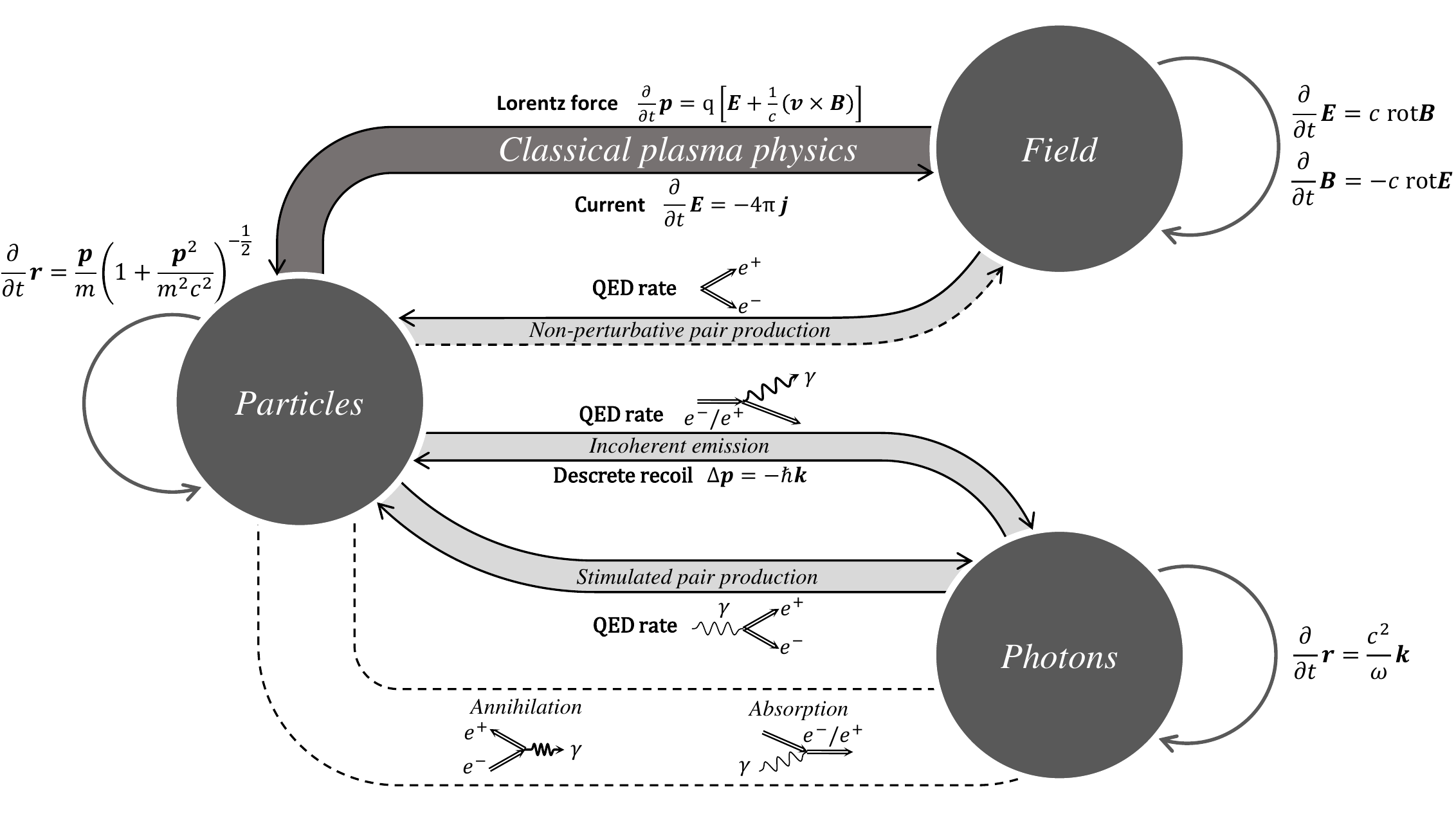}
\caption{Extension of the PIC approach for taking into account novel channels of energy transformation that could be triggered by laser fields of extreme intensity. Adapted from Fig.~7 in \citet{Gonoskov15}}
\label{Gonoskov15Fig7}
\end{figure}

Their arguments so far have been general and have not relied on any specific geometry. 
In order to demonstrate the validity of their arguments they have presented numerical results for a particular case. They considered the so-called giant attosecond pulse generation in the process of an over-dense plasma irradiation by an ultra-relativistic laser in the relativistic electronic spring regime \citep{Gonoskov11}.
They chose this example because it is a `worst case' scenario; a very wide (far beyond $I \sim \omega^{-8/3}$) spectrum of coherent emission is generated, which means that the two radiation regions could potentially have significant overlap, invalidating their results; they saw that this did not happen.

Despite the common origin for all energy deposition in Fig.~\ref{Gonoskov15Fig1}, they used the word field to mean coherent, low frequency radiation which can be resolved on the simulation grid, and the word photons to refer to the incoherent, high frequency radiation given by an ensemble of photons. Using this notation in Fig.~\ref{Gonoskov15Fig7} they have shown, schematically, three qualitatively different forms of energy allocation and the possible channels for conversion of energy between 
them. 
Conventional PIC simulations is indicated by the darker solid line connected between Particles and Field. One of basic interests in laser-plasma physics is electron-positron pair production. 

High-energy photons can interact with an arbitrary number of laser photons and produce real electron-positron pairs.
This generalization of the Breit-Wheeler process is called ``stimulated pair production.'' 
It is a perturbative process in the interaction between the stimulating photon and the pair. 
However, the interaction between the photon and the background can have both stimulated (perturbative) and non-perturbative dependencies on field strength and kinematics as shown in Fig.~\ref{Gonoskov15Fig7} \citep[e.g.,][]{Torgrimsson18,Nousch16}.
In the PIC simulations with QED effect, nonlinear Compton scattering is also implemented where the electron is excited by a laser and emits a photon.
When a charged particle is accelerated by an external field, it emits radiation. 
In the case that the electron is excited by a laser and emits a photon, it is called ``nonlinear Compton scattering''. 


The dashed lines in Fig.~\ref{Gonoskov15Fig7} indicate, for completeness, the processes which can be neglected. 
This is annihilation, absorption and the negligible loss of energy in Sauter--Schwinger pair creation. 
They remarked that neither spin nor polarisation are included in the code.

As one of laser-plasma physics applications the Inertial Confinement Fusion (ICF) has been 
extensively studied \citep[e.g.,][]{Olson2020}.



Contemporary lasers allow us to create shocks in the laboratory that propagate at a speed that matches that of energetic astrophysical shocks like those that ensheath supernova blast shells. 
The rapid growth time of the shocks and the spatio-temporal resolution, with which they can be sampled, allow us to identify the processes that are involved in their formation and evolution.
Some laser-generated unmagnetized shocks are mediated by collective electrostatic forces and effects caused by binary collisions between particles can be neglected.
Hydrodynamic models, which are valid for many large-scale astrophysical shocks, assume that collisions enforce a local thermodynamic equilibrium in the medium; laser-generated shocks are thus not always representative for astrophysical shocks. 
Laboratory studies of shocks can improve the understanding of their astrophysical counterparts if one can identify processes  that affect electrostatic shocks and hydrodynamic shocks alike. 
%

\begin{figure}[htb]
\includegraphics[width=\textwidth]{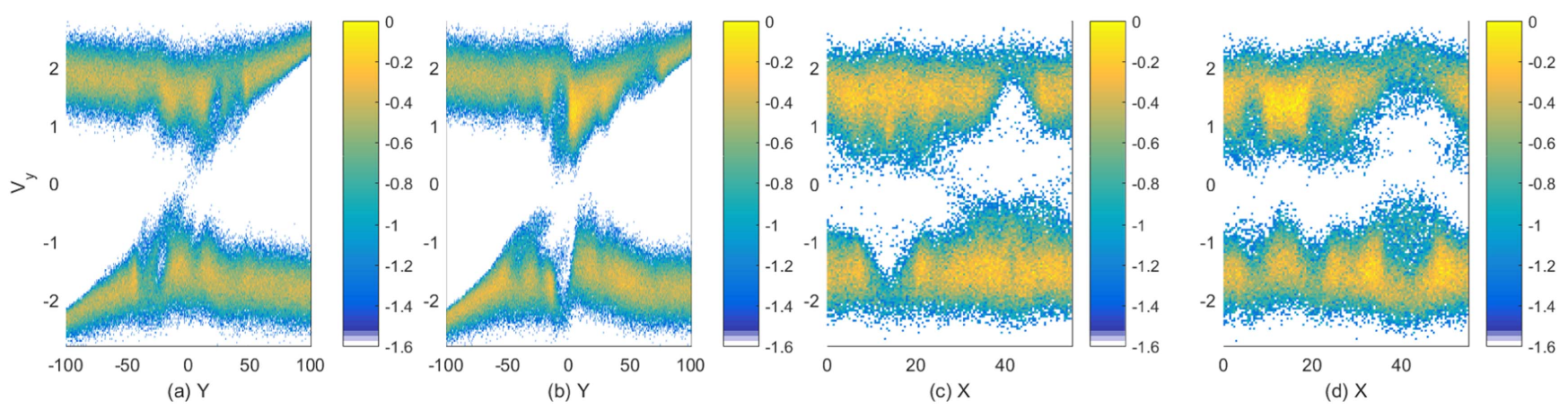}
\caption{The proton phase space density distribution for a laser-induced shock after $t = 1600$.
Panel (a) shows the velocity distribution along the slice $x=0$ and 
(b) shows it along $x = 13.75$. Panel (c) shows the distribution along $y=0$ and 
(d) that along $y=10$. The colour scale is logarithmic and to the peak value in the pair of panels. Image reproduced with permission from \citet{Dieckmann2018PPCF}, copyright by IOP.
} \label{Dieckmann18Fig6}
\end{figure}

\cite{Dieckmann2018PPCF} performed simulations that resolve a two-dimensional domain with the length $L_{\rm x}$ along $x$, which changes between simulations, and the fixed length $L_{\rm y}$ along $y$. 
The initially unmagnetized plasma consists of electrons and protons with the spatially uniform number density $n_{0}$. 
They use the correct proton-to-electron mass ratio with $L_{\rm x}= 55\lambda_{\rm D}= 125\Delta$ (Debye length $\lambda_{\rm D} = v_{\rm te}/\omega_{\rm pe}$. 
Figure~\ref{Dieckmann18Fig6} shows slices of the proton phase space density distribution along $x$ and along $y$ which sheds light on the cause of these density structures.
The distributions of the counterstreaming clouds are symmetric in Fig.~\ref{Dieckmann18Fig6}a and they are not symmetric in Figure~\ref{Dieckmann18Fig6}b. This symmetry break can be linked to the motion of the thin shell in the simulation frame. 
The thin shell at
$x = 13.75$ propagates from $y \approx  2.2$ at $t_{2}=177$ to $y \approx 15$ at $t_{4}=1600$, which corresponds to an average speed of $0.56c_{\rm s}$. 
The motion of the thin shell's electric potential to increasing $y$ implies that it affects the protons, which move in the opposite direction, less than the comoving ones. The counter-streaming proton clouds are well separated along $v_{\rm y}$, while 
this time a fully developed shock is observed. 
The modulation of the solitary wave by the NTSI apparently inhibits or delays the shock formation.
The slice of the proton phase space density distribution along $x=0$ in Fig.~\ref{Dieckmann18Fig6}c shows a depletion of the beam with $v_{\rm y} \approx 1.9c_{\rm s}$ at $x \approx 43$ and of the oppositely propagating proton beam at $x \approx 13.75$. The accumulation of protons at these values of $x$ and $y \approx 0$ seen in Fig.~\ref{Dieckmann18Fig6}d drives ambipolar electric fields, which are tilted relative to the $y$ axis. 
The protons that move to increasing $y$ are scattered by the oblique electric fields at $x=43$ and $y < 0$, which explains their depletion at $y = 0$ in Fig.~\ref{Dieckmann18Fig6}c. 
The scattering of the protons will broaden this depletion with increasing $y > 0$, which is confirmed by Fig.~\ref{Dieckmann18Fig6}d.
The proton density distribution in Fig.~\ref{Dieckmann18Fig6}d is caused by localised tilted electric fields and the ballistic motion of protons in the field-free intervals.

\vspace*{-0.01cm}
\citet{Dieckmann2018PPCF} have illustrated with a PIC simulation how a thin shell of dense plasma forms close to a planar collision boundary between two plasma clouds, which collide head on and orthogonal to the boundary, and how a pair of electrostatic shocks grows that separates the inflowing upstream protons from those in the thin shell. 
The absence of binary collisions between particles allowed some of the protons to traverse the thin shell and escape from it at the other side, while being re-accelerated by a double layer. 
The collisionless NTSI drove bending modes and breathing modes, which is also the case for the hydrodynamic one. 
The breathing modes involved large electric fields if the wavelength of the seed perturbation was short and they recovered here the complex spatial plasma density distribution from \cite{Dieckmann2015}. 
These structures could not develop in the case of the large initial seed wavelength due to the weaker electric field and a slower growth and they obtained the final proton density distribution from \citep{Dieckmann2015,Ahmed_2017}.

\rv{In the future, new numerical methods developed for laser plasma will be utilized for astrophysical plasma.}
\section{Summary and future outcome}
\label{sec:6}

Since the 1950s, PIC simulations -- initiated by a few scientists including Buneman, Hockney, Birdsall, and Dawson -- have been an alternative approach for the numerical investigation of plasma phenomena which is employed to solve plasma kinetic processes. 
In this review, we have described thoroughly the methods and applications of PIC simulations to laboratory, space, and astrophysical plasma associated to relativistic jets \rv{and various instabilities} providing useful insights into the evolution and associated phenomena such as the particle acceleration due to shocks as well as magnetic reconnection and radiation. 
\rv{In addition to the particle acceleration within relativistic jets, we have also discussed the physics of pulsars as well as discussed the particle acceleration in forced magnetic field turbulence. 
In contrast to astrophysical plasmas, contemporary PIC codes are also able to study the interaction of plasmas and laser beams. 
Present-day lasers are intense enough to create a density wake whose electric field 
accelerates charged particles making the laser-plasma interaction the ideal subject for PIC codes.}

\rv{Moreover, we have described future PIC simulations associated to the emission of electromagnetic radiation from binary mergers of neutron stars, which introduced the era of the multi-messenger physics of binary mergers. 
The merge of two colliding neutron stars not only emits gravitational waves, but also a significant electromagnetic component potentially capable of accelerating particle beams to relativistic energies which can be studied with relativistic PIC codes. 
Generally, PIC simulations become more and more powerful forming an excellent tool of studying the formation and physics of relativistic jets in the accretion 
disk system of rotating black holes and neutron stars.}

\rv{A future development of PIC simulations is to implement the full dynamic metrics in GRPIC codes including solving Einstein's equations self-consistently \citep{Baumgarte10NR}.}

Although, microscopic plasma particle simulations are computationally heavier than macroscopic plasma fluid simulations, they allow to study the individual particle behaviour in relativistic jets and are therefore suitable to understand jets in a more fundamental manner than by fluid simulations, or to even capture events which cannot be observed with fluid simulations.
Some of the key features that can be investigated with PIC, but not with fluid codes include, but are not limited to the following:

\begin {enumerate}
\item PIC models allow more sophisticated models of particle beams in shear-/shock-systems including Weibel, kinetic Kelvin--Helmholtz and mushroom instabilities.
\item PIC simulations involving Weibel instabilities have shown that some jet electrons which are injected into an ambient plasma, are thermalized whilst some electrons of the ambient plasma are heated by the jet electrons.
\item In addition to the Fermi process at shocks, magnetic reconnection provides an additional source of particle acceleration.
\item In turbulent astrophysical plasmas, there are incidents where ions partly energize more efficiently than electrons. 
\item Turbulent magnetic field accelerations occur in the nonlinear stage.
\item The magnetic reconnection in jets has the capability of accelerating particles up to energies of TeV which is beyond energies which can be explained by the WI, kKHI or MI.
\item PIC simulation can be used to understand the jet formation associated to mergers of neutrons stars, black holes or systems of neutron stars and black holes.
\item \rv{PIC codes are capable of simulating the formation and properties of relativistic jets in the accretion disk systems of black holes and neutron stars.}
\item PIC simulations have been used for laser-plasmas including QED effects.
\item \rv{GRPIC simulations will be further developed and provide new insights into the kinetic processes beyond the GRMHD simulations.}
\end {enumerate}

We would like to emphasize that according to Moore's original law \citep{moore_1965}, currently computational power doubles every 18 months such that performing relativistic PIC simulations will become a real alternative to fluid simulations in the distant future. 
Even though, Moore's law might fade these days, this does not mean that the development of computational power stagnates. 
Even without fulfilling Moore's law, computer chips are currently still becoming more powerful (even though more slowly) \citep{waldrop_2016}; additionally, Neven's law predicts that we might be on the edge to quantum supremacy \citep{mohseni_2017,arute_2019} which will raise computational power even more.  
However, even if the development of quantum computer systems might take decades before becoming realistic for PIC simulations, this year (2020) the barrier of exascale computing with $\gtrsim 10^{18}$ FLOPs has been broken by \href{https://www.techspot.com/news/84561-foldinghome-exceeds-15-exaflops-battle-against-covid-19.html}{the {\it Folding@Home} project}.
In the recent, exascale models for beam-plasma simulations 
\citep{vay_2018}, space weather prediction \citep{lapenta_2012,innocenti_2016} or multiscale systems \citep{alowayyed_2017} have been reported whilst PIC-plasma methods are currently being developed  \citep{vasileska_2020}.

Hence, in the future, both, exascale or quantum computing, will give us the opportunity to study relativistic jets in the vicinity of black holes or neutron stars  with much more precision than these days.
We believe that the further progresses of numerical methods of PIC simulations with advancing computing power ought to revise the contents in this review.

\label{ref}

%
%

\begin{acknowledgements}
We appreciate Seiji Zenitani, Keita Seto, Kenji Yoshida, Kouichi Hirotani, Yoshiharu Omura\id{,} and Bruno Giacomazzo for their 
critical reading and fruitful suggestions that improved the contents 
of this review. KI was supported by NSF AST-0908040,  NASA-NNX12AH06G, NNX13AP-21G, and
NNX13AP14G grants. The recent work is also provided by the NASA through Chandra Award Number 
GO7-18118X (PI: Ming Sun at UAH) issued by the Chandra X-ray Center, which 
is operated by the SAO 
for and on behalf of the NASA under contract NAS8-03060.  YM is supported by the ERC Synergy 
Grant `BlackHoleCam: Imaging the Event Horizon of Black Holes' (Grant No.\ 610058). 
The work of ID has been supported by the NUCLEU project. Simulations were performed using 
Pleiades and Endeavor facilities at NASA Advanced Supercomputing (NAS: s2004), using Comet 
at The San Diego Supercomputer Center (SDSC),  Bridges at The Pittsburgh 
Supercomputing Center, and Frontera at The Texas Advanced Computing Center,
which are supported by the NSF.
\end{acknowledgements}

\bibliographystyle{spbasic-FS}      
\bibliography{refs}   

\begin{thebibliography}{345}
\expandafter\ifx\csname url\endcsname\relax
 \def\url#1{\burl{#1}}\fi
\expandafter\ifx\csname urlprefix\endcsname\relax\def\urlprefix{URL }\fi
\providecommand{\bibinfo}[2]{#2}
\providecommand{\eprint}[2][]{\url{#2}}
\providecommand{\doi}[1]{\urlstyle{rm}\url{https://doi.org/#1}}

\bibitem[{{Abbott} et~al.(2020){Abbott}, et~al., {Kagra Collaboration}, and
  {VIRGO Collaboration}}]{Abbott2020LRR}
{Abbott} BP, et~al, {Kagra Collaboration} LSC, {VIRGO Collaboration} (2020)
  {Prospects for observing and localizing gravitational-wave transients with
  Advanced LIGO, Advanced Virgo and KAGRA}. Living Reviews in Relativity
  23(1):3. \doi{10.1007/s41114-020-00026-9}

\bibitem[{Abbott et~al.(2016{\natexlab{a}})}]{Abbott2016PhRvL.116f1102A}
Abbott BP, et~al. (2016{\natexlab{a}}) {Observation of Gravitational Waves from
  a Binary Black Hole Merger}. \prl 116(6):061102.
  \doi{10.1103/PhysRevLett.116.061102}.
  {\href{https://arxiv.org/abs/1602.03837}{{arXiv:1602.03837}}} {[gr-qc]}

\bibitem[{Abbott et~al.(2016{\natexlab{b}})}]{AbbottPhysRevLett.116.241102}
Abbott BP, et~al. (2016{\natexlab{b}}) Properties of the binary black hole
  merger {GW150914}. Phys Rev Lett 116:241102.
  \doi{10.1103/PhysRevLett.116.241102}

\bibitem[{Abbott et~al.(2017{\natexlab{a}})}]{abbott17c}
Abbott BP, et~al. (2017{\natexlab{a}}) {Gravitational Waves and Gamma-Rays from
  a Binary Neutron Star Merger: GW170817 and GRB 170817A}. \apjl 848(2):L13.
  \doi{10.3847/2041-8213/aa920c}.
  {\href{https://arxiv.org/abs/1710.05834}{{arXiv:1710.05834}}} {[astro-ph.HE]}

\bibitem[{Abbott et~al.(2017{\natexlab{b}})}]{abbott17b}
Abbott BP, et~al. (2017{\natexlab{b}}) {GW170817}: Observation of gravitational
  waves from a binary neutron star inspiral. Phys Rev Lett 119:161101.
  \doi{10.1103/PhysRevLett.119.161101}

\bibitem[{Abbott et~al.(2017{\natexlab{c}})}]{abbott17a}
Abbott BP, et~al. (2017{\natexlab{c}}) {Multi-messenger Observations of a
  Binary Neutron Star Merger}. \apjl 848(2):L12.
  \doi{10.3847/2041-8213/aa91c9}.
  {\href{https://arxiv.org/abs/1710.05833}{{arXiv:1710.05833}}} {[astro-ph.HE]}

\bibitem[{Abdalla et~al.(2017)}]{abdalla17}
Abdalla H, et~al. (2017) {TeV Gamma-Ray Observations of the Binary Neutron Star
  Merger GW170817 with H.E.S.S.} \apjl 850(2):L22.
  \doi{10.3847/2041-8213/aa97d2}.
  {\href{https://arxiv.org/abs/1710.05862}{{arXiv:1710.05862}}} {[astro-ph.HE]}

\bibitem[{{Abdo} and {Fermi GBM Collaboration}(2009)}]{Abdo09Sci}
{Abdo} AA, {Fermi GBM Collaboration} (2009) {Fermi Observations of High-Energy
  Gamma-Ray Emission from {GRB 080916C}}. Science 323(5922):1688.
  \doi{10.1126/science.1169101}

\bibitem[{{Aharonian} et~al.(2006)}]{Aharonian06}
{Aharonian} F, et~al. (2006) {Fast Variability of Tera-Electron Volt
  {\ensuremath{\gamma}} Rays from the Radio Galaxy {M87}}. Science
  314(5804):1424--1427. \doi{10.1126/science.1134408}.
  {\href{https://arxiv.org/abs/astro-ph/0612016}{{arXiv:astro-ph/0612016}}}
  {[astro-ph]}

\bibitem[{{Aharonian} et~al.(2007)}]{Aharonian07}
{Aharonian} F, et~al. (2007) {An Exceptional Very High Energy Gamma-Ray Flare
  of {PKS 2155-304}}. \apjl 664(2):L71--L74. \doi{10.1086/520635}.
  {\href{https://arxiv.org/abs/0706.0797}{{arXiv:0706.0797}}} {[astro-ph]}

\bibitem[{Ahmed et~al.(2017)Ahmed, Doria, Dieckmann, Sarri, Romagnani, Bret,
  Cerchez, Giesecke, Ianni, Kar, Notley, Prasad, Quinn, Willi, and
  Borghesi}]{Ahmed_2017}
Ahmed H, Doria D, Dieckmann ME, Sarri G, Romagnani L, Bret A, Cerchez M,
  Giesecke AL, Ianni E, Kar S, Notley M, Prasad R, Quinn K, Willi O, Borghesi M
  (2017) Experimental observation of thin-shell instability in a collisionless
  plasma. Astrophys J 834(2):L21. \doi{10.3847/2041-8213/834/2/l21}

\bibitem[{{Alowayyed} et~al.(2017){Alowayyed}, {Groen}, {Coveney}, and
  {Hoekstra}}]{alowayyed_2017}
{Alowayyed} S, {Groen} D, {Coveney} P, {Hoekstra} A (2017) {Multiscale
  computing in the exascale era}. J Comp Sci 22:15--25.
  \doi{10.1016/j.jocs.2017.07.004}

\bibitem[{{Aloy} et~al.(2000){Aloy}, {G{\'o}mez}, {Ib{\'a}{\~n}ez},
  {Mart{\'{\i}}}, and {M{\"u}ller}}]{aloy00}
{Aloy} MA, {G{\'o}mez} JL, {Ib{\'a}{\~n}ez} JM, {Mart{\'{\i}}} JM, {M{\"u}ller}
  E (2000) {Radio Emission from Three-dimensional Relativistic Hydrodynamic
  Jets: Observational Evidence of Jet Stratification}. Astrophys J Lett
  528:L85--L88. \doi{10.1086/312436}.
  {\href{https://arxiv.org/abs/astro-ph/9911153}{{astro-ph/9911153}}}

\bibitem[{Alves(2010)}]{alves10}
Alves EP (2010) Inductive and electrostatic acceleration in relativistic
  jet-plasma interactions. Master's thesis, Universidade T\'ecnica de Lisboa

\bibitem[{{Alves} et~al.(2012){Alves}, {Grismayer}, {Martins}, {Fi{\'u}za},
  {Fonseca}, and {Silva}}]{alves12}
{Alves} EP, {Grismayer} T, {Martins} SF, {Fi{\'u}za} F, {Fonseca} RA, {Silva}
  LO (2012) {Large-scale Magnetic Field Generation via the Kinetic
  Kelvin-Helmholtz Instability in Unmagnetized Scenarios}. \apjl 746(2):L14.
  \doi{10.1088/2041-8205/746/2/L14}.
  {\href{https://arxiv.org/abs/1107.6037}{{arXiv:1107.6037}}} {[astro-ph.HE]}

\bibitem[{{Alves} et~al.(2015{\natexlab{a}}){Alves}, {Grismayer}, {Fonseca},
  and {Silva}}]{alves14}
{Alves} EP, {Grismayer} T, {Fonseca} RA, {Silva} LO (2015{\natexlab{a}})
  {Transverse electron-scale instability in relativistic shear flows}. \pre
  92(2):021101. \doi{10.1103/PhysRevE.92.021101}.
  {\href{https://arxiv.org/abs/1505.06016}{{arXiv:1505.06016}}}
  {[physics.plasm-ph]}

\bibitem[{{Alves} et~al.(2015{\natexlab{b}}){Alves}, {Grismayer}, {Fonseca},
  and {Silva}}]{alves15}
{Alves} EP, {Grismayer} T, {Fonseca} RA, {Silva} LO (2015{\natexlab{b}})
  {Transverse electron-scale instability in relativistic shear flows}. PhRvE
  92(2):021101. \doi{10.1103/PhysRevE.92.021101}.
  {\href{https://arxiv.org/abs/1505.06016}{{arXiv:1505.06016}}}
  {[physics.plasm-ph]}

\bibitem[{Alves et~al.(2018)Alves, Zrake, and Fiuza}]{alves18eff}
Alves EP, Zrake J, Fiuza F (2018) Efficient nonthermal particle acceleration by
  the kink instability in relativistic jets. Phys Rev Lett 121:245101.
  \doi{10.1103/PhysRevLett.121.245101}

\bibitem[{{Alves} et~al.(2019){Alves}, {Zrake}, and {Fiuza}}]{Alves19ion}
{Alves} EP, {Zrake} J, {Fiuza} F (2019) {Nonthermal ion acceleration by the
  kink instability in nonrelativistic jets}. Phys Plasmas 26(7):072105.
  \doi{10.1063/1.5098478}.
  {\href{https://arxiv.org/abs/1907.11693}{{arXiv:1907.11693}}}
  {[physics.plasm-ph]}

\bibitem[{{Amano} and {Hoshino}(2007)}]{amano07}
{Amano} T, {Hoshino} M (2007) {Electron Injection at High Mach Number
  Quasi-perpendicular Shocks: Surfing and Drift Acceleration}. \apj
  661(1):190--202. \doi{10.1086/513599}.
  {\href{https://arxiv.org/abs/astro-ph/0612204}{{arXiv:astro-ph/0612204}}}
  {[astro-ph]}

\bibitem[{{Amano} and {Hoshino}(2009)}]{amano09}
{Amano} T, {Hoshino} M (2009) {Electron Shock Surfing Acceleration in
  Multidimensions: Two-Dimensional Particle-in-Cell Simulation of Collisionless
  Perpendicular Shock}. \apj 690(1):244--251.
  \doi{10.1088/0004-637X/690/1/244}.
  {\href{https://arxiv.org/abs/0805.1098}{{arXiv:0805.1098}}} {[astro-ph]}

\bibitem[{Arber et~al.(2015)Arber, Bennett, Brady, Lawrence-Douglas, Ramsay,
  Sircombe, Gillies, Evans, Schmitz, Bell, and Ridgers}]{Arber_2015}
Arber TD, Bennett K, Brady CS, Lawrence-Douglas A, Ramsay MG, Sircombe NJ,
  Gillies P, Evans RG, Schmitz H, Bell AR, Ridgers CP (2015) Contemporary
  particle-in-cell approach to laser-plasma modelling. Plasma Phys Control
  Fusion 57(11):113001. \doi{10.1088/0741-3335/57/11/113001}

\bibitem[{{Ardaneh} et~al.(2015){Ardaneh}, {Cai}, {Nishikawa}, and
  {Lemb{\'e}ge}}]{ardaneh15}
{Ardaneh} K, {Cai} D, {Nishikawa} KI, {Lemb{\'e}ge} B (2015) {Collisionless
  Weibel Shocks and Electron Acceleration in Gamma-Ray Bursts}. \apj 811(1):57.
  \doi{10.1088/0004-637X/811/1/57}.
  {\href{https://arxiv.org/abs/1507.05374}{{arXiv:1507.05374}}} {[astro-ph.HE]}

\bibitem[{Ardaneh et~al.(2016)Ardaneh, Cai, and Nishikawa}]{ardaneh16}
Ardaneh K, Cai D, Nishikawa KI (2016) Collisionless electron-ion shocks in
  relativistic unmagnetized jet-ambient interactions: Non-thermal electron
  injection by double layer. Astrophys J 827(2):124.
  \doi{10.3847/0004-637X/827/2/124}

\bibitem[{{Arnold} et~al.(2019){Arnold}, {Drake}, {Swisdak}, and
  {Dahlin}}]{arnold19}
{Arnold} H, {Drake} JF, {Swisdak} M, {Dahlin} J (2019) {Large-scale parallel
  electric fields and return currents in a global simulation model}. Phys
  Plasmas 26(10):102903. \doi{10.1063/1.5120373}.
  {\href{https://arxiv.org/abs/1907.07554}{{arXiv:1907.07554}}}
  {[physics.plasm-ph]}

\bibitem[{Arute and et~al.(2019)}]{arute_2019}
Arute F, et~al (2019) {Quantum supremacy using a programmable superconducting
  processor}. Nature 574(7779):505--510. \doi{10.1038/s41586-019-1666-5}.
  {\href{https://arxiv.org/abs/1910.11333}{{arXiv:1910.11333}}} {[quant-ph]}

\bibitem[{{Bai} et~al.(2015){Bai}, {Caprioli}, {Sironi}, and
  {Spitkovsky}}]{Bai2015}
{Bai} XN, {Caprioli} D, {Sironi} L, {Spitkovsky} A (2015)
  {Magnetohydrodynamic-particle-in-cell Method for Coupling Cosmic Rays with a
  Thermal Plasma: Application to Non-relativistic Shocks}. \apj 809(1):55.
  \doi{10.1088/0004-637X/809/1/55}.
  {\href{https://arxiv.org/abs/1412.1087}{{arXiv:1412.1087}}} {[astro-ph.HE]}

\bibitem[{{Baiotti} and {Rezzolla}(2017)}]{baiotti17}
{Baiotti} L, {Rezzolla} L (2017) {Binary neutron star mergers: a review of
  Einstein{\textquoteright}s richest laboratory}. Reports on Progress in
  Physics 80(9):096901. \doi{10.1088/1361-6633/aa67bb}.
  {\href{https://arxiv.org/abs/1607.03540}{{arXiv:1607.03540}}} {[gr-qc]}

\bibitem[{Barniol~Duran et~al.(2016)Barniol~Duran, Leng, and
  Giannios}]{barniol16}
Barniol~Duran R, Leng M, Giannios D (2016) An anisotropic minijets model for
  the grb prompt emission. Mon Not R Astron Soc Lett 455(1):L6--L10.
  \doi{10.1093/mnrasl/slv140}

\bibitem[{Barniol~Duran et~al.(2017)Barniol~Duran, Tchekhovskoy, and
  Giannios}]{barniol17}
Barniol~Duran R, Tchekhovskoy A, Giannios D (2017) Simulations of {AGN} jets:
  magnetic kink instability versus conical shocks. Mon Not R Astron Soc
  469(4):4957--4978. \doi{10.1093/mnras/stx1165}

\bibitem[{{Baty} et~al.(2013){Baty}, {Petri}, and {Zenitani}}]{Baty13}
{Baty} H, {Petri} J, {Zenitani} S (2013) {Explosive reconnection of double
  tearing modes in relativistic plasmas: application to the {Crab} flares}. Mon
  Not R Astron Soc 436:L20--L24. \doi{10.1093/mnrasl/slt104}.
  {\href{https://arxiv.org/abs/1308.0906}{{arXiv:1308.0906}}} {[astro-ph.HE]}

\bibitem[{{Baumgarte} and {Shapiro}(2010)}]{Baumgarte10NR}
{Baumgarte} TW, {Shapiro} SL (2010) {Numerical Relativity: Solving Einstein's
  Equations on the Computer}. Cambridge University Press.
  \doi{10.1017/CBO9781139193344}

\bibitem[{{Bird} et~al.(2021){Bird}, {Tan}, {Luedtke}, {Harrell}, {Taufer}, and
  {Albright}}]{Bird2021x}
{Bird} R, {Tan} N, {Luedtke} SV, {Harrell} SL, {Taufer} M, {Albright} B (2021)
  {VPIC 2.0: Next Generation Particle-in-Cell Simulations}. arXiv e-prints
  arXiv:2102.13133.
  {\href{https://arxiv.org/abs/2102.13133}{{arXiv:2102.13133}}} {[cs.DC]}

\bibitem[{Birdsall and Langdon(1991)}]{birdsall91}
Birdsall CK, Langdon AB (1991) Plasma physics via computer simulation. CRC
  Press 1(1):1--504

\bibitem[{{Birn} and {Priest}(2007)}]{Birn2007}
{Birn} J, {Priest} ER (2007) {Reconnection of magnetic fields:
  magnetohydrodynamics and collisionless theory and observations}. Cambridge
  University Press

\bibitem[{{Birn} et~al.(2001){Birn}, {Drake}, {Shay}, {Rogers}, {Denton},
  {Hesse}, {Kuznetsova}, {Ma}, {Bhattacharjee}, {Otto}, and
  {Pritchett}}]{birn01}
{Birn} J, {Drake} JF, {Shay} MA, {Rogers} BN, {Denton} RE, {Hesse} M,
  {Kuznetsova} M, {Ma} ZW, {Bhattacharjee} A, {Otto} A, {Pritchett} PL (2001)
  {Geospace Environmental Modeling (GEM) magnetic reconnection challenge}. J
  Geophys Res 106:3715--3720. \doi{10.1029/1999JA900449}

\bibitem[{{Biscoveanu} et~al.(2020){Biscoveanu}, {Thrane}, and
  {Vitale}}]{Biscoveanu2020ApJ}
{Biscoveanu} S, {Thrane} E, {Vitale} S (2020) {Constraining Short Gamma-Ray
  Burst Jet Properties with Gravitational Waves and Gamma-Rays}. \apj
  893(1):38. \doi{10.3847/1538-4357/ab7eaf}.
  {\href{https://arxiv.org/abs/1911.01379}{{arXiv:1911.01379}}} {[astro-ph.HE]}

\bibitem[{{Blandford} et~al.(2017){Blandford}, {Yuan}, {Hoshino}, and
  {Sironi}}]{Blandford2017}
{Blandford} R, {Yuan} Y, {Hoshino} M, {Sironi} L (2017) {Magnetoluminescence}.
  Space Sci Rev 207(1-4):291--317. \doi{10.1007/s11214-017-0376-2}.
  {\href{https://arxiv.org/abs/1705.02021}{{arXiv:1705.02021}}} {[astro-ph.HE]}

\bibitem[{{Blandford} et~al.(2019){Blandford}, {Meier}, and
  {Readhead}}]{Blandford2019}
{Blandford} R, {Meier} D, {Readhead} A (2019) {Relativistic Jets from Active
  Galactic Nuclei}. \araa 57:467--509.
  \doi{10.1146/annurev-astro-081817-051948}.
  {\href{https://arxiv.org/abs/1812.06025}{{arXiv:1812.06025}}} {[astro-ph.HE]}

\bibitem[{{Blandford} and {Znajek}(1977)}]{blandford1977}
{Blandford} RD, {Znajek} RL (1977) {Electromagnetic extraction of energy from
  Kerr black holes.} \mnras 179:433--456. \doi{10.1093/mnras/179.3.433}

\bibitem[{{Bogovalov}(1999)}]{Bogovalov1999}
{Bogovalov} SV (1999) {On the physics of cold MHD winds from oblique rotators}.
  \aap 349:1017--1026.
  {\href{https://arxiv.org/abs/astro-ph/9907051}{{arXiv:astro-ph/9907051}}}
  {[astro-ph]}

\bibitem[{{Borissov} et~al.(2020){Borissov}, {Neukirch}, {Kontar}, {Threlfall},
  and {Parnell}}]{Borissov2020}
{Borissov} A, {Neukirch} T, {Kontar} EP, {Threlfall} J, {Parnell} CE (2020)
  {Particle acceleration with anomalous pitch angle scattering in 3D separator
  reconnection}. \aap 635:A63. \doi{10.1051/0004-6361/201936977}.
  {\href{https://arxiv.org/abs/2001.07548}{{arXiv:2001.07548}}} {[astro-ph.SR]}

\bibitem[{{Bowers} et~al.(2008){Bowers}, {Albright}, {Yin}, {Bergen}, and
  {Kwan}}]{Bowers2008PhPl}
{Bowers} KJ, {Albright} BJ, {Yin} L, {Bergen} B, {Kwan} TJT (2008) {Ultrahigh
  performance three-dimensional electromagnetic relativistic kinetic plasma
  simulationa)}. Phys Plasmas 15(5):055703. \doi{10.1063/1.2840133}

\bibitem[{{Bret}(2009)}]{Bret2009}
{Bret} A (2009) {{Weibel}, Two-Stream, Filamentation, Oblique, Bell,
  Buneman...Which One Grows Faster?} \apj 699(2):990--1003.
  \doi{10.1088/0004-637X/699/2/990}.
  {\href{https://arxiv.org/abs/0903.2658}{{arXiv:0903.2658}}} {[astro-ph.HE]}

\bibitem[{{Broderick} and {Loeb}(2009)}]{broderick09}
{Broderick} AE, {Loeb} A (2009) {Imaging the Black Hole Silhouette of M87:
  Implications for Jet Formation and Black Hole Spin}. \apj 697(2):1164--1179.
  \doi{10.1088/0004-637X/697/2/1164}.
  {\href{https://arxiv.org/abs/0812.0366}{{arXiv:0812.0366}}} {[astro-ph]}

\bibitem[{Buneman(1993)}]{buneman93}
Buneman O (1993) Tristan: The 3-d electromagnetic particle code. In: Matsumoto
  H, Omura Y (eds) Computer Space Plasma Physics: Simulation Techniques and
  Software. Terra Scientific Publishing, Tokyo, pp 67--84.
  \urlprefix\url{http://www.terrapub.co.jp/e-library/cspp/index.html}

\bibitem[{{Buneman} et~al.(1992){Buneman}, {Neubert}, and
  {Nishikawa}}]{buneman92}
{Buneman} O, {Neubert} T, {Nishikawa} KI (1992) {Solar wind-magnetosphere
  interaction as simulated by a {3-D} {EM} particle code}. IEEE Trans Plasma
  Sci 20:810--816. \doi{10.1109/27.199533}

\bibitem[{Bussmann et~al.(2013)Bussmann, Burau, Cowan, Debus, Huebl, Juckeland,
  Kluge, Nagel, Pausch, Schmitt, Schramm, Schuchart, and
  Widera}]{Schuchart2020}
Bussmann M, Burau H, Cowan TE, Debus A, Huebl A, Juckeland G, Kluge T, Nagel
  WE, Pausch R, Schmitt F, Schramm U, Schuchart J, Widera R (2013) Radiative
  signatures of the relativistic kelvin-helmholtz instability. In: Proceedings
  of the International Conference on High Performance Computing, Networking,
  Storage and Analysis. SC '13. Association for Computing Machinery, New York,
  NY, USA. \doi{10.1145/2503210.2504564}

\bibitem[{Cai et~al.(2002)Cai, Li, Nishikawa, Xiao, and Yan}]{cai02}
Cai D, Li Y, Nishikawa Ki, Xiao C, Yan X (2002) Three-dimensional
  electromagnetic particle-in-cell code using high performance {Fortran} on
  {PC} cluster. In: Zima HP, Joe K, Sato M, Seo Y, Shimasaki M (eds) High
  Performance Computing. Springer, Berlin Heidelberg, pp 515--525.
  \doi{10.1007/3-540-47847-7_48}

\bibitem[{{Cai} et~al.(2003){Cai}, {Li}, {Nishikawa}, and {et al.}}]{cai03}
{Cai} D, {Li} Y, {Nishikawa} KI, {et al} (2003) {Parallel {3-D} Electromagnetic
  Particle Code Using High Performance {FORTRAN}: Parallel {TRISTAN}}. In:
  {B{\"u}chner} J, {Dum} C, {Scholer} M (eds) Space Plasma Simulation. Lecture
  Notes in Physics, vol 615. Springer, Berlin, pp 25--53.
  \doi{10.1007/3-540-36530-3_2}

\bibitem[{{Cai} et~al.(2015){Cai}, {Esmaeili}, {Lemb{\`e}ge}, and
  {Nishikawa}}]{cai15}
{Cai} D, {Esmaeili} A, {Lemb{\`e}ge} B, {Nishikawa} KI (2015) {Cusp dynamics
  under northward {IMF} using three-dimensional global particle-in-cell
  simulations}. J Geophys Res 120(10):8368--8386. \doi{10.1002/2015JA021230}

\bibitem[{{Cai} et~al.(2001){Cai}, {Li}, {Xiao}, and {Yan}}]{cai01}
{Cai} DS, {Li} YT, {Xiao} CJ, {Yan} XY (2001) {Three-dimensional
  electro-magnetic particle simulations of the solar wind-magnetosphere
  interaction with time-varying {IMF} using {HPF} {TRISTAN} code}. In:
  {B{\"u}chner} J, {Dum} CT, {Scholer} M (eds) Space Plasma Simulation. p~54

\bibitem[{Casse et~al.(2017)Casse, van Marle, and Marcowith}]{Casse_2017}
Casse F, van Marle AJ, Marcowith A (2017) On magnetic field amplification and
  particle acceleration near non-relativistic collisionless shocks: Particles
  in {MHD} cells simulations. Plasma Physics and Controlled Fusion
  60(1):014017. \doi{10.1088/1361-6587/aa8482},
  \urlprefix\url{https://doi.org/10.1088/1361-6587/aa8482}

\bibitem[{Cerutti and Beloborodov(2017)}]{Cerutti2017r}
Cerutti B, Beloborodov AM (2017) Electrodynamics of pulsar magnetospheres.
  Space Sci Rev 207(1):111--136. \doi{10.1007/s11214-016-0315-7}

\bibitem[{{Cerutti} and {Philippov}(2017{\natexlab{a}})}]{Cerutti2017}
{Cerutti} B, {Philippov} AA (2017{\natexlab{a}}) {Dissipation of the striped
  pulsar wind}. \aap 607:A134. \doi{10.1051/0004-6361/201731680}.
  {\href{https://arxiv.org/abs/1710.07320}{{arXiv:1710.07320}}} {[astro-ph.HE]}

\bibitem[{{Cerutti} and {Philippov}(2017{\natexlab{b}})}]{Cerutti17}
{Cerutti} B, {Philippov} AA (2017{\natexlab{b}}) {Dissipation of the striped
  pulsar wind}. Astron Astrophys 607:A134. \doi{10.1051/0004-6361/201731680}.
  {\href{https://arxiv.org/abs/1710.07320}{{arXiv:1710.07320}}} {[astro-ph.HE]}

\bibitem[{Cerutti et~al.(2013)Cerutti, Werner, Uzdensky, and
  Begelman}]{Cerutti_2013}
Cerutti B, Werner GR, Uzdensky DA, Begelman MC (2013) Simulations of particle
  acceleration beyond the classical synchrotron burnoff limit in magnetic
  reconnection: an explanation of the {Crab} flares. Astrophys J 770(2):147.
  \doi{10.1088/0004-637x/770/2/147}

\bibitem[{Cerutti et~al.(2014)Cerutti, Werner, Uzdensky, and
  Begelman}]{Cerutti_2014}
Cerutti B, Werner GR, Uzdensky DA, Begelman MC (2014) Three-dimensional
  relativistic pair plasma reconnection with radiative feedback in the {Crab}
  nebula. Astrophys J 782(2):104. \doi{10.1088/0004-637x/782/2/104}

\bibitem[{{Cerutti} et~al.(2016){Cerutti}, {Philippov}, and
  {Spitkovsky}}]{Cerutti2016}
{Cerutti} B, {Philippov} AA, {Spitkovsky} A (2016) {Modelling high-energy
  pulsar light curves from first principles}. \mnras 457(3):2401--2414.
  \doi{10.1093/mnras/stw124}.
  {\href{https://arxiv.org/abs/1511.01785}{{arXiv:1511.01785}}} {[astro-ph.HE]}

\bibitem[{{Cerutti} et~al.(2020){Cerutti}, {Philippov}, and
  {Dubus}}]{Cerutti2020A&A}
{Cerutti} B, {Philippov} AA, {Dubus} G (2020) {Dissipation of the striped
  pulsar wind and non-thermal particle acceleration: {3D} {PIC} simulations}.
  \aap 642:A204. \doi{10.1051/0004-6361/202038618}.
  {\href{https://arxiv.org/abs/2008.11462}{{arXiv:2008.11462}}} {[astro-ph.HE]}

\bibitem[{{Chang} et~al.(2008){Chang}, {Spitkovsky}, and {Arons}}]{chang08}
{Chang} P, {Spitkovsky} A, {Arons} J (2008) {Long-Term Evolution of Magnetic
  Turbulence in Relativistic Collisionless Shocks: Electron-Positron Plasmas}.
  \apj 674(1):378--387. \doi{10.1086/524764}.
  {\href{https://arxiv.org/abs/0704.3832}{{arXiv:0704.3832}}} {[astro-ph]}

\bibitem[{{Choi} et~al.(2014){Choi}, {Min}, {Nishikawa}, and {Choi}}]{choi14}
{Choi} EJ, {Min} K, {Nishikawa} KI, {Choi} CR (2014) {A study of the
  early-stage evolution of relativistic electron-ion shock using
  three-dimensional particle-in-cell simulations}. Phys Plasmas 21(7):072905.
  \doi{10.1063/1.4890479}

\bibitem[{Christie et~al.(2019)Christie, Petropoulou, Sironi, and
  Giannios}]{christie19}
Christie IM, Petropoulou M, Sironi L, Giannios D (2019) Radiative signatures of
  plasmoid-dominated reconnection in blazar jets. Mon Not R Astron Soc
  482(1):65--82. \doi{10.1093/mnras/sty2636}

\bibitem[{Ciolfi et~al.(2017)Ciolfi, Kastaun, Giacomazzo, Endrizzi, Siegel, and
  Perna}]{ciolfi17}
Ciolfi R, Kastaun W, Giacomazzo B, Endrizzi A, Siegel DM, Perna R (2017)
  General relativistic magnetohydrodynamic simulations of binary neutron star
  mergers forming a long-lived neutron star. Phys Rev D 95:063016.
  \doi{10.1103/PhysRevD.95.063016}

\bibitem[{{Clausen-Brown} et~al.(2011){Clausen-Brown}, {Lyutikov}, and
  {Kharb}}]{clausen11}
{Clausen-Brown} E, {Lyutikov} M, {Kharb} P (2011) {Signatures of large-scale
  magnetic fields in active galactic nuclei jets: transverse asymmetries}. Mon
  Not R Astron Soc 415:2081--2092. \doi{10.1111/j.1365-2966.2011.18757.x}

\bibitem[{{Comisso} and {Sironi}(2019)}]{Comisso19}
{Comisso} L, {Sironi} L (2019) {The Interplay of Magnetically Dominated
  Turbulence and Magnetic Reconnection in Producing Nonthermal Particles}.
  Astrophys J 886(2):122. \doi{10.3847/1538-4357/ab4c33}.
  {\href{https://arxiv.org/abs/1909.01420}{{arXiv:1909.01420}}} {[astro-ph.HE]}

\bibitem[{{Coroniti}(1990)}]{Coroniti1990}
{Coroniti} FV (1990) {Magnetically Striped Relativistic Magnetohydrodynamic
  Winds: The {Crab} Nebula Revisited}. \apj 349:538. \doi{10.1086/168340}

\bibitem[{{Crinquand} et~al.(2020){Crinquand}, {Cerutti}, {Philippov},
  {Parfrey}, and {Dubus}}]{Crinquand20x}
{Crinquand} B, {Cerutti} B, {Philippov} A, {Parfrey} K, {Dubus} G (2020)
  {Multidimensional Simulations of Ergospheric Pair Discharges around Black
  Holes}. \prl 124(14):145101. \doi{10.1103/PhysRevLett.124.145101}.
  {\href{https://arxiv.org/abs/2003.03548}{{arXiv:2003.03548}}} {[astro-ph.HE]}

\bibitem[{Daldorff et~al.(2014)Daldorff, T\'oth, Gombosi, Lapenta, Amaya,
  Markidis, and Brackbill}]{DALDORFF2014}
Daldorff LK, T\'oth G, Gombosi TI, Lapenta G, Amaya J, Markidis S, Brackbill JU
  (2014) Two-way coupling of a global hall magnetohydrodynamics model with a
  local implicit particle-in-cell model. J Comput Phys 268:236 -- 254.
  \doi{10.1016/j.jcp.2014.03.009}

\bibitem[{{D'Angelo}(1965)}]{d'amgelo65}
{D'Angelo} N (1965) {Kelvin}-{Helmholtz} instability in a fully ionized plasma
  in a magnetic field. Phys Fluids 8(9):1748--1750. \doi{10.1063/1.1761496}

\bibitem[{{d'Ascoli} et~al.(2018){d'Ascoli}, {Noble}, {Bowen}, {Campanelli},
  {Krolik}, and {Mewes}}]{d'Ascoli18}
{d'Ascoli} S, {Noble} SC, {Bowen} DB, {Campanelli} M, {Krolik} JH, {Mewes} V
  (2018) {Electromagnetic Emission from Supermassive Binary Black Holes
  Approaching Merger}. \apj 865(2):140. \doi{10.3847/1538-4357/aad8b4}.
  {\href{https://arxiv.org/abs/1806.05697}{{arXiv:1806.05697}}} {[astro-ph.HE]}

\bibitem[{{Daughton}(1999)}]{daughton99}
{Daughton} W (1999) {The unstable eigenmodes of a neutral sheet}. Phys Plasmas
  6:1329--1343. \doi{10.1063/1.873374}

\bibitem[{{Daughton} et~al.(2011){Daughton}, {Roytershteyn}, {Karimabadi},
  {Yin}, {Albright}, {Bergen}, and {Bowers}}]{Daughton11}
{Daughton} W, {Roytershteyn} V, {Karimabadi} H, {Yin} L, {Albright} BJ,
  {Bergen} B, {Bowers} KJ (2011) {Role of electron physics in the development
  of turbulent magnetic reconnection in collisionless plasmas}. Nature Physics
  7:539--542. \doi{10.1038/nphys1965}

\bibitem[{{Davelaar} et~al.(2020){Davelaar}, {Philippov}, {Bromberg}, and
  {Singh}}]{Davelaar2020}
{Davelaar} J, {Philippov} AA, {Bromberg} O, {Singh} CB (2020) {Particle
  Acceleration in Kink-unstable Jets}. \apjl 896(2):L31.
  \doi{10.3847/2041-8213/ab95a2}.
  {\href{https://arxiv.org/abs/1910.13370}{{arXiv:1910.13370}}} {[astro-ph.HE]}

\bibitem[{{Davis} and {Tchekhovskoy}(2020)}]{Davis2020}
{Davis} SW, {Tchekhovskoy} A (2020) {Magnetohydrodynamic Simulations of Active
  Galactic Nucleus Disks and Jets}. \araa 58(1):annurev.
  \doi{10.1146/annurev-astro-081817-051905}

\bibitem[{Dawson(1983)}]{dawson83}
Dawson JM (1983) Particle simulation of plasmas. Rev Mod Phys 55(2):403--447.
  \doi{10.1103/RevModPhys.55.403}

\bibitem[{{Dieckmann} et~al.(2008){Dieckmann}, {Shukla}, and
  {Drury}}]{dieckmann08}
{Dieckmann} ME, {Shukla} PK, {Drury} LOC (2008) {The Formation of a
  Relativistic Partially Electromagnetic Planar Plasma Shock}. Astrophys J
  675:586--595. \doi{10.1086/525516}.
  {\href{https://arxiv.org/abs/astro-ph/0702055}{{astro-ph/0702055}}}

\bibitem[{Dieckmann et~al.(2014)Dieckmann, Sarri, Doria, Ahmed, and
  Borghesi}]{Dieckmann_2014}
Dieckmann ME, Sarri G, Doria D, Ahmed H, Borghesi M (2014) Evolution of slow
  electrostatic shock into a plasma shock mediated by electrostatic turbulence.
  New J Phys 16(7):073001. \doi{10.1088/1367-2630/16/7/073001}

\bibitem[{{Dieckmann} et~al.(2015){Dieckmann}, {Ahmed}, {Doria}, {Sarri},
  {Walder}, {Folini}, {Bret}, {Ynnerman}, and {Borghesi}}]{Dieckmann2015}
{Dieckmann} ME, {Ahmed} H, {Doria} D, {Sarri} G, {Walder} R, {Folini} D, {Bret}
  A, {Ynnerman} A, {Borghesi} M (2015) {Thin-shell instability in collisionless
  plasma}. \pre 92(3):031101. \doi{10.1103/PhysRevE.92.031101}.
  {\href{https://arxiv.org/abs/1509.04504}{{arXiv:1509.04504}}} {[astro-ph.HE]}

\bibitem[{{Dieckmann} et~al.(2017){Dieckmann}, {Folini}, and
  {Walder}}]{Dieckmann2017}
{Dieckmann} ME, {Folini} D, {Walder} R (2017) {The interplay of the
  collisionless non-linear thin-shell instability with the ion acoustic
  instability}. \mnras 465(4):4240--4248. \doi{10.1093/mnras/stw3014}.
  {\href{https://arxiv.org/abs/1701.04051}{{arXiv:1701.04051}}}
  {[physics.plasm-ph]}

\bibitem[{{Dieckmann} et~al.(2018){Dieckmann}, {Doria}, {Sarri}, {Romagnani},
  {Ahmed}, {Folini}, {Walder}, {Bret}, and {Borghesi}}]{Dieckmann2018PPCF}
{Dieckmann} ME, {Doria} D, {Sarri} G, {Romagnani} L, {Ahmed} H, {Folini} D,
  {Walder} R, {Bret} A, {Borghesi} M (2018) {Electrostatic shock waves in the
  laboratory and astrophysics: similarities and differences}. Plasma Phys
  Control Fusion 60(1):014014. \doi{10.1088/1361-6587/aa8c8f}

\bibitem[{{Dieckmann} et~al.(2019){Dieckmann}, {Folini}, {Hotz}, {Nordman},
  {Dell'Acqua}, {Ynnerman}, and {Walder}}]{Dieckmann2019}
{Dieckmann} ME, {Folini} D, {Hotz} I, {Nordman} A, {Dell'Acqua} P, {Ynnerman}
  A, {Walder} R (2019) {Structure of a collisionless pair jet in a magnetized
  electron-proton plasma: flow-aligned magnetic field}. \aap 621:A142.
  \doi{10.1051/0004-6361/201834393}.
  {\href{https://arxiv.org/abs/1810.05415}{{arXiv:1810.05415}}} {[astro-ph.HE]}

\bibitem[{{Dong} et~al.(2020){Dong}, {Zhang}, and {Giannios}}]{Dong20}
{Dong} L, {Zhang} H, {Giannios} D (2020) {Kink Instabilities In Relativistic
  Jets Can Drive Quasi-Periodic Radiation Signatures}. Mon Not R Astron Soc
  \doi{10.1093/mnras/staa773}.
  {\href{https://arxiv.org/abs/2003.07765}{{arXiv:2003.07765}}} {[astro-ph.HE]}

\bibitem[{{Drake} et~al.(2019){Drake}, {Arnold}, {Swisdak}, and
  {Dahlin}}]{drakefe19}
{Drake} JF, {Arnold} H, {Swisdak} M, {Dahlin} JT (2019) {A computational model
  for exploring particle acceleration during reconnection in macroscale
  systems}. Phys Plasmas 26(1):012901. \doi{10.1063/1.5058140}.
  {\href{https://arxiv.org/abs/1809.04568}{{arXiv:1809.04568}}} {[astro-ph.SR]}

\bibitem[{{Du{\c{t}}an} et~al.(2016){Du{\c{t}}an}, Nishikawa, Mizuno, Niemiec,
  Kobzar, Pohl, G{\'o}mez, Pe'er, Frederiksen, Nordlund, and et~al.}]{dutan17}
{Du{\c{t}}an} I, Nishikawa KI, Mizuno Y, Niemiec J, Kobzar O, Pohl M, G{\'o}mez
  JL, Pe'er A, Frederiksen JT, Nordlund {\AA}, et~al (2016) Particle-in-cell
  simulations of global relativistic jets with helical magnetic fields.
  Proceedings of the International Astronomical Union 12(S324):199--202.
  \doi{10.1017/S1743921316012722}

\bibitem[{East et~al.(2016)East, Paschalidis, Pretorius, and Shapiro}]{east16}
East WE, Paschalidis V, Pretorius F, Shapiro SL (2016) Relativistic simulations
  of eccentric binary neutron star mergers: One-arm spiral instability and
  effects of neutron star spin. Phys Rev D 93:024011.
  \doi{10.1103/PhysRevD.93.024011}

\bibitem[{Esarey et~al.(2009)Esarey, Schroeder, and Leemans}]{Esarey09}
Esarey E, Schroeder CB, Leemans WP (2009) Physics of laser-driven plasma-based
  electron accelerators. Rev Mod Phys 81:1229--1285.
  \doi{10.1103/RevModPhys.81.1229}

\bibitem[{{Esirkepov}(2001)}]{Esirkepov2001}
{Esirkepov} TZ (2001) {Exact charge conservation scheme for Particle-in-Cell
  simulation with an arbitrary form-factor}. Comput Phys Commun
  135(2):144--153. \doi{10.1016/S0010-4655(00)00228-9}

\bibitem[{{Etienne} et~al.(2017){Etienne}, {Wan}, {Babiuc}, {McWilliams}, and
  {Choudhary}}]{etienne17}
{Etienne} ZB, {Wan} MB, {Babiuc} MC, {McWilliams} ST, {Choudhary} A (2017)
  {GiRaFFE: an open-source general relativistic force-free electrodynamics
  code}. Classical and Quantum Gravity 34(21):215001.
  \doi{10.1088/1361-6382/aa8ab3}.
  {\href{https://arxiv.org/abs/1704.00599}{{arXiv:1704.00599}}} {[gr-qc]}

\bibitem[{{Event Horizon Telescope Collaboration}(2019)}]{EHT2019_6}
{Event Horizon Telescope Collaboration} (2019) {First {M87} Event Horizon
  Telescope Results. {VI}. {T}he Shadow and Mass of the Central Black Hole}.
  Astrophys J Lett 875(1):L6. \doi{10.3847/2041-8213/ab1141}.
  {\href{https://arxiv.org/abs/1906.11243}{{arXiv:1906.11243}}} {[astro-ph.GA]}

\bibitem[{{Frederiksen} et~al.(2004){Frederiksen}, {Hededal}, {Haugb{\o}lle},
  and {Nordlund}}]{frederiksen04}
{Frederiksen} JT, {Hededal} CB, {Haugb{\o}lle} T, {Nordlund} {\r{A}} (2004)
  {Magnetic Field Generation in Collisionless Shocks: Pattern Growth and
  Transport}. \apjl 608(1):L13--L16. \doi{10.1086/421262}.
  {\href{https://arxiv.org/abs/astro-ph/0308104}{{arXiv:astro-ph/0308104}}}
  {[astro-ph]}

\bibitem[{{Frederiksen} et~al.(2010){Frederiksen}, {Haugb{\o}lle}, {Medvedev},
  and {Nordlund}}]{Frederiksen2010ApJ}
{Frederiksen} JT, {Haugb{\o}lle} T, {Medvedev} MV, {Nordlund} {\AA} (2010)
  {Radiation Spectral Synthesis of Relativistic Filamentation}. \apjl
  722(1):L114--L119. \doi{10.1088/2041-8205/722/1/L114}.
  {\href{https://arxiv.org/abs/1003.1140}{{arXiv:1003.1140}}} {[astro-ph.HE]}

\bibitem[{{Fried}(1959)}]{Fried1959}
{Fried} BD (1959) {Mechanism for Instability of Transverse Plasma Waves}.
  Physics of Fluids 2(3):337--337. \doi{10.1063/1.1705933}

\bibitem[{{Giacomazzo} et~al.(2012){Giacomazzo}, {Baker}, {Miller}, {Reynolds},
  and {van Meter}}]{Giacomazzo12}
{Giacomazzo} B, {Baker} JG, {Miller} MC, {Reynolds} CS, {van Meter} JR (2012)
  {General Relativistic Simulations of Magnetized Plasmas around Merging
  Supermassive Black Holes}. \apjl 752(1):L15.
  \doi{10.1088/2041-8205/752/1/L15}.
  {\href{https://arxiv.org/abs/1203.6108}{{arXiv:1203.6108}}} {[astro-ph.HE]}

\bibitem[{Giannios(2010)}]{giannios10}
Giannios D (2010) Uhecrs from magnetic reconnection in relativistic jets. Mon
  Not R Astron Soc Lett 408(1):L46--L50. \doi{10.1111/j.1745-3933.2010.00925.x}

\bibitem[{{Giannios}(2013)}]{Giannios2013}
{Giannios} D (2013) {Reconnection-driven plasmoids in blazars: fast flares on a
  slow envelope}. \mnras 431(1):355--363. \doi{10.1093/mnras/stt167}.
  {\href{https://arxiv.org/abs/1211.0296}{{arXiv:1211.0296}}} {[astro-ph.HE]}

\bibitem[{Giannios and Uzdensky(2019)}]{Giannios19}
Giannios D, Uzdensky DA (2019) {{GRB} and blazar jets shining through their
  stripes}. Mon Not R Astron Soc 484(1):1378--1389. \doi{10.1093/mnras/stz082}

\bibitem[{Giannios et~al.(2009)Giannios, Uzdensky, and Begelman}]{giannios09}
Giannios D, Uzdensky DA, Begelman MC (2009) Fast tev variability in blazars:
  jets in a jet. Mon Not R Astron Soc Lett 395(1):L29--L33.
  \doi{10.1111/j.1745-3933.2009.00635.x}

\bibitem[{Godfrey(1974)}]{GODFREY1974}
Godfrey BB (1974) Numerical cherenkov instabilities in electromagnetic particle
  codes. J Comput Phys 15(4):504--521. \doi{10.1016/0021-9991(74)90076-X}

\bibitem[{{Goldstein} et~al.(2017){Goldstein}, {Veres}, {Burns}, {Briggs},
  {Hamburg}, {Kocevski}, {Wilson-Hodge}, {Preece}, {Poolakkil}, {Roberts},
  {Hui}, {Connaughton}, {Racusin}, {von Kienlin}, {Dal Canton}, {Christensen},
  {Littenberg}, {Siellez}, {Blackburn}, {Broida}, {Bissaldi}, {Cleveland},
  {Gibby}, {Giles}, {Kippen}, {McBreen}, {McEnery}, {Meegan}, {Paciesas}, and
  {Stanbro}}]{goldstein17}
{Goldstein} A, {Veres} P, {Burns} E, {Briggs} MS, {Hamburg} R, {Kocevski} D,
  {Wilson-Hodge} CA, {Preece} RD, {Poolakkil} S, {Roberts} OJ, {Hui} CM,
  {Connaughton} V, {Racusin} J, {von Kienlin} A, {Dal Canton} T, {Christensen}
  N, {Littenberg} T, {Siellez} K, {Blackburn} L, {Broida} J, {Bissaldi} E,
  {Cleveland} WH, {Gibby} MH, {Giles} MM, {Kippen} RM, {McBreen} S, {McEnery}
  J, {Meegan} CA, {Paciesas} WS, {Stanbro} M (2017) {An Ordinary Short
  Gamma-Ray Burst with Extraordinary Implications: Fermi-GBM Detection of GRB
  170817A}. \apjl 848(2):L14. \doi{10.3847/2041-8213/aa8f41}.
  {\href{https://arxiv.org/abs/1710.05446}{{arXiv:1710.05446}}} {[astro-ph.HE]}

\bibitem[{{Gonoskov} et~al.(2015){Gonoskov}, {Bastrakov}, {Efimenko},
  {Ilderton}, {Marklund}, {Meyerov}, {Muraviev}, {Sergeev}, {Surmin}, and
  {Wallin}}]{Gonoskov15}
{Gonoskov} A, {Bastrakov} S, {Efimenko} E, {Ilderton} A, {Marklund} M,
  {Meyerov} I, {Muraviev} A, {Sergeev} A, {Surmin} I, {Wallin} E (2015)
  {Extended particle-in-cell schemes for physics in ultrastrong laser fields:
  Review and developments}. \pre 92(2):023305.
  \doi{10.1103/PhysRevE.92.023305}.
  {\href{https://arxiv.org/abs/1412.6426}{{arXiv:1412.6426}}}
  {[physics.plasm-ph]}

\bibitem[{{Gonoskov} et~al.(2011){Gonoskov}, {Korzhimanov}, {Kim}, {Marklund},
  and {Sergeev}}]{Gonoskov11}
{Gonoskov} AA, {Korzhimanov} AV, {Kim} AV, {Marklund} M, {Sergeev} AM (2011)
  {Ultrarelativistic nanoplasmonics as a route towards extreme-intensity
  attosecond pulses}. \pre 84(4):046403. \doi{10.1103/PhysRevE.84.046403}.
  {\href{https://arxiv.org/abs/1104.5375}{{arXiv:1104.5375}}}
  {[physics.plasm-ph]}

\bibitem[{Gonsalves et~al.(2019)Gonsalves, Nakamura, Daniels, Benedetti,
  Pieronek, de~Raadt, Steinke, Bin, Bulanov, van Tilborg, Geddes, Schroeder,
  T\'oth, Esarey, Swanson, Fan-Chiang, Bagdasarov, Bobrova, Gasilov, Korn,
  Sasorov, and Leemans}]{Gonsalves19}
Gonsalves AJ, Nakamura K, Daniels J, Benedetti C, Pieronek C, de~Raadt TCH,
  Steinke S, Bin JH, Bulanov SS, van Tilborg J, Geddes CGR, Schroeder CB,
  T\'oth C, Esarey E, Swanson K, Fan-Chiang L, Bagdasarov G, Bobrova N, Gasilov
  V, Korn G, Sasorov P, Leemans WP (2019) Petawatt laser guiding and electron
  beam acceleration to {8 GeV} in a laser-heated capillary discharge waveguide.
  Phys Rev Lett 122:084801. \doi{10.1103/PhysRevLett.122.084801}

\bibitem[{Granot(2012)}]{granot12}
Granot J (2012) Scaling relations between numerical simulations and physical
  systems they represent. Mon Not R Astron Soc 421(3):2610--2615.
  \doi{10.1111/j.1365-2966.2012.20489.x}

\bibitem[{Granot et~al.(2011)Granot, Komissarov, and Spitkovsky}]{granot11}
Granot J, Komissarov SS, Spitkovsky A (2011) Impulsive acceleration of strongly
  magnetized relativistic flows. Mon Not R Astron Soc 411(2):1323--1353.
  \doi{10.1111/j.1365-2966.2010.17770.x}

\bibitem[{Granot et~al.(2018)Granot, Gill, Guetta, and {De Colle}}]{granot18}
Granot J, Gill R, Guetta D, {De Colle} F (2018) Off-axis emission of short
  {GRB} jets from double neutron star mergers and {GRB 170817A}. Mon Not R
  Astron Soc 481(2):1597--1608. \doi{10.1093/mnras/sty2308}

\bibitem[{{Gravity Collaboration} et~al.(2017){Gravity Collaboration}, {Abuter}
  et~al.}]{Gravity17}
{Gravity Collaboration}, {Abuter} R, et~al. (2017) {First light for {GRAVITY}:
  Phase referencing optical interferometry for the Very Large Telescope
  Interferometer}. \aap 602:A94. \doi{10.1051/0004-6361/201730838}.
  {\href{https://arxiv.org/abs/1705.02345}{{arXiv:1705.02345}}} {[astro-ph.IM]}

\bibitem[{Greenwood et~al.(2004)Greenwood, Cartwright, Luginsland, and
  Baca}]{GREENWOOD2004}
Greenwood AD, Cartwright KL, Luginsland JW, Baca EA (2004) On the elimination
  of numerical cerenkov radiation in {PIC} simulations. J Comput Phys
  201(2):665 -- 684. \doi{10.1016/j.jcp.2004.06.021}

\bibitem[{Grismayer et~al.(2013)Grismayer, Alves, Fonseca, and
  Silva}]{grysmayer13a}
Grismayer T, Alves EP, Fonseca RA, Silva LO (2013) dc-magnetic-field generation
  in unmagnetized shear flows. Phys Rev Lett 111:015005.
  \doi{10.1103/PhysRevLett.111.015005}

\bibitem[{{Grismayer} et~al.(2013b){Grismayer}, {Alves}, {Fonseca}, and
  {Silva}}]{grysmayer13b}
{Grismayer} T, {Alves} EP, {Fonseca} RA, {Silva} LO (2013b) {Theory of
  multidimensional electron-scale instabilities in unmagnetized shear flows}.
  Plasma Physics and Controlled Fusion 55(12):124031.
  \doi{10.1088/0741-3335/55/12/124031}

\bibitem[{{Gruzinov}(2008)}]{gruzino08}
{Gruzinov} A (2008) {{GRB}: magnetic fields, cosmic rays, and emission from
  first principles?} ArXiv e-prints
  {\href{https://arxiv.org/abs/0803.1182}{{arXiv:0803.1182}}}

\bibitem[{Guo et~al.(2015)Guo, Liu, Daughton, and Li}]{Guo_2015}
Guo F, Liu YH, Daughton W, Li H (2015) Particle acceleration and plasma
  dynamics during magnetic reconnection in the magnetically dominated regime.
  Astrophys J 806(2):167. \doi{10.1088/0004-637x/806/2/167}

\bibitem[{Guo et~al.(2016)Guo, Li, Li, Daughton, Zhang, Lloyd-Ronning, Liu,
  Zhang, and Deng}]{Guo_2016}
Guo F, Li X, Li H, Daughton W, Zhang B, Lloyd-Ronning N, Liu YH, Zhang H, Deng
  W (2016) Efficient production of high-energy nonthermal particles during
  magnetic reconnection in a magnetically dominated ion--electron plasma.
  Astrophys J 818(1):L9. \doi{10.3847/2041-8205/818/1/l9}

\bibitem[{Guo et~al.(2019)Guo, Li, Daughton, Kilian, Li, Liu, Yan, and
  Ma}]{Guo_2019}
Guo F, Li X, Daughton W, Kilian P, Li H, Liu YH, Yan W, Ma D (2019) Determining
  the dominant acceleration mechanism during relativistic magnetic reconnection
  in large-scale systems. Astrophys J 879(2):L23.
  \doi{10.3847/2041-8213/ab2a15}

\bibitem[{{Guo} et~al.(2020){Guo}, {Liu}, {Li}, {Li}, {Daughton}, and
  {Kilian}}]{Guo2020PhPl}
{Guo} F, {Liu} YH, {Li} X, {Li} H, {Daughton} W, {Kilian} P (2020) {Recent
  progress on particle acceleration and reconnection physics during magnetic
  reconnection in the magnetically-dominated relativistic regime}. Phys Plasmas
  27(8):080501. \doi{10.1063/5.0012094}.
  {\href{https://arxiv.org/abs/2006.15288}{{arXiv:2006.15288}}} {[astro-ph.HE]}

\bibitem[{{Guo} et~al.(2014){Guo}, {Sironi}, and {Narayan}}]{guo14}
{Guo} X, {Sironi} L, {Narayan} R (2014) {Non-thermal Electron Acceleration in
  Low Mach Number Collisionless Shocks. I. Particle Energy Spectra and
  Acceleration Mechanism}. \apj 794(2):153. \doi{10.1088/0004-637X/794/2/153}.
  {\href{https://arxiv.org/abs/1406.5190}{{arXiv:1406.5190}}} {[astro-ph.HE]}

\bibitem[{{Harris}(1962)}]{harris62}
{Harris} EG (1962) {On a plasma sheath separating regions of oppositely
  directed magnetic field}. Nuovo Cimento 23:115--121. \doi{10.1007/BF02733547}

\bibitem[{Hawley et~al.(2015)Hawley, Fendt, Hardcastle, Nokhrima, and
  Tchekhovskoy}]{hawley15}
Hawley JF, Fendt C, Hardcastle M, Nokhrima E, Tchekhovskoy A (2015) Disks and
  jets. Space Sci Rev 191(S324):441--469. \doi{10.1007/s1121}

\bibitem[{{Hededal}(2005)}]{Hededal2005PhDT}
{Hededal} C (2005) {Gamma-Ray Bursts, Collisionless Shocks and Synthetic
  Spectra}. PhD thesis, -.
  \urlprefix\url{https://arxiv.org/pdf/astro-ph/0506559.pdf}

\bibitem[{{Hededal} and {Nishikawa}(2005)}]{hededakk05}
{Hededal} CB, {Nishikawa} KI (2005) {The Influence of an Ambient Magnetic Field
  on Relativistic collisionless Plasma Shocks}. \apjl 623(2):L89--L92.
  \doi{10.1086/430253}.
  {\href{https://arxiv.org/abs/astro-ph/0412317}{{arXiv:astro-ph/0412317}}}
  {[astro-ph]}

\bibitem[{{Hededal} et~al.(2004){Hededal}, {Haugb{\o}lle}, {Frederiksen}, and
  {Nordlund}}]{heledal04}
{Hededal} CB, {Haugb{\o}lle} T, {Frederiksen} JT, {Nordlund} {\r{A}} (2004)
  {Non-Fermi Power-Law Acceleration in Astrophysical Plasma Shocks}. \apjl
  617(2):L107--L110. \doi{10.1086/427387}.
  {\href{https://arxiv.org/abs/astro-ph/0408558}{{arXiv:astro-ph/0408558}}}
  {[astro-ph]}

\bibitem[{{Higuera} and {Cary}(2017)}]{Higuera17}
{Higuera} AV, {Cary} JR (2017) {Structure-preserving second-order integration
  of relativistic charged particle trajectories in electromagnetic fields}.
  Phys Plasmas 24(5):052104. \doi{10.1063/1.4979989}.
  {\href{https://arxiv.org/abs/1701.05605}{{arXiv:1701.05605}}}
  {[physics.plasm-ph]}

\bibitem[{{Hirabayashi} and {Hoshino}(2017)}]{Hirabayashi2017}
{Hirabayashi} K, {Hoshino} M (2017) {Stratified Simulations of Collisionless
  Accretion Disks}. \apj 842(1):36. \doi{10.3847/1538-4357/aa74b3}.
  {\href{https://arxiv.org/abs/1705.06507}{{arXiv:1705.06507}}} {[astro-ph.HE]}

\bibitem[{{Hirotani} et~al.(2021){Hirotani}, {Krasnopolsky}, {Shang},
  {Nishikawa}, and {Watson}}]{Hirotani2021}
{Hirotani} K, {Krasnopolsky} R, {Shang} H, {Nishikawa} Ki, {Watson} M (2021)
  {Two-dimensional Particle-in-cell Simulations of Axisymmetric Black Hole
  Magnetospheres}. \apj 908(1):88. \doi{10.3847/1538-4357/abd3a6}.
  {\href{https://arxiv.org/abs/2012.07229}{{arXiv:2012.07229}}} {[astro-ph.HE]}

\bibitem[{{Hockney} and {Eastwood}(1988)}]{hockney88}
{Hockney} RW, {Eastwood} JW (1988) {Computer simulation using particles}.
  Hilger, Bristol Philadelphia

\bibitem[{Hockney and Eastwood(1989)}]{hockney89}
Hockney RW, Eastwood JW (1989) Computer simulation using particles. CRC Press
  1(1):1--540

\bibitem[{{Hoshino}(2015)}]{Hoshino2015PhRvL}
{Hoshino} M (2015) {Angular Momentum Transport and Particle Acceleration During
  Magnetorotational Instability in a Kinetic Accretion Disk}. \prl
  114(6):061101. \doi{10.1103/PhysRevLett.114.061101}.
  {\href{https://arxiv.org/abs/1502.02452}{{arXiv:1502.02452}}} {[astro-ph.HE]}

\bibitem[{{Hoshino} and {Shimada}(2002)}]{hoshino02}
{Hoshino} M, {Shimada} N (2002) {Nonthermal Electrons at High Mach Number
  Shocks: Electron Shock Surfing Acceleration}. \apj 572(2):880--887.
  \doi{10.1086/340454}.
  {\href{https://arxiv.org/abs/astro-ph/0203073}{{arXiv:astro-ph/0203073}}}
  {[astro-ph]}

\bibitem[{{Innocenti} et~al.(2016){Innocenti}, {Johnson}, {Markidis}, {Amaya},
  {Deca}, {Olshesky}, and {Lapenta}}]{innocenti_2016}
{Innocenti} ME, {Johnson} A, {Markidis} S, {Amaya} J, {Deca} J, {Olshesky} V,
  {Lapenta} G (2016) {Progress towards physics-based space weather forecasting
  with exascale computing}. Adv Eng Software 111:3--17.
  \doi{10.1016/j.advengsoft.2016.06.011}

\bibitem[{Ioka and Nakamura(2018)}]{ioka18}
Ioka K, Nakamura T (2018) Can an off-axis gamma-ray burst jet in {GW170817}
  explain all the electromagnetic counterparts? Progress of Theoretical and
  Experimental Physics 2018(4):043E02. \doi{10.1093/ptep/pty036}

\bibitem[{{Jackson}(1999)}]{Jackson1999}
{Jackson} JD (1999) {Classical electrodynamics}. John Wiley \& Sons, Inc, New
  York, 3rd edition. \doi{10.1007/3-540-36530-3}

\bibitem[{{Jaroschek} et~al.(2005){Jaroschek}, {Lesch}, and
  {Treumann}}]{jaroschek05}
{Jaroschek} CH, {Lesch} H, {Treumann} RA (2005) {Ultrarelativistic Plasma Shell
  Collisions in {\ensuremath{\gamma}}-Ray Burst Sources: Dimensional Effects on
  the Final Steady State Magnetic Field}. \apj 618(2):822--831.
  \doi{10.1086/426066}

\bibitem[{J\"uttner(1911)}]{juettner11}
J\"uttner F (1911) Das {M}axwellsche {G}esetz der {G}eschwindigkeitsverteilung
  in der {R}elativtheorie. Ann Phys 339(5):856--882.
  \doi{10.1002/andp.19113390503}

\bibitem[{{Kagan} et~al.(2013){Kagan}, {Milosavljevi{\'c}}, and
  {Spitkovsky}}]{kagan13}
{Kagan} D, {Milosavljevi{\'c}} M, {Spitkovsky} A (2013) {A Flux Rope Network
  and Particle Acceleration in Three-dimensional Relativistic Magnetic
  Reconnection}. Astrophys J 774(1):41. \doi{10.1088/0004-637X/774/1/41}.
  {\href{https://arxiv.org/abs/1208.0849}{{arXiv:1208.0849}}} {[astro-ph.HE]}

\bibitem[{Kagan et~al.(2015)Kagan, Sironi, Cerutti, and Giannios}]{Kagan2015}
Kagan D, Sironi L, Cerutti B, Giannios D (2015) Relativistic magnetic
  reconnection in pair plasmas and its astrophysical applications. Space Sci
  Rev 191(1):545--573. \doi{10.1007/s11214-014-0132-9}

\bibitem[{{Kagan} et~al.(2016){Kagan}, {Nakar}, and {Piran}}]{Kagan2016}
{Kagan} D, {Nakar} E, {Piran} T (2016) {Beaming of Particles and Synchrotron
  Radiation in Relativistic Magnetic Reconnection}. \apj 826(2):221.
  \doi{10.3847/0004-637X/826/2/221}.
  {\href{https://arxiv.org/abs/1601.07349}{{arXiv:1601.07349}}} {[astro-ph.HE]}

\bibitem[{{Karimabadi} et~al.(1991){Karimabadi}, {Omidi}, and
  {Quest}}]{Karimabadi1991}
{Karimabadi} H, {Omidi} N, {Quest} KB (1991) {Two-dimensional simulations of
  the ion/ion acoustic instability and electrostatic shocks}. \grl
  18(10):1813--1816. \doi{10.1029/91GL02241}

\bibitem[{Karimabadi et~al.(2014)Karimabadi, Roytershteyn, Vu, Omelchenko,
  Scudder, Daughton, Dimmock, Nykyri, Wan, Sibeck, Tatineni, Majumdar, Loring,
  and Geveci}]{karimabadi14}
Karimabadi H, Roytershteyn V, Vu HX, Omelchenko YA, Scudder J, Daughton W,
  Dimmock A, Nykyri K, Wan M, Sibeck D, Tatineni M, Majumdar A, Loring B,
  Geveci B (2014) The link between shocks, turbulence, and magnetic
  reconnection in collisionless plasmas. Phys Plasmas 21(6):062308.
  \doi{10.1063/1.4882875}

\bibitem[{{Kato} and {Takabe}(2010)}]{Kato2010}
{Kato} TN, {Takabe} H (2010) {Electrostatic and electromagnetic instabilities
  associated with electrostatic shocks: Two-dimensional particle-in-cell
  simulation}. Phys Plasmas 17(3):032114--032114. \doi{10.1063/1.3372138}.
  {\href{https://arxiv.org/abs/1003.1217}{{arXiv:1003.1217}}}
  {[physics.plasm-ph]}

\bibitem[{Kawamura et~al.(2016)Kawamura, Giacomazzo, Kastaun, Ciolfi, Endrizzi,
  Baiotti, and Perna}]{kawamura16}
Kawamura T, Giacomazzo B, Kastaun W, Ciolfi R, Endrizzi A, Baiotti L, Perna R
  (2016) Binary neutron star mergers and short gamma-ray bursts: Effects of
  magnetic field orientation, equation of state, and mass ratio. Phys Rev D
  94:064012. \doi{10.1103/PhysRevD.94.064012}

\bibitem[{Kelly et~al.(2017)Kelly, Baker, Etienne, Giacomazzo, and
  Schnittman}]{kelly17}
Kelly BJ, Baker JG, Etienne ZB, Giacomazzo B, Schnittman J (2017) Prompt
  electromagnetic transients from binary black hole mergers. Phys Rev D
  96:123003. \doi{10.1103/PhysRevD.96.123003}

\bibitem[{{Kimura} et~al.(2019){Kimura}, {Tomida}, and {Murase}}]{Kimura2019}
{Kimura} SS, {Tomida} K, {Murase} K (2019) {Acceleration and escape processes
  of high-energy particles in turbulence inside hot accretion flows}. \mnras
  485(1):163--178. \doi{10.1093/mnras/stz329}.
  {\href{https://arxiv.org/abs/1812.03901}{{arXiv:1812.03901}}} {[astro-ph.HE]}

\bibitem[{{Kirk} et~al.(2009){Kirk}, {Lyubarsky}, and {Petri}}]{Kirk2009}
{Kirk} JG, {Lyubarsky} Y, {Petri} J (2009) {The Theory of Pulsar Winds and
  Nebulae}. In: {Becker} W (ed) Astrophysics and Space Science Library. vol
  357. Springer, Berlin Heidelberg, p 421. \doi{10.1007/978-3-540-76965-1_16}

\bibitem[{K{\"o}hn et~al.(2017)K{\"o}hn, Diniz, and Harakeh}]{koehn_2017}
K{\"o}hn C, Diniz G, Harakeh MN (2017) Production mechanisms of leptons,
  photons, and hadrons and their possible feedback close to lightning leaders.
  Journal of Geophysical Research: Atmospheres 122(2):1365--1383.
  \doi{https://doi.org/10.1002/2016JD025445},
  \urlprefix\url{https://agupubs.onlinelibrary.wiley.com/doi/abs/10.1002/2016JD025445}.
  {\href{https://arxiv.org/abs/https://agupubs.onlinelibrary.wiley.com/doi/pdf/10.1002/2016JD025445}{{https://agupubs.onlinelibrary.wiley.com/doi/pdf/10.1002/2016JD025445}}}

\bibitem[{{K{\"o}hn} et~al.(2018){K{\"o}hn}, {Enghoff}, and
  {Svensmark}}]{koehn_2018}
{K{\"o}hn} C, {Enghoff} MB, {Svensmark} H (2018) {A 3D particle Monte Carlo
  approach to studying nucleation}. Journal of Computational Physics
  363:30--38. \doi{10.1016/j.jcp.2018.02.032}

\bibitem[{K{\"o}hn et~al.(2020)K{\"o}hn, Chanrion, Nishikawa, Babich, and
  Neubert}]{koehn_2020}
K{\"o}hn C, Chanrion O, Nishikawa K, Babich L, Neubert T (2020) The emission of
  energetic electrons from the complex streamer corona adjacent to leader
  stepping. Plasma Sources Sci Technol 29(3):035023.
  \doi{10.1088/1361-6595/ab6e57}

\bibitem[{{Komissarov}(2007)}]{komissarov07}
{Komissarov} SS (2007) {Multidimensional numerical scheme for resistive
  relativistic magnetohydrodynamics}. Mon Not R Astron Soc 382(3):995--1004.
  \doi{10.1111/j.1365-2966.2007.12448.x}.
  {\href{https://arxiv.org/abs/0708.0323}{{arXiv:0708.0323}}} {[astro-ph]}

\bibitem[{Komissarov(2012)}]{komissarov12}
Komissarov SS (2012) Shock dissipation in magnetically dominated impulsive
  flows. Mon Not R Astron Soc 422(1):326--346.
  \doi{10.1111/j.1365-2966.2012.20609.x}

\bibitem[{{Komissarov} et~al.(2009){Komissarov}, {Vlahakis}, {K{\"o}nigl}, and
  {Barkov}}]{komissarov09}
{Komissarov} SS, {Vlahakis} N, {K{\"o}nigl} A, {Barkov} MV (2009) {Magnetic
  acceleration of ultrarelativistic jets in gamma-ray burst sources}. Mon Not R
  Astron Soc 394(3):1182--1212. \doi{10.1111/j.1365-2966.2009.14410.x}.
  {\href{https://arxiv.org/abs/0811.1467}{{arXiv:0811.1467}}} {[astro-ph]}

\bibitem[{Kumar et~al.(2020)Kumar, Kim, Kang, Hur, and Chung}]{kumar2020}
Kumar S, Kim YK, Kang T, Hur MS, Chung M (2020) Evolution of magnetic field in
  a weakly relativistic counterstreaming inhomogeneous $e^-/e^+$ plasmas. Laser
  Particle Beams 38(3):181--187. \doi{10.1017/S0263034620000233}

\bibitem[{{Kunz} et~al.(2016){Kunz}, {Stone}, and {Quataert}}]{Kunz2016}
{Kunz} MW, {Stone} JM, {Quataert} E (2016) {Magnetorotational Turbulence and
  Dynamo in a Collisionless Plasma}. \prl 117(23):235101.
  \doi{10.1103/PhysRevLett.117.235101}.
  {\href{https://arxiv.org/abs/1608.07911}{{arXiv:1608.07911}}} {[astro-ph.HE]}

\bibitem[{Lalescu et~al.(2015)Lalescu, Shi, Eyink, Drivas, Vishniac, and
  Lazarian}]{Lalescu2015}
Lalescu CC, Shi YK, Eyink GL, Drivas TD, Vishniac ET, Lazarian A (2015)
  Inertial-range reconnection in magnetohydrodynamic turbulence and in the
  solar wind. Phys Rev Lett 115:025001. \doi{10.1103/PhysRevLett.115.025001}

\bibitem[{{Lapenta}(2012)}]{lapenta_2012}
{Lapenta} G (2012) {Particle simulations of space weather}. Journal of
  Computational Physics 231(3):795--821. \doi{10.1016/j.jcp.2011.03.035}

\bibitem[{Lapenta et~al.(2020{\natexlab{a}})Lapenta, Berchem, Alaoui, and
  Walker}]{Lapenta2020prl}
Lapenta G, Berchem J, Alaoui ME, Walker R (2020{\natexlab{a}}) Turbulent
  energization of electron power law tails during magnetic reconnection. Phys
  Rev Lett 125:225101. \doi{10.1103/PhysRevLett.125.225101},
  \urlprefix\url{https://link.aps.org/doi/10.1103/PhysRevLett.125.225101}

\bibitem[{Lapenta et~al.(2020{\natexlab{b}})Lapenta, El~Alaoui, Berchem, and
  Walker}]{Lapenta2020}
Lapenta G, El~Alaoui M, Berchem J, Walker R (2020{\natexlab{b}}) Multiscale
  mhd-kinetic pic study of energy fluxes caused by reconnection. Journal of
  Geophysical Research: Space Physics 125(3):e2019JA027276.
  \doi{https://doi.org/10.1029/2019JA027276},
  \urlprefix\url{https://agupubs.onlinelibrary.wiley.com/doi/abs/10.1029/2019JA027276},
  e2019JA027276 10.1029/2019JA027276.
  {\href{https://arxiv.org/abs/https://agupubs.onlinelibrary.wiley.com/doi/pdf/10.1029/2019JA027276}{{https://agupubs.onlinelibrary.wiley.com/doi/pdf/10.1029/2019JA027276}}}

\bibitem[{Lautenbach and Grauer(2018)}]{Lautenbach2018}
Lautenbach S, Grauer R (2018) Multiphysics simulations of collisionless
  plasmas. Frontiers in Physics 6:113. \doi{10.3389/fphy.2018.00113}

\bibitem[{{Lazarian} et~al.(2012){Lazarian}, {Vlahos}, {Kowal}, {Yan},
  {Beresnyak}, and {de Gouveia Dal Pino}}]{lazarian12r}
{Lazarian} A, {Vlahos} L, {Kowal} G, {Yan} H, {Beresnyak} A, {de Gouveia Dal
  Pino} EM (2012) {Turbulence, Magnetic Reconnection in Turbulent Fluids and
  Energetic Particle Acceleration}. Space Sci Rev 173(1-4):557--622.
  \doi{10.1007/s11214-012-9936-7}.
  {\href{https://arxiv.org/abs/1211.0008}{{arXiv:1211.0008}}} {[astro-ph.SR]}

\bibitem[{{Lazarian} et~al.(2020){Lazarian}, {Eyink}, {Jafari}, {Kowal}, {Li},
  {Xu}, and {Vishniac}}]{Lazarian2020}
{Lazarian} A, {Eyink} GL, {Jafari} A, {Kowal} G, {Li} H, {Xu} S, {Vishniac} ET
  (2020) {3D turbulent reconnection: Theory, tests, and astrophysical
  implications}. Phys Plasmas 27(1):012305. \doi{10.1063/1.5110603}.
  {\href{https://arxiv.org/abs/2001.00868}{{arXiv:2001.00868}}} {[astro-ph.HE]}

\bibitem[{{Lazzati} et~al.(2018){Lazzati}, {Perna}, {Morsony}, {Lopez-Camara},
  {Cantiello}, {Ciolfi}, {Giacomazzo}, and {Workman}}]{Lazzati18}
{Lazzati} D, {Perna} R, {Morsony} BJ, {Lopez-Camara} D, {Cantiello} M, {Ciolfi}
  R, {Giacomazzo} B, {Workman} JC (2018) {Late Time Afterglow Observations
  Reveal a Collimated Relativistic Jet in the Ejecta of the Binary Neutron Star
  Merger {GW170817}}. \prl 120(24):241103.
  \doi{10.1103/PhysRevLett.120.241103}.
  {\href{https://arxiv.org/abs/1712.03237}{{arXiv:1712.03237}}} {[astro-ph.HE]}

\bibitem[{Lehe et~al.(2016)Lehe, Kirchen, Andriyash, Godfrey, and
  Vay}]{LEHE2016}
Lehe R, Kirchen M, Andriyash IA, Godfrey BB, Vay JL (2016) A spectral,
  quasi-cylindrical and dispersion-free particle-in-cell algorithm. Computer
  Physics Communications 203:66 -- 82. \doi{10.1016/j.cpc.2016.02.007}

\bibitem[{{Liang} et~al.(2013a){Liang}, {Boettcher}, and {Smith}}]{liang13a}
{Liang} E, {Boettcher} M, {Smith} I (2013a) {Magnetic Field Generation and
  Particle Energization at Relativistic Shear Boundaries in Collisionless
  Electron-Positron Plasmas}. \apjl 766(2):L19.
  \doi{10.1088/2041-8205/766/2/L19}.
  {\href{https://arxiv.org/abs/1111.3326}{{arXiv:1111.3326}}} {[astro-ph.HE]}

\bibitem[{{Liang} et~al.(2013b){Liang}, {Fu}, {Boettcher}, {Smith}, and
  {Roustazadeh}}]{liang13b}
{Liang} E, {Fu} W, {Boettcher} M, {Smith} I, {Roustazadeh} P (2013b)
  {Relativistic Positron-Electron-Ion Shear Flows and Application to Gamma-Ray
  Bursts}. \apjl 779(2):L27. \doi{10.1088/2041-8205/779/2/L27}.
  {\href{https://arxiv.org/abs/1303.2153}{{arXiv:1303.2153}}} {[astro-ph.HE]}

\bibitem[{{Liang} et~al.(2017){Liang}, {Fu}, and {B{\"o}ttcher}}]{liang17}
{Liang} E, {Fu} W, {B{\"o}ttcher} M (2017) {Relativistic Shear Flow between
  Electron-Ion and Electron-Positron Plasmas and Astrophysical Applications}.
  Astrophys J 847(2):90. \doi{10.3847/1538-4357/aa8772}.
  {\href{https://arxiv.org/abs/1612.07418}{{arXiv:1612.07418}}} {[astro-ph.HE]}

\bibitem[{{Liang} et~al.(2018){Liang}, {Fu}, {B{\"o}ttcher}, and
  {Roustazadeh}}]{liang18}
{Liang} E, {Fu} W, {B{\"o}ttcher} M, {Roustazadeh} P (2018) {Scaling of
  Relativistic Shear Flows with the Bulk Lorentz Factor}. Astrophys J
  854(2):129. \doi{10.3847/1538-4357/aaa7f5}.
  {\href{https://arxiv.org/abs/1703.06135}{{arXiv:1703.06135}}} {[astro-ph.HE]}

\bibitem[{Liang(1981)}]{liang81}
Liang EPT (1981) Inverse comptonization and the nature of the march 1979
  {$\gamma$}-ray burst event. Nature 292:319--321. \doi{10.1038/292319a0}

\bibitem[{{Liska} et~al.(2019){Liska}, {Chatterjee}, {Tchekhovskoy}, {Yoon},
  {van Eijnatten}, {Hesp}, {Markoff}, {Ingram}, and {van der
  Klis}}]{Liska2019x}
{Liska} M, {Chatterjee} K, {Tchekhovskoy} A, {Yoon} D, {van Eijnatten} D,
  {Hesp} C, {Markoff} S, {Ingram} A, {van der Klis} M (2019) {H-AMR: A New
  GPU-accelerated GRMHD Code for Exascale Computing With 3D Adaptive Mesh
  Refinement and Local Adaptive Time-stepping}. arXiv e-prints
  arXiv:1912.10192.
  {\href{https://arxiv.org/abs/1912.10192}{{arXiv:1912.10192}}} {[astro-ph.HE]}

\bibitem[{{Liska} et~al.(2020{\natexlab{a}}){Liska}, {Hesp}, {Tchekhovskoy},
  {Ingram}, {van der Klis}, {Markoff}, and {Van Moer}}]{Liska2020m}
{Liska} M, {Hesp} C, {Tchekhovskoy} A, {Ingram} A, {van der Klis} M, {Markoff}
  SB, {Van Moer} M (2020{\natexlab{a}}) {Disc Tearing and Bardeen-Petterson
  Alignment in GRMHD Simulations of Highly Tilted Thin Accretion Discs}. \mnras
  \doi{10.1093/mnras/staa099}.
  {\href{https://arxiv.org/abs/1904.08428}{{arXiv:1904.08428}}} {[astro-ph.HE]}

\bibitem[{{Liska} et~al.(2020{\natexlab{b}}){Liska}, {Tchekhovskoy}, and
  {Quataert}}]{Liska2020L}
{Liska} M, {Tchekhovskoy} A, {Quataert} E (2020{\natexlab{b}}) {Large-scale
  poloidal magnetic field dynamo leads to powerful jets in GRMHD simulations of
  black hole accretion with toroidal field}. \mnras 494(3):3656--3662.
  \doi{10.1093/mnras/staa955}.
  {\href{https://arxiv.org/abs/1809.04608}{{arXiv:1809.04608}}} {[astro-ph.HE]}

\bibitem[{{Liu} et~al.(2018){Liu}, {Hesse}, {Guo}, {Li}, and
  {Nakamura}}]{liu18}
{Liu} YH, {Hesse} M, {Guo} F, {Li} H, {Nakamura} TKM (2018) {Strongly localized
  magnetic reconnection by the super-Alfv{\'e}nic shear flow}. Phys Plasmas
  25(8):080701. \doi{10.1063/1.5042539}.
  {\href{https://arxiv.org/abs/1806.00533}{{arXiv:1806.00533}}}
  {[physics.plasm-ph]}

\bibitem[{{Lyubarsky}(2019)}]{Lyubarsky2019}
{Lyubarsky} Y (2019) {Radio emission of the {Crab} and {Crab}-like pulsars}.
  \mnras 483(2):1731--1736. \doi{10.1093/mnras/sty3233}.
  {\href{https://arxiv.org/abs/1811.11122}{{arXiv:1811.11122}}} {[astro-ph.HE]}

\bibitem[{{Lyutikov} et~al.(2019){Lyutikov}, {Temim}, {Komissarov}, {Slane},
  {Sironi}, and {Comisso}}]{lyutikov_2019}
{Lyutikov} M, {Temim} T, {Komissarov} S, {Slane} P, {Sironi} L, {Comisso} L
  (2019) {Interpreting Crab Nebula's synchrotron spectrum: two acceleration
  mechanisms}. \mnras 489(2):2403--2416. \doi{10.1093/mnras/stz2023}.
  {\href{https://arxiv.org/abs/1811.01767}{{arXiv:1811.01767}}} {[astro-ph.HE]}

\bibitem[{{MacDonald} and {Marscher}(2018)}]{nick18}
{MacDonald} NR, {Marscher} AP (2018) {Faraday Conversion in Turbulent Blazar
  Jets}. \apj 862(1):58. \doi{10.3847/1538-4357/aacc62}.
  {\href{https://arxiv.org/abs/1611.09954}{{arXiv:1611.09954}}} {[astro-ph.HE]}

\bibitem[{Makwana et~al.(2017)Makwana, Keppens, and Lapenta}]{MAKWANA2017}
Makwana K, Keppens R, Lapenta G (2017) Two-way coupling of magnetohydrodynamic
  simulations with embedded particle-in-cell simulations. Computer Physics
  Communications 221:81 -- 94. \doi{10.1016/j.cpc.2017.08.003}

\bibitem[{{Marcowith} et~al.(2020){Marcowith}, {Ferrand}, {Grech}, {Meliani},
  {Plotnikov}, and {Walder}}]{Marcowith2020}
{Marcowith} A, {Ferrand} G, {Grech} M, {Meliani} Z, {Plotnikov} I, {Walder} R
  (2020) {Multi-scale simulations of particle acceleration in astrophysical
  systems}. Living Rev Comput Astrophys 6:1. \doi{10.1007/s41115-020-0007-6}.
  {\href{https://arxiv.org/abs/2002.09411}{{arXiv:2002.09411}}} {[astro-ph.HE]}

\bibitem[{Markidis et~al.(2010)Markidis, Lapenta, and
  Rizwan-uddin}]{MARKIDIS2010}
Markidis S, Lapenta G, Rizwan-uddin (2010) Multi-scale simulations of plasma
  with {iPIC3D}. Mathematics and Computers in Simulation 80(7):1509--1519.
  \doi{10.1016/j.matcom.2009.08.038}, multiscale modeling of moving interfaces
  in materials

\bibitem[{Markidis et~al.(2014)Markidis, Lapenta, Delzanno, Henri, Goldman,
  Newman, Intrator, and Laure}]{Markidis_2014}
Markidis S, Lapenta G, Delzanno GL, Henri P, Goldman MV, Newman DL, Intrator T,
  Laure E (2014) Signatures of secondary collisionless magnetic reconnection
  driven by kink instability of a flux rope. Plasma Phys Control Fusion
  56(6):064010. \doi{10.1088/0741-3335/56/6/064010}

\bibitem[{{Mart{\'\i}}(2019)}]{Marti2019}
{Mart{\'\i}} JM (2019) {Numerical Simulations of Jets from Active Galactic
  Nuclei}. Galaxies 7(1):24. \doi{10.3390/galaxies7010024}

\bibitem[{Mart{\'i} and M{\"u}ller(2015)}]{marti15}
Mart{\'i} JM, M{\"u}ller E (2015) Grid-based methods in relativistic
  hydrodynamics and magnetohydrodynamics. Living Rev Comput Astrophys 1:3.
  \doi{10.1007/lrca-2015-3}

\bibitem[{{Martins} et~al.(2009){Martins}, {Fonseca}, {Silva}, and
  {Mori}}]{martins09}
{Martins} SF, {Fonseca} RA, {Silva} LO, {Mori} WB (2009) {Ion Dynamics and
  Acceleration in Relativistic Shocks}. \apjl 695(2):L189--L193.
  \doi{10.1088/0004-637X/695/2/L189}.
  {\href{https://arxiv.org/abs/0903.3573}{{arXiv:0903.3573}}} {[astro-ph.HE]}

\bibitem[{Matsumoto et~al.(2017)Matsumoto, Amano, Kato, and
  Hoshino}]{matsumoto17}
Matsumoto Y, Amano T, Kato TN, Hoshino M (2017) Electron surfing and drift
  accelerations in a {Weibel}-dominated high-{Mach}-number shock. Phys Rev Lett
  119:105101. \doi{10.1103/PhysRevLett.119.105101}

\bibitem[{McKinney and Uzdensky(2012)}]{mckinney12}
McKinney JC, Uzdensky DA (2012) A reconnection switch to trigger gamma-ray
  burst jet dissipation. Mon Not R Astron Soc 419(1):573--607.
  \doi{10.1111/j.1365-2966.2011.19721.x}

\bibitem[{{Medvedev} and {Loeb}(1999)}]{medvedev99}
{Medvedev} MV, {Loeb} A (1999) {Generation of Magnetic Fields in the
  Relativistic Shock of Gamma-Ray Burst Sources}. \apj 526(2):697--706.
  \doi{10.1086/308038}.
  {\href{https://arxiv.org/abs/astro-ph/9904363}{{arXiv:astro-ph/9904363}}}
  {[astro-ph]}

\bibitem[{{Medvedev} and {Zakutnyaya}(2009)}]{medvedev09}
{Medvedev} MV, {Zakutnyaya} OV (2009) {Magnetic Fields and Cosmic Rays in GRBs:
  A Self-Similar Collisionless Foreshock}. \apj 696(2):2269--2274.
  \doi{10.1088/0004-637X/696/2/2269}.
  {\href{https://arxiv.org/abs/0812.1906}{{arXiv:0812.1906}}} {[astro-ph]}

\bibitem[{Medvedev et~al.(2011)Medvedev, Frederiksen, Haugb{\o}lle, and
  Nordlund}]{Medvedev_2011}
Medvedev MV, Frederiksen JT, Haugb{\o}lle T, Nordlund {\AA} (2011) Radiation
  signatures of sub-{Larmor} scale magnetic fields. Astrophys J 737(2):55.
  \doi{10.1088/0004-637x/737/2/55}

\bibitem[{Meier(2008)}]{meier08}
Meier DL (2008) {Exhaust inspection}. Nature 452:pages 945--946.
  \doi{10.1038/452945b}

\bibitem[{Meier et~al.(2001)Meier, Koide, and Uchida}]{Meier01}
Meier DL, Koide S, Uchida Y (2001) Magnetohydrodynamic production of
  relativistic jets. Science 291(5501):84--92.
  \doi{10.1126/science.291.5501.84}

\bibitem[{{Meli} et~al.(2021){Meli}, {Nishikawa}, {Dutan}, {Mizuno}, {Niemiec},
  {Gomez}, {Pohl}, {K{\"o}hn}, and {MacDonald}}]{Meli2021}
{Meli} A, {Nishikawa} K, {Dutan} I, {Mizuno} Y, {Niemiec} J, {Gomez} JL, {Pohl}
  M, {K{\"o}hn} C, {MacDonald} N (2021) {Particle Acceleration in Relativistic
  Electron-positron Jets with Helical Magnetic Fields}. arXiv e-prints
  arXiv:2009.04158.
  {\href{https://arxiv.org/abs/2009.04158}{{arXiv:2009.04158}}} {[astro-ph.HE]}

\bibitem[{{Melzani} et~al.(2013){Melzani}, {Winisdoerffer}, {Walder}, {Folini},
  {Favre}, {Krastanov}, and {Messmer}}]{Melzani2013}
{Melzani} M, {Winisdoerffer} C, {Walder} R, {Folini} D, {Favre} JM, {Krastanov}
  S, {Messmer} P (2013) {Apar-T: code, validation, and physical interpretation
  of particle-in-cell results}. \aap 558:A133.
  \doi{10.1051/0004-6361/201321557}.
  {\href{https://arxiv.org/abs/1308.5892}{{arXiv:1308.5892}}} {[astro-ph.HE]}

\bibitem[{{Melzani} et~al.(2014{\natexlab{a}}){Melzani}, {Walder}, {Folini},
  {Winisdoerffer}, and {Favre}}]{Melzani2014a}
{Melzani} M, {Walder} R, {Folini} D, {Winisdoerffer} C, {Favre} JM
  (2014{\natexlab{a}}) {Relativistic magnetic reconnection in collisionless
  ion-electron plasmas explored with particle-in-cell simulations}. \aap
  570:A111. \doi{10.1051/0004-6361/201424083}.
  {\href{https://arxiv.org/abs/1404.7366}{{arXiv:1404.7366}}} {[astro-ph.HE]}

\bibitem[{{Melzani} et~al.(2014{\natexlab{b}}){Melzani}, {Walder}, {Folini},
  {Winisdoerffer}, and {Favre}}]{Melzani2014b}
{Melzani} M, {Walder} R, {Folini} D, {Winisdoerffer} C, {Favre} JM
  (2014{\natexlab{b}}) {The energetics of relativistic magnetic reconnection:
  ion-electron repartition and particle distribution hardness}. \aap 570:A112.
  \doi{10.1051/0004-6361/201424193}.
  {\href{https://arxiv.org/abs/1405.2938}{{arXiv:1405.2938}}} {[astro-ph.HE]}

\bibitem[{Messmer(2001)}]{Messmer2001T}
Messmer P (2001) {Observations and simulations of particle acceleration in
  solar flares}. PhD Thesis, Insstute of Astronomy 14412.
  \doi{https://doi.org/10.3929/ethz-a-004225437}

\bibitem[{{Messmer}(2002)}]{Messmer2002}
{Messmer} P (2002) {Temperature isotropization in solar flare plasmas due to
  the electron firehose instability}. \aap 382:301--311.
  \doi{10.1051/0004-6361:20011583}.
  {\href{https://arxiv.org/abs/astro-ph/0111303}{{arXiv:astro-ph/0111303}}}
  {[astro-ph]}

\bibitem[{{Metzger}(2017{\natexlab{a}})}]{Metzger2017r}
{Metzger} BD (2017{\natexlab{a}}) Kilonovae. Living Rev Relativ 20:3.
  \doi{10.1007/s41114-017-0006-z}

\bibitem[{{Metzger}(2017{\natexlab{b}})}]{metzger17x}
{Metzger} BD (2017{\natexlab{b}}) {Welcome to the Multi-Messenger Era! Lessons
  from a Neutron Star Merger and the Landscape Ahead}. ArXiv e-prints
  {\href{https://arxiv.org/abs/1710.05931}{{arXiv:1710.05931}}} {[astro-ph.HE]}

\bibitem[{{Metzger} and {Berger}(2012)}]{Metzger12}
{Metzger} BD, {Berger} E (2012) {What is the Most Promising Electromagnetic
  Counterpart of a Neutron Star Binary Merger?} \apj 746(1):48.
  \doi{10.1088/0004-637X/746/1/48}.
  {\href{https://arxiv.org/abs/1108.6056}{{arXiv:1108.6056}}} {[astro-ph.HE]}

\bibitem[{{Michel}(1971)}]{Michel1971}
{Michel} FC (1971) {Coherent Neutral Sheet Radiation from Pulsars}. Comments on
  Astrophysics and Space Physics 3:80

\bibitem[{{Michel}(1973)}]{Michel1973}
{Michel} FC (1973) {Rotating Magnetospheres: an Exact 3-D Solution}. \apjl
  180:L133. \doi{10.1086/181169}

\bibitem[{Miniati and Elyiv(2013)}]{Miniati_2013}
Miniati F, Elyiv A (2013) Relaxation of blazar-induced pair beams in cosmic
  voids. Astrophys J 770(1):54. \doi{10.1088/0004-637x/770/1/54}

\bibitem[{{Mizuno}(2013)}]{mizuno13}
{Mizuno} Y (2013) {The Role of the Equation of State in Resistive Relativistic
  Magnetohydrodynamics}. Astrophys J Suppl Ser 205(1):7.
  \doi{10.1088/0067-0049/205/1/7}.
  {\href{https://arxiv.org/abs/1301.6052}{{arXiv:1301.6052}}} {[astro-ph.HE]}

\bibitem[{{Mizuno} et~al.(2007){Mizuno}, {Hardee}, and {Nishikawa}}]{mizuno07}
{Mizuno} Y, {Hardee} P, {Nishikawa} KI (2007) {Three-dimensional Relativistic
  Magnetohydrodynamic Simulations of Magnetized Spine-Sheath Relativistic
  Jets}. Astrophys J 662:835--850. \doi{10.1086/518106}.
  {\href{https://arxiv.org/abs/astro-ph/0703190}{{astro-ph/0703190}}}

\bibitem[{{Mizuno} et~al.(2009){Mizuno}, {Lyubarsky}, {Nishikawa}, and
  {Hardee}}]{mizuno09}
{Mizuno} Y, {Lyubarsky} Y, {Nishikawa} KI, {Hardee} PE (2009)
  {Three-Dimensional Relativistic Magnetohydrodynamic Simulations of
  Current-Driven Instability. I. Instability of a Static Column}. \apj
  700(1):684--693. \doi{10.1088/0004-637X/700/1/684}.
  {\href{https://arxiv.org/abs/0903.5358}{{arXiv:0903.5358}}} {[astro-ph.HE]}

\bibitem[{{Mizuno} et~al.(2014){Mizuno}, {Hardee}, and {Nishikawa}}]{mizuno14}
{Mizuno} Y, {Hardee} PE, {Nishikawa} KI (2014) {Spatial Growth of the
  Current-driven Instability in Relativistic Jets}. \apj 784(2):167.
  \doi{10.1088/0004-637X/784/2/167}.
  {\href{https://arxiv.org/abs/1402.2370}{{arXiv:1402.2370}}} {[astro-ph.HE]}

\bibitem[{{Mizuno} et~al.(2015){Mizuno}, {G{\'o}mez}, {Nishikawa}, {Meli},
  {Hardee}, and {Rezzolla}}]{mizuno15}
{Mizuno} Y, {G{\'o}mez} JL, {Nishikawa} KI, {Meli} A, {Hardee} PE, {Rezzolla} L
  (2015) {Recollimation Shocks in Magnetized Relativistic Jets}. \apj
  809(1):38. \doi{10.1088/0004-637X/809/1/38}.
  {\href{https://arxiv.org/abs/1505.00933}{{arXiv:1505.00933}}} {[astro-ph.HE]}

\bibitem[{Mohseni et~al.(2017)Mohseni, Read, Neven, Boixo, Denchev, Babbush,
  Fowler, Smelyanskiy, and Martinis}]{mohseni_2017}
Mohseni M, Read P, Neven H, Boixo S, Denchev V, Babbush R, Fowler A,
  Smelyanskiy V, Martinis J (2017) Commercialize quantum technologies in five
  years. Nature 543:171--174.
  \urlprefix\url{https://www.nature.com/news/commercialize-quantum-technologies-in-five-years-1.21583}

\bibitem[{Moore(1965)}]{moore_1965}
Moore GE (1965) Cramming more components onto integrated circuits. Electronics
  38:114--117.
  \urlprefix\url{https://www.cs.utexas.edu/~fussell/courses/cs352h/papers/moore.pdf}

\bibitem[{Mo\'scibrodzka et~al.(2017)Mo\'scibrodzka, Dexter, Davelaar, and
  Falcke}]{moscibrodzka17}
Mo\'scibrodzka M, Dexter J, Davelaar J, Falcke H (2017) Faraday rotation in
  {GRMHD} simulations of the jet launching zone of {M87}. Mon Not R Astron Soc
  468(2):2214--2221. \doi{10.1093/mnras/stx587}

\bibitem[{{M{\"o}sta} et~al.(2020){M{\"o}sta}, {Radice}, {Haas}, {Schnetter},
  and {Bernuzzi}}]{Moesta2020ApJ}
{M{\"o}sta} P, {Radice} D, {Haas} R, {Schnetter} E, {Bernuzzi} S (2020) {A
  Magnetar Engine for Short {GRB}s and Kilonovae}. \apjl 901(2):L37.
  \doi{10.3847/2041-8213/abb6ef}.
  {\href{https://arxiv.org/abs/2003.06043}{{arXiv:2003.06043}}} {[astro-ph.HE]}

\bibitem[{Na et~al.(2020)Na, Nicolini, Lee, Borges, Omelchenko, and
  Teixeira}]{NA2020}
Na DY, Nicolini J, Lee R, Borges BH, Omelchenko Y, Teixeira F (2020) Diagnosing
  numerical cherenkov instabilities in relativistic plasma simulations based on
  general meshes. J Comput Phys 402:108880. \doi{10.1016/j.jcp.2019.108880}

\bibitem[{{Nagakura} et~al.(2014){Nagakura}, {Hotokezaka}, {Sekiguchi},
  {Shibata}, and {Ioka}}]{nagakura14}
{Nagakura} H, {Hotokezaka} K, {Sekiguchi} Y, {Shibata} M, {Ioka} K (2014) {Jet
  Collimation in the Ejecta of Double Neutron Star Mergers: A New Canonical
  Picture of Short Gamma-Ray Bursts}. \apjl 784(2):L28.
  \doi{10.1088/2041-8205/784/2/L28}.
  {\href{https://arxiv.org/abs/1403.0956}{{arXiv:1403.0956}}} {[astro-ph.HE]}

\bibitem[{{Nakar}(2020)}]{Nakar2020PhR}
{Nakar} E (2020) {The electromagnetic counterparts of compact binary mergers}.
  \physrep 886:1--84. \doi{10.1016/j.physrep.2020.08.008}.
  {\href{https://arxiv.org/abs/1912.05659}{{arXiv:1912.05659}}} {[astro-ph.HE]}

\bibitem[{Ng and Noble(2006)}]{ng06}
Ng JST, Noble RJ (2006) Inductive and electrostatic acceleration in
  relativistic jet-plasma interactions. Phys Rev Lett 96:115006.
  \doi{10.1103/PhysRevLett.96.115006}

\bibitem[{{Niemiec} et~al.(2008){Niemiec}, {Pohl}, {Stroman}, and
  {Nishikawa}}]{niemiec08}
{Niemiec} J, {Pohl} M, {Stroman} T, {Nishikawa} KI (2008) {Production of
  Magnetic Turbulence by Cosmic Rays Drifting Upstream of Supernova Remnant
  Shocks}. \apj 684(2):1174--1189. \doi{10.1086/590054}.
  {\href{https://arxiv.org/abs/0802.2185}{{arXiv:0802.2185}}} {[astro-ph]}

\bibitem[{{Nishikawa} et~al.(2020){Nishikawa}, {Mizuno}, {G{\'o}mez},
  {Du{\c{t}}an}, {Niemiec}, {Kobzar}, {MacDonald}, {Meli}, {Pohl}, and
  {Hirotani}}]{Nishikawa2020}
{Nishikawa} K, {Mizuno} Y, {G{\'o}mez} JL, {Du{\c{t}}an} I, {Niemiec} J,
  {Kobzar} O, {MacDonald} N, {Meli} A, {Pohl} M, {Hirotani} K (2020) {Rapid
  particle acceleration due to recollimation shocks and turbulent magnetic
  fields in injected jets with helical magnetic fields}. Mon Not R Astron Soc
  493(2):2652--2658. \doi{10.1093/mnras/staa421}.
  {\href{https://arxiv.org/abs/1906.10302}{{arXiv:1906.10302}}} {[astro-ph.HE]}

\bibitem[{{Nishikawa}(1997)}]{nishikawa97}
{Nishikawa} KI (1997) {Particle entry into the magnetosphere with a southward
  interplanetary magnetic field studied by a three-dimensional electromagnetic
  particle code}. J Geophys Res 102(A8):17631--17642. \doi{10.1029/97JA00826}

\bibitem[{{Nishikawa} et~al.(2003){Nishikawa}, {Hardee}, {Richardson},
  {Preece}, {Sol}, and {Fishman}}]{nishikawa03}
{Nishikawa} KI, {Hardee} P, {Richardson} G, {Preece} R, {Sol} H, {Fishman} GJ
  (2003) {Particle Acceleration in Relativistic Jets Due to Weibel
  Instability}. \apj 595(1):555--563. \doi{10.1086/377260}.
  {\href{https://arxiv.org/abs/astro-ph/0305091}{{arXiv:astro-ph/0305091}}}
  {[astro-ph]}

\bibitem[{{Nishikawa} et~al.(2005){Nishikawa}, {Hardee}, {Richardson},
  {Preece}, {Sol}, and {Fishman}}]{nishikawa05}
{Nishikawa} KI, {Hardee} P, {Richardson} G, {Preece} R, {Sol} H, {Fishman} GJ
  (2005) {Particle Acceleration and Magnetic Field Generation in
  Electron-Positron Relativistic Shocks}. \apj 622(2):927--937.
  \doi{10.1086/428394}.
  {\href{https://arxiv.org/abs/astro-ph/0409702}{{arXiv:astro-ph/0409702}}}
  {[astro-ph]}

\bibitem[{Nishikawa et~al.(2006a)Nishikawa, Hardee, Hededal, Richardson,
  Preece, Sol, and Fishman}]{nishikawa06}
Nishikawa KI, Hardee P, Hededal CB, Richardson G, Preece R, Sol H, Fishman GJ
  (2006a) Particle acceleration, magnetic field generation, and emission in
  relativistic shocks. Adv Space Res 38(7):1316--1319.
  \doi{10.1016/j.asr.2005.01.036}

\bibitem[{{Nishikawa} et~al.(2006b){Nishikawa}, {Hardee}, {Hededal}, and
  {Fishman}}]{nishikawa06a}
{Nishikawa} KI, {Hardee} PE, {Hededal} CB, {Fishman} GJ (2006b) {Acceleration
  Mechanics in Relativistic Shocks by the Weibel Instability}. \apj
  642(2):1267--1274. \doi{10.1086/501426}.
  {\href{https://arxiv.org/abs/astro-ph/0510590}{{arXiv:astro-ph/0510590}}}
  {[astro-ph]}

\bibitem[{{Nishikawa} et~al.(2008){Nishikawa}, {Mizuno}, {Fishman}, and
  {Hardee}}]{nishikawa08}
{Nishikawa} KI, {Mizuno} Y, {Fishman} GJ, {Hardee} P (2008) {Particle
  Acceleration, Magnetic Field Generation, and Associated Emission in
  Collisionless Relativistic Jets}. Int J Mod Phys D 17:1761--1767.
  \doi{10.1142/S0218271808013388}.
  {\href{https://arxiv.org/abs/0801.4390}{{arXiv:0801.4390}}}

\bibitem[{{Nishikawa} et~al.(2009){Nishikawa}, {Niemiec}, {Hardee}, {Medvedev},
  {Sol}, {Mizuno}, {Zhang}, {Pohl}, {Oka}, and {Hartmann}}]{nishikawa09}
{Nishikawa} KI, {Niemiec} J, {Hardee} PE, {Medvedev} M, {Sol} H, {Mizuno} Y,
  {Zhang} B, {Pohl} M, {Oka} M, {Hartmann} DH (2009) {Weibel Instability and
  Associated Strong Fields in a Fully Three-Dimensional Simulation of a
  Relativistic Shock}. \apjl 698(1):L10--L13.
  \doi{10.1088/0004-637X/698/1/L10}.
  {\href{https://arxiv.org/abs/0904.0096}{{arXiv:0904.0096}}} {[astro-ph.HE]}

\bibitem[{{Nishikawa} et~al.(2011){Nishikawa}, {Niemiec}, {Medvedev}, {Zhang},
  {Hardee}, {Nordlund}, {Frederiksen}, {Mizuno}, {Sol}, {Pohl}, {Hartmann}, and
  {Fishman}}]{Nishikawa11AIPC}
{Nishikawa} KI, {Niemiec} J, {Medvedev} M, {Zhang} B, {Hardee} P, {Nordlund}
  {\AA}, {Frederiksen} J, {Mizuno} Y, {Sol} H, {Pohl} M, {Hartmann} DH,
  {Fishman} GJ (2011) {Simulation of Relativistic Shocks and Associated
  Self-consistent Radiation}. In: {Florinski} V, {Heerikhuisen} J, {Zank} GP,
  {Gallagher} DL (eds) Partially ionized plasmas throughout the cosmos. AIP
  Conference Series, vol 1366. American Institute of Physics, pp 163--171.
  \doi{10.1063/1.3625602}

\bibitem[{Nishikawa et~al.(2011)Nishikawa, Niemiec, Medvedev, Zhang, Hardee,
  Nordlund, Frederiksen, Mizuno, Sol, Pohl, Hartmann, Oka, and
  Fishman}]{nishikawa11ad}
Nishikawa KI, Niemiec J, Medvedev M, Zhang B, Hardee P, Nordlund A, Frederiksen
  J, Mizuno Y, Sol H, Pohl M, Hartmann DH, Oka M, Fishman GJ (2011) Radiation
  from relativistic shocks in turbulent magnetic fields. Adv Space Res
  47(8):1434--1440. \doi{10.1016/j.asr.2011.01.032}

\bibitem[{{Nishikawa} et~al.(2012){Nishikawa}, {Choi}, {Min}, {Niemiec},
  {Zhang}, {Hardee}, {Mizuno}, {Medvedev}, {Nordlund}, {Frederiksen}, {Sol},
  {Pohl}, {Hartmann}, and {Fishman}}]{Nishikawa2012grb}
{Nishikawa} KI, {Choi} EJ, {Min} KW, {Niemiec} J, {Zhang} B, {Hardee} P,
  {Mizuno} Y, {Medvedev} M, {Nordlund} {\AA}, {Frederiksen} J, {Sol} H, {Pohl}
  M, {Hartmann} DH, {Fishman} GJ (2012) {Radiation from shock-accelerated
  particles}. In: Gamma-Ray Bursts 2012 Conference ({GRB} 2012). p~28

\bibitem[{{Nishikawa} et~al.(2013a){Nishikawa}, {Hardee}, {Mizuno},
  {Du{\c{t}}an}, {Zhang}, {Medvedev}, {Choi}, {Min}, {Niemiec}, {Nordlund},
  {Frederiksen}, {Sol}, {Pohl}, {Hartmann}, {Marscher}, and
  {G{\'o}mez}}]{Nishikawa2013EPJ}
{Nishikawa} KI, {Hardee} P, {Mizuno} Y, {Du{\c{t}}an} I, {Zhang} B, {Medvedev}
  M, {Choi} EJ, {Min} KW, {Niemiec} J, {Nordlund} {\AA}, {Frederiksen} J, {Sol}
  H, {Pohl} M, {Hartmann} DH, {Marscher} A, {G{\'o}mez} JL (2013a) {Radiation
  from accelerated particles in relativistic jets with shocks, shear-flow, and
  reconnection}. Eur Phys J Web Conf 61:02003.
  \doi{10.1051/epjconf/20136102003}

\bibitem[{Nishikawa et~al.(2013b)Nishikawa, Hardee, Zhang, {Du{\c{t}}an},
  Medvedev, Choi, Min, Niemiec, Mizuno, Nordlund, Frederiksen, Sol, Pohl, and
  Hartmann}]{nishikawa13}
Nishikawa KI, Hardee P, Zhang B, {Du{\c{t}}an} I, Medvedev M, Choi EJ, Min KW,
  Niemiec J, Mizuno Y, Nordlund A, Frederiksen JT, Sol H, Pohl M, Hartmann DH
  (2013b) Magnetic field generation in a jet-sheath plasma via the kinetic
  {Kelvin}-{Helmholtz} instability. Ann Geophys 31(9):1535--1541.
  \doi{10.5194/angeo-31-1535-2013}

\bibitem[{{Nishikawa} et~al.(2014a){Nishikawa}, {Hardee}, {Du{\c{t}}an},
  {Zhang}, {Meli}, {Choi}, {Min}, {Niemiec}, {Mizuno}, {Medvedev}, {Nordlund},
  {Frederiksen}, {Sol}, {Pohl}, and {Hartmann}}]{nishikawa14c}
{Nishikawa} KI, {Hardee} P, {Du{\c{t}}an} I, {Zhang} B, {Meli} A, {Choi} EJ,
  {Min} K, {Niemiec} J, {Mizuno} Y, {Medvedev} M, {Nordlund} A, {Frederiksen}
  JT, {Sol} H, {Pohl} M, {Hartmann} D (2014a) {Radiation from Particles
  Accelerated in Relativistic Jet Shocks and Shear-flows}. ArXiv e-prints
  {\href{https://arxiv.org/abs/1412.7064}{{arXiv:1412.7064}}} {[astro-ph.HE]}

\bibitem[{{Nishikawa} et~al.(2014b){Nishikawa}, {Hardee}, {Du{\c{t}}an},
  {Niemiec}, {Medvedev}, {Mizuno}, {Meli}, {Sol}, {Zhang}, {Pohl}, and
  {Hartmann}}]{nishikawa14b}
{Nishikawa} KI, {Hardee} PE, {Du{\c{t}}an} I, {Niemiec} J, {Medvedev} M,
  {Mizuno} Y, {Meli} A, {Sol} H, {Zhang} B, {Pohl} M, {Hartmann} DH (2014b)
  {Magnetic Field Generation in Core-sheath Jets via the Kinetic
  Kelvin-Helmholtz Instability}. \apj 793(1):60.
  \doi{10.1088/0004-637X/793/1/60}.
  {\href{https://arxiv.org/abs/1405.5247}{{arXiv:1405.5247}}} {[astro-ph.HE]}

\bibitem[{{Nishikawa} et~al.(2016a){Nishikawa}, {Frederiksen}, {Nordlund},
  {Mizuno}, {Hardee}, {Niemiec}, {G{\'o}mez}, {Pe'er}, {Du{\c{t}}an}, {Meli},
  {Sol}, {Pohl}, and {Hartmann}}]{nishikawa16a}
{Nishikawa} KI, {Frederiksen} JT, {Nordlund} {\r{A}}, {Mizuno} Y, {Hardee} PE,
  {Niemiec} J, {G{\'o}mez} JL, {Pe'er} A, {Du{\c{t}}an} I, {Meli} A, {Sol} H,
  {Pohl} M, {Hartmann} DH (2016a) {Evolution of Global Relativistic Jets:
  Collimations and Expansion with kKHI and the Weibel Instability}. \apj
  820(2):94. \doi{10.3847/0004-637X/820/2/94}.
  {\href{https://arxiv.org/abs/1511.03581}{{arXiv:1511.03581}}} {[astro-ph.HE]}

\bibitem[{Nishikawa et~al.(2016b)Nishikawa, Mizuno, Niemiec, Kobzar, Pohl,
  G{\'o}mez, {Du{\c{t}}an}, Pe'er, Frederiksen, Nordlund, Meli, Sol, Hardee,
  and Hartmann}]{nishikawa16b}
Nishikawa KI, Mizuno Y, Niemiec J, Kobzar O, Pohl M, G{\'o}mez JL,
  {Du{\c{t}}an} I, Pe'er A, Frederiksen JT, Nordlund {\AA}, Meli A, Sol H,
  Hardee PE, Hartmann DH (2016b) Microscopic processes in global relativistic
  jets containing helical magnetic fields. Galaxies 4(4):38.
  \doi{10.3390/galaxies4040038}

\bibitem[{Nishikawa et~al.(2017)Nishikawa, Mizuno, G{\'o}mez, {Du{\c{t}}an},
  Meli, White, Niemiec, Kobzar, Pohl, Pe'er, Frederiksen, Nordlund, Sol,
  Hardee, and Hartmann}]{nishikawa17}
Nishikawa KI, Mizuno Y, G{\'o}mez JL, {Du{\c{t}}an} I, Meli A, White C, Niemiec
  J, Kobzar O, Pohl M, Pe'er A, Frederiksen JT, Nordlund {\AA}, Sol H, Hardee
  PE, Hartmann DH (2017) Microscopic processes in global relativistic jets
  containing helical magnetic fields: Dependence on jet radius. Galaxies
  5(4):58. \doi{10.3390/galaxies5040058}

\bibitem[{Nishikawa et~al.(2019)Nishikawa, Mizuno, G{\'o}mez, {Du{\c{t}}an},
  Meli, Niemiec, Kobzar, Pohl, Sol, MacDonald, and Hartmann}]{nishikawa19gal}
Nishikawa KI, Mizuno Y, G{\'o}mez JL, {Du{\c{t}}an} I, Meli A, Niemiec J,
  Kobzar O, Pohl M, Sol H, MacDonald N, Hartmann DH (2019) Relativistic jet
  simulations of the {Weibel} instability in the slab model to cylindrical jets
  with helical magnetic fields. Galaxies 7(1):29. \doi{10.3390/galaxies7010029}

\bibitem[{Nousch et~al.(2016)Nousch, Seipt, K{\"a}mpfer, and Titov}]{Nousch16}
Nousch T, Seipt D, K{\"a}mpfer B, Titov AI (2016) Spectral caustics in laser
  assisted {Breit}--{Wheeler} process. Phys Lett B 755:162--167.
  \doi{10.1016/j.physletb.2016.01.062}

\bibitem[{{Oka} et~al.(2008){Oka}, {Fujimoto}, {Nakamura}, {Shinohara}, and
  {Nishikawa}}]{oka08}
{Oka} M, {Fujimoto} M, {Nakamura} TKM, {Shinohara} I, {Nishikawa} KI (2008)
  {Magnetic Reconnection by a Self-Retreating X Line}. Phys Rev Lett
  101(20):205004. \doi{10.1103/PhysRevLett.101.205004}.
  {\href{https://arxiv.org/abs/0808.2179}{{arXiv:0808.2179}}} {[astro-ph]}

\bibitem[{Olson et~al.(2020)Olson, Leeper, Batha, Peterson, Bradley, Stygar,
  LeChien, Robey, Young, and Meezan}]{Olson2020}
Olson RE, Leeper RJ, Batha SH, Peterson RR, Bradley PA, Stygar WA, LeChien KR,
  Robey HF, Young CV, Meezan NB (2020) Pulsed power indirect drive approach to
  inertial confinement fusion. High Energy Density Phys 36:100749.
  \doi{10.1016/j.hedp.2020.100749}

\bibitem[{{Omura}(2007)}]{omura07}
{Omura} Y (2007) {One-dimensional Electromagnetic Particle Code: {KEMPO1}. A
  Tutorial on Microphysics in Space Plasmas}. In: Advanced Methods for Space
  Simulations. Terrapub, Tokyo, pp 1--21.
  \urlprefix\url{https://www.terrapub.co.jp/e-library/amss/}

\bibitem[{{O'Sullivan} et~al.(2013){O'Sullivan}, {McClure-Griffiths}, {Feain},
  {Gaensler}, and {Sault}}]{O'Sullivan13}
{O'Sullivan} SP, {McClure-Griffiths} NM, {Feain} IJ, {Gaensler} BM, {Sault} RJ
  (2013) {Broad-band radio circular polarization spectrum of the relativistic
  jet in {PKS B2126-158}}. Mon Not R Astron Soc 435:311--319.
  \doi{10.1093/mnras/stt1298}.
  {\href{https://arxiv.org/abs/1307.5121}{{arXiv:1307.5121}}} {[astro-ph.HE]}

\bibitem[{Palastro et~al.(2020)Palastro, Shaw, Franke, Ramsey, Simpson, and
  Froula}]{Palastro2020}
Palastro JP, Shaw JL, Franke P, Ramsey D, Simpson TT, Froula DH (2020)
  Dephasingless laser wakefield acceleration. Phys Rev Lett 124:134802.
  \doi{10.1103/PhysRevLett.124.134802}

\bibitem[{{Palenzuela} et~al.(2010){Palenzuela}, {Lehner}, and
  {Liebling}}]{Palenzuela10}
{Palenzuela} C, {Lehner} L, {Liebling} SL (2010) {Dual Jets from Binary Black
  Holes}. Science 329(5994):927--930. \doi{10.1126/science.1191766}.
  {\href{https://arxiv.org/abs/1005.1067}{{arXiv:1005.1067}}} {[astro-ph.HE]}

\bibitem[{{Palmroth} et~al.(2018){Palmroth}, {Ganse}, {Pfau-Kempf},
  {Battarbee}, {Turc}, {Brito}, {Grandin}, {Hoilijoki}, {Sandroos}, and {von
  Alfthan}}]{palmroth18}
{Palmroth} M, {Ganse} U, {Pfau-Kempf} Y, {Battarbee} M, {Turc} L, {Brito} T,
  {Grandin} M, {Hoilijoki} S, {Sandroos} A, {von Alfthan} S (2018) {Vlasov
  methods in space physics and astrophysics}. Living Rev Comput Astrophys 4:1.
  \doi{10.1007/s41115-018-0003-2}.
  {\href{https://arxiv.org/abs/1808.05885}{{arXiv:1808.05885}}}
  {[physics.space-ph]}

\bibitem[{{Parfrey} et~al.(2019){Parfrey}, {Philippov}, and
  {Cerutti}}]{Parfrey19}
{Parfrey} K, {Philippov} A, {Cerutti} B (2019) {First-Principles Plasma
  Simulations of Black-Hole Jet Launching}. Phys Rev Lett 122(3):035101.
  \doi{10.1103/PhysRevLett.122.035101}.
  {\href{https://arxiv.org/abs/1810.03613}{{arXiv:1810.03613}}} {[astro-ph.HE]}

\bibitem[{{Parker} et~al.(1999){Parker}, {Petschek}, {Otto}, and {et
  al.}}]{Buechner1999}
{Parker} E, {Petschek} H, {Otto} A, {et al} (1999) {Plasma Astrophysics And
  Space Physics}. Kluwer Academic Publishers. \doi{10.1007/978-94-011-4203-8}

\bibitem[{Paschalidis(2017)}]{paschalidis17}
Paschalidis V (2017) General relativistic simulations of compact binary mergers
  as engines for short gamma-ray bursts. Class Quantum Grav 34(8):084002

\bibitem[{Paschalidis and Stergioulas(2017)}]{paschalidisST2017}
Paschalidis V, Stergioulas N (2017) Rotating stars in relativity. Living Rev
  Relativ 20:7. \doi{10.1007/s41114-017-0008-x}

\bibitem[{Pe'er(2014)}]{peer14}
Pe'er A (2014) Energetic and broad band spectral distribution of emission from
  astronomical jets. Space Sci Rev 183(1):371--403.
  \doi{10.1007/s11214-013-0001-y}

\bibitem[{{Perucho} and {Lobanov}(2008)}]{perucho08}
{Perucho} M, {Lobanov} AP (2008) {Kelvin-Helmholtz Modes Revealed by the
  Transversal Structure of the Jet in 0836+710}. In: {Rector} TA, {De Young} DS
  (eds) Extragalactic Jets: Theory and Observation from Radio to Gamma Ray.
  Astronomical Society of the Pacific Conference Series, vol 386. p 381.
  {\href{https://arxiv.org/abs/0707.2887}{{arXiv:0707.2887}}}

\bibitem[{{Petropoulou} and {Sironi}(2018)}]{Petropoulou2018}
{Petropoulou} M, {Sironi} L (2018) {The steady growth of the high-energy
  spectral cut-off in relativistic magnetic reconnection}. \mnras
  481(4):5687--5701. \doi{10.1093/mnras/sty2702}.
  {\href{https://arxiv.org/abs/1808.00966}{{arXiv:1808.00966}}} {[astro-ph.HE]}

\bibitem[{Petropoulou et~al.(2016)Petropoulou, Giannios, and
  Sironi}]{petropoulou16}
Petropoulou M, Giannios D, Sironi L (2016) Blazar flares powered by plasmoids
  in relativistic reconnection. Mon Not R Astron Soc 462(3):3325--3343.
  \doi{10.1093/mnras/stw1832}

\bibitem[{Philippov et~al.(2019)Philippov, Uzdensky, Spitkovsky, and
  Cerutti}]{Philippov_2019}
Philippov A, Uzdensky DA, Spitkovsky A, Cerutti B (2019) Pulsar radio emission
  mechanism: Radio nanoshots as a low-frequency afterglow of relativistic
  magnetic reconnection. Astrophys J 876(1):L6. \doi{10.3847/2041-8213/ab1590}

\bibitem[{Philippov and Spitkovsky(2018)}]{Philippov_2018}
Philippov AA, Spitkovsky A (2018) Ab-initio pulsar magnetosphere: Particle
  acceleration in oblique rotators and high-energy emission modeling. Astrophys
  J 855(2):94. \doi{10.3847/1538-4357/aaabbc}

\bibitem[{Piran(2005)}]{piran05}
Piran T (2005) {Magnetic Fields in Gamma-Ray Bursts: A Short Overview}. AIP
  Conference Proceedings 784:164. \doi{10.1063/1.2077181}

\bibitem[{Pohl et~al.(2020)Pohl, Hoshino, and Niemiec}]{Pohl2020r}
Pohl M, Hoshino M, Niemiec J (2020) {PIC} simulation methods for cosmic
  radiation and plasma instabilities. Prog Part Nucl Phys 111:103751.
  \doi{10.1016/j.ppnp.2019.103751}

\bibitem[{Popruzhenko et~al.(2019)Popruzhenko, Liseykina, and
  Macchi}]{Popruzhenko_2019}
Popruzhenko SV, Liseykina TV, Macchi A (2019) Efficiency of radiation friction
  losses in laser-driven `hole boring' of dense targets. New J Phys
  21(3):033009. \doi{10.1088/1367-2630/ab0119}

\bibitem[{{Porth} et~al.(2017){Porth}, {Olivares}, {Mizuno}, {Younsi},
  {Rezzolla}, {Moscibrodzka}, {Falcke}, and {Kramer}}]{oliber17}
{Porth} O, {Olivares} H, {Mizuno} Y, {Younsi} Z, {Rezzolla} L, {Moscibrodzka}
  M, {Falcke} H, {Kramer} M (2017) {The black hole accretion code}. Comput
  Astrophys Cosmol 4:1. \doi{10.1186/s40668-017-0020-2}.
  {\href{https://arxiv.org/abs/1611.09720}{{arXiv:1611.09720}}} {[gr-qc]}

\bibitem[{{Pozdnyakov} et~al.(1977){Pozdnyakov}, {Sobol}, and
  {Syunyaev}}]{Pozdnyakov77}
{Pozdnyakov} LA, {Sobol} IM, {Syunyaev} RA (1977) {Effect of the multiple
  Compton scatterings on an x-ray emission spectrum. Calculation by the Monte
  Carlo method}. Sov Astron 21:708--714

\bibitem[{{Pozdnyakov} et~al.(1983){Pozdnyakov}, {Sobol}, and
  {Syunyaev}}]{Pozdnyakov83}
{Pozdnyakov} LA, {Sobol} IM, {Syunyaev} RA (1983) {Comptonization and the
  shaping of {X-ray} source spectra - {Monte Carlo} calculations}. Sov Sci Rev
  2:189--331

\bibitem[{{Pritchett} et~al.(2003){Pritchett}, {Cai}, {Lamb\'ege}, and {et
  al.}}]{Buechner2003}
{Pritchett} PL, {Cai} D, {Lamb\'ege} B, {et al} (2003) {Space Plasma
  Simulation}. Springer-Verlag Berlin Heidelberg. \doi{10.1007/3-540-36530-3}

\bibitem[{{Punsly}(1996)}]{Punsly1996}
{Punsly} B (1996) {Fast Waves and the Causality of Black Hole Dynamos}. \apj
  467:105. \doi{10.1086/177588}

\bibitem[{{Quataert} and {Gruzinov}(1999)}]{quataert_1999}
{Quataert} E, {Gruzinov} A (1999) {Turbulence and Particle Heating in
  Advection-dominated Accretion Flows}. \apj 520(1):248--255.
  \doi{10.1086/307423}.
  {\href{https://arxiv.org/abs/astro-ph/9803112}{{arXiv:astro-ph/9803112}}}
  {[astro-ph]}

\bibitem[{{Ramirez-Ruiz} et~al.(2007){Ramirez-Ruiz}, {Nishikawa}, and
  {Hededal}}]{ramirez07}
{Ramirez-Ruiz} E, {Nishikawa} KI, {Hededal} CB (2007) {e$^{+/-}$ Pair Loading
  and the Origin of the Upstream Magnetic Field in GRB Shocks}. \apj
  671(2):1877--1885. \doi{10.1086/522072}.
  {\href{https://arxiv.org/abs/0707.4381}{{arXiv:0707.4381}}} {[astro-ph]}

\bibitem[{{Reville} and {Bell}(2013)}]{Reville2013}
{Reville} B, {Bell} AR (2013) {Universal behaviour of shock precursors in the
  presence of efficient cosmic ray acceleration}. \mnras 430(4):2873--2884.
  \doi{10.1093/mnras/stt100}.
  {\href{https://arxiv.org/abs/1301.3173}{{arXiv:1301.3173}}} {[astro-ph.HE]}

\bibitem[{{Reville} and {Kirk}(2010)}]{Reville2010ApJ}
{Reville} B, {Kirk} JG (2010) {Computation of Synthetic Spectra from
  Simulations of Relativistic Shocks}. \apj 724(2):1283--1295.
  \doi{10.1088/0004-637X/724/2/1283}.
  {\href{https://arxiv.org/abs/1010.0872}{{arXiv:1010.0872}}} {[astro-ph.HE]}

\bibitem[{{Rezzolla} and {Zanotti}(2013)}]{Rezzolla13}
{Rezzolla} L, {Zanotti} O (2013) {Relativistic Hydrodynamics}. Oxford
  University Press. \doi{10.1093/acprof:oso/9780198528906.001.0001}

\bibitem[{{Rezzolla} et~al.(2011){Rezzolla}, {Giacomazzo}, {Baiotti}, {Granot},
  {Kouveliotou}, and {Aloy}}]{rezzolla11}
{Rezzolla} L, {Giacomazzo} B, {Baiotti} L, {Granot} J, {Kouveliotou} C, {Aloy}
  MA (2011) {The Missing Link: Merging Neutron Stars Naturally Produce Jet-like
  Structures and Can Power Short Gamma-ray Bursts}. \apjl 732(1):L6.
  \doi{10.1088/2041-8205/732/1/L6}.
  {\href{https://arxiv.org/abs/1101.4298}{{arXiv:1101.4298}}} {[astro-ph.HE]}

\bibitem[{{Rieke} et~al.(2015){Rieke}, {Trost}, and {Grauer}}]{Rieke2015}
{Rieke} M, {Trost} T, {Grauer} R (2015) {Coupled Vlasov and two-fluid codes on
  {GPU}s}. J Comput Phys 283:436--452. \doi{10.1016/j.jcp.2014.12.016}.
  {\href{https://arxiv.org/abs/1406.5445}{{arXiv:1406.5445}}}
  {[physics.plasm-ph]}

\bibitem[{{Rincon}(2019)}]{Rincon2019}
{Rincon} F (2019) {Dynamo theories}. Journal of Plasma Physics 85(4):205850401.
  \doi{10.1017/S0022377819000539}.
  {\href{https://arxiv.org/abs/1903.07829}{{arXiv:1903.07829}}}
  {[physics.plasm-ph]}

\bibitem[{{Ripperda} et~al.(2018){Ripperda}, {Bacchini}, {Teunissen}, {Xia},
  {Porth}, {Sironi}, {Lapenta}, and {Keppens}}]{Ripperda18}
{Ripperda} B, {Bacchini} F, {Teunissen} J, {Xia} C, {Porth} O, {Sironi} L,
  {Lapenta} G, {Keppens} R (2018) {A Comprehensive Comparison of Relativistic
  Particle Integrators}. Astrophys J Suppl Ser 235(1):21.
  \doi{10.3847/1538-4365/aab114}.
  {\href{https://arxiv.org/abs/1710.09164}{{arXiv:1710.09164}}} {[astro-ph.IM]}

\bibitem[{Robinson et~al.(2009)Robinson, Gibbon, Zepf, Kar, Evans, and
  Bellei}]{Robinson_2009}
Robinson APL, Gibbon P, Zepf M, Kar S, Evans RG, Bellei C (2009)
  Relativistically correct hole-boring and ion acceleration by circularly
  polarized laser pulses. Plasma Physics and Controlled Fusion 51(2):024004.
  \doi{10.1088/0741-3335/51/2/024004},
  \urlprefix\url{https://doi.org/10.1088/0741-3335/51/2/024004}

\bibitem[{Rubino and Tuffin(2009)}]{rubino_2009}
Rubino G, Tuffin B (eds)  (2009) Rare Event Simulation Using {Monte Carlo}
  Methods. John Wiley \& Sons, West Sussex. \doi{10.1002/9780470745403}

\bibitem[{Ruiz and Shapiro(2017)}]{ruiz17}
Ruiz M, Shapiro SL (2017) General relativistic magnetohydrodynamics simulations
  of prompt-collapse neutron star mergers: The absence of jets. Phys Rev D
  96:084063. \doi{10.1103/PhysRevD.96.084063}

\bibitem[{{Ruiz} et~al.(2016){Ruiz}, {Lang}, {Paschalidis}, and
  {Shapiro}}]{ruiz16}
{Ruiz} M, {Lang} RN, {Paschalidis} V, {Shapiro} SL (2016) {Binary Neutron Star
  Mergers: A Jet Engine for Short Gamma-Ray Bursts}. \apjl 824(1):L6.
  \doi{10.3847/2041-8205/824/1/L6}.
  {\href{https://arxiv.org/abs/1604.02455}{{arXiv:1604.02455}}} {[astro-ph.HE]}

\bibitem[{Ruiz et~al.(2018)Ruiz, Shapiro, and Tsokaros}]{ruiz18}
Ruiz M, Shapiro SL, Tsokaros A (2018) {GW170817}, general relativistic
  magnetohydrodynamic simulations, and the neutron star maximum mass. Phys Rev
  D 97:021501. \doi{10.1103/PhysRevD.97.021501}

\bibitem[{{Ruiz} et~al.(2020a){Ruiz}, {Tsokaros}, and {Shapiro}}]{Ruiz2020a}
{Ruiz} M, {Tsokaros} A, {Shapiro} SL (2020a) {Magnetohydrodynamic simulations
  of binary neutron star mergers in general relativity: Effects of magnetic
  field orientation on jet launching}. \prd 101(6):064042.
  \doi{10.1103/PhysRevD.101.064042}.
  {\href{https://arxiv.org/abs/2001.09153}{{arXiv:2001.09153}}} {[astro-ph.HE]}

\bibitem[{{Ruiz} et~al.(2020b){Ruiz}, {Tsokaros}, {Shapiro}, {Nelli}, and
  {Qunell}}]{Ruiz2020b}
{Ruiz} M, {Tsokaros} A, {Shapiro} SL, {Nelli} KC, {Qunell} S (2020b) {Magnetic
  ergostars, jet formation, and gamma-ray bursts: Ergoregions versus horizons}.
  \prd 102(10):104022. \doi{10.1103/PhysRevD.102.104022}.
  {\href{https://arxiv.org/abs/2009.08982}{{arXiv:2009.08982}}} {[astro-ph.HE]}

\bibitem[{{Ruiz} et~al.(2020c){Ruiz}, {Paschalidis}, {Tsokaros}, and
  {Shapiro}}]{Ruiz2020c}
{Ruiz} M, {Paschalidis} V, {Tsokaros} A, {Shapiro} SL (2020c) {Black
  hole-neutron star coalescence: Effects of the neutron star spin on jet
  launching and dynamical ejecta mass}. \prd 102(12):124077.
  \doi{10.1103/PhysRevD.102.124077}.
  {\href{https://arxiv.org/abs/2011.08863}{{arXiv:2011.08863}}} {[astro-ph.HE]}

\bibitem[{{Ruiz} et~al.(2021){Ruiz}, {Shapiro}, and {Tsokaros}}]{Ruiz2021Multi}
{Ruiz} M, {Shapiro} SL, {Tsokaros} A (2021) {Multimessenger Binary Mergers
  Containing Neutron Stars: Gravitational Waves, Jets, and
  $\boldsymbol{\gamma}$-Ray Bursts}. arXiv e-prints arXiv:2102.03366.
  {\href{https://arxiv.org/abs/2102.03366}{{arXiv:2102.03366}}} {[astro-ph.HE]}

\bibitem[{{Ryan} et~al.(2018){Ryan}, {Ressler}, {Dolence}, {Gammie}, and
  {Quataert}}]{ryan_2018}
{Ryan} BR, {Ressler} SM, {Dolence} JC, {Gammie} C, {Quataert} E (2018)
  {Two-temperature GRRMHD Simulations of M87}. \apj 864(2):126.
  \doi{10.3847/1538-4357/aad73a}.
  {\href{https://arxiv.org/abs/1808.01958}{{arXiv:1808.01958}}} {[astro-ph.HE]}

\bibitem[{{Sari} and {Piran}(1999)}]{sari99}
{Sari} R, {Piran} T (1999) {Predictions for the Very Early Afterglow and the
  Optical Flash}. \apj 520(2):641--649. \doi{10.1086/307508}.
  {\href{https://arxiv.org/abs/astro-ph/9901338}{{arXiv:astro-ph/9901338}}}
  {[astro-ph]}

\bibitem[{{Savchenko} et~al.(2017){Savchenko}, {Ferrigno}, {Kuulkers},
  {Bazzano}, {Bozzo}, {Brandt}, {Chenevez}, {Courvoisier}, {Diehl}, {Domingo},
  {Hanlon}, {Jourdain}, {von Kienlin}, {Laurent}, {Lebrun}, {Lutovinov},
  {Martin-Carrillo}, {Mereghetti}, {Natalucci}, {Rodi}, {Roques}, {Sunyaev},
  and {Ubertini}}]{savchenko17}
{Savchenko} V, {Ferrigno} C, {Kuulkers} E, {Bazzano} A, {Bozzo} E, {Brandt} S,
  {Chenevez} J, {Courvoisier} TJL, {Diehl} R, {Domingo} A, {Hanlon} L,
  {Jourdain} E, {von Kienlin} A, {Laurent} P, {Lebrun} F, {Lutovinov} A,
  {Martin-Carrillo} A, {Mereghetti} S, {Natalucci} L, {Rodi} J, {Roques} JP,
  {Sunyaev} R, {Ubertini} P (2017) {INTEGRAL Detection of the First Prompt
  Gamma-Ray Signal Coincident with the Gravitational-wave Event GW170817}.
  \apjl 848(2):L15. \doi{10.3847/2041-8213/aa8f94}.
  {\href{https://arxiv.org/abs/1710.05449}{{arXiv:1710.05449}}} {[astro-ph.HE]}

\bibitem[{Schlickeiser et~al.(2013)Schlickeiser, Krakau, and
  Supsar}]{Schlickeiser_2013}
Schlickeiser R, Krakau S, Supsar M (2013) Plasma effects on fast pair beams.
  {II}. {R}eactive versus kinetic instability of parallel electrostatic waves.
  Astrophys J 777(1):49. \doi{10.1088/0004-637x/777/1/49}

\bibitem[{{Sgattoni} et~al.(2015{\natexlab{a}}){Sgattoni}, {Fedeli},
  {Sinigardi}, {Marocchino}, {Macchi}, {Weinberg}, and
  {Karmakar}}]{Sgattoni2015x}
{Sgattoni} A, {Fedeli} L, {Sinigardi} S, {Marocchino} A, {Macchi} A, {Weinberg}
  V, {Karmakar} A (2015{\natexlab{a}}) {Optimising {PICCANTE} - an Open Source
  Particle-in-Cell Code for Advanced Simulations on Tier-0 Systems}. arXiv
  e-prints {\href{https://arxiv.org/abs/1503.02464}{{arXiv:1503.02464}}}
  {[cs.DC]}

\bibitem[{{Sgattoni} et~al.(2015{\natexlab{b}}){Sgattoni}, {Sinigardi},
  {Fedeli}, {Pegoraro}, and {Macchi}}]{Sgattoni2015PhRvE}
{Sgattoni} A, {Sinigardi} S, {Fedeli} L, {Pegoraro} F, {Macchi} A
  (2015{\natexlab{b}}) {Laser-driven Rayleigh-Taylor instability: Plasmonic
  effects and three-dimensional structures}. \pre 91(1):013106.
  \doi{10.1103/PhysRevE.91.013106}.
  {\href{https://arxiv.org/abs/1404.1260}{{arXiv:1404.1260}}}
  {[physics.plasm-ph]}

\bibitem[{Shapiro(2017)}]{shapiro17}
Shapiro SL (2017) Black holes, disks, and jets following binary mergers and
  stellar collapse: The narrow range of electromagnetic luminosities and
  accretion rates. Phys Rev D 95:101303. \doi{10.1103/PhysRevD.95.101303}

\bibitem[{{Shapiro} et~al.(1976){Shapiro}, {Lightman}, and
  {Eardley}}]{shapiro_1976}
{Shapiro} SL, {Lightman} AP, {Eardley} DM (1976) {A two-temperature accretion
  disk model for Cygnus X-1: structure and spectrum.} \apj 204:187--199.
  \doi{10.1086/154162}

\bibitem[{Shibata et~al.(2017)Shibata, Fujibayashi, Hotokezaka, Kiuchi,
  Kyutoku, Sekiguchi, and Tanaka}]{shibata17}
Shibata M, Fujibayashi S, Hotokezaka K, Kiuchi K, Kyutoku K, Sekiguchi Y,
  Tanaka M (2017) Modeling {GW170817} based on numerical relativity and its
  implications. Phys Rev D 96:123012. \doi{10.1103/PhysRevD.96.123012}

\bibitem[{{Silva} et~al.(2003){Silva}, {Fonseca}, {Tonge}, {Dawson}, {Mori},
  and {Medvedev}}]{silva03}
{Silva} LO, {Fonseca} RA, {Tonge} JW, {Dawson} JM, {Mori} WB, {Medvedev} MV
  (2003) {Interpenetrating Plasma Shells: Near-equipartition Magnetic Field
  Generation and Nonthermal Particle Acceleration}. \apjl 596(1):L121--L124.
  \doi{10.1086/379156}.
  {\href{https://arxiv.org/abs/astro-ph/0307500}{{arXiv:astro-ph/0307500}}}
  {[astro-ph]}

\bibitem[{{Singh} et~al.(2016){Singh}, {Mizuno}, and {de Gouveia Dal
  Pino}}]{singh16}
{Singh} CB, {Mizuno} Y, {de Gouveia Dal Pino} EM (2016) {Spatial Growth of
  Current-driven Instability in Relativistic Rotating Jets and the Search for
  Magnetic Reconnection}. \apj 824(1):48. \doi{10.3847/0004-637X/824/1/48}.
  {\href{https://arxiv.org/abs/1603.03276}{{arXiv:1603.03276}}} {[astro-ph.HE]}

\bibitem[{Sironi and Cerutti(2017)}]{sironi17}
Sironi L, Cerutti B (2017) Particle acceleration in pulsar wind nebulae: {PIC}
  modelling. In: Torres DF (ed) Modelling Pulsar Wind Nebulae. Astrophysics and
  Space Science Library, vol 446. Springer, Cham, pp 247--277.
  \doi{10.1007/978-3-319-63031-1_11}

\bibitem[{{Sironi} and {Giannios}(2014)}]{sironi14}
{Sironi} L, {Giannios} D (2014) {Relativistic Pair Beams from TeV Blazars: A
  Source of Reprocessed GeV Emission rather than Intergalactic Heating}. \apj
  787(1):49. \doi{10.1088/0004-637X/787/1/49}.
  {\href{https://arxiv.org/abs/1312.4538}{{arXiv:1312.4538}}} {[astro-ph.HE]}

\bibitem[{Sironi and Spitkovsky(2009)}]{Sironi_2009syn}
Sironi L, Spitkovsky A (2009) Synthetic spectra from particle-in-cell
  simulations of relativistic collisionless shocks. Astrophys J
  707(1):L92--L96. \doi{10.1088/0004-637x/707/1/l92}

\bibitem[{Sironi and Spitkovsky(2011)}]{Sironi_2011}
Sironi L, Spitkovsky A (2011) Acceleration of particles at the termination
  shock of a relativistic striped wind. Astrophys J 741(1):39.
  \doi{10.1088/0004-637x/741/1/39}

\bibitem[{{Sironi} and {Spitkovsky}(2014)}]{Sironi2014}
{Sironi} L, {Spitkovsky} A (2014) {Relativistic Reconnection: An Efficient
  Source of Non-thermal Particles}. \apjl 783(1):L21.
  \doi{10.1088/2041-8205/783/1/L21}.
  {\href{https://arxiv.org/abs/1401.5471}{{arXiv:1401.5471}}} {[astro-ph.HE]}

\bibitem[{{Sironi} et~al.(2015){Sironi}, {Keshet}, and {Lemoine}}]{Sironi15r}
{Sironi} L, {Keshet} U, {Lemoine} M (2015) {Relativistic Shocks: Particle
  Acceleration and Magnetization}. Space Sci Rev 191(1-4):519--544.
  \doi{10.1007/s11214-015-0181-8}.
  {\href{https://arxiv.org/abs/1506.02034}{{arXiv:1506.02034}}} {[astro-ph.HE]}

\bibitem[{Sironi et~al.(2015)Sironi, Petropoulou, and Giannios}]{sironi15}
Sironi L, Petropoulou M, Giannios D (2015) Relativistic jets shine through
  shocks or magnetic reconnection? Mon Not R Astron Soc 450(1):183--191.
  \doi{10.1093/mnras/stv641}

\bibitem[{Sironi et~al.(2016)Sironi, Giannios, and Petropoulou}]{sironi16}
Sironi L, Giannios D, Petropoulou M (2016) Plasmoids in relativistic
  reconnection, from birth to adulthood: first they grow, then they go. Mon Not
  R Astron Soc 462(1):48--74. \doi{10.1093/mnras/stw1620}

\bibitem[{{Spitkovsky}(2005)}]{Spitkovsky2005}
{Spitkovsky} A (2005) {Simulations of relativistic collisionless shocks: shock
  structure and particle acceleration}. In: {Bulik} T, {Rudak} B, {Madejski} G
  (eds) Astrophysical Sources of High Energy Particles and Radiation. American
  Institute of Physics Conference Series, vol 801. pp 345--350.
  \doi{10.1063/1.2141897}.
  {\href{https://arxiv.org/abs/astro-ph/0603211}{{arXiv:astro-ph/0603211}}}
  {[astro-ph]}

\bibitem[{{Spitkovsky}(2008a)}]{spit08a}
{Spitkovsky} A (2008a) {On the Structure of Relativistic Collisionless Shocks
  in Electron-Ion Plasmas}. \apjl 673(1):L39. \doi{10.1086/527374}.
  {\href{https://arxiv.org/abs/0706.3126}{{arXiv:0706.3126}}} {[astro-ph]}

\bibitem[{{Spitkovsky}(2008b)}]{spit08b}
{Spitkovsky} A (2008b) {Particle Acceleration in Relativistic Collisionless
  Shocks: Fermi Process at Last?} \apjl 682(1):L5. \doi{10.1086/590248}.
  {\href{https://arxiv.org/abs/0802.3216}{{arXiv:0802.3216}}} {[astro-ph]}

\bibitem[{Stone et~al.(2008)Stone, Gardiner, Teuben, Hawley, and
  Simon}]{Stone_2008}
Stone JM, Gardiner TA, Teuben P, Hawley JF, Simon JB (2008) Athena: A new code
  for astrophysical {MHD}. Astrophys J Suppl Ser 178(1):137--177.
  \doi{10.1086/588755}

\bibitem[{Sun et~al.(2017)Sun, Paschalidis, Ruiz, and Shapiro}]{sun17}
Sun L, Paschalidis V, Ruiz M, Shapiro SL (2017) Magnetorotational collapse of
  supermassive stars: Black hole formation, gravitational waves, and jets. Phys
  Rev D 96:043006. \doi{10.1103/PhysRevD.96.043006}

\bibitem[{{Synge}(1957)}]{Synge57}
{Synge} JL (1957) {The Relativistic Gas}. North-Holland

\bibitem[{{Takahashi} et~al.(2011){Takahashi}, {Kudoh}, {Masada}, and
  {Matsumoto}}]{takahashi11b}
{Takahashi} HR, {Kudoh} T, {Masada} Y, {Matsumoto} J (2011) {Scaling Law of
  Relativistic Sweet-Parker-type Magnetic Reconnection}. Astrophys J
  739(2):L53. \doi{10.1088/2041-8205/739/2/L53}.
  {\href{https://arxiv.org/abs/1108.3891}{{arXiv:1108.3891}}} {[astro-ph.HE]}

\bibitem[{{Takahashi} et~al.(2009){Takahashi}, {Kishishita}, {Uchiyama},
  {Tanaka}, {Yamaoka}, {Khangulyan}, {Aharonian}, {Bosch-Ramon}, and
  {Hinton}}]{takahashi_2009}
{Takahashi} T, {Kishishita} T, {Uchiyama} Y, {Tanaka} T, {Yamaoka} K,
  {Khangulyan} D, {Aharonian} FA, {Bosch-Ramon} V, {Hinton} JA (2009) {Study of
  the Spectral and Temporal Characteristics of X-Ray Emission of the Gamma-Ray
  Binary LS 5039 with Suzaku}. \apj 697(1):592--600.
  \doi{10.1088/0004-637X/697/1/592}.
  {\href{https://arxiv.org/abs/0812.3358}{{arXiv:0812.3358}}} {[astro-ph]}

\bibitem[{{Takamoto} et~al.(2019){Takamoto}, {Matsumoto}, and
  {Kato}}]{Takamoto2019}
{Takamoto} M, {Matsumoto} Y, {Kato} TN (2019) {Evolution of Three-dimensional
  Relativistic Ion {Weibel} Instability: Competition with Kink Instability}.
  \apj 877(2):137. \doi{10.3847/1538-4357/ab1911}.
  {\href{https://arxiv.org/abs/1904.07008}{{arXiv:1904.07008}}} {[astro-ph.HE]}

\bibitem[{{Tamburini} et~al.(2010){Tamburini}, {Pegoraro}, {Di Piazza},
  {Keitel}, and {Macchi}}]{Tamburini2010}
{Tamburini} M, {Pegoraro} F, {Di Piazza} A, {Keitel} CH, {Macchi} A (2010)
  {Radiation reaction effects on radiation pressure acceleration}. New J Phys
  12(12):123005. \doi{10.1088/1367-2630/12/12/123005}.
  {\href{https://arxiv.org/abs/1008.1685}{{arXiv:1008.1685}}}
  {[physics.plasm-ph]}

\bibitem[{Tchekhovskoy(2015)}]{sasha15}
Tchekhovskoy A (2015) {Launching of Active Galactic Nuclei Jets}. The Formation
  and Disruption of Black Hole Jets 414:45--82

\bibitem[{{Thompson}(2003)}]{thompson03}
{Thompson} TA (2003) {Magnetic Protoneutron Star Winds and r-Process
  Nucleosynthesis}. \apjl 585(1):L33--L36. \doi{10.1086/374261}.
  {\href{https://arxiv.org/abs/astro-ph/0302132}{{arXiv:astro-ph/0302132}}}
  {[astro-ph]}

\bibitem[{{Thurgood} and {Tsiklauri}(2015)}]{Thurgood2015}
{Thurgood} JO, {Tsiklauri} D (2015) {Self-consistent particle-in-cell
  simulations of fundamental and harmonic plasma radio emission mechanisms}.
  \aap 584:A83. \doi{10.1051/0004-6361/201527079}.
  {\href{https://arxiv.org/abs/1509.07004}{{arXiv:1509.07004}}} {[astro-ph.SR]}

\bibitem[{Toggweiler et~al.(2014)Toggweiler, Adelmann, Arbenz, and
  Yang}]{toggweiler14}
Toggweiler M, Adelmann A, Arbenz P, Yang J (2014) A novel adaptive time
  stepping variant of the {Boris}--{Buneman} integrator for the simulation of
  particle accelerators with space charge. J Comput Phys 273:255--267.
  \doi{10.1016/j.jcp.2014.05.008}

\bibitem[{Torgrimsson et~al.(2018)Torgrimsson, Schneider, and
  Sch\"utzhold}]{Torgrimsson18}
Torgrimsson G, Schneider C, Sch\"utzhold R (2018) Sauter-schwinger pair
  creation dynamically assisted by a plane wave. Phys Rev D 97:096004.
  \doi{10.1103/PhysRevD.97.096004}

\bibitem[{{Umeda} et~al.(2003){Umeda}, {Omura}, {Tominaga}, and
  {Matsumoto}}]{umeda03}
{Umeda} T, {Omura} Y, {Tominaga} T, {Matsumoto} H (2003) {A new charge
  conservation method in electromagnetic particle-in-cell simulations}. Comput
  Phys Commun 156:73--85. \doi{10.1016/S0010-4655(03)00437-5}

\bibitem[{Uzdensky(2011)}]{Uzdensky11}
Uzdensky DA (2011) Magnetic reconnection in extreme astrophysical environments.
  Space Sci Rev 160(1):45--71. \doi{10.1007/s11214-011-9744-5}

\bibitem[{{Vafin} et~al.(2018){Vafin}, {Rafighi}, {Pohl}, and
  {Niemiec}}]{Vafin2018}
{Vafin} S, {Rafighi} I, {Pohl} M, {Niemiec} J (2018) {The Electrostatic
  Instability for Realistic Pair Distributions in Blazar/{EBL} Cascades}. \apj
  857(1):43. \doi{10.3847/1538-4357/aab552}.
  {\href{https://arxiv.org/abs/1803.02990}{{arXiv:1803.02990}}} {[astro-ph.HE]}

\bibitem[{{van Marle} et~al.(2019){van Marle}, {Casse}, and
  {Marcowith}}]{vanMarle2019}
{van Marle} AJ, {Casse} F, {Marcowith} A (2019) {Three-dimensional simulations
  of non-resonant streaming instability and particle acceleration near
  non-relativistic astrophysical shocks}. \mnras 490(1):1156--1165.
  \doi{10.1093/mnras/stz2624}.
  {\href{https://arxiv.org/abs/1909.06931}{{arXiv:1909.06931}}} {[astro-ph.HE]}

\bibitem[{Vanthieghem et~al.(2020)Vanthieghem, Lemoine, Plotnikov, Grassi,
  Grech, Gremillet, and Pelletier}]{Vanthieghem2020}
Vanthieghem A, Lemoine M, Plotnikov I, Grassi A, Grech M, Gremillet L,
  Pelletier G (2020) Physics and phenomenology of weakly magnetized,
  relativistic astrophysical shock waves. Galaxies 8(2):33.
  \doi{10.3390/galaxies8020033}

\bibitem[{{Vasileska} et~al.(2020){Vasileska}, {Tom\v{s}i\v{c}}, and
  {Kos}}]{vasileska_2020}
{Vasileska} I, {Tom\v{s}i\v{c}} P, {Kos} L (2020) {Modernization of the {PIC}
  codes for exascale plasma simulation}. In: 2020 43rd International Convention
  on Information, Communication and Electronic Technology (MIPRO). pp 209--213.
  \doi{10.23919/MIPRO48935.2020.9245299},
  \urlprefix\url{http://docs.mipro-proceedings.com/dsbe/07_DSBE_6210.pdf}

\bibitem[{{Vay}(2008)}]{Vay08}
{Vay} JL (2008) {Simulation of beams or plasmas crossing at relativistic
  velocity}. Phys Plasmas 15(5):056701. \doi{10.1063/1.2837054}

\bibitem[{{Vay} and {et al.}(2018)}]{vay_2018}
{Vay} JL, {et al} (2018) {{Warp-X}: A new exascale computing platform for
  beam--plasma simulations}. Nucl Instr Methods Phys 909:476--479.
  \doi{10.1016/j.nima.2018.01.035}

\bibitem[{{Villasenor} and {Buneman}(1992)}]{villasenor92}
{Villasenor} J, {Buneman} O (1992) {Rigorous charge conservation for local
  electromagnetic field solvers}. Comput Phys Commun 69:306--316.
  \doi{10.1016/0010-4655(92)90169-Y}

\bibitem[{{Vishniac}(1983)}]{Vishniac1983}
{Vishniac} ET (1983) {The dynamic and gravitational instabilities of spherical
  shocks}. \apj 274:152--167. \doi{10.1086/161433}

\bibitem[{Waldrop(2016)}]{waldrop_2016}
Waldrop MM (2016) The chips are down for {Moore}'s law. Nature 530:144--147.
  \urlprefix\url{https://www.nature.com/news/the-chips-are-down-for-moore-s-law-1.19338}

\bibitem[{{Walker} et~al.(2019){Walker}, {Lapenta}, {Berchem}, {El-Alaoui}, and
  {Schriver}}]{Walker2019}
{Walker} RJ, {Lapenta} G, {Berchem} J, {El-Alaoui} M, {Schriver} D (2019)
  {Embedding particle-in-cell simulations in global magnetohydrodynamic
  simulations of the magnetosphere}. Journal of Plasma Physics 85(1):905850109.
  \doi{10.1017/S0022377819000072}

\bibitem[{Watson and Nishikawa(2010)}]{watson10}
Watson M, Nishikawa KI (2010) A method for incorporating the {Kerr}--{Schild}
  metric in electromagnetic particle-in-cell code. Comput Phys Commun
  181(10):1750--1757. \doi{10.1016/j.cpc.2010.06.034}

\bibitem[{Weibel(1959)}]{weibel59}
Weibel ES (1959) Spontaneously growing transverse waves in a plasma due to an
  anisotropic velocity distribution. Phys Rev Lett 2:83--84.
  \doi{10.1103/PhysRevLett.2.83}

\bibitem[{Werner et~al.(2018)Werner, Philippov, and Uzdensky}]{Werner18}
Werner GR, Philippov AA, Uzdensky DA (2018) {Particle acceleration in
  relativistic magnetic reconnection with strong inverse-Compton cooling in
  pair plasmas}. Mon Not R Astron Soc Lett 482(1):L60--L64.
  \doi{10.1093/mnrasl/sly157}

\bibitem[{{Winkel} et~al.(2015){Winkel}, {Speck}, and {Ruprecht}}]{Winkel15}
{Winkel} M, {Speck} R, {Ruprecht} D (2015) {A high-order {Boris} integrator}. J
  Comput Phys 295:456--474. \doi{10.1016/j.jcp.2015.04.022}.
  {\href{https://arxiv.org/abs/1409.5677}{{arXiv:1409.5677}}} {[math.NA]}

\bibitem[{Yano et~al.(2019)Yano, Zhidkov, Koga, Hosokai, and Kodama}]{Yano19}
Yano M, Zhidkov A, Koga JK, Hosokai T, Kodama R (2019) Effects of hole-boring
  and relativistic transparency on particle acceleration in overdense plasma
  irradiated by short multi-{PW} laser pulses. Phys Plasmas 26(9):093108.
  \doi{10.1063/1.5120068}

\bibitem[{Yee(1966)}]{yee66}
Yee K (1966) Numerical solution of initial boundary value problems involving
  {Maxwell}'s equations in isotropic media. IEEE Trans Antennas Propagation
  14(3):302--307. \doi{10.1109/TAP.1966.1138693}

\bibitem[{Yin et~al.(2008)Yin, Daughton, Karimabadi, Albright, Bowers, and
  Margulies}]{yin08}
Yin L, Daughton W, Karimabadi H, Albright BJ, Bowers KJ, Margulies J (2008)
  Three-dimensional dynamics of collisionless magnetic reconnection in
  large-scale pair plasmas. Phys Rev Lett 101:125001.
  \doi{10.1103/PhysRevLett.101.125001}

\bibitem[{Zamaninasab et~al.(2014)Zamaninasab, Clausen-Brown, Savolainen, and
  Tchekhovskoy}]{zaman14}
Zamaninasab M, Clausen-Brown E, Savolainen T, Tchekhovskoy A (2014) Dynamically
  important magnetic fields near accreting supermassive black holes. Nature
  510(1):126--128. \doi{10.1038/nature13399}

\bibitem[{Zenitani(2015)}]{zenitani15}
Zenitani S (2015) Loading relativistic maxwell distributions in particle
  simulations. Phys Plasmas 22(4):042116. \doi{10.1063/1.4919383}

\bibitem[{{Zenitani} and {Hoshino}(2001)}]{Zenitani01}
{Zenitani} S, {Hoshino} M (2001) {The Generation of Nonthermal Particles in the
  Relativistic Magnetic Reconnection of Pair Plasmas}. Astrophys J Lett
  562(1):L63--L66. \doi{10.1086/337972}.
  {\href{https://arxiv.org/abs/1402.7139}{{arXiv:1402.7139}}} {[astro-ph.HE]}

\bibitem[{{Zenitani} and {Hoshino}(2005)}]{zenitani05prl}
{Zenitani} S, {Hoshino} M (2005) {Three-Dimensional Evolution of a Relativistic
  Current Sheet: Triggering of Magnetic Reconnection by the Guide Field}. Phys
  Rev Lett 95(9):095001. \doi{10.1103/PhysRevLett.95.095001}.
  {\href{https://arxiv.org/abs/astro-ph/0505493}{{arXiv:astro-ph/0505493}}}
  {[astro-ph]}

\bibitem[{{Zenitani} and {Hoshino}(2007)}]{Zenitani07}
{Zenitani} S, {Hoshino} M (2007) {Particle Acceleration and Magnetic
  Dissipation in Relativistic Current Sheet of Pair Plasmas}. Astrophys J
  670(1):702--726. \doi{10.1086/522226}.
  {\href{https://arxiv.org/abs/0708.1000}{{arXiv:0708.1000}}} {[astro-ph]}

\bibitem[{{Zenitani} and {Hoshino}(2008)}]{zenitani08}
{Zenitani} S, {Hoshino} M (2008) {The Role of the Guide Field in Relativistic
  Pair Plasma Reconnection}. Astrophys J 677(1):530--544. \doi{10.1086/528708}.
  {\href{https://arxiv.org/abs/0712.2016}{{arXiv:0712.2016}}} {[astro-ph]}

\bibitem[{{Zenitani} and {Kato}(2018)}]{Zenitani18j}
{Zenitani} S, {Kato} T (2018) {Numerical methods for charged particles in
  relativistic particle-in-cell simulation}. Sustainable humanosphere: Bulletin
  of Research Institute for Sustainable Humanosphere) 14:62--77.
  \urlprefix\url{http://hdl.handle.net/2433/235378}, in Japanese

\bibitem[{Zenitani and Kato(2020)}]{Zenitani2020}
Zenitani S, Kato TN (2020) Multiple {Boris} integrators for particle-in-cell
  simulation. Comput Phys Commun 247:106954. \doi{10.1016/j.cpc.2019.106954}

\bibitem[{Zenitani and Umeda(2018)}]{Zenitani18}
Zenitani S, Umeda T (2018) On the {Boris} solver in particle-in-cell
  simulation. Phys Plasmas 25(11):112110. \doi{10.1063/1.5051077}

\bibitem[{{Zenitani} et~al.(2009{\natexlab{a}}){Zenitani}, {Hesse}, and
  {Klimas}}]{2009ApJ...705..907Z}
{Zenitani} S, {Hesse} M, {Klimas} A (2009{\natexlab{a}}) {Relativistic
  Two-fluid Simulations of Guide Field Magnetic Reconnection}. Astrophys J
  705(1):907--913. \doi{10.1088/0004-637X/705/1/907}.
  {\href{https://arxiv.org/abs/0909.1955}{{arXiv:0909.1955}}} {[astro-ph.HE]}

\bibitem[{{Zenitani} et~al.(2009{\natexlab{b}}){Zenitani}, {Hesse}, and
  {Klimas}}]{2009ApJ...696.1385Z}
{Zenitani} S, {Hesse} M, {Klimas} A (2009{\natexlab{b}}) {Two-Fluid
  Magnetohydrodynamic Simulations of Relativistic Magnetic Reconnection}.
  Astrophys J 696(2):1385--1401. \doi{10.1088/0004-637X/696/2/1385}.
  {\href{https://arxiv.org/abs/0902.2074}{{arXiv:0902.2074}}} {[astro-ph.HE]}

\bibitem[{{Zenitani} et~al.(2010){Zenitani}, {Hesse}, and
  {Klimas}}]{zenitani10}
{Zenitani} S, {Hesse} M, {Klimas} A (2010) {Resistive Magnetohydrodynamic
  Simulations of Relativistic Magnetic Reconnection}. Astrophys J
  716(2):L214--L218. \doi{10.1088/2041-8205/716/2/L214}.
  {\href{https://arxiv.org/abs/1005.4485}{{arXiv:1005.4485}}} {[astro-ph.HE]}

\bibitem[{Zhang and Yan(2010)}]{Zhang_2010}
Zhang B, Yan H (2010) The internal-collision-induced magnetic reconnection and
  turbulence ({ICMART}) model of gamma-ray bursts. Astrophys J 726(2):90.
  \doi{10.1088/0004-637x/726/2/90}

\bibitem[{Zhang et~al.(2020)Zhang, Li, Giannios, Guo, Liu, and
  Dong}]{Zhang_2020}
Zhang H, Li X, Giannios D, Guo F, Liu YH, Dong L (2020) Radiation and
  polarization signatures from magnetic reconnection in relativistic jets. {I}.
  {A} systematic study. Astrophys J 901(2):149. \doi{10.3847/1538-4357/abb1b0}

\bibitem[{{Zhang} et~al.(2009){Zhang}, {MacFadyen}, and {Wang}}]{zhang09}
{Zhang} W, {MacFadyen} A, {Wang} P (2009) {Three-Dimensional Relativistic
  Magnetohydrodynamic Simulations of the Kelvin-Helmholtz Instability: Magnetic
  Field Amplification by a Turbulent Dynamo}. \apjl 692(1):L40--L44.
  \doi{10.1088/0004-637X/692/1/L40}.
  {\href{https://arxiv.org/abs/0811.3638}{{arXiv:0811.3638}}} {[astro-ph]}

\bibitem[{Zhdankin et~al.(2019)Zhdankin, Uzdensky, Werner, and
  Begelman}]{Zhdankin19}
Zhdankin V, Uzdensky DA, Werner GR, Begelman MC (2019) Electron and ion
  energization in relativistic plasma turbulence. Phys Rev Lett 122:055101.
  \doi{10.1103/PhysRevLett.122.055101}

\bibitem[{{Ziebell} et~al.(2012){Ziebell}, {Yoon}, {Gaelzer}, and
  {Pavan}}]{Ziebell2012}
{Ziebell} LF, {Yoon} PH, {Gaelzer} R, {Pavan} J (2012) {Langmuir condensation
  by spontaneous scattering off electrons in two dimensions}. Plasma Physics
  and Controlled Fusion 54(5):055012. \doi{10.1088/0741-3335/54/5/055012}

\end{thebibliography}

\end{document}